\documentclass[11pt]{article}

\pdfoutput=1

\usepackage{verbatim,amsmath,amssymb}
\usepackage{geometry}
\usepackage[export]{adjustbox}
\usepackage{subfig}
\usepackage{natbib}
\usepackage{placeins}
\usepackage{amsthm,mathrsfs,amsfonts,dsfont} 
\usepackage{tikz}
	\usetikzlibrary{arrows,matrix,positioning,angles,quotes,calc,math}
\usepackage{pgfplots}
	\pgfplotsset{compat=newest}
\usepackage{scrextend}
\usepackage{relsize}
\usepackage{exscale}
\usepackage{nicefrac}
\usepackage{changepage}

\newcommand{\norm}[1]{\left\lVert#1\right\rVert}
\newcommand\tightdots{\hbox to 1em{.\hss.\hss.}}

\usepackage[font=small,labelfont=bf]{caption}

\definecolor{darkblue}{rgb}{0,0,1}
\definecolor{dgreen}{rgb}{0,0.5,0}
\usepackage{hyperref}
\hypersetup{colorlinks=true, breaklinks=true, linkcolor=darkblue, menucolor=darkblue, citecolor=darkblue, urlcolor=darkblue}

\DeclareRobustCommand{\rchi}{{\mathpalette\irchi\relax}}
\newcommand{\irchi}[2]{\raisebox{\depth}{$#1\chi$}}

\newcommand{\mrin}{\mathrm{in}}
\newcommand{\mrel}{\mathrm{el}}
\newcommand{\bmrg}{\boldsymbol{\mathrm{g}}}

\newcommand {\dpa}[2]{\dfrac{\partial{#1}}{\partial{#2}}}

\newcommand{\mrvisc}{\mathrm{visc}}

\newcommand{\ab}{{\alpha\beta}}

\newcommand{\gd}{{\gamma\delta}}

\newcommand{\ez}{{\epsilon\zeta}}

\newcommand{\ca}{{,\alpha}}
\newcommand{\cb}{{,\beta}}
\newcommand{\cg}{{,\gamma}}
\newcommand{\cd}{{,\delta}}

\newcommand{\mrT}{\mathrm{T}}

\newcommand {\eqb}[1]{\begin{equation}\begin{array}{#1}}
\newcommand {\eqe}{\end{array}\end{equation}}

\newcommand {\esb}[1]{\begin{equation*}\begin{array}{#1}}
\newcommand {\ese}{\end{array}\end{equation*}}
\newcommand {\ds}{\displaystyle}

\newcommand {\back}{\! \! \!}
\newcommand {\is}{\back &=& \back}
\newcommand {\dis}{\back &:=& \back}

\newcommand {\dif}{\mathrm{d}}

\newcommand {\II}{{I\kern-.3em I}}
\newcommand {\III}{{I\kern-.3em I\kern-.3em I}}

\newcommand {\mra}{\mathrm{a}}
\newcommand {\mrb}{\mathrm{b}}
\newcommand {\mrc}{\mathrm{c}}

\newcommand {\mrg}{\mathrm{g}}

\newcommand {\mrn}{\mathrm{n}}

\newcommand {\mrs}{\mathrm{s}}

\newcommand {\ma}{\mathbf{a}}
\newcommand {\mcc}{\mathbf{c}}
\newcommand {\md}{\mathbf{d}}

\newcommand {\mf}{\mathbf{f}}
\newcommand {\mg}{\mathbf{g}}

\newcommand {\mk}{\mathbf{k}}

\newcommand {\mr}{\mathbf{r}}

\newcommand {\mx}{\mathbf{x}}

\newcommand {\ba}{\boldsymbol{a}}

\newcommand {\be}{\boldsymbol{e}}
\newcommand {\bff}{\boldsymbol{f}}
\newcommand {\bg}{\boldsymbol{g}}

\newcommand {\bi}{\boldsymbol{i}}

\newcommand {\bn}{\boldsymbol{n}}

\newcommand {\bu}{\boldsymbol{u}}

\newcommand {\bx}{\boldsymbol{x}}

\newcommand {\bnu}{\mbox{\boldmath$\nu$}}

\newcommand {\mA}{\mathbf{A}}
\newcommand {\mB}{\mathbf{B}}
\newcommand {\mC}{\mathbf{C}}

\newcommand {\mI}{\mathbf{I}}

\newcommand {\mN}{\mathbf{N}}

\newcommand {\mX}{\mathbf{X}}

\newcommand {\bA}{\boldsymbol{A}}
\newcommand {\bB}{\boldsymbol{B}}
\newcommand {\bC}{\boldsymbol{C}}

\newcommand {\bE}{\boldsymbol{E}}
\newcommand {\bF}{\boldsymbol{F}}
\newcommand {\bG}{\boldsymbol{G}}

\newcommand {\bI}{\boldsymbol{I}}

\newcommand {\bM}{\boldsymbol{M}}
\newcommand {\bN}{\boldsymbol{N}}

\newcommand {\bS}{\boldsymbol{S}}
\newcommand {\bT}{\boldsymbol{T}}

\newcommand {\bX}{\boldsymbol{X}}

\newcommand {\bsig}{\mbox{\boldmath$\sigma$}}

\newcommand {\bone}{\mathbf{1}}

\newcommand {\IR}{{\rm\kern.24em
   \vrule width.02em height1.53ex depth-.05ex
   \kern-.3em R}}
\newcommand {\ic}{{\rm\kern.20em
   \vrule width.02em height1.0ex depth-.05ex
   \kern-.22em c}}
\newcommand {\ia}{{\rm\kern.20em
   \vrule width.02em height1.05ex depth-.0ex
   \kern-.25em a}}
\newcommand {\IC}{{\rm\kern.24em
   \vrule width.02em height1.4ex depth-.05ex
   \kern-.26em C}}
\newcommand {\ID}{{\rm\kern.34em
   \vrule width.02em height1.5ex depth-.05ex
   \kern-.36em D}}
\newcommand {\IS}{{\rm\kern.24em
   \vrule width.02em height1.6ex depth.05ex
   \kern-.26em S}}
\newcommand {\IT}{{\rm\kern.50em
   \vrule width.02em height1.55ex depth-.05ex
   \kern-.52em T}}

\newcommand {\IE}{{\rm\kern.24em
   \vrule width.02em height1.55ex depth-.05ex
   \kern-.33em E}}
\newcommand {\IEa}{{\rm\kern.24em
   \vrule width.02em height1.55ex depth-.05ex
   \kern-.33em E}^{1}_{ijkl}}
\newcommand {\IEb}{{\rm\kern.24em
   \vrule width.02em height1.55ex depth-.05ex
   \kern-.33em E}^{2}_{ijkl}}

\newcommand {\sD}{\mathcal{D}}

\newcommand {\sS}{\mathcal{S}}

\newcommand {\sU}{\mathcal{U}}

\newcommand {\Ass}[2]{\kern 0.9ex \vrule width0.45em height0.2ex depth0ex \kern -2.1ex \bigwedge_{#1}^{#2}}
\newcommand {\ASS}[2]{\kern 1.45ex \vrule width0.5em height0.2ex depth0ex \kern -2.65ex \bigwedge_{#1}^{#2}}

\newlength{\figwidth}
\newlength{\figheight}

\newcommand{\seclabel}[1]{\label{s:#1}}
\newcommand{\secref}[1]{Sec.~\ref{s:#1}}
\newcommand{\eqlabel}[1]{\label{e:#1}}
\newcommand{\eqsref}[1]{Eq.~\eqref{e:#1}}
\newcommand{\figlabel}[1]{\label{f:#1}}
\newcommand{\figref}[1]{Fig.~\ref{f:#1}}
\newcommand{\tablabel}[1]{\label{t:#1}}
\newcommand{\tabref}[1]{Table~\ref{t:#1}}
\newcommand{\applabel}[1]{\label{a:#1}}
\newcommand{\appref}[1]{Appendix~\ref{a:#1}}

\pgfplotsset{every axis/.append style={
                    label style={font=\small},
                    tick label style={font=\footnotesize}  
                    }}

\usepackage{titlesec}
\titleformat{\paragraph}
{\normalfont\normalsize\bfseries}{\theparagraph}{1em}{}
\titlespacing*{\paragraph}
{0pt}{3.25ex plus 1ex minus .2ex}{1.5ex plus .2ex}

\newtheoremstyle{rem}
{6pt}
{6pt}
{\small}
{}
{\bf}
{:}
{.5em}
{}

\theoremstyle{rem}
\newtheorem{remark}{Remark}[section]

\makeatletter
\newcommand\footnoteref[1]{\protected@xdef\@thefnmark{\ref{#1}}\@footnotemark}
\newcommand\footnoterefs[2]{\protected@xdef\@thefnmark{\ref{#1},\ref{#2}}\@footnotemark}
\makeatother

\begin{document}
\renewcommand{\thefootnote}{\fnsymbol{footnote}}
\begin{adjustwidth}{-2pt}{-2pt}
\begin{center}
\Large{\bf{An~isogeometric~finite~element~formulation~for~boundary~and~shell viscoelasticity~based~on~a~multiplicative~surface~deformation~split}}\\
\end{center}
\end{adjustwidth}

\begin{center}
\large{Karsten Paul$^\mra$ and Roger A. Sauer$^{\mra,\mrb,\mrc,}$\footnote{corresponding author, email: sauer@aices.rwth-aachen.de}}
\vspace{4mm}

\small{\textit{$^\mra$Aachen Institute for Advanced Study in Computational Engineering Science (AICES),\\RWTH Aachen University, Templergraben 55, 52062 Aachen, Germany}}

\small{\textit{$^\mrb$Faculty of Civil and Environmental Engineering, Gda\'nsk University of Technology, \\ul.~Gabriela Narutowicza 11/12, 80-233 Gda\'nsk, Poland}}

\small{\textit{$^\mrc$Department of Mechanical Engineering, Indian Institute of Technology Guwahati,\\Assam 781039, India}}

\end{center}
\vspace{-4mm}
\setcounter{footnote}{0}
\renewcommand{\thefootnote}{\arabic{footnote}}

\begin{center}
\small{Published\footnote{This pdf is the personal version of an article whose journal version is available at \href{https://doi.org/10.1002/nme.7080}{https:/\!/onlinelibrary.wiley.com}} in \textit{International Journal for Numerical Methods in Engineering},\\\href{https://doi.org/10.1002/nme.7080}{doi: 10.1002/nme.7080} \\
Submitted on March 21, 2022; Revised on June 23, 2022; Accepted on June 24, 2022}
\end{center}

\vspace{-3mm}

\rule{\linewidth}{.15mm}
{\bf Abstract:}
This work presents a numerical formulation to model isotropic viscoelastic material behavior for membranes and thin shells. The surface and the shell theory are formulated within a curvilinear coordinate system, which allows the representation of general surfaces and deformations. The kinematics follow from Kirchhoff-Love theory and the discretization makes use of isogeometric shape functions. A multiplicative split of the surface deformation gradient is employed, such that an intermediate surface configuration is introduced. The surface metric and curvature of this intermediate configuration follow from the solution of nonlinear evolution laws -- ordinary differential equations (ODEs) -- that stem from a generalized viscoelastic solid model. The evolution laws are integrated numerically with the implicit Euler scheme and linearized within the Newton-Raphson scheme of the nonlinear finite element framework. The implementation of membrane and bending viscosity is verified with the help of analytical solutions and shows ideal convergence behavior. The chosen numerical examples capture large deformations and typical viscoelasticity behavior, such as creep, relaxation, and strain rate dependence. It is also shown that the proposed formulation can be straightforwardly applied to model boundary viscoelasticity of 3D bodies.

{\bf Keywords:} Viscoelasticity, Kirchhoff-Love shells, multiplicative split, isogeometric analysis, nonlinear finite element methods, surface elasticity

\vspace{-4mm}
\rule{\linewidth}{.15mm}

\section{Introduction} \seclabel{intro}
Thin-walled structures are common to many engineering branches, as they have low weight combined with high strength. A suitable approach to model these structures is Kirchhoff-Love theory, especially for very slender structures. Many materials, such as polymers, bitumens, biological materials, and several metals or plastics at elevated temperatures, exhibit viscoelastic behavior, i.e.~they show both elastic and viscous material properties. Elastic materials return to their original shape instantaneously once the applied loads are removed, whereas viscoelastic materials return to their original shape gradually. For large deformations, the elastic and viscous deformations are in general modeled by a multiplicative split of the deformation gradient, which leads to an additive split of certain strains.

Kirchhoff-Love theory considers thin shells, such that out-of-plane shear strains can be neglected. Therefore, the cross-section always remains orthogonal to the shell's mid-plane during deformation. The governing equation of Kirchhoff-Love shells contains higher-order derivatives, as it describes bending in terms of deformation and not in terms of rotation. The weak form thus requires $C^1$-continuity, which poses a requirement on the numerical discretization. Isogeometric analysis (IGA), introduced by \cite{hughes2005}, offers the possibility of discretizations with higher continuity by making use of splines for the description of the geometry and the solution, as is discussed in detail in \cite{cottrell2009}. IGA has interesting mathematical properties \citep{evans2009} and can be integrated into standard finite element codes by performing the numerical integration of smooth basis functions on Bézier elements \citep{borden2011} or Lagrange elements \citep{schillinger2016}. \cite{nguyen2015} provide an overview on isogeometric analysis and discuss implementation aspects. Isogeometric Kirchhoff-Love shells have been intensively studied in the literature, starting with \cite{kiendl2009}. Further works on isogeometric shells followed, such as blended shells \citep{benson2013}, hierarchical shells \citep{echter2013}, and large-strain Kirchhoff-Love shells \citep{kiendl2015,duong2017}. For complex engineering problems, single patches are often not sufficient to represent shell geometries, such that multi-patch descriptions are required. The continuity of such discretizations is not preserved at patch interfaces and needs to be restored, see \cite{paul2020a} for a recent review on patch enforcement techniques in isogeometric analysis. Isogeometric analysis has been used within several shell and membrane applications, such as laminated composite shells \citep{thai2012,deng2015}, anisotropic shells \citep{nagy2013,faroughi2020}, shape optimization of shells \citep{kiendl2014,hirschler2019b}, liquid membranes \citep{sauer2014a,roohbakhshan2019}, biological tissues \citep{tepole2015,roohbakhshan2017}, topology optimization of shells \citep{kang2016,zhang2020}, shell fracture \citep{ambati2016b,kiendl2016}, lipid bilayers \citep{sauer2017b,bartezzaghi2019}, 2D materials \citep{ghaffari2018,ghaffari2019}, shell elastoplasticity \citep{ambati2018,huynh2020}, inverse analysis of shells \citep{vubac2018,borzeszkowski2022}, Cahn-Hilliard phase separations on deforming surfaces \citep{valizadeh2019,zimmermann2019}, adaptive surface refinement \citep{paul2020b,proserpio2020}, and fiber-reinforced shells \citep{schulte2020,duong2022}.

The rigorous treatment of linear viscoelasticity dates back to the works by \cite{coleman1961} and \cite{crochet1969}. Extensions to finite viscoelasticity have been provided by \cite{koh1963} and \cite{sidoroff1974}. Finite element implementations of viscoelasticity followed by \cite{simo1987} and \cite{gvondjee1992}. \cite{reese1998} present a theory for finite viscoelasticity based on a multiplicative split of the deformation gradient. Recent discussions on the origin, mathematics and application of the multiplicative split are provided by \cite{lubarda2004}, \cite{gupta2007}, and \cite{reina2018}. \cite{shaw1999} describe viscoelastic material models and discuss their numerical treatment including spatial adaptivity of the discretization. \cite{miehe2000} present a model for filled rubbery polymers considering a superimposed finite elastic-viscoelastic-plastoelastic stress response with damage and they discuss its numerical implementation. \cite{holzapfel2001} provide a model for viscoelastic fiber-reinforced composites at finite strains and compare its finite element results with experimental data. \cite{bonet2001} proposes a continuum formulation for viscoelastic materials and corresponding constitutive models considering finite deformations. \cite{adolfsson2003} make use of fractional derivatives to model large strain viscoelasticity. \cite{fancello2006} propose a general variational framework for finite viscoelastic models and compare several models with each other. \cite{amin2006} consider finite strain viscoelasticity of rubbers and investigate the effects of internal variables on viscous phenomena based on experiments and numerical models. \cite{hossain2010} focus on the modeling of polymer curing by considering viscoelasticity and shrinking at finite strains. \cite{kaestner2012} propose an experimental procedure to classify the nonlinear, inelastic mechanical behavior of polymers and derive a corresponding material model. \cite{marques2012} provide a summary on computational viscoelasticity. \cite{shutov2013} suggest an explicit solution for an implicit time stepping scheme used in finite strain viscoelasticity based on a multiplicative split of the deformation gradient. \cite{james2016} demonstrate the importance of the viscoelastic response in structural design optimization.

The extension of viscoelasticity to shells has been addressed in fewer works. \cite{evans1976} propose a theory for viscoelastic membranes under large deformations and apply it to red cell discocytes. \cite{neff2005} provide the first model for viscoelastic membranes and shells based on the classical multiplicative split of the deformation gradient. Such a split can also be formulated directly on the surface, as was shown in the recent work of \cite{sauer2019}. Additive decompositions are used for erythrocyte membranes \citep{lubarda2011}, for shell formulations based on three-dimensional viscoelasticity \citep{li2012}, for micropolar plates and shells \citep{altenbach2015}, and for fiber reinforced composite shells \citep{doerr2017}. \cite{liu2017} consider surface viscoelasticity coupled with three-dimensional viscoelastic bulks to model cells and cell aggregates. \cite{hernandez2019} study viscoelastic behavior of Reissner-Mindlin shells. \cite{dadgarrad2021} consider incompressible membranes and quasi-linear viscoelasticity. Following the work on surface elasticity by \cite{javili2009,javili2010}, \cite{dortdivanlioglu2021} develop a theory for nonlinear surface viscoelasticity at finite strains by accounting for strain-dependent boundary stresses.

Recently, viscoelasticity has also been considered in isogeometric analysis. The work of \cite{dortdivanlioglu2021} uses IGA, while \cite{shafei2021} make use of isogeometric analysis for the analysis of nonlinear vibration of viscoelastic plates. Further, \cite{sun2021} employ the isogeometric boundary element method to investigate viscoelastic materials.

The present work makes use of the new multiplicative surface deformation split of \cite{sauer2019} in order to model isotropic viscoelastic material behavior of thin shell structures. Here, the flexibility and consistency of this split are demonstrated and the corresponding computational formulation is described. The nonlinear isogeometric thin shell formulation follows the work of \cite{duong2017}, see also Remark~\ref{r:shell}. The proposed formulation thus benefits from efficiency and accuracy gains provided by rotation-free isogeometric finite element formulations. A generalized viscoelastic solid is considered to model both membrane and bending viscosity, see also Remark~\ref{r:surface}. Several material models are outlined and a time integration scheme is presented to solve the evolution laws. Several material models are outlined and a time integration scheme is presented to solve the evolution laws. For several special cases, analytical solutions are derived. It is further shown that the proposed formulation can be applied to model boundary viscoelasticity of 3D bodies without any further modifications. In summary, the formulation in the present work
\begin{itemize}	
	\item employs a multiplicative split of the surface deformation gradient for large strains,
	\item allows to flexibly describe in-plane and out-of-plane viscoelasticity,
	\item captures typical viscoelastic effects, such as creep, relaxation, and strain rate dependency,
	\item uses an isogeometric finite element implementation,
	\item uses numerical time integration to solve the evolution laws,
	\item is verified by several analytical solutions,
	\item exhibits ideal convergence rates, and
	\item can model boundary viscoelasticity of 3D bodies without any further modifications.
\end{itemize}
\begin{remark} \label{r:shell}
	The employed shell formulation is based on a direct surface approach, which directly formulates the kinematical quantities, balance laws and constitutive relations (in particular) for the surface, without resorting to 3D counterparts. The approach is consistent with 3D continuum mechanics, because the underlying kinematical assumptions, balance laws and constitutive principles are the same. It is therefore possible to show their equivalency for certain constitutive models \citep{duong2017}, see also Remark~\ref{r:3dConstitution}. The direct surface approach enables a straightforward decomposition of the elastic energy density into membrane and bending contributions, which makes the constitutive modeling more flexible. Thus, the proposed approach admits any hyperelastic and viscous material that is consistent with the principles of material modeling such as the second law of thermodynamics.
\end{remark}
\begin{remark} \label{r:surface}
The proposed formulation captures two separate cases in a unified manner: shell viscoelasticity (that includes membrane viscoelasticity as a special case) and boundary viscoelasticity of 3D bodies -- a case which is sometimes referred to as \textit{surface viscoelasticity} in the literature \citep{liu2017,dortdivanlioglu2021}. In the two cases, the term \textit{surface} therefore either refers to the mid-plane of the shell or the boundary of the 3D solid. The separation into the two cases is only conceptual. Mathematically and computationally the two cases become identical in the proposed formulation. For brevity, terms like \textit{surface viscoelasticity} are therefore simply used to subsume both cases.
\end{remark}

The remainder of this paper is structured as follows: \secref{cntVisc} summarizes the surface description and kinematics considering the multiplicative split of the surface deformation gradient. \secref{cntViscConst} describes constitutive relations and the extension of hyperelasticity to account for membrane and bending viscosity. In \secref{cmpVisc}, the computational formulation including the time integration scheme and the linearization is outlined. \secref{numExVisc} presents several numerical examples that highlight viscoelastic material behavior and ideal convergence rates. \secref{concl} concludes this paper with a summary and an outlook.

\section{The multiplicative surface deformation split} \seclabel{cntVisc}
This section summarizes the continuum formulation of viscoelastic shells based on a curvilinear surface description and the multiplicative surface deformation split of \cite{sauer2019}. A concise summary of the mathematical background of the thin shell theory can be found in \cite{sauer2018}.

\subsection{Surface description} \seclabel{cntViscDscr}
The surface description is based on the four different domains shown in \figref{cntViscMappings}. The parameter domain is denoted by $\mathcal{P}$, while the reference, intermediate and current surface are denoted by $\sS_0$, $\hat\sS$, and $\sS$, respectively. The introduction of the intermediate configuration is required to account for deformations composed of two separate components, i.e.~elastic and inelastic components. The mapping $\bX(\xi^\alpha)$ describes the reference surface and it associates the covariant basis $\{\bA_\alpha,\bN\}$ to each surface point. Here, $\xi^\alpha$ ($\alpha=1,2$) denote the convective coordinates, $\bA_\alpha$ are the tangent vectors and $\bN$ is the surface normal. In analogy, the current surface $\sS$ follows from $\bx(\xi^\alpha,t)$ with covariant basis $\{\ba_\alpha,\bn\}$. Time is denoted $t$. In the intermediate configuration, the covariant basis is denoted $\{\hat\ba_\alpha,\hat\bn\}$. The deformation $\sS_0\rightarrow\hat\sS$ is taken as the inelastic part (`in'), and $\hat\sS\rightarrow\sS$ as the elastic part (`el').
\begin{figure}[!ht]
	\centering
		\includegraphics[width=0.81\textwidth]{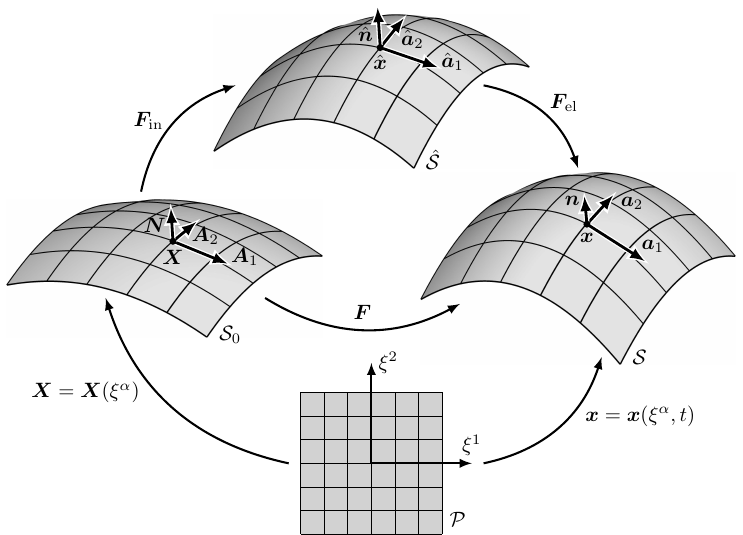}
	\caption{Surface mappings with the multiplicative surface deformation split into inelastic and elastic components \citep{sauer2019}.} \figlabel{cntViscMappings}
\end{figure}

Given the mapping $\bx(\xi^\alpha,t)$, the tangent vectors and surface normal follow as
\eqb{l}
	\ba_\alpha=\dfrac{\partial\bx}{\partial\xi^\alpha}\,,\quad\mathrm{and}\quad\bn=\dfrac{\ba_1\times\ba_2}{||\ba_1\times\ba_2||}\,, \eqlabel{cntViscDscrCoVariantBasis}
\eqe
such that in-plane and out-of-plane surface objects can be characterized. The tangent vectors define the covariant surface metric with components
\eqb{l}
	a_\ab=\ba_\alpha\cdot\ba_\beta\,,  \eqlabel{cntViscDscrSurfaceMetric}
\eqe
which describe length and angle changes. In general, the basis $\{\ba_\alpha,\bn\}$ is not orthonormal. Orthonormality is restored by introducing a contravariant basis $\{\ba^\alpha,\bn\}$ through
\eqb{l}
	\ba^\alpha=a^\ab\,\ba_\beta\,,\quad\mathrm{with}\quad\bigl[a^\ab\bigr]=\bigl[a_\ab\bigr]^{-1}\,, \eqlabel{cntViscDscrContraVariantBasis}
\eqe
where $a^\ab$ refers to the contravariant surface metric. Note that index notation is used here, such that summation from $1$ to $2$ is implied on all terms with repeated Greek indices. Now, $\ba^\alpha\cdot\ba_\beta=\delta^\alpha_\beta$ with Kronecker delta $\delta^\alpha_\beta$.

The second parametric derivative $\bx_{,\ab}=\ba_{\alpha,\beta}=\partial\ba_\alpha/\partial\xi^\beta$ is introduced to define the surface curvature
\eqb{l}
	b_\ab=\ba_{\alpha,\beta}\cdot\bn\,.  \eqlabel{cntViscDscrCurvature}
\eqe
It can be used to determine the mean and Gaussian curvature
\eqb{l}
	H=1/2\,a^\ab\,b_\ab\,,\quad\mathrm{and}\quad\kappa=\det\bigl[{b^\alpha}_{\!\beta}\bigr]\,,\eqlabel{cntViscDscrMeanGaussianCurvature}
\eqe
respectively, where $[{b^\alpha}_{\!\beta}]=[a^{\alpha\gamma}b_{\gamma\beta}]$. In contrast to the parametric derivative $\ba_{\alpha,\beta}$, the covariant derivative of $\ba_\alpha$ can be defined as $\ba_{\alpha;\beta}:=(\bn\otimes\bn)\,\ba_{\alpha,\beta}$.

The description of the reference surface $\sS_0$ follows in analogy to the surface description of $\sS$ presented above. The intermediate configuration $\hat\sS$ is described by the unknown tangent vectors $\hat\ba_\alpha$. Given these tangent vectors, the intermediate surface quantities can be characterized analogously to $\sS_0$ and $\sS$. For the reference surface, capital letters or the index `0' are used, whereas quantities on the intermediate configuration are denoted with a hat, `$\hat\bullet$'. The intermediate surface metric $\hat{a}_\ab$ and curvature $\hat{b}_\ab$ are particularly important to describe viscoelasticity. These quantities follow in analogy to Eqs.~\eqref{e:cntViscDscrSurfaceMetric} and \eqref{e:cntViscDscrCurvature}.

\subsection{Surface kinematics} \seclabel{cntViscKin}
\figref{cntViscMappings} shows that the mappings between the different surface configurations are characterized by the surface deformation gradient $\bF$, or its elastic or inelastic part, $\bF_{\!\mrel}$ and $\bF_{\!\mrin}$, respectively. They are given by
\eqb{l}
	\bF=\ba_\alpha\otimes\bA^\alpha\,,\quad\bF_{\!\mrel}=\ba_\alpha\otimes\hat{\ba}^\alpha\,,\quad\mathrm{and}\quad\bF_{\!\mrin}=\hat{\ba}_\alpha\otimes\bA^\alpha\,, \eqlabel{cntViscKinSurfaceDeformationTensor}
\eqe
such that the deformation gradient is multiplicatively split into its elastic and inelastic part, i.e.~$\bF=\bF_{\!\mrel}\,\bF_{\!\mrin}$. Based on this split, the tangent vectors can be expressed as
\eqb{rlrlrl}
	\bA_\alpha&\!\!\!=\bF_{\!\mrin}^{-1}\hat\ba_\alpha=\bF^{-1}\ba_\alpha\,,& \quad\hat\ba_\alpha &\!\!\!=\bF_{\!\mrin}\,\bA_\alpha=\bF_{\!\mrel}^{-1}\ba_\alpha\,,& \quad \ba_\alpha&\!\!\!=\bF_{\!\mrel}\,\hat\ba_\alpha=\bF\bA_\alpha\,,\\[2mm]
	\bA^\alpha&\!\!\!=\bF_{\!\mrin}^\mrT\,\hat\ba^\alpha=\bF^\mrT\ba^\alpha\,,& \hat\ba^\alpha&\!\!\!=\bF_{\!\mrin}^{-\mrT}\bA^\alpha=\bF_{\!\mrel}^\mrT\,\ba^\alpha\,,& \ba^\alpha&\!\!\!=\bF_{\!\mrel}^{-\mrT}\hat\ba^\alpha=\bF^{-\mrT}\!\bA^\alpha\,. \eqlabel{cntViscTangentVectors}
\eqe
Based on the surface deformation gradient $\bF$ in \eqsref{cntViscKinSurfaceDeformationTensor}, the surface Cauchy-Green tensors can be determined, i.e.
\eqb{l}
	\bC=\bF^\mrT\bF=a_\ab\,\bA^\alpha\otimes\bA^\beta\,,\quad\mathrm{and}\quad\bB=\bF\bF^\mrT=A^\ab\,\ba_\alpha\otimes\ba_\beta\,. \eqlabel{cntViscKinCauchyGreenTensorOnS}
\eqe
They have the two invariants
\eqb{l}
	I_1=\bI:\bC=\bi:\bB=A^\ab\,a_\ab\,,\quad\mathrm{and}\quad J=\dfrac{\sqrt{\det[a_\ab]}}{\sqrt{\det[A_\ab]}}\,.\eqlabel{cntViscKinInvariantsOnS}
\eqe
The second invariant in \eqsref{cntViscKinInvariantsOnS} characterizes the surface stretch between $\sS_0$ and $\sS$.
Based on the surface identities $\bI=\bA_\alpha\otimes\bA^\alpha$ on $\sS_0$ and $\bi=\ba_\alpha\otimes\ba^\alpha$ on $\sS$ and \eqsref{cntViscKinCauchyGreenTensorOnS}, the surface Green-Lagrange and surface Almansi strain tensors follow as
\eqb{l}
	\bE=\dfrac{1}{2}\,\bigl(\bC-\bI\bigr)=\varepsilon_\ab\,\bA^\alpha\otimes\bA^\beta\,,\quad\mathrm{and}\quad\be=\dfrac{1}{2}\,\bigl(\bi-\bB^{-1}\bigr)=\varepsilon_\ab\,\ba^\alpha\otimes\ba^\beta\,, \eqlabel{cntViscKinStrainTensorsOnS}
\eqe
respectively. In \eqsref{cntViscKinStrainTensorsOnS}, the strain components are given by
\eqb{l}
	\varepsilon_\ab=\dfrac{1}{2}\,\bigl(a_\ab-A_\ab\bigr)\,. \eqlabel{cntViscKinStrainComponents}
\eqe

Now, the right surface Cauchy-Green tensor can be pushed forward to the intermediate configuration, and the inverse of the left surface Cauchy-Green tensor can be pulled back to the intermediate configuration, i.e.
\eqb{rll}
	\bC_\mrel \dis \bF_{\!\mrin}^{-\mrT}\,\bC\,\bF_{\!\mrin}^{-1}=\bF_{\!\mrel}^\mrT\,\bF_{\!\mrel}= a_\ab\,\hat{\ba}^\alpha\otimes\hat{\ba}^\beta\,,\quad\mathrm{and}\\[2mm]
	\bB_\mrin^{-1}\dis\bF_{\!\mrel}^\mrT\,\bB^{-1}\,\bF_{\!\mrel}=\bF_{\!\mrin}^{-\mrT}\bF_{\!\mrin}^{-1}=A_\ab\,\hat\ba^\alpha\otimes\hat\ba^\beta\,. \eqlabel{cntViscKinCelBinvin}
\eqe
In analogy to the push forward and pull back operations in \eqsref{cntViscKinCelBinvin}, the surface Green-Lagrange strain tensor $\bE$ can be pushed forward, or equivalently, the surface Almansi strain tensor $\be$ can be pulled back to the intermediate configuration, i.e.
\eqb{l}
	\hat\be = \bF_{\!\mrin}^{-\mrT}\,\bE\,\bF_{\!\mrin}^{-1}=\bF_{\!\mrel}^\mrT\,\be\,\bF_{\!\mrel}\,, \eqlabel{cntViscKinStrainTensorHatDerivation}
\eqe
which results in
\eqb{l}
	\hat\be = \dfrac{1}{2}\Bigl(\bC_\mrel-\bB_\mrin^{-1}\Bigr) = \varepsilon_\ab\,\hat\ba^\alpha\otimes\hat\ba^\beta\,, \eqlabel{cntViscKinStrainTensorHat}
\eqe
due to Eqs.~\eqref{e:cntViscTangentVectors}, \eqref{e:cntViscKinStrainTensorsOnS} and \eqref{e:cntViscKinCelBinvin}.
\eqsref{cntViscKinStrainTensorHat} shows that the multiplicative split of the deformation tensor, see \eqsref{cntViscKinSurfaceDeformationTensor}, leads to the additive split of the strains
\eqb{l}
	\hat\be=\hat\be_\mrel+\hat\be_\mrin\,, \eqlabel{cntViscKinStrainTensorSplit}
\eqe
where
\eqb{l}
	\hat\be_\mrel=\dfrac{1}{2}\Bigl(\bC_\mrel-\hat\bi\Bigr)=\varepsilon_\ab^\mrel\,\hat\ba^\alpha\otimes\hat\ba^\beta\,, \quad\mathrm{and}\quad \hat\be_\mrin=\dfrac{1}{2}\Bigl(\hat\bi-\bB_\mrin^{-1}\Bigr)=\varepsilon_\ab^\mrin\,\hat\ba^\alpha\otimes\hat\ba^\beta\,, \eqlabel{cntViscKinStrainTensorSplitComponents}
\eqe
with the surface identity $\hat\bi=\hat\ba_\alpha\otimes\hat\ba^\alpha$ on $\hat\sS$. The strain components in \eqsref{cntViscKinStrainComponents} are then also additively split, i.e.
\eqb{l}
	\varepsilon_\ab=\varepsilon_\ab^\mrel+\varepsilon_\ab^\mrin\,, \eqlabel{cntViscKinStrainSplit}
\eqe
with
\eqb{l}
	\varepsilon_\ab^\mrel=\dfrac{1}{2}\,\bigl(a_\ab-\hat{a}_\ab\bigr)\,,\quad\mathrm{and}\quad\varepsilon_\ab^\mrin=\dfrac{1}{2}\bigl(\hat{a}_\ab-A_\ab\bigr)\,. \eqlabel{cntViscKinStrainSplitComponents}
\eqe
Based on the multiplicative split of $\bF$, the surface stretch in Eq.~(\ref{e:cntViscKinInvariantsOnS}.2) becomes
\eqb{l}
	J=J_\mrel\,J_\mrin\,, \eqlabel{cntViscKinSurfaceStretchSplit}
\eqe
with
\eqb{l}
	J_\mrel=\dfrac{\sqrt{\det[a_\ab]}}{\sqrt{\det[\hat{a}_\ab]}}\,,\quad\mathrm{and}\quad J_\mrin=\dfrac{\sqrt{\det[\hat{a}_\ab]}}{\sqrt{\det[A_\ab]}}\,. \eqlabel{cntViscKinSurfaceStretchSplitComponents}
\eqe
Further, the first invariant of $\bC_\mrel$ is given by
\eqb{l}
	I^\mrel_1 = \hat{a}^\ab a_\ab\,. \eqlabel{cntViscKinFirstInvariantElastic}
\eqe

In analogy to \eqsref{cntViscKinStrainSplit}, the additive curvature decomposition
\eqb{l}
	\kappa_\ab=\kappa_\ab^\mrel+\kappa_\ab^\mrin\,, \eqlabel{cntViscKinCurvatureSplit}
\eqe
with
\eqb{l}
	\kappa_\ab^\mrel=b_\ab-\hat{b}_\ab\,,\quad\mathrm{and}\quad\kappa_\ab^\mrin=\hat{b}_\ab-B_\ab\,, \eqlabel{cntViscInCurvatureSplitComponents}
\eqe
is introduced. In analogy to \eqsref{cntViscDscrMeanGaussianCurvature}, the mean and Gaussian curvature of the intermediate surface are given by
\eqb{l}
	\hat{H}=\dfrac{1}{2}\,\hat{a}^\ab\,\hat{b}_\ab\,,\quad\mathrm{and}\quad\hat\kappa=\det\bigl[{\hat{b}^\alpha}_{\,\,\beta}\bigr]\,. \eqlabel{cntViscKinMeanCurvatures}
\eqe
The influences of the split of the surface deformation gradient on the surface motion, stresses, moments and balance laws are further discussed in \cite{sauer2019}.

\subsection{Weak form} \seclabel{cntViscWeak}
The quasi-static weak form for deforming shells is given by \citep{sauer2017a}
\eqb{l}
	G_\mathrm{int}-G_\mathrm{ext}=0\,,\quad\forall\,\delta\bx\in\mathcal{U}\,, \eqlabel{cntViscWeakWeakForm}
\eqe
with variation $\delta\bx$ taken from some suitable space $\mathcal{U}$. The internal and external virtual work in \eqsref{cntViscWeakWeakForm} are given by
\eqb{rll}
	G_\mathrm{int} \dis \ds\int_{\sS_0}\dfrac{1}{2}\delta a_\ab\,\tau^\ab\,\dif A+\int_{\sS_0}\delta b_\ab\,M_0^\ab\,\dif A\,,\quad\mathrm{and}\\[4mm]
	G_\mathrm{ext} \dis \ds\int_\sS\delta\bx\cdot\bff\,\dif A+\int_{\partial\sS}\delta\bx\cdot\bT\,\dif s+\int_{\partial\sS}\delta\bn\cdot\bM\,\dif s\,, \eqlabel{cntViscWeakIntExtVirtualWork}
\eqe
where $\delta a_\ab$ and $\delta b_\ab$ refer to the variation of the surface metric and curvature, respectively. The external forces $\bff$, $\bT$, and $\bM$ describe prescribed body forces, boundary tractions and boundary moments, respectively. The stress components $\tau^\ab$ and moment components $M_0^\ab$ in Eq.~(\ref{e:cntViscWeakIntExtVirtualWork}.1) follow from constitution, which is discussed in the subsequent section.

\section{Surface constitution} \seclabel{cntViscConst}
This section is concerned with the constitutive relations for viscoelastic shells. \secref{cntViscConstElasticity} focuses on elastic materials, which is then extended to viscoelastic materials in \secref{cntViscConstViscoelasticity}.

\subsection{Surface elasticity} \seclabel{cntViscConstElasticity}
For hyperelastic shells, it is advantageous to consider a decomposition of the elastic energy density into in-plane and out-of-plane contributions, i.e.
\eqb{l}
	\Psi_\mrel(a_\ab,b_\ab) = \Psi_\mathrm{mem}(a_\ab)+\Psi_\mathrm{bend}(a_\ab,b_\ab)\,, \eqlabel{cntViscConstPsiEl}
\eqe
where $\Psi_\mathrm{mem}$ refers to the elastic energy density associated with in-plane membrane deformations, while $\Psi_\mathrm{bend}$ refers to the elastic energy density associated with out-of-plane bending. Given the elastic energy density, the membrane stress and moment components can be computed via
\eqb{l}
	\sigma^\ab=\dfrac{2}{J}\,\dfrac{\partial\Psi_\mrel}{\partial a_\ab}\,,\quad\mathrm{and}\quad M^\ab=\dfrac{1}{J}\,\dfrac{\partial\Psi_\mrel}{\partial b_\ab}\,. \eqlabel{cntViscConstSigmaM}
\eqe
The corresponding Kirchhoff stress and moment components (w.r.t.~the reference configuration) follow as $\tau^\ab=J\,\sigma^\ab$ and $M_0^\ab=J\,M^\ab$. The symmetric stress $\sigma^\ab$ appearing in Eqs.~(\ref{e:cntViscWeakIntExtVirtualWork}.1) and (\ref{e:cntViscConstSigmaM}.1) is denoted \textit{effective stress} by \cite{simo1989}. It is generally not the same as the physical (Cauchy) stress
\eqb{l}
	N^\ab=\sigma^\ab+b^\beta_\gamma\,M^{\gamma\alpha}\,, \eqlabel{cntViscConstCauchyStress}
\eqe
that appears in the equilibrium equation. Only for pure membranes or flat shells $\sigma^\ab=N^\ab$. Given $N^\ab$, the surface tension and the deviatoric membrane stresses are given by
\eqb{l}
	\gamma=\dfrac{1}{2}\,N^\ab\,a_\ab\,,\quad\mathrm{and}\quad N^\ab_\mathrm{dev}=N^\ab-\gamma\,a^\ab\,. \eqlabel{cntViscConstGammaNabDev}
\eqe
In Secs.~\ref{s:KoiterMembraneModel}--\ref{s:HelfrichBendingModel}, several choices for $\Psi_\mathrm{mem}$ and $\Psi_\mathrm{bend}$ and their resulting stresses and moments are presented.

\begin{remark} \label{r:3dConstitution}
As pointed out in Remark~\ref{r:shell}, the employed shell formulation is based on a direct surface approach, for which the elastic energy density in \eqsref{cntViscConstPsiEl} is an energy per surface area. Instead of providing this energy function directly, it can also be extracted from classical three-dimensional constitutive models. In that case, the 3D constitutive model is projected onto the surface via thickness integration, as is briefly described subsequently \citep{duong2017}. The in-plane components of the three-dimensional second Piola-Kirchhoff stress tensor $\tilde\bS$ are
\eqb{l}
	\tilde\tau^\ab=\bG^\alpha\cdot\tilde\bS\,\bG^\beta\,, \eqlabel{cntViscStress3D}
\eqe
where $\bG^\alpha$ denotes the tangent vector of the shell layer at position $\bX+\xi_0\,\bN$ \citep{duong2017}, with thickness coordinate $\xi_0\in[-T/2,T/2]$ and initial shell thickness $T$. The tensor $\tilde\bS$ can then be taken from classical three-dimensional hyperelasticity and viscoelasticity models. The resulting stress and moment components of the shell then follow from the thickness integration \citep{duong2017}
\eqb{l}
	\tau^\ab\approx\ds\int_{-\frac{T}{2}}^{\frac{T}{2}}\tilde\tau^\ab\,\dif\xi_0\,,\quad\mathrm{and}\quad M_0^\ab\approx-\int_{-\frac{T}{2}}^{\frac{T}{2}}\xi_0\,\tilde\tau^\ab\,\dif\xi_0\,. \eqlabel{cntViscConstStressMomentsThicknessIntegation}
\eqe
The numerical integration that is generally required here makes the projection approach less efficient than the direct surface approach.
\end{remark}

\subsubsection{Koiter membrane model} \label{s:KoiterMembraneModel}
Given the membrane energy density of the Koiter model \citep{ciarlet2005,sauer2017a}
\eqb{l}
	\Psi_\mathrm{mem}(a_\ab)=\dfrac{1}{8}\bigl(a_\ab-A_\ab\bigr)\,c^{\ab\gd}\,\bigl(a_\gd-A_\gd\bigr)\,, \eqlabel{cntViscConstElasticityPsiMemKoiter}
\eqe
with
\eqb{l}
	c^{\ab\gd}=\Lambda\,A^\ab\,A^\gd+\mu\,(A^{\alpha\gamma}A^{\beta\delta}+A^{\alpha\delta}A^{\beta\gamma})\,, \eqlabel{cntViscConstElasticityCabgdKoiter}
\eqe
the stresses follow from Eq.~(\ref{e:cntViscConstSigmaM}.1) as 
\eqb{l}
	\sigma^\ab\bigl(a_\gd\bigr)=\dfrac{1}{2\,J}\,c^{\ab\gd}\,\bigl(a_\gd-A_\gd)=\dfrac{1}{J}\,\biggl[\dfrac{\Lambda}{2}\,\bigl(I_1-2\bigr)\,a^\ab+\mu\,\bigl(A^{\alpha\gamma}\,a_\gd\,A^{\beta\delta}-a^\ab\bigr)\biggr]\,. \eqlabel{cntViscConstElasticitySigmaABkoiter}
\eqe

\subsubsection{Neo-Hookean membrane model}
Analogous to the classical three-dimensional Neo-Hookean material model, the Neo-Hookean membrane energy density is given by \citep{sauer2017a}
\eqb{l}
	\Psi_\mathrm{mem}\bigl(a_\ab\bigr) = \dfrac{\Lambda}{4}\,\bigl(J^2-1-2\,\ln J\bigr)+\dfrac{\mu}{2}\Bigl(I_1-2-2\,\ln J\Bigr)\,. \eqlabel{cntViscConstElasticityPsiMemClassicalNeoHookean}
\eqe
Its resulting stresses are
\eqb{l}
	\sigma^\ab\bigl(a_\gd\bigr)=\dfrac{1}{J}\,\biggl[\dfrac{\Lambda}{2}\,\bigl(J^2-1\bigr)\,a^\ab+\mu\,\bigl(A^\ab-a^\ab\bigr)\biggr]\,. \eqlabel{cntViscConstElasticitySigmaABclassicalNeoHookean}
\eqe
Note that in this work, $\Lambda$ may be taken as zero in \eqsref{cntViscConstElasticitySigmaABclassicalNeoHookean}, as it allows for an analytical solution of the evolution laws, see also \secref{cntViscConstViscoelasticity}.

\subsubsection{Neo-Hookean membrane model with dilatational/deviatoric split}
The membrane energy density in \eqsref{cntViscConstElasticityPsiMemClassicalNeoHookean} does not properly split the energy into pure dilatational and deviatoric components. A proper split is achieved by the following membrane energy density \citep{sauer2017b}
\eqb{l}
	\Psi_\mathrm{mem}(a_\ab)=\dfrac{K}{4}\,\bigl(J^2-1-2\,\ln J\bigr)+\dfrac{\mu}{2}\,\Bigl(\dfrac{I_1}{J}-2\Bigr)\,, \eqlabel{cntViscConstElasticityPsiMemNeoHookean}
\eqe
which yields the stresses
\eqb{l}
	\sigma^\ab\bigl(a_\gd\bigr)=\dfrac{1}{J}\,\biggl[\dfrac{K}{2}\,\bigl(J^2-1)\,a^\ab+\dfrac{\mu}{2\,J}\,\bigl(2\,A^\ab-I_1\,a^\ab\bigr)\biggr]\,. \eqlabel{cntViscConstElasticitySigmaABneoHookean}
\eqe

\subsubsection{Incompressible Neo-Hookean membrane model}
The classical 3D incompressible Neo-Hookean material model is described by the stress \citep{odgen1997}
\eqb{l}
	\tilde\bsig=\tilde\mu\,\tilde\bB+q\,\bone\,. \eqlabel{cntViscConstElasticityTildeSigmaIncompressibleNeoHookean}
\eqe
Here, $\tilde\mu$ denotes the 3D shear modulus and $q$ is the Lagrange multiplier associated with the incompressibility constraint. For membranes and thin shells, $\tilde\bB=\bB+\lambda_3^2\,(\bn\otimes\bn)$ and $\tilde\bsig=\bsig/\tilde{t}+\sigma_{33}\,(\bn\otimes\bn)$, where $\lambda_3$ denotes the out-of-plane stretch and $\tilde{t}$ is the shell thickness.\footnote{A tilde is placed on the current thickness to distinguish it from the time $t$.} Based on \eqsref{cntViscConstElasticityTildeSigmaIncompressibleNeoHookean}, the membrane stresses follow as \citep{sauer2014b}
\eqb{l}
	\sigma^\ab\bigl(a_\gd\bigr)=\dfrac{\mu}{J}\,\biggl(A^\ab-\dfrac{a^\ab}{J^2}\biggr)\,. \eqlabel{cntViscConstElasticitySigmaABincompressibleNeoHookean}
\eqe

\subsubsection{Membranes with constant surface tension}
A constant surface tension $\gamma$, see Eq.~(\ref{e:cntViscConstGammaNabDev}.1), can be imposed with the energy density \citep{sauer2016}
\eqb{l}
	\Psi_\mathrm{mem}(a_\ab)=\gamma\,J\,, \eqlabel{cntViscConstElasticityPsiMemSurfaceTensionDriven}
\eqe
which yields the stresses
\eqb{l}
	\sigma^\ab\bigl(a_\gd\bigr)=\gamma\,a^\ab\,. \eqlabel{cntViscConstElasticitySigmaABsurfaceTensionDriven}
\eqe

\subsubsection{Koiter bending model} 
Similar to \eqsref{cntViscConstElasticityPsiMemKoiter}, the Koiter model for bending is defined via \citep{ciarlet2005,sauer2017a}
\eqb{l}
	\Psi_\mathrm{bend}(b_\ab)=\dfrac{1}{2}\bigl(b_\ab-B_\ab\bigr)\,f^{\ab\gd}\,\bigl(b_\gd-B_\gd\bigr)\,, \eqlabel{cntViscConstElasticityPsiBendKoiter}
\eqe
with for instance $f^{\ab\gd}=c/2\,(A^{\alpha\gamma}A^{\beta\delta}+A^{\alpha\delta}A^{\beta\gamma})$, where $c$ is the bending modulus. The resulting moment components follow from Eq.~(\ref{e:cntViscConstSigmaM}.2) as
\eqb{l}
	M^\ab(b_\gd)=\dfrac{1}{J}\,f^{\ab\gd}\,\bigl(b_\gd-B_\gd\bigr)\,. \eqlabel{cntViscConstElasticityMabKoiter}
\eqe

\subsubsection{Helfrich bending model} \label{s:HelfrichBendingModel}
The Helfrich bending energy density is given by \citep{helfrich1973,sauer2017a}
\eqb{l}
	\Psi_\mathrm{bend}(a_\ab,b_\ab)=J\,\Bigl(k\,\bigl(H-H_0\bigr)^2+k^\star\,\kappa\Bigr)\,, \eqlabel{cntViscConstElasticityPsiBendHelfrich}
\eqe
which leads to the stresses
\eqb{l}
	\sigma^\ab(a_\gd,b_\gd)=\Bigl(k\,\bigl(H-H_0\bigr)^2-\kappa^\star\,\kappa\Bigr)\,a^\ab-2\,k\,\bigl(H-H_0)\,b^\ab\,, \eqlabel{cntViscConstElasticitySigmaABhelfrich}
\eqe
and moment components
\eqb{l}
	M^\ab(a_\gd,b_\gd)=\Bigl(k\,\bigl(H-H_0\bigr)+2\,k^\star\,H\Bigr)\,a^\ab-k^\star\,b^\ab\,, \eqlabel{cntViscConstElasticityMabHelfrich}
\eqe
with material parameters $k$ and $k^\star$, and the so-called spontaneous curvature $H_0$, which can be used to impose an initial stress-free mean curvature. Note that in this work, $k^\star$ is taken as zero in Eqs.~\eqref{e:cntViscConstElasticityPsiBendHelfrich}--\eqref{e:cntViscConstElasticityMabHelfrich}.

\subsection{Surface viscoelasticity} \seclabel{cntViscConstViscoelasticity}
In order to model viscoelastic material behavior, a generalized viscoelastic solid as shown by the rheological model in \figref{cntViscConstSrfSurfaceRheology} is considered.
\begin{figure}[!ht]
	\centering
		\includegraphics[scale=1]{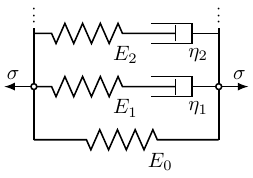}
	\caption{Surface rheology: Generalized viscoelastic solid. Here, $E_\bullet$ denotes the Young's modulus of a spring element, and $\eta_\bullet$ refers to the damping coefficient of a dashpot element.} \figlabel{cntViscConstSrfSurfaceRheology}
\end{figure}
It is composed of one branch containing a spring element (`elastic branch'), and several Maxwell branches that contain a spring and dashpot element. For simplicity, here, only one Maxwell branch is considered, such that the total stresses are given by
\eqb{l}
	\sigma^\ab\bigl(a^\gd,\hat{a}^\gd\bigr) = \sigma^\ab_0\bigl(a^\gd\bigr)+\sigma^\ab_1\bigl(\hat{a}^\gd\bigr)\,, \eqlabel{cntViscTotalStress}
\eqe
where $\sigma^\ab_0$ are the stresses in the elastic branch, and $\sigma_1^\ab$ are the stresses in the Maxwell branch. Due to the multiplicative split of the surface deformation gradient and the resulting additive split of the strains, the strains in the spring and dashpot element of a Maxwell branch are not equal. But the stresses are equal, i.e.
\eqb{l}
	\sigma_{1(\mrel)}^\ab\bigl(\hat{a}^\gd\bigr)=\sigma^\ab_{1(\mrin)}\bigl(\hat{a}^\gd,\dot{\hat{a}}^\gd\bigr)\,. \eqlabel{cntViscMaxwellStressCondition}
\eqe
\eqsref{cntViscMaxwellStressCondition} resembles three generally nonlinear ODEs for the components $\hat{a}^\gd$. The indices `$(\mrel)$' and `$(\mrin)$' refer to the spring and dashpot element of a Maxwell branch, respectively.\footnote{As the elastic and inelastic stresses are equal in the Maxwell model, the indices `$(\mrel)$' and `$(\mrin)$' can be omitted. But they are kept here in order to emphasize that these are the stresses in the spring and dashpot elements.} Once $\hat{a}^\gd$ is determined, the stresses $\sigma_1^\ab(\hat{a}^\gd)$ in \eqsref{cntViscTotalStress} can be computed, and the total stresses $\sigma^\ab\bigl(a^\gd,\hat{a}^\gd\bigr) $ are then used in the weak form, see Eq.~(\ref{e:cntViscWeakIntExtVirtualWork}.1).

A simple shear viscosity model for the inelastic stresses is given by \citep{sauer2019}
\eqb{l}
	\sigma^\ab_{1(\mrin)}\bigl(\dot{\hat{a}}^\gd\bigr)=-\dfrac{1}{J_\mrel}\,\eta_\mrs\,\dot{\hat{a}}^\ab\,, \eqlabel{cntViscSigma1in}
\eqe
where $\eta_\mrs\geq0$ denotes the in-plane shear viscosity. Note that the model in \eqsref{cntViscSigma1in} causes both shear and dilatation as the example in \secref{numExVisc2DmemStressRelaxation} shows. This is due to the fact that it leads to nonzero surface tension $\gamma$, see Eq.~(\ref{e:cntViscConstGammaNabDev}.1). Another simple viscosity model is given by \citep{sauer2019}
\eqb{l}
	\sigma^\ab_{1(\mrin)}\bigl(\dot{\hat{a}}^\gd\bigr)=\dfrac{1}{J_\mrel}\,\eta_\mrs\,\dot{J}_\mrin\,\hat{a}^\ab\,, \eqlabel{cntViscSigma1in2}
\eqe
with $\eta_\mrs\geq0$. In general, \eqsref{cntViscSigma1in2} leads to nonzero shear stresses, such that the model is not purely dilatational.

In all cases, the dissipated energy can be computed via
\eqb{l}
	\sD = \ds\int_0^t\int_{\sS}\sigma_1^\ab\,\dot{\varepsilon}_\ab^\mrin\,\dif a\,\dif t\,. \eqlabel{cntViscDissipatedEnergy}
\eqe
Given \eqsref{cntViscSigma1in} and Maxwell stresses $\sigma_{1(\mrel)}^\ab$, \eqsref{cntViscMaxwellStressCondition} can be rewritten in the form
\eqb{l}
	\dot{\hat{a}}^\ab=-\dfrac{J_\mrel}{\eta_\mrs}\,\sigma^\ab_{1(\mrel)}\bigl(\hat{a}^\gd\bigr)\,, \eqlabel{cntViscSurfaceODEgeneral}
\eqe
with the initial condition $\hat{a}^\ab\big|_{t=0}=A^\ab$. In analogy to Eq.~(\ref{e:cntViscConstSigmaM}.1), the Maxwell stresses $\hat\sigma_{1(\mrel)}^\ab$ follow from an energy density $\hat\Psi_\mathrm{mem}(\hat{a}_\ab,a_\ab)$, i.e.
\eqb{l}
	\hat\sigma_{1(\mrel)}^\ab=\dfrac{\partial\hat\Psi_\mathrm{mem}}{\partial\varepsilon^\mrel_\ab}=\dfrac{\partial\hat\Psi_\mathrm{mem}}{\partial\varepsilon_\ab}\,, \eqlabel{cntViscConstSrfHatSigma}
\eqe
where the elastic energy density $\hat\Psi_\mathrm{mem}(\hat{a}_\ab,a_\ab)$ follows analogously to the membrane energy densities presented in \secref{cntViscConstElasticity}, see also \cite{sauer2019}. The stresses $\hat\sigma^\ab_{1(\mrel)}$ can then be referred to the current surface via $\sigma^\ab_{1(\mrel)}=1/J_\mrel\,\hat\sigma_{1(\mrel)}^\ab$.
\begin{remark} \label{r:hatPsiMem}
	To obtain the elastic energy density $\hat\Psi_\mathrm{el}$ from one of the energy densities $\Psi_\mathrm{el}$ presented in \secref{cntViscConstElasticity}, the variables $A_\ab$ and $B_\ab$ need to be replaced by $\hat{a}_\ab$ and $\hat{b}_\ab$. Thus, $\hat\Psi_\mathrm{el}$ will be dependent on $\hat{a}_\ab$, $\hat{b}_\ab$, $a_\ab$, and $b_\ab$. In contrast, $\Psi_\mathrm{el}$ is expressed in terms of $A_\ab$, $B_\ab$, $a_\ab$, and $b_\ab$.
\end{remark}
In general, time integration schemes are needed to solve the evolution laws in \eqsref{cntViscSurfaceODEgeneral}, which are discussed in \secref{cmpViscODEs}. In some special cases, however, the ODEs in \eqsref{cntViscSurfaceODEgeneral} can be solved analytically. For instance, using the Neo-Hookean material model from \eqsref{cntViscConstElasticitySigmaABclassicalNeoHookean} with $K=0$ for the Maxwell branch, the ODEs simplify to
\eqb{l}
	\dot{\hat{a}}^\ab=\dfrac{\mu_1}{\eta_\mrs}\bigl(a^\ab-\hat{a}^\ab\bigr)\,, \eqlabel{cntViscMemODEneoHookeanSimple}
\eqe
where the contravariant surface metric generally depends on time, i.e.~$a^\ab=a^\ab(t)$. \eqsref{cntViscMemODEneoHookeanSimple} resembles three independent, linear, first-order, inhomogeneous ODEs, which can be solved analytically for specific choices of $a^\ab(t)$, e.g.~see the numerical examples in Secs.~\ref{s:numExViscInflMem} and \ref{s:numExViscInflSphr}.

In analogy to \eqsref{cntViscTotalStress}, the total moment components for the generalized viscoelastic solid with one Maxwell branch are\footnote{Here, the brackets in the index are used in order to distinguish the moments in the elastic branch, $M^\ab_{(0)}$, from the moments components w.r.t.~the reference configuration, $M^\ab_0$, see also Eq.~(\ref{e:cntViscConstSigmaM}.2).}
\eqb{l}
	M^\ab\bigl(b_\gd,\hat{b}_\gd\bigr)= M^\ab_{(0)}\bigl(b_\gd\bigr)+M^\ab_{(1)}\bigl(\hat{b}_\gd\bigr)\,. \eqlabel{cntViscTotalMoment}
\eqe
Similar to the stress equality condition for membrane viscosity in \eqsref{cntViscMaxwellStressCondition}, bending viscosity requires that the moments in the spring and dashpot element of the Maxwell branch, see \figref{cntViscConstSrfSurfaceRheology}, are equal, i.e.
\eqb{l}
	M^\ab_{(1)(\mrel)}\bigl(\hat{b}_\gd\bigr)=M^\ab_{(1)(\mrin)}\bigl(\hat{b}_\gd,\dot{\hat{b}}_\gd\bigr)\,, \eqlabel{cntViscMaxwellMomentCondition}
\eqe
where `$(1)$' refers to the \textit{first} Maxwell branch. In analogy to \eqsref{cntViscSigma1in}, a simple model for the moment components in the dashpot element is \citep{sauer2019}
\eqb{l}
	M_\ab^{(1),(\mrin)}\bigl(\dot{\hat{b}}_\gd\bigr)=\dfrac{\eta_\mrb}{J_\mrel}\,\dot{\hat{b}}_\ab\,, \eqlabel{cntViscMoment1in}
\eqe 
with $\eta_\mrb\geq0$. Combining Eqs.~\eqref{e:cntViscMaxwellMomentCondition}--\eqref{e:cntViscMoment1in}, the evolution laws can be written as
\eqb{l}
	\dot{\hat{b}}_\ab=\dfrac{J_\mrel}{\eta_\mrb}\,M_\ab^{(1)(\mrel)}\bigl(\hat{b}_\gd\bigr)\,, \eqlabel{cntViscBendingODEgeneral}
\eqe
with initial condition $\hat{b}_\ab\big|_{t=0}=B_\ab$. In analogy to Eq.~(\ref{e:cntViscConstSigmaM}.2), the Maxwell moments $\hat{M}_{1(\mrel)}^\ab$ follow from a bending energy density $\hat\Psi_\mathrm{bend}(\hat{a}_\ab,\hat{b}_\ab,a_\ab,b_\ab)$, i.e.
\eqb{l}
	\hat{M}_{(1)(\mrel)}^\ab=\dfrac{\partial\hat\Psi_\mathrm{bend}}{\partial\kappa^\mrel_\ab}=\dfrac{\partial\hat\Psi_\mathrm{bend}}{\partial\kappa_\ab}\,, \eqlabel{cntViscConstSrfHatM}
\eqe
where a material model from \secref{cntViscConstElasticity} can be chosen for $\hat{\Psi}_\mathrm{bend}(\hat{a}_\ab,\hat{b}_\ab,a_\ab,b_\ab)$ (by replacing $A_\ab$ with $\hat{a}_\ab$ and $B_\ab$ with $\hat{b}_\ab$, see also Remark~\ref{r:hatPsiMem}). \eqsref{cntViscConstSrfHatM} can then be referred to the current surface via $M_{(1)(\mrel)}^\ab=1/J_\mrel\,\hat{M}_{(1)(\mrel)}^\ab$. Note that in case of the Koiter bending model from \eqsref{cntViscConstElasticityMabKoiter}, \eqsref{cntViscBendingODEgeneral} leads to the evolution laws
\eqb{l}
	\dot{\hat{b}}_\ab = \dfrac{c_1}{\eta_\mrb}\,\bigl(b_\ab-\hat{b}_\ab\bigr)\,, \eqlabel{cntViscBendODEKoiter}
\eqe
which resembles three independent linear, first-order, inhomogeneous ODEs, similar to \eqsref{cntViscMemODEneoHookeanSimple}. For specific choices of $b_\ab(t)$, these ODEs are analytically solvable, e.g.~see the numerical examples in Secs.~\ref{s:numExViscPureBend}--\ref{s:numExViscInflSphr}. If no analytical solution is possible, numerical time integration needs to be used to solve the evolution laws, which is discussed in \secref{cmpViscODEs}.

\section{Computational formulation} \seclabel{cmpVisc}
This section briefly summarizes isogeometric surface discretizations in \secref{cmpViscIGA} and a corresponding computational formulation for modeling thin shells in Secs.~\ref{s:cmpViscDiscrPrm}--\ref{s:cmpViscDiscrWeak}, following \cite{duong2017}. \secref{cmpViscODEs} presents the numerical treatment for viscoelastic shells together with a time integration scheme for the evolution laws. The linearization within the finite element framework is discussed in \secref{cmpViscSolutionNES} and a summary of the governing equations is provided in \secref{cmpViscSummary}.

\subsection{Isogeometric surface discretization} \seclabel{cmpViscIGA}
The solution of the weak form of thin shells in Eqs.~\eqref{e:cntViscWeakWeakForm}--\eqref{e:cntViscWeakIntExtVirtualWork} requires $C^1$-continuous surface discretizations. Isogeometric analysis (IGA), proposed by \cite{hughes2005}, provides such surface discretizations, while offering many additional advantages. The basis functions of IGA are based on splines. \figref{cmpViscIGABasisFunctionsSpline} visualizes these spline basis functions for cubic order ($p=3$) and knot vector $\Xi=[0,0,0,0,0.25,0.5,0.75,1,1,1,1]$. Each basis function spans over $p+1$ elements, which is different from standard finite element methods based on Lagrangian basis functions.
\begin{figure}[!ht]
	\setlength{\figheight}{0.35\textwidth}
	\centering
	\subfloat[Spline basis function\figlabel{cmpViscIGABasisFunctionsSpline}]{
		\begin{tikzpicture}
			\def\cdot{\times}
			\begin{axis}[
			grid=both,xlabel={$\xi$},ylabel={},width=0.6\textwidth,height=\figheight,
			xmin=0,xmax=1,ymin=0,ymax=1,
			xtick={0,0.25,0.5,0.75,1},
			tick label style={font=\footnotesize},legend style={at={(0.5,0.7)},anchor=south,nodes={scale=0.75, transform shape}},legend cell align={left},legend columns=4,
			]
				\addplot[black,solid,line width=1]table [x index = {0}, y index = {5},col sep=comma]{fig/main/basisFunctions.csv};
					\addlegendentry{$\hat{N}_1$};
				\addplot[blue,solid,line width=1]table [x index = {0}, y index = {6},col sep=comma]{fig/main/basisFunctions.csv};
					\addlegendentry{$\hat{N}_2$};
				\addplot[red,solid,line width=1]table [x index = {0}, y index = {7},col sep=comma]{fig/main/basisFunctions.csv};
					\addlegendentry{$\hat{N}_3$};
				\addplot[cyan,solid,line width=1]table [x index = {0}, y index = {8},col sep=comma]{fig/main/basisFunctions.csv};
					\addlegendentry{$\hat{N}_4$};
				\addplot[purple,solid,line width=1]table [x index = {0}, y index = {9},col sep=comma]{fig/main/basisFunctions.csv};
					\addlegendentry{$\hat{N}_5$};
				\addplot[brown,solid,line width=1]table [x index = {0}, y index = {10},col sep=comma]{fig/main/basisFunctions.csv};
					\addlegendentry{$\hat{N}_6$};
				\addplot[black!40!green,solid,line width=1]table [x index = {0}, y index = {11},col sep=comma]{fig/main/basisFunctions.csv};
					\addlegendentry{$\hat{N}_7$};
			\end{axis}
		\end{tikzpicture}
	}
	\quad
	\subfloat[Bernstein polynomials\figlabel{cmpViscIGABasisFunctionsBernstein}]{
		\begin{tikzpicture}
			\def\cdot{\times}
			\begin{axis}[
			grid=both,xlabel={$\xi$},width=0.3\textwidth,height=\figheight,
			xmin=0,xmax=1,ymin=0,ymax=1,
			tick label style={font=\footnotesize},legend style={at={(0.5,0.58)},anchor=south,nodes={scale=0.75, transform shape}},legend cell align={left},
			]
				\addplot[black,solid,line width=1]table [x index = {0}, y index = {1},col sep=comma]{fig/main/basisFunctions.csv};
					\addlegendentry{$B_{0,3}$};
				\addplot[blue,solid,line width=1]table [x index = {0}, y index = {2},col sep=comma]{fig/main/basisFunctions.csv};
					\addlegendentry{$B_{1,3}$};
				\addplot[red,solid,line width=1]table [x index = {0}, y index = {3},col sep=comma]{fig/main/basisFunctions.csv};
					\addlegendentry{$B_{2,3}$};
				\addplot[cyan,solid,line width=1]table [x index = {0}, y index = {4},col sep=comma]{fig/main/basisFunctions.csv};
					\addlegendentry{$B_{3,3}$};
			\end{axis}
		\end{tikzpicture}
	}
	\caption{Isogeometric discretization: (a) Cubic spline basis functions $\hat{N}_A$ for the knot vector $\Xi=[0,0,0,0,0.25,0.5,0.75,1,1,1,1]$ and (b) cubic Bernstein polynomials $B_{i,p}$ $(i=0,\tightdots,p)$.} \figlabel{cmpViscIGABasisFunctions}
\end{figure}
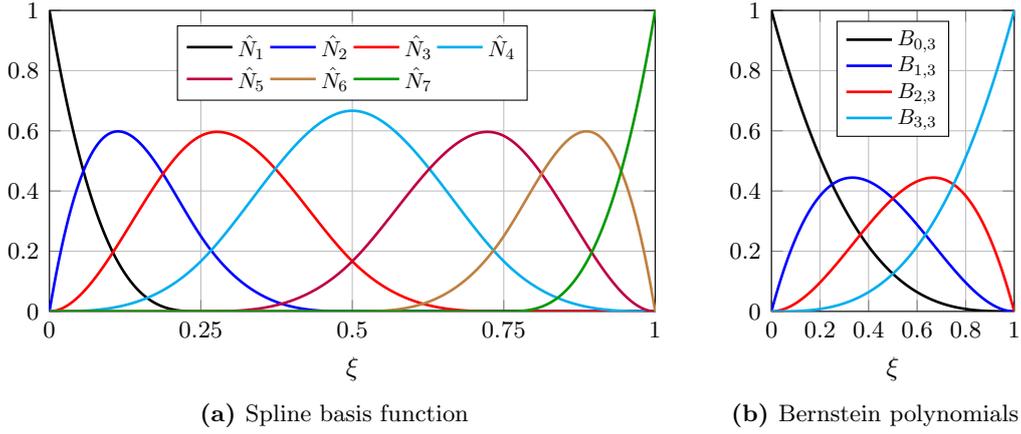

For two parametric directions, as requried for the surface description presented in \secref{cntViscDscr}, the basis functions follow from the tensor product of the basis functions in each parametric dimension. In order to embed isogeometric analysis into a standard finite element code, the Bézier extraction operator of \cite{borden2011} can be used. For a given element $\Omega^e$, there are $n$ spline basis function that have support on $\Omega^e$, i.e.~that are non-vanishing on this element. The NURBS (Non-Uniform Rational B-Splines) basis functions $\{N_A\}_{A=1}^n$ on $\Omega^e$ at the parametric coordinate $(\xi,\eta)$ are given by
\eqb{l}
N_A(\xi,\eta)=\dfrac{w_A\,\hat{N}_A(\xi,\eta)}{\sum_{\tilde{A}=1}^n w_{\tilde{A}}\,\hat{N}_{\tilde{A}}(\xi,\eta)}\,, \eqlabel{cmpViscIGANA}
\eqe
with weights $w_A$ and B-spline basis functions $\{\hat{N}_A\}_{A=1}^n$. This B-spline basis is mapped to a Bézier basis, which does not span over multiple elements. Cubic Bézier polynomials $B_{i,p}$ $(i=0,\tightdots,p)$ are illustrated in \figref{cmpViscIGABasisFunctionsBernstein}. Considering the Bézier extraction operators $\mC_\xi^e$ and $\mC_\eta^e$ for element $\Omega^e$ in $\xi$- and $\eta$-direction, respectively, yields the B-spline basis functions
\eqb{l}
	\hat{\mN}^e(\xi,\eta)=\mC_\xi^e\,\mB(\xi)\otimes\mC_\eta^e\,\mB(\eta)\,. \eqlabel{cmpViscIGAHatN}
\eqe
In \eqsref{cmpViscIGAHatN}, $\hat{\mN}^e$ is of size $n\times 1$ and contains the basis functions $\hat{N}_A$, $A=1,\tightdots n$. The Bézier extraction operators are of size $n\times n$, and $\mB$ is of size $n\times 1$ and contains the Bernstein polynomials in the corresponding parametric direction. Note that instead of Bézier extraction, also Lagrange extraction can be used \citep{schillinger2016}.

\subsection{Discretization of the primary field} \seclabel{cmpViscDiscrPrm}
Given are the $n_e$ spline basis functions on parametric element $\Omega^e$ with global indices $i_1,\dots,i_{n_e}$. The surface representation is then given by
\eqb{l}
	\bX\approx\bX^h = \mN^e\,\mX_e\,,\quad\mathrm{and}\quad\bx\approx\bx^h = \mN^e\,\mx_e\,, \eqlabel{cmpShllDiscretizationXx}
\eqe
for the reference and current surface, respectively. Here, the shape function array is
\eqb{l}
	\mN^e := [N_{i_1}\bone,\,N_{i_2}\bone,\,...,\,N_{i_{n_e}}\bone]\,, \eqlabel{cmpShllShapeFunctionArray}
\eqe
with dimension $3\times3\,n_e$. The $(3\times3)$-identity matrix is denoted $\bone$ and the element-level vectors for the nodal displacements are denoted $\mX_e$ and $\mx_e$. Analogously to \eqsref{cmpShllDiscretizationXx}, the variations of the nodal displacements on element $\Omega^e$ are
\eqb{l}
	\delta\bX^h = \mN^e\,\delta\mX_e\,,\quad\mathrm{and}\quad\delta\bx^h = \mN^e\,\delta\mx_e\,. \eqlabel{cmpShllDiscretizationDeltaXx}
\eqe
The discretized covariant tangent vectors, see also Eq.~(\ref{e:cntViscDscrCoVariantBasis}.1), follow as
\eqb{l}
	\bA_\alpha^h = \mN^e_{\!,\alpha}\,\mX_e\,,\quad\mathrm{and}\quad \ba_\alpha^h = \mN^e_{\!,\alpha}\,\mx_e\,, \eqlabel{cmpShllDiscretizationTangentVector}
\eqe
with $\mN^e_{\!,\alpha}=\partial\mN^e/\partial\xi^\alpha$. The discretized surface normals $\bN^h$ and $\bn^h$ follow in analogy to Eq.~(\ref{e:cntViscDscrCoVariantBasis}.2). The metric and curvature tensor components are then given by
\eqb{lll}
	A_{\alpha\beta}^h \is \mX_e^\mrT\,\bigl(\mN^e_{\!,\alpha}\bigr)^{\!\mrT}\,\mN^e_{\!,\beta}\,\mX_e\,, \quad\mathrm{and}\quad B_{\alpha\beta}^h = \bN^h\cdot\mN^e_{\!,\alpha\beta}\,\mX_e\,, \eqlabel{cmpShllDiscretizationMetricCurvatureReference}
\eqe 
and
\eqb{lll}
	a_{\alpha\beta}^h \is \mx_e^\mrT\,\bigl(\mN^e_{\!,\alpha}\bigr)^{\!\mrT}\,\mN^e_{\!,\beta}\,\mx_e\,,\quad\mathrm{and}\quad b_{\alpha\beta}^h = \bn^h\cdot\mN^e_{\!,\alpha\beta}\,\mx_e\,. \eqlabel{cmpShllDiscretizationMetricCurvatureCurrent}
\eqe
Eqs.~\eqref{e:cmpShllDiscretizationMetricCurvatureReference}--\eqref{e:cmpShllDiscretizationMetricCurvatureCurrent} can then be used to compute the discretized contravariant surface metrics, $[A^{\alpha\beta}_h]=[A_{\alpha\beta}^h]^{-1}$ and $[a^{\alpha\beta}_h]=[a_{\alpha\beta}^h]^{-1}$. In analogy, the discretized variations of the surface metric and curvature are given by
\eqb{lll}
	\delta a_{\alpha\beta}^h \is \delta\mx_e^\mrT\Bigl(\bigl(\mN^e_{\!,\alpha}\bigr)^{\!\mrT}\,\mN^e_{\!,\beta}+\bigl(\mN^e_{\!,\beta}\bigr)^{\!\mrT}\,\mN^e_{\!,\alpha}\Bigr)\,\mx_e\,, \quad\mathrm{and}\quad\delta b_{\alpha\beta}^h =\delta\mx_e^\mrT\,\bigl(\mN^e_{\!;\alpha\beta}\bigr)^{\!\mrT}\bn^h\,, \eqlabel{cmpShllDiscretizationDeltaMetricCurvature}
\eqe
with
\eqb{l}
	\mN^e_{\!;\alpha\beta} := \mN^e_{\!,\alpha\beta} - \Gamma^\gamma_{\alpha\beta}\,\mN^e_{\!,\gamma}\,. \eqlabel{cmpShllCovariantDerivativeOfShapeFunctionArray}
\eqe
Here, the discretized Christoffel symbols of the second kind on $\sS$ are
\eqb{l}
	\Gamma^\gamma_{\alpha\beta} = \mx_e^\mrT\,\bigl(\mN^e_{\!,\alpha\beta}\bigr)^{\!\mrT}a^{\gamma\delta}_h\,\mN^e_{\!,\delta}\,\mx_e\,. \eqlabel{cmpShllDiscretizedChristoffelSymbolCurrent}
\eqe
Note that subsequently, the superscript `h' may be omitted for notational simplicity.

Since the evolution laws for membrane and bending viscosity, see Eqs.~\eqref{e:cntViscSurfaceODEgeneral} and \eqref{e:cntViscBendingODEgeneral}, are purely temporal, no spatial discretization is required for $\hat{a}^\ab$ and $\hat{b}_\ab$. Thus, $\hat{a}^\ab$ and $\hat{b}_\ab$ can be treated as history variables that are evolved and stored at each quadrature point. This effectively eliminates them from the set of unknowns leading to a dependency of $\hat{a}^\ab$ and $\hat{b}_\ab$ on the primary unknown $\bx$ that affects the linearization, see also \secref{cmpViscSolutionNES}.

\subsection{Discretized mechanical weak form} \seclabel{cmpViscDiscrWeak}
Inserting the discretization from above into Eqs.~\eqref{e:cntViscWeakWeakForm}--\eqref{e:cntViscWeakIntExtVirtualWork} yields the discretized mechanical weak form
\eqb{l}
	\delta\mx^\mrT\,\big[\mf_\mathrm{int} - \mf_\mathrm{ext}\big] = 0\,, \quad \forall\,\delta\mx\in\sU^h\,, \eqlabel{cmpShllDiscretizedWeakForm}
\eqe
with the finite-dimensional space $\sU^h\subset\sU$, and the global force vectors $\mf_\mathrm{int}$ and $\mf_\mathrm{ext}$. These are assembled from their element-level contributions
\eqb{rll}	
	\mf^e_\mathrm{int} \dis \ds\int_{\Omega^e}\sigma^\ab\,\bigl(\mN^e_{\!,\alpha}\bigr)^{\!\mrT}\ba^h_\beta\,\dif a
	+ \int_{\Omega^e}M^\ab\,\bigl(\mN^e_{\!;\alpha\beta}\bigr)^{\!\mrT}\bn^h\,\dif a\,,\quad\mathrm{and} \\[4mm] 
	\mf^e_\mathrm{ext} \dis \ds\int_{\Omega^e}\bigl(\mN^e\bigr)^{\!\mrT}p\,\bn^h\,\dif a + \ds\int_{\Omega^e}\bigl(\mN^e\bigr)^{\!\mrT}f^\alpha\,\ba^h_\alpha\,\dif a\,, \eqlabel{cmpShllForceVectors}
\eqe
where $\Omega^e$ denotes the domain of element $e$ in the current configuration. In $\mf^e_\mathrm{ext}$, the boundary loads $\bT$ and $\bM$ acting on $\partial\sS$ are assumed to be zero. The extension to boundary loads can be found in \cite{duong2017}. The computation of the stresses $\sigma^\ab$ and moments $M^\ab$ is outlined in \secref{cntViscConst}.

From \eqsref{cmpShllDiscretizedWeakForm} follows the equation of motion at the free nodes (where no Dirichlet boundary conditions are prescribed)
\eqb{l}
	\mf(\mx) =\mf_\mathrm{int}(\mx) - \mf_\mathrm{ext}(\mx) = \mathbf{0}\,,\eqlabel{cmpShllODE}
\eqe
where $\mx$ is the global unknown, similar to the element-level unknowns $\mx_e$.

\subsection{Solution of the evolution laws} \seclabel{cmpViscODEs}
In \secref{cntViscConst}, ODEs for $\hat{a}^\ab$ and $\hat{b}_\ab$ are derived, see Eqs.~\eqref{e:cntViscSurfaceODEgeneral} and \eqref{e:cntViscBendingODEgeneral}. In general, they need to be solved numerically as no analytical solution exists. The temporal integration of $\hat{a}^\ab$ based on \eqsref{cntViscSurfaceODEgeneral} is described in \secref{cmpViscODEsMembrane}, and the numerical treatment of $\hat{b}_\ab$ is described in \secref{cmpViscODEsBending}.

\subsubsection{Membrane material models} \seclabel{cmpViscODEsMembrane}
Given all quantities at time step $n$ and the time step size $\Delta t_{n+1}:=t_{n+1}-t_n$, \eqsref{cntViscSurfaceODEgeneral} needs to be solved for $\hat{a}^\ab_{n+1}$. Here, an implicit/backward Euler method is employed, such that the ODEs in \eqsref{cntViscSurfaceODEgeneral} reduce to the nonlinear algebraic equations
\eqb{l}
	\hat{\mrg}^\ab_\mrs\bigl(\hat{a}^\gd_{n+1}\bigr):=\dfrac{\hat{a}^\ab_{n+1}-\hat{a}^\ab_n}{\Delta t_{n+1}}+\dfrac{J_{\mrel,n+1}}{\eta_\mrs}\,\sigma_{1(\mrel)}^\ab\bigl(\hat{a}^\gd_{n+1}\bigr)=0\,, \eqlabel{cmpViscDiscretizedSurfaceODEgeneral}
\eqe
which need to be solved for $\hat{a}^\ab_{n+1}$. The initial condition is $\hat{a}^\ab\big|_{t=0}=A^\ab$. \eqsref{cmpViscDiscretizedSurfaceODEgeneral} contains four equations, which need to be solved for the four unknowns $\hat{a}^\ab$. As $\hat{a}^{12}=\hat{a}^{21}$, the unknown $\hat{a}^{21}$ and the corresponding equation $\hat{\mrg}_\mrs^{21}=0$ can be eliminated. Thus, there remain three equations and unknowns, which are arranged in the vectors
\eqb{l}
	\hat{\bmrg}_\mrs:=\begin{bmatrix}\hat{\mrg}_\mrs^{11}\\\hat{\mrg}_\mrs^{12}\\\hat{\mrg}_\mrs^{22}\end{bmatrix}=\boldsymbol{0}\,,\quad\mathrm{and}\quad\hat\ma:=\begin{bmatrix}\hat{a}^{11}\\\hat{a}^{12}\\\hat{a}^{22}\end{bmatrix}. \eqlabel{cmpViscAssembledEqns}
\eqe
In general, it is not possible to solve Eq.~(\ref{e:cmpViscAssembledEqns}.1) analytically, and thus a local Newton-Raphson iteration is used. Using Taylor expansion, Eq.~(\ref{e:cmpViscAssembledEqns}.1) is approximated around $\hat\ma_{n+1}\big|_i$ (where $i$ denotes the $i$'th Newton-Raphson step) by
\eqb{l}
	\hat{\bmrg}_\mrs\big|_{i+1}\approx\hat{\bmrg}_\mrs\big|_i+\dpa{\hat{\bmrg}_\mrs}{\hat{\ma}_{n+1}}\bigg|_{i}\,\Delta\hat{\ma}_{n+1}\big|_{i+1}=\boldsymbol{0}\,, \eqlabel{cmpViscAssembledEqnsTaylorExpansion}
\eqe
 where
\eqb{l}
	\dpa{\hat{\bmrg}_\mrs}{\hat{\ma}_{n+1}}=\begin{bmatrix}
		\nicefrac{\partial\hat{\mrg}_\mrs^{11}}{\partial\hat{a}_{n+1}^{11}} & \nicefrac{\partial\hat{\mrg}_\mrs^{11}}{\partial\hat{a}_{n+1}^{12}}+\nicefrac{\partial\hat{\mrg}_\mrs^{11}}{\partial\hat{a}_{n+1}^{21}} & \nicefrac{\partial\hat{\mrg}_\mrs^{11}}{\partial\hat{a}_{n+1}^{22}}\\[2mm]
		\nicefrac{\partial\hat{\mrg}_\mrs^{12}}{\partial\hat{a}_{n+1}^{11}} & \nicefrac{\partial\hat{\mrg}_\mrs^{12}}{\partial\hat{a}_{n+1}^{12}}+\nicefrac{\partial\hat{\mrg}_\mrs^{12}}{\partial\hat{a}_{n+1}^{21}} & \nicefrac{\partial\hat{\mrg}_\mrs^{12}}{\partial\hat{a}_{n+1}^{22}}\\[2mm]
		\nicefrac{\partial\hat{\mrg}_\mrs^{22}}{\partial\hat{a}_{n+1}^{11}} & \nicefrac{\partial\hat{\mrg}_\mrs^{22}}{\partial\hat{a}_{n+1}^{12}}+\nicefrac{\partial\hat{\mrg}_\mrs^{22}}{\partial\hat{a}_{n+1}^{21}} & \nicefrac{\partial\hat{\mrg}_\mrs^{22}}{\partial\hat{a}_{n+1}^{22}}
	\end{bmatrix}\,. \eqlabel{cmpViscAssembledEqnsLin}
\eqe
Note that the sum in the second column of \eqsref{cmpViscAssembledEqnsLin} occurs because the variable $\hat{a}^{21}$ is eliminated, as mentioned above. Provided the starting guess $\hat{\ma}_{n+1}\big|_0=\hat{\ma}_n$, the iteration ($i=0,1,2,\tightdots$)
\eqb{l}
	\mathrm{solve}\quad\dpa{\hat{\bmrg}_\mrs\bigl(\hat{\ma}_{n+1}\bigr)}{\hat{\ma}_{n+1}}\bigg|_i\,\Delta\hat{\ma}_{n+1}\big|_{i+1}=-\hat{\bmrg}_\mrs\bigl(\hat{\ma}_{n+1}\bigr)\big|_i\quad\mathrm{for}\:\:\Delta\hat{\ma}_{n+1}\big|_{i+1}\,,\\[5mm]
	\mathrm{update}\quad \hat{\ma}_{n+1}\big|_{i+1}=\hat{\ma}_{n+1}\big|_{i}+\Delta\hat{\ma}_{n+1}\big|_{i+1}\,, \eqlabel{cmpViscNRiteration}
\eqe
is repeated until convergence is obtained. Convergence is monitored by checking if
\eqb{l}
	\norm{\Delta\hat{\ma}_{n+1}\big|_{i+1}}_2\leq10^{-10}\,. \eqlabel{cmpViscNRconvergence}
\eqe
Subsequently, the algrebraic equations following from \eqsref{cmpViscDiscretizedSurfaceODEgeneral} are specified for all the membrane material models of \secref{cntViscConst}. The index ‘$n+1$’ is omitted for notational simplicity in the subsequent sections. Every quantity without index ‘$n$’ is evaluated at the current time step.

\paragraph{Koiter membrane model} 
Given the material model from \eqsref{cntViscConstElasticityPsiMemKoiter},\footnote{with $A_\ab$ replaced by $\hat{a}_\ab$, see also Remark~\ref{r:hatPsiMem}\label{footnote:Aab}} \eqsref{cmpViscDiscretizedSurfaceODEgeneral} becomes
\eqb{l}
	\hat{\mrg}_\mrs^\ab\bigl(\hat{a}^\gd\bigr)=\dfrac{\hat{a}^\ab-\hat{a}^\ab_n}{\Delta t} + \dfrac{\Lambda_1}{2\,\eta_\mrs}\,\bigl(I_1^{\mrel}-2\bigr)\,\hat{a}^\ab+\dfrac{\mu_1}{\eta_\mrs}\,\bigl(\hat{a}^{\alpha\gamma}\,a_\gd\,\hat{a}^{\beta\delta}-\hat{a}^\ab\bigr)=0\,. \eqlabel{cmpViscDiscretizedODEKoiter}
\eqe

\paragraph{Neo-Hookean membrane model} 
Considering the material model from \eqsref{cntViscConstElasticityPsiMemClassicalNeoHookean} with $\Lambda=0$,\footnoteref{footnote:Aab} \eqsref{cmpViscDiscretizedSurfaceODEgeneral} becomes
\eqb{l}
	\hat{\mrg}_\mrs^\ab\bigl(\hat{a}^\gd\bigr)=\dfrac{\hat{a}^\ab-\hat{a}^\ab_n}{\Delta t} + \dfrac{\mu_1}{\eta_\mrs}\,\bigl(\hat{a}^\ab-a^\ab\bigr)=0\,. \eqlabel{cmpviscDiscretizedODENeoHookeanSimple}
\eqe
This is a linear model that can be solved directly for $\hat{a}^\ab$ (without using \eqsref{cmpViscNRiteration}) giving
\eqb{l}
	\hat{a}^\ab = \dfrac{\eta_\mrs\,\hat{a}_n^\ab+\mu_1\,\Delta t\,a^\ab}{\eta_\mrs+\mu_1\,\Delta t}\,. \eqlabel{cmpviscDiscretizedODENeoHookeanSimple2}
\eqe

\paragraph{Neo-Hookean membrane model with dilatational/deviatoric split}
For the material model from \eqsref{cntViscConstElasticityPsiMemNeoHookean},\footnoteref{footnote:Aab}  \eqsref{cmpViscDiscretizedSurfaceODEgeneral} becomes
\eqb{l}
	\hat{\mrg}_\mrs^\ab\bigl(\hat{a}^\gd\bigr)=\dfrac{\hat{a}^\ab-\hat{a}_n^\ab}{\Delta t}+\dfrac{K_1}{2\,\eta_\mrs}\,\bigl(J_{\mrel}^2-1\bigr)\,a^\ab+\dfrac{\mu_1}{2\,\eta_\mrs\,J_{\mrel}}\,\bigl(2\,\hat{a}^\ab-I_1^{\mrel}a^\ab\bigr)=0\,. \eqlabel{cmpViscDiscretizedODENeoHookean}
\eqe

\paragraph{Incompressible Neo-Hookean membrane model}
For the material model from \eqsref{cntViscConstElasticitySigmaABincompressibleNeoHookean},\footnoteref{footnote:Aab} \eqsref{cmpViscDiscretizedSurfaceODEgeneral} becomes
\eqb{l}
	\hat{\mrg}_\mrs^\ab\bigl(\hat{a}^\gd\bigr)=\dfrac{\hat{a}^\ab-\hat{a}_n^\ab}{\Delta t}+\dfrac{\mu_1}{\eta_\mrs}\,\biggl(\hat{a}^\ab-\dfrac{a^\ab}{J_{\mrel}^2}\biggr)=0\,. \eqlabel{cmpViscDiscretizedODEIncompressibleNeoHookean}
\eqe

\paragraph{Membranes with constant surface tension}
Considering the material model from \eqsref{cntViscConstElasticityPsiMemSurfaceTensionDriven},\footnoteref{footnote:Aab}  \eqsref{cmpViscDiscretizedSurfaceODEgeneral} becomes
\eqb{l}
	\hat{\mrg}_\mrs^\ab\bigl(\hat{a}^\gd\bigr)=\dfrac{\hat{a}^\ab-\hat{a}^\ab_n}{\Delta t} + \dfrac{\hat\gamma}{\eta_\mrs}\,J_{\mrel}\,a^\ab=0\,, \eqlabel{cmpviscDiscretizedODEsurfaceTensionDriven}
\eqe
where $\hat\gamma$ denotes the prescribed surface tension w.r.t. the intermediate configuration.

\subsubsection{Bending material models} \seclabel{cmpViscODEsBending}
The numerical time integration of the evolution laws for bending viscosity follows in analogy to the presented scheme for the evolution laws for membrane viscosity. In analogy to \eqsref{cmpViscDiscretizedSurfaceODEgeneral}, the ODEs in \eqsref{cntViscBendingODEgeneral} reduce to the nonlinear algebraic equations
\eqb{l}
	\hat{\mrg}_\ab^\mrb\bigl(\hat{b}_\gd\bigr):=\dfrac{\hat{b}_\ab-\hat{b}_\ab^n}{\Delta t}-\dfrac{J_{\mrel}}{\eta_\mrb}\,M_\ab^{(1)(\mrel)}\bigl(\hat{b}_\gd\bigr)=0\,, \eqlabel{cmpViscDiscretizedBendingODEgeneral}
\eqe
with initial condition $\hat{b}_\ab\big|_{t=0}=B_\ab$.

\paragraph{Koiter bending model}
The algebraic equations, see \eqsref{cmpViscDiscretizedBendingODEgeneral},\footnote{with $B_\ab$ replaced by $\hat{b}_\ab$, see also Remark~\ref{r:hatPsiMem}\label{footnote:Bab}} for the material model from \eqsref{cntViscConstElasticityPsiBendKoiter} become
\eqb{l}
	\hat{\mrg}^\mrb_\ab\bigl(\hat{b}_\gd\bigr)=\dfrac{\hat{b}_\ab-\hat{b}_\ab^n}{\Delta t}+\dfrac{c_1}{\eta_\mrb}\,\bigl(\hat{b}_\ab-b_\ab\bigr)=0\,. \eqlabel{cmpViscDiscretizedODEKoiterBending}
\eqe
This is also a linear model that can be directly solved for $\hat{b}_\ab$, similar to \eqsref{cmpviscDiscretizedODENeoHookeanSimple2}.

\paragraph{Helfrich bending model} 
For the material model from \eqsref{cntViscConstElasticityPsiBendHelfrich} with $k^\star=0$,\footnoterefs{footnote:Aab}{footnote:Bab} Eqs.~\eqref{e:cmpViscDiscretizedSurfaceODEgeneral} and \eqref{e:cmpViscDiscretizedBendingODEgeneral} become
\eqb{l}
	\hat{\mrg}_\mrs^\ab\bigl(\hat{a}^\gd\bigr)=\dfrac{\hat{a}^\ab-\hat{a}^\ab_n}{\Delta t}+\dfrac{k_1\,J_{\mrel}}{\eta_\mrb}\,\Bigl[\bigl(H-\hat{H}\bigr)^2\,a^\ab-2\,\bigl(H-\hat{H}\bigr)\,b^\ab\Bigr]=0\,, \eqlabel{cmpViscDiscretizedODEHelfrichBendingGs}
\eqe
and
\eqb{l}
	\hat{\mrg}^\mrb_\ab\bigl(\hat{b}_\gd\bigr)=\dfrac{\hat{b}_\ab-\hat{b}_\ab^n}{\Delta t}-\dfrac{k_1\,J_{\mrel}}{\eta_\mrb}\,\bigl(H-\hat{H}\bigr)\,a_\ab=0\,. \eqlabel{cmpViscDiscretizedODEHelfrichBendingGb}
\eqe

The derivatives of all these material models, which are required in \eqsref{cmpViscAssembledEqnsLin}, are reported in \appref{viscLinearizationImplicitEuler}.

\begin{remark} \label{r:coupledViscosity}
For coupled membrane and bending viscosity, iteration \eqref{e:cmpViscNRiteration} has to be performed for the unknowns $\hat{a}^\ab$ and $\hat{b}_\ab$ simultaneously. The linear equation system in Eq.~(\ref{e:cmpViscNRiteration}.1) then becomes
\eqb{l}
	\renewcommand*{\arraystretch}{1.5}
	\begin{bmatrix}
		\partial\hat{\bmrg}_\mrs\bigl(\hat\ma_{n+1},\hat{\mathbf{b}}_{n+1}\bigr)/\partial\hat{\ma}_{n+1} & \partial\hat{\bmrg}_\mrs\bigl(\hat\ma_{n+1},\hat{\mathbf{b}}_{n+1}\bigr)/\partial\hat{\mathbf{b}}_{n+1} \\
		\partial\hat{\bmrg}^\mrb\bigl(\hat\ma_{n+1},\hat{\mathbf{b}}_{n+1}\bigr)/\partial\hat{\ma}_{n+1} & \partial\hat{\bmrg}^\mrb\bigl(\hat\ma_{n+1},\hat{\mathbf{b}}_{n+1}\bigr)/\partial\hat{\mathbf{b}}_{n+1}
	\end{bmatrix}_i\,\begin{bmatrix}\Delta\hat\ma_{n+1}\\\Delta\hat{\mathbf{b}}_{n+1}\end{bmatrix}_{i+1}=-\begin{bmatrix}\hat\bmrg_\mrs\bigl(\hat\ma_{n+1},\hat{\mathbf{b}}_{n+1}\bigr)\\\hat\bmrg_\mrb\bigl(\hat\ma_{n+1},\hat{\mathbf{b}}_{n+1}\bigr)\end{bmatrix}_i\,,
\eqe
where the vector $\hat{\mathbf{b}}$ arranges the terms $\hat{b}_\ab$ analogously to Eq.~(\ref{e:cmpViscAssembledEqns}.2). For example, the discretized evolution laws for the Helfrich model, see Eqs.~\eqref{e:cmpViscDiscretizedODEHelfrichBendingGs}--\eqref{e:cmpViscDiscretizedODEHelfrichBendingGb}, lead to coupled membrane and bending viscosity, see also Eqs.~\eqref{e:viscLinearizationImplicitEulerHelfrichGs}--\eqref{e:viscLinearizationImplicitEulerHelfrichGb}.
\end{remark}

\begin{remark} \label{r:implEuler}
Examining \eqsref{cmpviscDiscretizedODENeoHookeanSimple2} shows that the absolute value of the factor in front of $\hat{a}^\ab_n$ is $\eta_\mrs/(\eta_\mrs+\mu_1\,\Delta t)$, which is smaller to one and thus, the method is stable. For the explicit Euler scheme, this factor would be $|(\eta_\mrs-\mu_1\,\Delta t)/\eta_\mrs|$, which might become larger to one, such that the method is not stable. Thus, the implicit Euler scheme is used in this work.
\end{remark}

\begin{remark} \label{r:reformulatingODEs}
The presented formulation of membrane viscosity leads to ODEs for the intermediate surface metric $\hat{a}^\ab$. For some discretizations, e.g.~based on unstructured splines, solving for $\hat{a}^\ab$ might lead to numerical ill-conditioning. For instance, at the extraordinary points of an unstructured spline sphere \citep{toshniwal2017b}, the parametrization becomes singular, i.e.~one tangent vector $\ba_\alpha$ approaches $\boldsymbol{0}$. The computation of the inverse $[a^\ab]=[a_\ab]^{-1}$ then might lead to numerical ill-conditioning. The same problem pertains to the intermediate surface metric $\hat{a}^\ab$. In order to resolve this issue, the ODEs for $\hat{a}^\ab$ may be reformulated, e.g.~as ODEs for $I_1^\mrel$, $J_\mrin$ and $J_\mrel$, but also other options could be possible, for instance based on the Cauchy-Green tensor $\bC_\mrel$.
\end{remark}

\subsection{Solution of the nonlinear equation system} \seclabel{cmpViscSolutionNES}
The nonlinear equation system in \eqsref{cmpShllODE} is solved using a global Newton-Raphson iteration. This scheme requires the linearization of the force vector in \eqsref{cmpShllODE}. This leads to the element-level stiffness matrix $\mk^e:=\mk^e_{\tau\tau}+\mk^e_{\tau M}+\mk^e_{M\tau}+\mk^e_{MM}+\mk^e_{\tau}+\mk^e_{M}$ with the element-level material stiffness matrices 
\eqb{rllrll}
	\mk_{\tau\tau}^e\dis\ds\int_{\Omega_0^e}c^{\ab\gd}\,\bigl(\mN^e_\ca\bigr)^{\!\mrT}\bigl(\ba_\beta\otimes\ba_\gamma\bigr)\,\mN^e_\cd\,\dif A\,,&\:\:\:\mk_{\tau M}^e\dis\ds\int_{\Omega_0^e}d^{\ab\gd}\,\bigl(\mN^e_\ca\bigr)^{\!\mrT}\bigl(\ba_\beta\otimes\bn\bigr)\,\mN^e_{;\gd}\,\dif A\,,\\[4mm]
	\mk_{M\tau}^e\dis\ds\int_{\Omega_0^e}e^{\ab\gd}\,\bigl(\mN^e_{;\ab}\bigr)^{\!\mrT}\bigl(\bn\otimes\ba_\gamma\bigr)\,\mN^e_\cd\,\dif A\,,&\:\:\:\mk_{MM}^e\dis\ds\int_{\Omega_0^e}f^{\ab\gd}\,\bigl(\mN^e_{;\ab}\bigr)^{\!\mrT}\bigl(\bn\otimes\bn\bigr)\,\mN^e_{;\gd}\,\dif A\,,\eqlabel{cmpViscSolutionNESmaterialStiffnessMatrices}
\eqe
and the element-level geometric stiffness matrices
\eqb{l}
	\mk_{\tau}^e:=\ds\int_{\Omega_0^e}\bigl(\mN^e_\ca\bigr)^{\!\mrT}\tau^\ab\,\mN^e_\cb\,\dif A\,,\quad\mathrm{and}\quad\mk_M^e:=\mk_{M1}^e+\mk_{M2}^e+(\mk_{M2}^e)^\mrT\,,\eqlabel{cmpViscSolutionNESgeometricStiffnessMatrices1}
\eqe
with
\eqb{rll}
	\mk_{M1}^e\dis\ds -\int_{\Omega_0^e}b_\ab\,M_0^\ab\,a^\gd\,\bigl(\mN^e_\cg\bigr)^{\!\mrT}\bigl(\bn\otimes\bn\bigr)\,\mN^e_\cd\,\dif A\,,\quad\mathrm{and}\\[4mm]
	\mk_{M2}^e\dis\ds -\int_{\Omega_0^e}M_0^\ab\,\bigl(\mN^e_\cg\bigr)^{\!\mrT}\bigl(\bn\otimes\ba^\gamma\bigr)\,\mN^e_{;\ab}\,\dif A\,,\eqlabel{cmpViscSolutionNESgeometricStiffnessMatrices2}
\eqe
see \cite{duong2017} for more details and for the linearization of the external element-level force vector $\mf_\mathrm{ext}^e$ in Eq.~(\ref{e:cmpShllForceVectors}.2). In \eqsref{cmpViscSolutionNESmaterialStiffnessMatrices}, the material tangents
\eqb{l}
	c^{\ab\gd}=2\,\dpa{\tau^\ab}{a_\gd}\,,\quad d^{\ab\gd}=\dpa{\tau^\ab}{b_\gd}\,,\quad e^{\ab\gd}=2\,\dpa{M_0^\ab}{a_\gd}\,,\quad\mathrm{and}\quad f^{\ab\gd}=\dpa{M_0^\ab}{b_\gd}\,, \eqlabel{cmpViscTangentsFE}
\eqe
need to be defined for the employed material model. As the stresses and moments are composed of elastic and Maxwell components, see Eqs.~\eqref{e:cntViscTotalStress} and \eqref{e:cntViscTotalMoment}, the material tangents in \eqsref{cmpViscTangentsFE} will also be composed of the two contributions
\eqb{l}
	\tilde{c}^{\ab\gd} = \tilde{c}^{\ab\gd}_0+\tilde{c}^{\ab\gd}_1\,,\quad\tilde{c}=c,d,e,f\,, \eqlabel{cmpViscMaterialTangents}
\eqe
where again the index `0' refers to the elastic branch and `1' refers to the Maxwell branch. The Maxwell material tangents $\tilde{c}_1^{\ab\gd}$ in \eqsref{cmpViscMaterialTangents} are derived in \appref{viscLinearizationFEM}.

\begin{remark} \label{r:eliminationHataabHatbab}
In \cite{sauer2019}, the variables $\hat{a}_\ab$ and $\hat{b}_\ab$ are assumed to be independent variables from $a_\ab$ and $b_\ab$. In this work here, the unknowns $\hat{a}_\ab$ and $\hat{b}_\ab$ are locally eliminated by solving Eqs.~\eqref{e:cntViscSurfaceODEgeneral} and \eqref{e:cntViscBendingODEgeneral} at each quadrature point, see \secref{cmpViscODEs}. This elimination makes $\hat{a}_\ab$ and $\hat{b}_\ab$ a function of $a_\ab$ and $b_\ab$, which leads to additional derivatives in the linearization that are not appearing in the theory of \cite{sauer2019}.
\end{remark}

\subsection{Summary of the computational formulation for viscoelastic shells} \seclabel{cmpViscSummary}
A concise summary of the presented computational formulation for viscoelastic shells is given in \tabref{cmpViscSummary}.
\begin{table}[!ht]
	\caption{Summary of the computational formulation for viscoelastic shells} \tablabel{cmpViscSummary}
	\centering
	\setlength\fboxrule{0.6pt}
	\fbox{%
 	\parbox{0.9\textwidth}{%
    		\small
    		The governing nonlinear equation system for the shell deformation is\vspace{-2mm}
    			\esb{l}
    				\mf(\mx) =\mf_\mathrm{int}(\mx) - \mf_\mathrm{ext}(\mx) = \mathbf{0}\,,
    			\ese
    		which is assembled from the element-level contributions
    			\esb{rll}
				\mf^e_\mathrm{int} \dis \ds\int_{\Omega^e}\sigma^\ab\,\mN_{\!,\alpha}^\mrT\,\ba^h_\beta\,\dif a
				+ \int_{\Omega^e}M^\ab\,\mN^\mrT_{\!;\alpha\beta}\,\bn^h\,\dif a\,,\quad\mathrm{and}\\[4mm]
				\mf^e_\mathrm{ext} \dis \ds\int_{\Omega^e}\mN^\mrT\,p\,\bn^h\,\dif a + \ds\int_{\Omega^e}\mN^\mrT\,f^\alpha\,\ba^h_\alpha\,\dif a\,.
			\ese
		The stresses and moments (considering one Maxwell branch) are given by\vspace{-2mm}
			\esb{l}
				\sigma^\ab = \sigma^\ab_0\bigl(a^\gd\bigr)+\sigma^\ab_1\bigl(\hat{a}^\gd\bigr)\,,\quad\mathrm{and}\quad M^\ab = M^\ab_{(0)}\bigl(b^\gd\bigr)+M^\ab_{(1)}\bigl(\hat{b}^\gd\bigr)\,.
			\ese		
		Here, $\sigma^\ab_0$, $\sigma_1^\ab=\sigma^\ab_{1(\mrel)}$, $M^\ab_{(0)}$, and $M^\ab_{(1)}=M^\ab_{(1)(\mrel)}$ follow from a specific choice of the elastic energy density, see \secref{cntViscConstElasticity} for examples. The Maxwell stress $\sigma^\ab_1$ and moment $M^\ab_{(1)}$ follow from the conditions in Eqs.~\eqref{e:cntViscMaxwellStressCondition} and \eqref{e:cntViscMaxwellMomentCondition}. For the simple shear viscosity models in Eqs.~\eqref{e:cntViscSigma1in} and \eqref{e:cntViscMoment1in}, the evolution laws for $\hat{a}^\ab$ and $\hat{b}_\ab$ are given by\vspace{-2mm}
			\esb{l}
				\dot{\hat{a}}^\ab=-\dfrac{J_\mrel}{\eta_\mrs}\,\sigma^\ab_{1(\mrel)}\bigl(\hat{a}^\gd\bigr)\,,\quad\mathrm{and}\quad\dot{\hat{b}}_\ab=\dfrac{J_\mrel}{\eta_\mrb}\,M_\ab^{(1)(\mrel)}\bigl(\hat{b}_\gd\bigr)\,,
			\ese
		with initial conditions $\hat{a}^\ab|_{t=0}=A^\ab$ and $\hat{b}_\ab|_{t=0}=B_\ab$. The resulting ODEs are solved with the implicit Euler scheme, which leads to the temporal discretized nonlinear equations\vspace{-2mm}
			\esb{rll}
				\hat{\mrg}^\ab_\mrs\bigl(\hat{a}^\gd\bigr)\dis\dfrac{\hat{a}^\ab-\hat{a}^\ab_n}{\Delta t}+\dfrac{J_\mrel}{\eta_\mrs}\,\sigma_{1(\mrel)}^\ab\bigl(\hat{a}^\gd\bigr)=0\,,\quad\mathrm{and}\\[4mm]
				\hat{\mrg}_\ab^\mrb\bigl(\hat{b}_\gd\bigr)\dis\dfrac{\hat{b}_\ab-\hat{b}_\ab^n}{\Delta t}-\dfrac{J_\mrel}{\eta_\mrb}\,M_\ab^{(1)(\mrel)}\bigl(\hat{b}_\gd\bigr)=0\,,
			\ese
		which are solved with a local Newton-Raphson method, see \eqsref{cmpViscNRiteration}.\\		
		To solve $\mf(\mx)=\boldsymbol{0}$, a global Newton-Raphson method is employed. The required stiffness matrix is\vspace{-2mm}
			\esb{l}
				\mk^e:=\mk^e_{\tau\tau}+\mk^e_{\tau M}+\mk^e_{M\tau}+\mk^e_{MM}+\mk^e_{\tau}+\mk^e_{M}\,,
			\ese
		with the individual terms given in Eqs.~\eqref{e:cmpViscSolutionNESmaterialStiffnessMatrices}--\eqref{e:cmpViscSolutionNESgeometricStiffnessMatrices2}. The material tangents in the material stiffness matrices are computed from
			\esb{l}
				\tilde{c}^{\ab\gd} = \tilde{c}^{\ab\gd}_0+\tilde{c}^{\ab\gd}_1\,,\quad\tilde{c}=c,d,e,f\,,
			\ese
		where the index `0' refers to the elastic branch and `1' refers to the Maxwell branch. The material tangent $\tilde{c}^{\ab\gd}_1$ is derived in \appref{viscLinearizationFEM} for various material models.
    		}%
	}
\end{table}

\section{Numerical examples} \seclabel{numExVisc}
This section presents several numerical examples that illustrate viscoelastic behavior of shells. In \secref{numExVisc2Dmem}, typical viscoelastic behavior is investigated on two-dimensional square membranes. The implementation of membrane, bending, and coupled membrane and bending viscosity is then verified in Secs.~\ref{s:numExViscInflMem}--\ref{s:numExViscInflSphr}. In \secref{numExViscScordelisLos}, a viscoelastic Scordelis-Lo roof is investigated to illustrate inhomogeneous deformations. \secref{numExViscCube} shows that the presented formulation is also capable of modeling boundary viscoelasticity of 3D bodies. For all examples in this section, the surface is discretized by bi-quadratic NURBS, if not stated otherwise. Numerical integration on the bi-unit parent element is performed using Gaussian quadrature with $(p+1)\times(q+1)$ quadrature points, which represents a very conservative approach. Here, $p$ and $q$ refer to the polynomial orders of the surface discretization in the two parametric directions. To post-process the surface quantities $\hat{a}_\ab$ and $\hat{b}_\ab$ at any point on the surface, an $L^2$-projection is employed to map these values from the quadrature point level to the control point level. All quantities in this section are non-dimensionalized by the introduction of a reference length $L_0$, time $T_0$, and stiffness $\mu_0$, $K_0$, or $c_0$. For all examples, the shear viscosity model from \eqsref{cntViscSigma1in} is employed.

\subsection{2D viscoelastic membrane} \seclabel{numExVisc2Dmem}
The first examples study two-dimensional viscoelasticity of an initially square membrane to demonstrate typical viscoelastic behavior, such as stress relaxation, creep, and strain rate dependence. \figref{numExVisc2DmemSetups} shows the two setups that are used in this section to model pure shear and pure dilatation. The black crosses in the figures mark the positions where the surface quantities are evaluated for visualization. Since the deformations are homogeneous, a single finite element is used for the computation, if not stated otherwise.
\begin{figure}[!ht]
	\centering
		\subfloat[Pure shear\figlabel{numExVisc2DmemSetupPureShear}]{\includegraphics[scale=1]{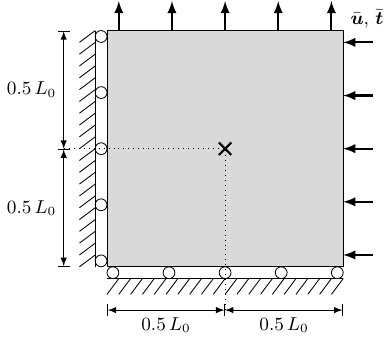}}
	\qquad
		\subfloat[Pure dilatation\figlabel{numExVisc2DmemSetupPureDilatation}]{\includegraphics[scale=1]{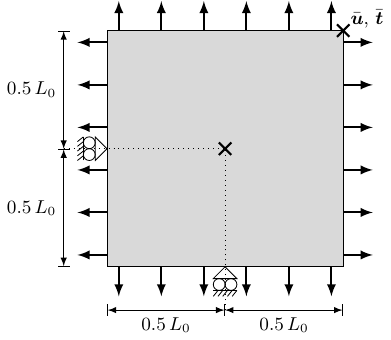}}
	\caption{2D viscoelastic membrane: Geometry, loading and boundary conditions for (a) pure shear and (b) pure dilatation. The black crosses mark the positions where surface quantities are evaluated.} \figlabel{numExVisc2DmemSetups}
\end{figure}

\subsubsection{Stress relaxation} \seclabel{numExVisc2DmemStressRelaxation}
First, the setup in \figref{numExVisc2DmemSetupPureShear} is considered and the Neo-Hookean material model is employed for both the elastic and Maxwell branch, see \eqsref{cntViscConstElasticityPsiMemNeoHookean}. The material parameters in the elastic branch are $\mu=3\,\mu_0$ and $K=0$; the ones in the Maxwell branch are $\mu_1=\mu$ and $K_1\in\{0,\mu/3\}$. The end time is $t_\mathrm{end}=5\,T_0$ and the constant time step size is chosen as $\Delta t=0.1\,T_0$.

On the top edge, the displacement profile as shown in \figref{numExVisc2DmemStressRelaxationPureShearDisplacement} is imposed, which leads to the stretch $\lambda_y=1+\bar{u}/L_0$. On the right edge, the stretch $\lambda_x=1/\lambda_y$ is imposed to ensure pure shear.
       
The resulting stress $\sigma^2_2$,\footnote{$\sigma^\alpha_\beta=\sigma^{\alpha\gamma}a_{\gamma\beta}$} which is composed of elastic and Maxwell stresses, is shown over time in \figref{numExVisc2DmemStressRelaxationPureShearStress} for three different values of the in-plane shear viscosity $\eta_\mrs$ and the purely elastic case. The stress exhibits jumps whenever there is a jump of the imposed displacement. In the viscoelastic case, the magnitude of the stress is higher because the Maxwell stress is added to the total stress, see \eqsref{cntViscTotalStress}. Over time, the Maxwell stress relaxes such that the elastic stress level is approached. In \figref{numExVisc2DmemStressRelaxationPureShearI1}, the invariants $I_1$ and $I_1^\mrel$ from Eqs.~(\ref{e:cntViscKinInvariantsOnS}.1) and \eqref{e:cntViscKinFirstInvariantElastic} are visualized over time. For $K_1=0$, the elastic invariant $I_1^\mrel$ is monotonically increasing. In contrast, if $K_1$ is not vanishing, $I_1^\mrel$ decreases to its initial value. This happens faster for lower values of $\eta_\mrs$. This behavior happens because the employed shear viscosity model from \eqsref{cntViscSigma1in} is not a pure shear model. Instead, it causes both shear and dilatation.
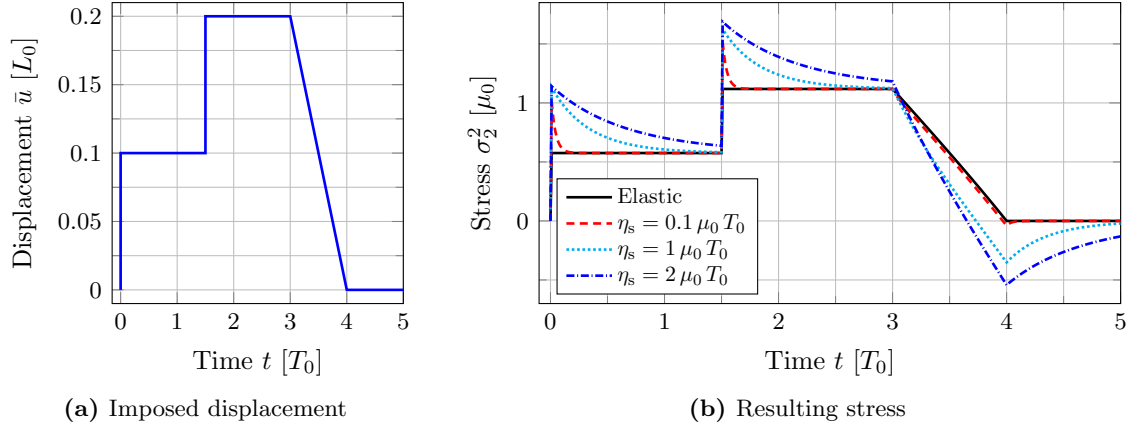
\begin{figure}[!ht]
	\setlength{\figwidth}{0.45\textwidth}
	\setlength{\figheight}{0.35\textwidth}
	\centering
		\subfloat[Imposed displacement\figlabel{numExVisc2DmemStressRelaxationPureShearDisplacement}]{
			\begin{tikzpicture}
				\def\cdot{\times}
				\begin{axis}[grid=both,xlabel={Time $t$ $[T_0]$},ylabel={Displacement $\bar{u}$ $[L_0]$},width=0.34\textwidth,height=\figheight,
				legend cell align={left},legend style={nodes={scale=0.75, transform shape}},tick label style={font=\footnotesize},
				xmin=-0.15,xmax=5,ymin=-0.01,ymax=0.21,xtick={0,1,2,3,4,5},minor ytick={0.025,0.075,0.125,0.175},
				yticklabel style={/pgf/number format/fixed,/pgf/number format/precision=5},scaled y ticks=false]
					\draw[blue,line width=1] (0,0) -- (0,0.1) -- (1.5,0.1) -- (1.5,0.2) -- (3,0.2) -- (4,0) -- (5,0);
				\end{axis}
			\end{tikzpicture}}
	\quad
		\subfloat[Resulting stress\figlabel{numExVisc2DmemStressRelaxationPureShearStress}]{
			\begin{tikzpicture}
				\def\cdot{\times}
				\begin{axis}[grid=both,xlabel={Time $t$ $[T_0]$},ylabel={Stress $\sigma^2_2$ $[\mu_0]$},width=0.58\textwidth,height=\figheight,
				legend cell align={left},legend style={nodes={scale=0.75, transform shape}},tick label style={font=\footnotesize},legend pos=south west,
				xmax=5,xmin=-0.1,ymin=-0.7,ymax=1.85,minor xtick={0.5,1.5,2.5,3.5,4.5},minor ytick={-0.5,0.5,1.5}]
					\addplot[black,solid,line width=1]table [x index = {0}, y index = {1},col sep=comma,]{fig/2Dmem/2DmemPureShearRelaxationSigma22.csv};
						\addlegendentry{Elastic};
					\addplot[red,densely dashed,line width=1]table [x index = {0}, y index = {2},col sep=comma,]{fig/2Dmem/2DmemPureShearRelaxationSigma22.csv};
						\addlegendentry{$\eta_\mrs=0.1\,\mu_0\,T_0$};
					\addplot[cyan,densely dotted,line width=1]table [x index = {0}, y index = {3},col sep=comma,]{fig/2Dmem/2DmemPureShearRelaxationSigma22.csv};
						\addlegendentry{$\eta_\mrs=1\,\mu_0\,T_0$};
					\addplot[blue,densely dash dot,line width=1]table [x index = {0}, y index = {4},col sep=comma,]{fig/2Dmem/2DmemPureShearRelaxationSigma22.csv};
						\addlegendentry{$\eta_\mrs=2\,\mu_0\,T_0$};
				\end{axis}
			\end{tikzpicture}
		}
	\caption{2D viscoelastic membrane: Stress relaxation for pure shear (according to \figref{numExVisc2DmemSetupPureShear}).} \figlabel{numExVisc2DmemStressRelaxationPureShearStressStrain}
\end{figure}           
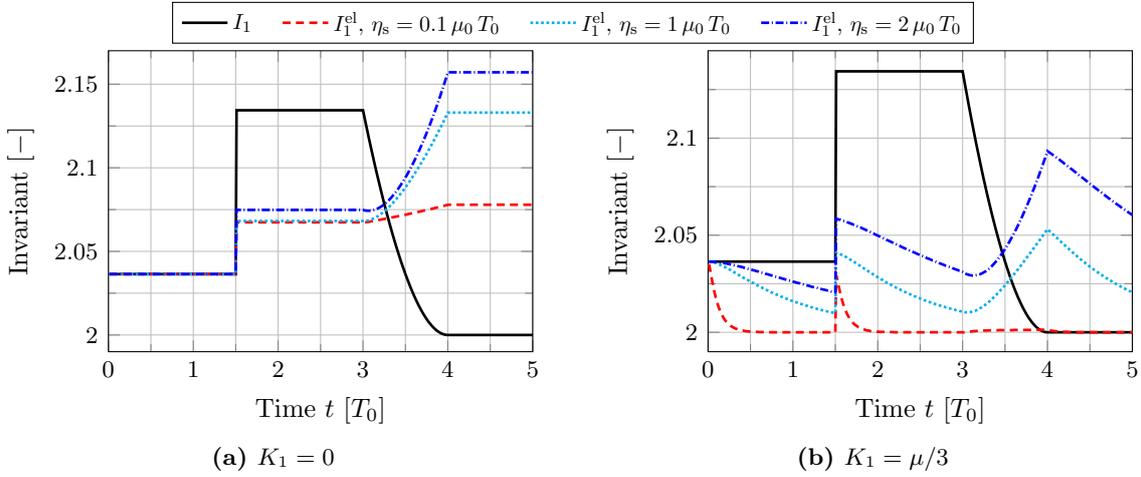
\begin{figure}[!ht]
	\setlength{\figwidth}{0.45\textwidth}
	\setlength{\figheight}{0.35\textwidth}
	\centering
	\begin{tikzpicture}
		\pgfplotsset{width=1\textwidth,height=0.2\textwidth,compat=newest,}
		\begin{axis}[hide axis,xmin=0,xmax=0.00001,ymin=0,ymax=0.00001,legend cell align={left},
					 legend columns=4,legend style={/tikz/every even column/.append style={column sep=2ex}},
					 tick label style={font=\footnotesize},legend style={nodes={scale=0.75, transform shape}},legend cell align={left}]
  			  	\addlegendimage{black,solid,line width=1,}
  			  	\addlegendentry{$I_1$};
  			  	\addlegendimage{red,densely dashed,line width=1,}
  			  	\addlegendentry{$I_1^\mrel$, $\eta_\mrs=0.1\,\mu_0\,T_0$};
  			  	\addlegendimage{cyan,densely dotted,line width=1,}
  			  	\addlegendentry{$I_1^\mrel$, $\eta_\mrs=1\,\mu_0\,T_0$};
  			  	\addlegendimage{blue,densely dash dot,line width=1,}
  			  	\addlegendentry{$I_1^\mrel$, $\eta_\mrs=2\,\mu_0\,T_0$};
  			 \end{axis}
	\end{tikzpicture}
	\\ \vspace{-3mm}
	\subfloat[$K_1=0$\figlabel{numExVisc2DmemStressRelaxationPureShearI1noLame}]{
			\begin{tikzpicture}
				\def\cdot{\times}
				\begin{axis}[grid=both,xlabel={Time $t$ $[T_0]$},ylabel={Invariant $[-]$},width=\figwidth,height=\figheight,
				legend cell align={left},legend style={nodes={scale=0.75, transform shape}},tick label style={font=\footnotesize},legend pos=outer north east ,
				xmin=0,xmax=5,ymin=1.99,ymax=2.17,minor xtick={0.5,1.5,2.5,3.5,4.5},minor ytick={2.025,2.075,2.125,2.175},
				]
					\addplot[black,solid,line width=1]table [x index = {0}, y index = {1},col sep=comma,]{fig/2Dmem/2DmemPureShearRelaxationI1noLame.csv};
					\addplot[red,densely dashed,line width=1]table [x index = {0}, y index = {2},col sep=comma,]{fig/2Dmem/2DmemPureShearRelaxationI1noLame.csv};
					\addplot[cyan,densely dotted,line width=1]table [x index = {0}, y index = {4},col sep=comma,]{fig/2Dmem/2DmemPureShearRelaxationI1noLame.csv};
					\addplot[blue,densely dash dot,line width=1]table [x index = {0}, y index = {6},col sep=comma,]{fig/2Dmem/2DmemPureShearRelaxationI1noLame.csv};
				\end{axis}
			\end{tikzpicture}
		}
		\quad		
		\subfloat[$K_1=\mu/3$\figlabel{numExVisc2DmemStressRelaxationPureShearI1withLame}]{
			\begin{tikzpicture}
				\def\cdot{\times}
				\begin{axis}[grid=both,xlabel={Time $t$ $[T_0]$},ylabel={Invariant $[-]$},width=\figwidth,height=\figheight,
				legend cell align={left},legend style={nodes={scale=0.75, transform shape}},tick label style={font=\footnotesize},legend pos=outer north east ,
				xmin=0,xmax=5,ymin=1.99,ymax=2.145,minor xtick={0.5,1.5,2.5,3.5,4.5},minor ytick={2.025,2.075,2.125},
				]
					\addplot[black,solid,line width=1]table [x index = {0}, y index = {1},col sep=comma,]{fig/2Dmem/2DmemPureShearRelaxationI1withLame.csv};
					\addplot[red,densely dashed,line width=1]table [x index = {0}, y index = {2},col sep=comma,]{fig/2Dmem/2DmemPureShearRelaxationI1withLame.csv};
					\addplot[cyan,densely dotted,line width=1]table [x index = {0}, y index = {4},col sep=comma,]{fig/2Dmem/2DmemPureShearRelaxationI1withLame.csv};
					\addplot[blue,densely dash dot,line width=1]table [x index = {0}, y index = {6},col sep=comma,]{fig/2Dmem/2DmemPureShearRelaxationI1withLame.csv};
				\end{axis}
			\end{tikzpicture}
		}
	\caption{2D viscoelastic membrane: Stress relaxation for pure shear: First invariant for the imposed displacement profile from \figref{numExVisc2DmemStressRelaxationPureShearDisplacement} and two different areal bulk moduli $K_1$.} \figlabel{numExVisc2DmemStressRelaxationPureShearI1}
\end{figure}

Next, pure dilatation is considered, see \figref{numExVisc2DmemSetupPureDilatation}, and the Koiter material model is employed, see \eqsref{cntViscConstElasticityPsiMemKoiter}. The material parameters in the elastic branch are $\mu=0$ and $\Lambda=K=3\,K_0$; the ones in the Maxwell branch are $\mu_1=0$, $\Lambda_1=K_1$, and $\eta_\mrs=0.33\,K_0\,T_0$. The time stepping is the same as in the previous example. The displacement profile from \figref{numExVisc2DmemStressRelaxationPureDilatationDisplacement} is imposed, and the surface quantities are evaluated at the center of the sheet, see \figref{numExVisc2DmemSetupPureDilatation}. The resulting stress $\sigma^2_2$ is shown in \figref{numExVisc2DmemStressRelaxationPureDilatationStress}, which exhibits similar behavior as in the previous example. For larger values of $K_1$, the total stress is larger and the relaxation time decreases. The inelastic and elastic surface stretches, $J_\mrin$ and $J_\mrel$, respectively, are visualized in \figref{numExVisc2DmemStressRelaxationPureDilatationJ}. For larger values of $K_1$, the inelastic surface stretch is also larger, whereas the elastic surface stretch decreases. Note that the total surface stretch $J=J_\mrin\,J_\mrel$ is equal to $J_\mrel$ in the purely elastic case.
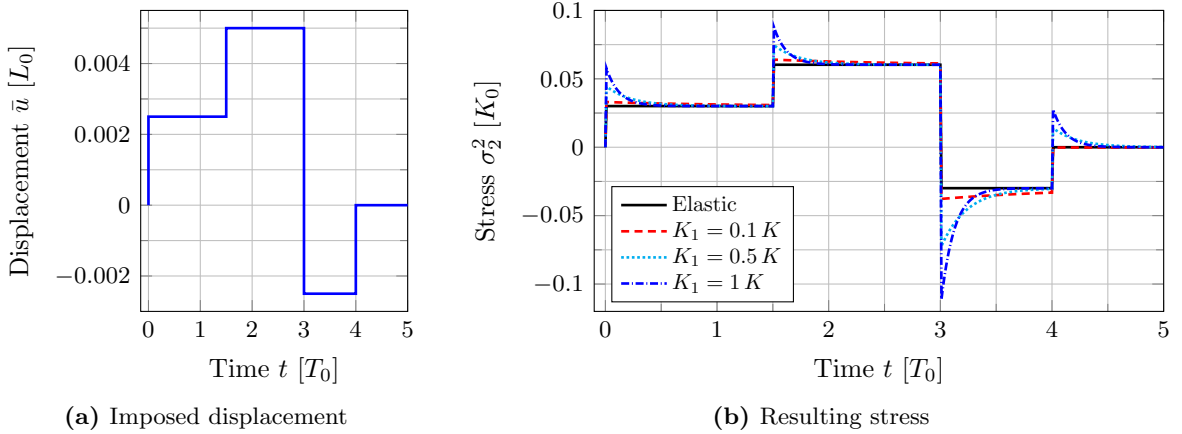
\begin{figure}[!ht]
	\setlength{\figwidth}{0.45\textwidth}
	\setlength{\figheight}{0.35\textwidth}
	\centering
		\subfloat[Imposed displacement\figlabel{numExVisc2DmemStressRelaxationPureDilatationDisplacement}]{
			\begin{tikzpicture}
				\def\cdot{\times}
				\begin{axis}[grid=both,xlabel={Time $t$ $[T_0]$},ylabel={Displacement $\bar{u}$ $[L_0]$},width=0.32\textwidth,height=\figheight,
				legend cell align={left},legend style={nodes={scale=0.75, transform shape}},tick label style={font=\footnotesize},
				xmin=-0.15,xmax=5,ymin=-0.003,ymax=0.0055,xtick={0,1,2,3,4,5},minor ytick={-0.001,0.001,0.003,0.005},
				yticklabel style={/pgf/number format/fixed,/pgf/number format/precision=5},scaled y ticks=false
				]
					\draw[blue,line width=1] (0,0) -- (0,0.0025) -- (1.5,0.0025) -- (1.5,0.005) -- (3,0.005) -- (3,-0.0025) -- (4,-0.0025) -- (4,0) -- (5,0);
				\end{axis}
			\end{tikzpicture}}
	\quad
		\subfloat[Resulting stress\figlabel{numExVisc2DmemStressRelaxationPureDilatationStress}]{
			\begin{tikzpicture}
				\def\cdot{\times}
				\begin{axis}[grid=both,xlabel={Time $t$ $[T_0]$},ylabel={Stress $\sigma^2_2$ $[K_0]$},width=0.57\textwidth,height=\figheight,
				legend cell align={left},legend style={nodes={scale=0.75, transform shape}},tick label style={font=\footnotesize},legend pos=south west,
				xmax=5,xmin=-0.1,ymin=-0.12,ymax=0.1,minor xtick={0.5,1.5,2.5,3.5,4.5},minor ytick={-0.075,-0.025,0.025,0.075},
				yticklabel style={/pgf/number format/fixed,/pgf/number format/precision=5},scaled y ticks=false]
					\addplot[black,solid,line width=1]table [x index = {0}, y index = {1},col sep=comma,]{fig/2Dmem/2DmemPureDilatationRelaxationSigma22.csv};
						\addlegendentry{Elastic};
					\addplot[red,densely dashed,line width=1]table [x index = {0}, y index = {2},col sep=comma,]{fig/2Dmem/2DmemPureDilatationRelaxationSigma22.csv};
						\addlegendentry{$K_1=0.1\,K$};
					\addplot[cyan,densely dotted,line width=1]table [x index = {0}, y index = {3},col sep=comma,]{fig/2Dmem/2DmemPureDilatationRelaxationSigma22.csv};
						\addlegendentry{$K_1=0.5\,K$};
					\addplot[blue,densely dash dot,line width=1]table [x index = {0}, y index = {4},col sep=comma,]{fig/2Dmem/2DmemPureDilatationRelaxationSigma22.csv};
						\addlegendentry{$K_1=1\,K$};
				\end{axis}
			\end{tikzpicture}
		}
	\caption{2D viscoelastic membrane: Stress relaxation for pure dilatation (according to \figref{numExVisc2DmemSetupPureDilatation}).} \figlabel{numExVisc2DmemStressRelaxationPureDilatationStressStrain}
\end{figure}  
\begin{figure}[!ht]
	\setlength{\figwidth}{0.45\textwidth}
	\setlength{\figheight}{0.35\textwidth}
	\centering
	\begin{tikzpicture}
		\pgfplotsset{width=1\textwidth,height=0.2\textwidth,compat=newest,}
		\begin{axis}[hide axis,xmin=0,xmax=0.00001,ymin=0,ymax=0.00001,legend cell align={left},
					 legend columns=4,legend style={/tikz/every even column/.append style={column sep=2ex}},
					 tick label style={font=\footnotesize},legend style={nodes={scale=0.75, transform shape}},legend cell align={left}]
  			  	\addlegendimage{black,solid,line width=1,}
  			  	\addlegendentry{Elastic};
  			  	\addlegendimage{red,densely dashed,line width=1,}
  			  	\addlegendentry{$K_1=0.1\,K$};
  			  	\addlegendimage{cyan,densely dotted,line width=1,}
  			  	\addlegendentry{$K_1=0.5\,K$};
  			  	\addlegendimage{blue,densely dash dot,line width=1,}
  			  	\addlegendentry{$K_1=1\,K$};
  			 \end{axis}
	\end{tikzpicture}
	\\ \vspace{-3mm}
	\subfloat[Inelastic surface stretch $J_\mrin$\figlabel{numExVisc2DmemStressRelaxationPureDilatationJin}]{
			\begin{tikzpicture}
				\def\cdot{\times}
				\begin{axis}[grid=both,xlabel={Time $t$ $[T_0]$},ylabel={Surface stretch $J_\mrin$ $[-]$},width=\figwidth,height=\figheight,
				legend cell align={left},legend style={nodes={scale=0.75, transform shape}},tick label style={font=\footnotesize},legend pos=outer north east ,
				xmin=0,xmax=5,ymin=0.988,ymax=1.022,minor xtick={0.5,1.5,2.5,3.5,4.5},minor ytick={0.995,1.005,1.015},
				]
					\addplot[black,solid,line width=1,domain=0:5,samples=2]{1};
					\addplot[red,densely dashed,line width=1]table [x index = {0}, y index = {2},col sep=comma,]{fig/2Dmem/2DmemPureDilatationRelaxationJ.csv};
					\addplot[cyan,densely dotted,line width=1]table [x index = {0}, y index = {3},col sep=comma,]{fig/2Dmem/2DmemPureDilatationRelaxationJ.csv};
					\addplot[blue,densely dash dot,line width=1]table [x index = {0}, y index = {4},col sep=comma,]{fig/2Dmem/2DmemPureDilatationRelaxationJ.csv};
				\end{axis}
			\end{tikzpicture}
		}
		\quad		
		\subfloat[Elastic surface stretch $J_\mrel$\figlabel{numExVisc2DmemStressRelaxationPureDilatationJel}]{
			\begin{tikzpicture}
				\def\cdot{\times}
				\begin{axis}[grid=both,xlabel={Time $t$ $[T_0]$},ylabel={Surface stretch $J_\mrel$ $[-]$},width=\figwidth,height=\figheight,
				legend cell align={left},legend style={nodes={scale=0.75, transform shape}},tick label style={font=\footnotesize},legend pos=outer north east ,
				xmin=0,xmax=5,ymin=0.97,ymax=1.023,minor xtick={0.5,1.5,2.5,3.5,4.5},minor ytick={0.97,0.99,1.01},
				]
					\addplot[black,solid,line width=1]table [x index = {0}, y index = {1},col sep=comma,]{fig/2Dmem/2DmemPureDilatationRelaxationJ.csv};
					\addplot[red,densely dashed,line width=1]table [x index = {0}, y index = {5},col sep=comma,]{fig/2Dmem/2DmemPureDilatationRelaxationJ.csv};
					\addplot[cyan,densely dotted,line width=1]table [x index = {0}, y index = {6},col sep=comma,]{fig/2Dmem/2DmemPureDilatationRelaxationJ.csv};
					\addplot[blue,densely dash dot,line width=1]table [x index = {0}, y index = {7},col sep=comma,]{fig/2Dmem/2DmemPureDilatationRelaxationJ.csv};
				\end{axis}
			\end{tikzpicture}
		}
	\caption{2D viscoelastic membrane: Stress relaxation for pure dilatation: Surface stretches for the imposed displacement profile from \figref{numExVisc2DmemStressRelaxationPureDilatationDisplacement}. Note that the total surface stretch $J=J_\mrin\,J_\mrel$ is equal to $J_\mrel$ in the purely elastic case.} \figlabel{numExVisc2DmemStressRelaxationPureDilatationJ}
\end{figure}
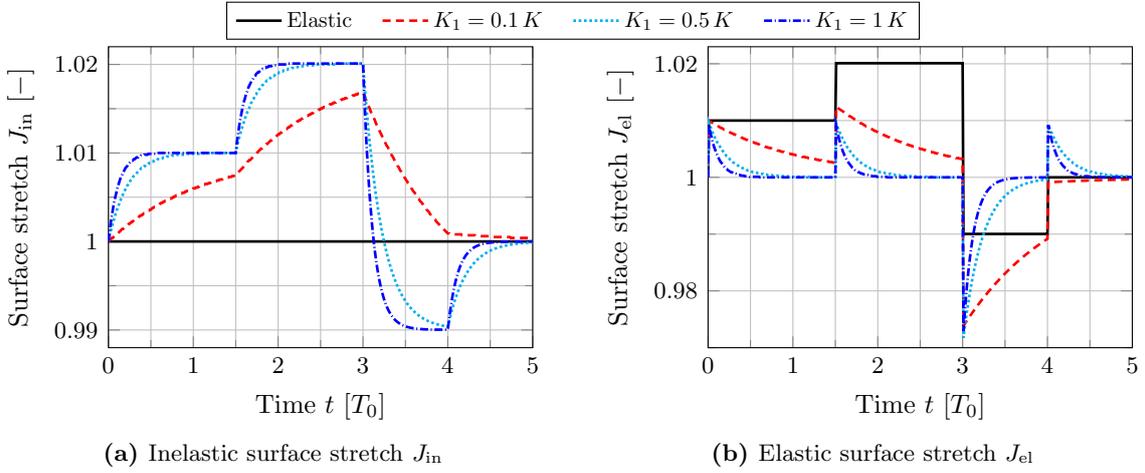

\subsubsection{Creep} \seclabel{numExVisc2DmemCreep}
Second, creep under pure dilatation (according to \figref{numExVisc2DmemSetupPureDilatation}) is considered using the Neo-Hookean material model from \eqsref{cntViscConstElasticityPsiMemNeoHookean} with material parameters $\mu=3\,\mu_0$, $K=\mu$, $\mu_1=\mu$ and $K_1=K$. The end time is $t_\mathrm{end}=5\,T_0$ and the constant time step size is chosen as $\Delta t=0.1\,T_0$. \figref{numExVisc2DmemCreepPureDilatationTraction} shows the imposed traction profile. The surface quantities are evaluated at the top right corner of the membrane, see the top right cross in \figref{numExVisc2DmemSetupPureDilatation}. \figref{numExVisc2DmemCreepPureDilatationDisplacement} shows the resulting vertical displacement for different values of the in-plane shear viscosity $\eta_\mrs$. A jump of the imposed traction leads to an instantaneous elastic response, i.e.~a jump of the displacement. Over time, the displacement magnitude increases further, which is known as creep.
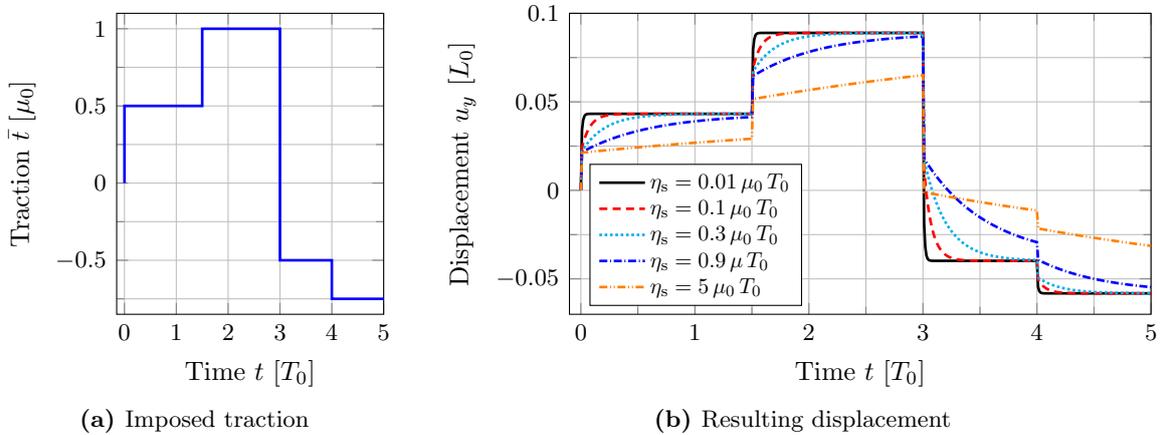
\begin{figure}[!ht]
	\setlength{\figwidth}{0.45\textwidth}
	\setlength{\figheight}{0.35\textwidth}
	\centering
		\subfloat[Imposed traction\figlabel{numExVisc2DmemCreepPureDilatationTraction}]{
			\begin{tikzpicture}
				\def\cdot{\times}
				\begin{axis}[grid=both,xlabel={Time $t$ $[T_0]$},ylabel={Traction $\bar{t}$ $[\mu_0]$},width=0.32\textwidth,height=\figheight,
				legend cell align={left},legend style={nodes={scale=0.75, transform shape}},tick label style={font=\footnotesize},
				xmin=-0.15,xmax=5,ymin=-0.85,ymax=1.1,xtick={0,1,2,3,4,5},minor ytick={-0.75,-0.25,0.25,0.75},
				yticklabel style={/pgf/number format/fixed,/pgf/number format/precision=5},scaled y ticks=false]
					\draw[blue,line width=1] (0,0) -- (0,0.5) -- (1.5,0.5) -- (1.5,1) -- (3,1) -- (3,-0.5) -- (4,-0.5) -- (4,-0.75) -- (5,-0.75);
				\end{axis}
			\end{tikzpicture}}
	\quad
		\subfloat[Resulting displacement\figlabel{numExVisc2DmemCreepPureDilatationDisplacement}]{
			\begin{tikzpicture}
				\def\cdot{\times}
				\begin{axis}[grid=both,xlabel={Time $t$ $[T_0]$},ylabel={Displacement $u_y$ $[L_0]$},width=0.58\textwidth,height=\figheight,
				legend cell align={left},legend style={nodes={scale=0.75, transform shape}},tick label style={font=\footnotesize},legend style={at={(0.4,0.5)}},
				xmin=-0.1,xmax=5,ymin=-0.07,ymax=0.1,minor xtick={0.5,1.5,2.5,3.5,4.5},minor ytick={-0.025,0.025,0.075},
				yticklabel style={/pgf/number format/fixed,/pgf/number format/precision=5},scaled y ticks=false]
					\addplot[black,solid,line width=1]table [x index = {0}, y index = {1},col sep=comma,]{fig/2Dmem/2DmemPureDilatationCreepU.csv};
						\addlegendentry{$\eta_\mrs=0.01\,\mu_0\,T_0$};
					\addplot[red,densely dashed,line width=1]table [x index = {0}, y index = {2},col sep=comma,]{fig/2Dmem/2DmemPureDilatationCreepU.csv};
						\addlegendentry{$\eta_\mrs=0.1\,\mu_0\,T_0$};
					\addplot[cyan,densely dotted,line width=1]table [x index = {0}, y index = {3},col sep=comma,]{fig/2Dmem/2DmemPureDilatationCreepU.csv};
						\addlegendentry{$\eta_\mrs=0.3\,\mu_0\,T_0$};
					\addplot[blue,densely dash dot,line width=1]table [x index = {0}, y index = {4},col sep=comma,]{fig/2Dmem/2DmemPureDilatationCreepU.csv};
						\addlegendentry{$\eta_\mrs=0.9\,\mu\,T_0$};
					\addplot[orange,densely dash dot dot,line width=1]table [x index = {0}, y index = {5},col sep=comma,]{fig/2Dmem/2DmemPureDilatationCreepU.csv};
						\addlegendentry{$\eta_\mrs=5\,\mu_0\,T_0$};
				\end{axis}
			\end{tikzpicture}
		}
	\caption{2D viscoelastic membrane: Creep for pure dilatation (according to \figref{numExVisc2DmemSetupPureDilatation}).}\figlabel{numExVisc2DmemCreepPureDilatationStressStrain}
\end{figure}   

\subsubsection{Strain rate dependence} \seclabel{numExVisc2DmemStrainRateDependence}
Third, the setup in \figref{numExVisc2DmemSetupPureShear} is used again to show the influences of different strain rates and cyclic loading. The Neo-Hookean material model from \eqsref{cntViscConstElasticityPsiMemNeoHookean} is used with $K_1=\mu_1=\mu=1\,\mu_0$, $K=0$, and the in-plane shear viscosity is set to $\eta_\mrs=0.25\,\mu_0\,T_0$. The imposed displacement over time is given by the function
\eqb{l}
	\bar{u}_y(t)=0.25\,L_0\,\sin(\omega\,t)\,, \eqlabel{numExVisc2DmemStrainRateDependenceImposedDiplacement}
\eqe
with excitation frequency $\omega$. At first, the time span $t\in[0,\pi/\omega]$ is considered such that only one loading-unloading cycle is computed. The time step size is constant and $500$ time steps are used. \figref{numExVisc2DmemStrainRateDependence} shows the total, elastic, and Maxwell stress over the displacement. In the elastic case, the stress-displacement curve is identical for loading and unloading, see \figref{numExVisc2DmemStrainRateDependenceElastic}. But in the viscoelastic case, these curves are not coinciding. Instead, the unloading occurs at lower stress than the loading, which can be seen for both the total stress in \figref{numExVisc2DmemStrainRateDependenceTotal} and the Maxwell stress in \figref{numExVisc2DmemStrainRateDependenceViscous}. This indicates that energy is dissipated during the loading-unloading cycle, which is further studied in the next section. \figref{numExVisc2DmemStrainRateDependence} shows that with increasing excitation frequency, and thus with increasing strain rate, the Maxwell stress is larger. Further, for very low and very high strain rates, the loading and unloading curves come closer to each other such that less energy is dissipated.
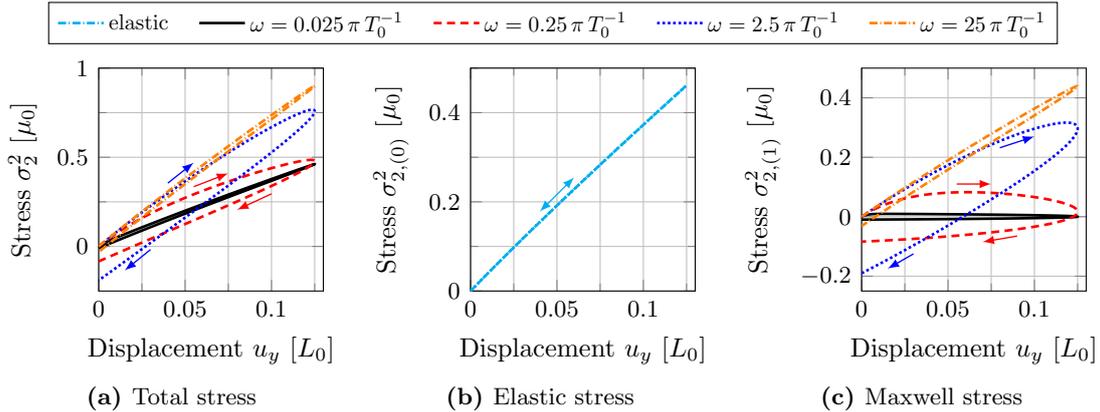
\begin{figure}[!ht]
	\setlength{\figwidth}{0.285\textwidth}
	\centering
	\begin{tikzpicture}
		\pgfplotsset{width=1\textwidth,height=0.2\textwidth,compat=newest,}
		\begin{axis}[hide axis,xmin=0,xmax=0.00001,ymin=0,ymax=0.00001,legend cell align={left},
					 legend columns=5,legend style={/tikz/every even column/.append style={column sep=2ex}},
					 tick label style={font=\footnotesize},legend style={nodes={scale=0.75, transform shape}},legend cell align={left}]
  			  	\addlegendimage{cyan,densely dash dot,line width=1,}
  			  	\addlegendentry{elastic};
  			  	\addlegendimage{black,solid,line width=1,}
  			  	\addlegendentry{$\omega=0.025\,\pi\,T_0^{-1}$};
  			  	\addlegendimage{red,densely dashed,line width=1,}
  			  	\addlegendentry{$\omega=0.25\,\pi\,T_0^{-1}$};
  			  	\addlegendimage{blue,densely dotted,line width=1,}
  			  	\addlegendentry{$\omega=2.5\,\pi\,T_0^{-1}$};
  			  	\addlegendimage{orange,densely dash dot,line width=1,}
  			  	\addlegendentry{$\omega=25\,\pi\,T_0^{-1}$};
  			 \end{axis}
	\end{tikzpicture}
	\\ \vspace{-3mm}
	\subfloat[Total stress\figlabel{numExVisc2DmemStrainRateDependenceTotal}]{
		\begin{tikzpicture}
			\begin{axis}[grid=both,xlabel={Displacement $u_y$ $[L_0]$},ylabel={Stress $\sigma^2_2$ $[\mu_0]$},width=\figwidth,height=\figwidth,
			xmin=0,xmax=0.13,
			ymin=-0.25,ymax=1,
			minor xtick={0.025,0.075,0.125},
			minor ytick={0.25,0.75},
			tick label style={font=\footnotesize},legend style={nodes={scale=0.75, transform shape}},legend cell align={left},
			xticklabel style={/pgf/number format/fixed,/pgf/number format/precision=5},scaled x ticks=false,
			]
				\addplot[black,solid,line width=1]table [x index = {0}, y index = {3},col sep=comma]{fig/2Dmem/strainRateDependence.csv};
				\addplot[red,densely dashed,line width=1]table [x index = {4}, y index = {7},col sep=comma]{fig/2Dmem/strainRateDependence.csv};
				\addplot[blue,densely dotted,,line width=1]table [x index = {8}, y index = {11},col sep=comma]{fig/2Dmem/strainRateDependence.csv};
				\addplot[orange,densely dash dot,line width=1]table [x index = {12}, y index = {15},col sep=comma]{fig/2Dmem/strainRateDependence.csv};
				\draw[->,>=latex,line width=.5,red] (0.055,0.335) -- (0.075,0.41);
				\draw[->,>=latex,line width=.5,red] (0.1,0.295) -- (0.08,0.205);
				\draw[->,>=latex,line width=.5,blue] (0.04,0.355) -- (0.055,0.47);
				\draw[->,>=latex,line width=.5,blue] (0.03,-0.0175) -- (0.015,-0.135);
			\end{axis}
		\end{tikzpicture}
	}
	\subfloat[Elastic stress\figlabel{numExVisc2DmemStrainRateDependenceElastic}]{
		\begin{tikzpicture}
			\begin{axis}[grid=both,xlabel={Displacement $u_y$ $[L_0]$},ylabel={Stress $\sigma^2_{2,(0)}$ $[\mu_0]$},width=\figwidth,height=\figwidth,
			xmin=0,xmax=0.13,
			ymin=0,ymax=0.5,
			minor xtick={0.025,0.075,0.125},
			minor ytick={0.1,0.3,0.5},
			tick label style={font=\footnotesize},legend style={nodes={scale=0.75, transform shape}},legend cell align={left},
			xticklabel style={/pgf/number format/fixed,/pgf/number format/precision=5},scaled x ticks=false,
			]
				\addplot[cyan,densely dash dot dot,line width=1]table [x index = {16}, y index = {17},col sep=comma]{fig/2Dmem/strainRateDependence.csv};
				\draw[<->,>=latex,line width=.5,cyan] (0.04,0.178) -- (0.06,0.255);
			\end{axis}
		\end{tikzpicture}
	}
	\subfloat[Maxwell stress\figlabel{numExVisc2DmemStrainRateDependenceViscous}]{
		\begin{tikzpicture}
			\begin{axis}[grid=both,xlabel={Displacement $u_y$ $[L_0]$},ylabel={Stress $\sigma^2_{2,(1)}$ $[\mu_0]$},width=\figwidth,height=\figwidth,
			xmin=0,xmax=0.13,
			ymin=-0.25,ymax=0.5,
			minor xtick={0.025,0.075,0.125},
			tick label style={font=\footnotesize},
			legend style={nodes={scale=0.75, transform shape}},legend cell align={left},			
			xticklabel style={/pgf/number format/fixed,/pgf/number format/precision=5},scaled x ticks=false,
			]
				\addplot[black,solid,line width=1]table [x index = {0}, y index = {2},col sep=comma]{fig/2Dmem/strainRateDependence.csv};
				\addplot[red,densely dashed,line width=1]table [x index = {4}, y index = {6},col sep=comma]{fig/2Dmem/strainRateDependence.csv};
				\addplot[blue,densely dotted,line width=1]table [x index = {8}, y index = {10},col sep=comma]{fig/2Dmem/strainRateDependence.csv};
				\addplot[orange,densely dash dot,line width=1]table [x index = {12}, y index = {14},col sep=comma]{fig/2Dmem/strainRateDependence.csv};
				\draw[->,>=latex,line width=.5,red] (0.055,0.11) -- (0.075,0.11);
				\draw[->,>=latex,line width=.5,red] (0.09,-0.065) -- (0.07,-0.085);
				\draw[->,>=latex,line width=.5,blue] (0.08,0.232) -- (0.1,0.27);
				\draw[->,>=latex,line width=.5,blue] (0.03,-0.125) -- (0.015,-0.175);
			\end{axis}
		\end{tikzpicture}
	}
	\caption{2D viscoelastic membrane: Strain rate dependence for one loading-unloading cycle. The Maxwell branch exhibits hysteresis -- the loading and unloading curves do not coincide.} \figlabel{numExVisc2DmemStrainRateDependence}
\end{figure}

\subsubsection{Cyclic loading} \seclabel{numExVisc2DmemCyclicLoading}
Fourth, the same setup as in \secref{numExVisc2DmemStrainRateDependence} with the displacement profile from \eqsref{numExVisc2DmemStrainRateDependenceImposedDiplacement} is considered, but the time span is now extended to $t\in[0,20\,\pi/\omega]$, such that ten loading-unloading cycles occur. For this, $10,000$ time steps are used. The resulting stress-displacement curves are shown in \figref{numExVisc2DmemCyclicLoading} for three different excitation frequencies $\omega$ and $\eta_\mrs=1\,\mu_0\,T_0$. Note that the first loading-unloading cycle is slightly offset from the following cycles as the first one starts at zero stress, whereas the next ones start at non-vanishing stress. The areas within the hystereses are an indicator for the dissipated energy. This energy decreases for very low or very high excitation frequencies. Note that the shown results exhibit nonlinear and non-symmetric behavior in compression and tension. \figref{numExVisc2DmemCyclicLoadingDissipation} shows the dissipated energy, see \eqsref{cntViscDissipatedEnergy}, over the excitation frequencies for three different values of $\eta_\mrs$. For each setup, there exists an excitation frequency where the dissipation is maximum.
\begin{figure}[!ht]
	\setlength{\figwidth}{0.3\textwidth}
	\setlength{\figheight}{0.31\textwidth}
	\centering
	\subfloat{
		\begin{tikzpicture}
			\begin{axis}[grid=both,xlabel={Displacement $u_y$ $[L_0]$},ylabel={Stress $\sigma^2_2$ $[\mu_0]$},width=\figwidth,height=\figheight,
			tick label style={font=\footnotesize},legend style={nodes={scale=0.75, transform shape},at={(0,1.15)},anchor=west},legend cell align={left},
			minor xtick={-0.05,0.05},ymin=-0.65,ymax=0.5,ytick={-0.5,-0.25,0,0.25,0.5},minor ytick={-0.375,-0.125,0.125,0.375},
			]
				\addplot[black,solid,line width=1]table [x index = {0}, y index = {1},col sep=comma]{fig/2Dmem/hystereses.csv};
					\addlegendentry{$\omega=0.025\,\pi\,T_0^{-1}$};
			\end{axis}
		\end{tikzpicture}
		}
	\subfloat{
		\begin{tikzpicture}
			\begin{axis}[grid=both,xlabel={Displacement $u_y$ $[L_0]$},ylabel={Stress $\sigma^2_2$ $[\mu_0]$},width=\figwidth,height=\figheight,
			tick label style={font=\footnotesize},legend style={nodes={scale=0.75, transform shape},at={(0,1.15)},anchor=west},legend cell align={left},
			minor xtick={-0.05,0.05},ymin=-0.85,ymax=0.85,minor ytick={-0.75,-0.25,0.25,0.75},
			]
				\addplot[red,densely dashed,line width=1]table [x index = {2}, y index = {3},col sep=comma]{fig/2Dmem/hystereses.csv};
					\addlegendentry{$\omega=0.25\,\pi\,T_0^{-1}$};
			\end{axis}
		\end{tikzpicture}
		}
	\subfloat{
		\begin{tikzpicture}
			\begin{axis}[grid=both,xlabel={Displacement $u_y$ $[L_0]$},ylabel={Stress $\sigma^2_2$ $[\mu_0]$},width=\figwidth,height=\figheight,
			tick label style={font=\footnotesize},legend style={nodes={scale=0.75, transform shape},at={(0,1.15)},anchor=west},legend cell align={left},
			minor xtick={-0.05,0.05},ymin=-1.1,ymax=1.1,ytick={-1,-0.5,0,0.5,1},minor ytick={-0.75,-0.25,0.25,0.75,1.25},
			]
				\addplot[blue,densely dotted,line width=1]table [x index = {4}, y index = {5},col sep=comma]{fig/2Dmem/hystereses.csv};
					\addlegendentry{$\omega=1\,\pi\,T_0^{-1}$};
			\end{axis}
		\end{tikzpicture}
		}
	\caption{2D viscoelastic membrane: Hystereses for cyclic loading and different excitation frequencies.} \figlabel{numExVisc2DmemCyclicLoading}
\end{figure}
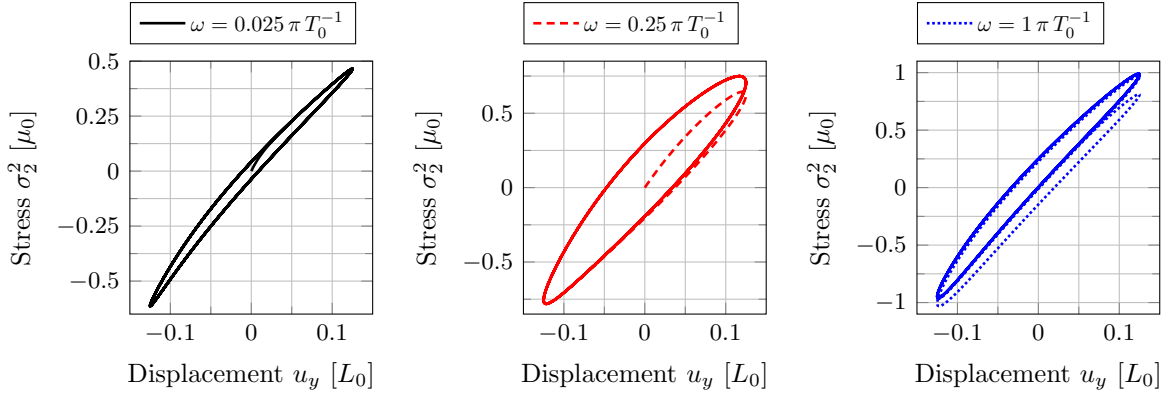

\begin{figure}[!ht]
	\centering
		\begin{tikzpicture}
			\begin{axis}[xmode=log,ymode=log,grid=both,xlabel={Excitation frequency $\omega$ $[\pi\,T_0^{-1}]$},ylabel={Dissipation $\sD$ $[\mu_0\,L_0^{2}]$},width=0.81\textwidth,height=0.35\textwidth,
			xmin=1e-4,xmax=1e3,
			tick label style={font=\footnotesize},legend pos = north west,legend style={nodes={scale=0.75, transform shape}},legend cell align={left},legend style={at={(0.2,0.325)}}
			]
				\addplot[black,solid,mark=o,mark size=1.25,line width=1,mark options={solid}]table [x index = {0}, y index = {1},col sep=comma]{fig/2Dmem/dissipation_etaS1em1.csv};
					\addlegendentry{$\eta_\mrs=0.1\,\mu_0\,T_0$};
				\addplot[red,densely dashed,mark=triangle*,mark size=1.5,line width=1,mark options={solid}]table [x index = {0}, y index = {1},col sep=comma]{fig/2Dmem/dissipation_etaS1.csv};
					\addlegendentry{$\eta_\mrs=1\,\mu_0\,T_0$};
				\addplot[blue,densely dotted,mark=square*,mark size=1.25,line width=1,mark options={solid}]table [x index = {0}, y index = {1},col sep=comma]{fig/2Dmem/dissipation_etaS10.csv};
					\addlegendentry{$\eta_\mrs=10\,\mu_0\,T_0$};
			\end{axis}
		\end{tikzpicture}
	\caption{2D viscoelastic membrane: Dissipation $\sD$, see \eqsref{cntViscDissipatedEnergy}, over the excitation frequency $\omega$ for cyclic loading.} \figlabel{numExVisc2DmemCyclicLoadingDissipation}
\end{figure}
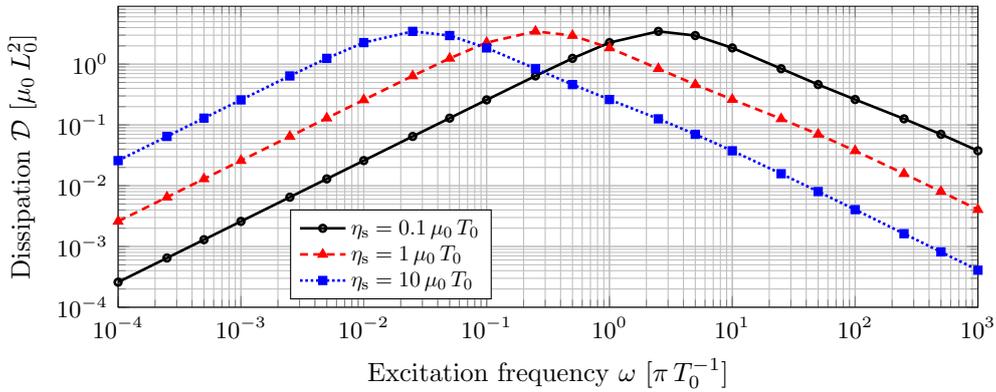

\subsection{Inflated membrane balloon} \seclabel{numExViscInflMem}
 This section deals with the inflation of a viscoelastic spherical rubber balloon, similar to the elastic counterpart considered in \cite{sauer2014b}. This example is used to verify the formulation and implementation for membrane viscosity as an analytical solution for this problem can be derived. The finite element model is shown in \figref{numExViscInflMemSetupGeometry}. Only a quarter of the sphere is used for the simulation and appropriate boundary conditions are provided to prevent rigid body motion and to maintain the symmetry of the inflating balloon across the gray marked planes. The bold black lines mark the patch interfaces between the four patches, of which the quarter mesh is composed. The finite element mesh contains $6m^2$ elements, where $2m$ $(m=1,2,\tightdots)$ denotes the number of elements along the equator of the quarter sphere.
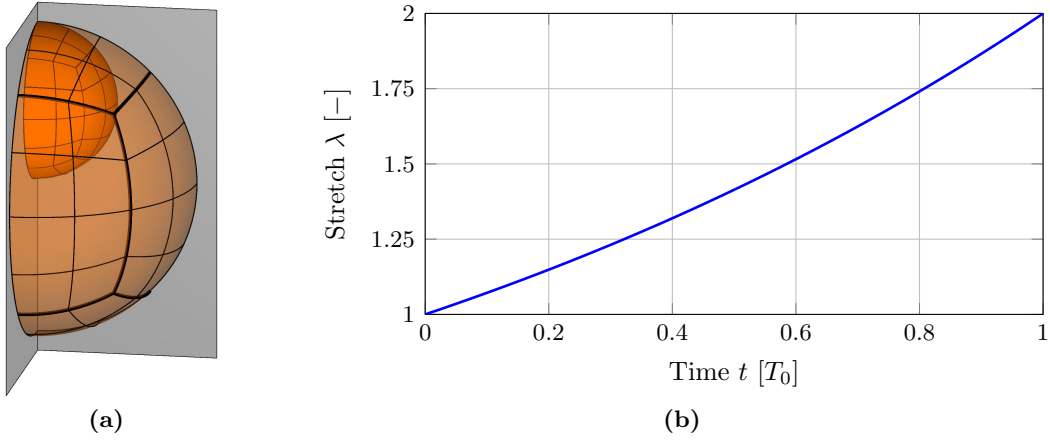
\begin{figure}[!ht]
	\centering
		\subfloat[\figlabel{numExViscInflMemSetupGeometry}]{\includegraphics[width=0.175\textwidth]{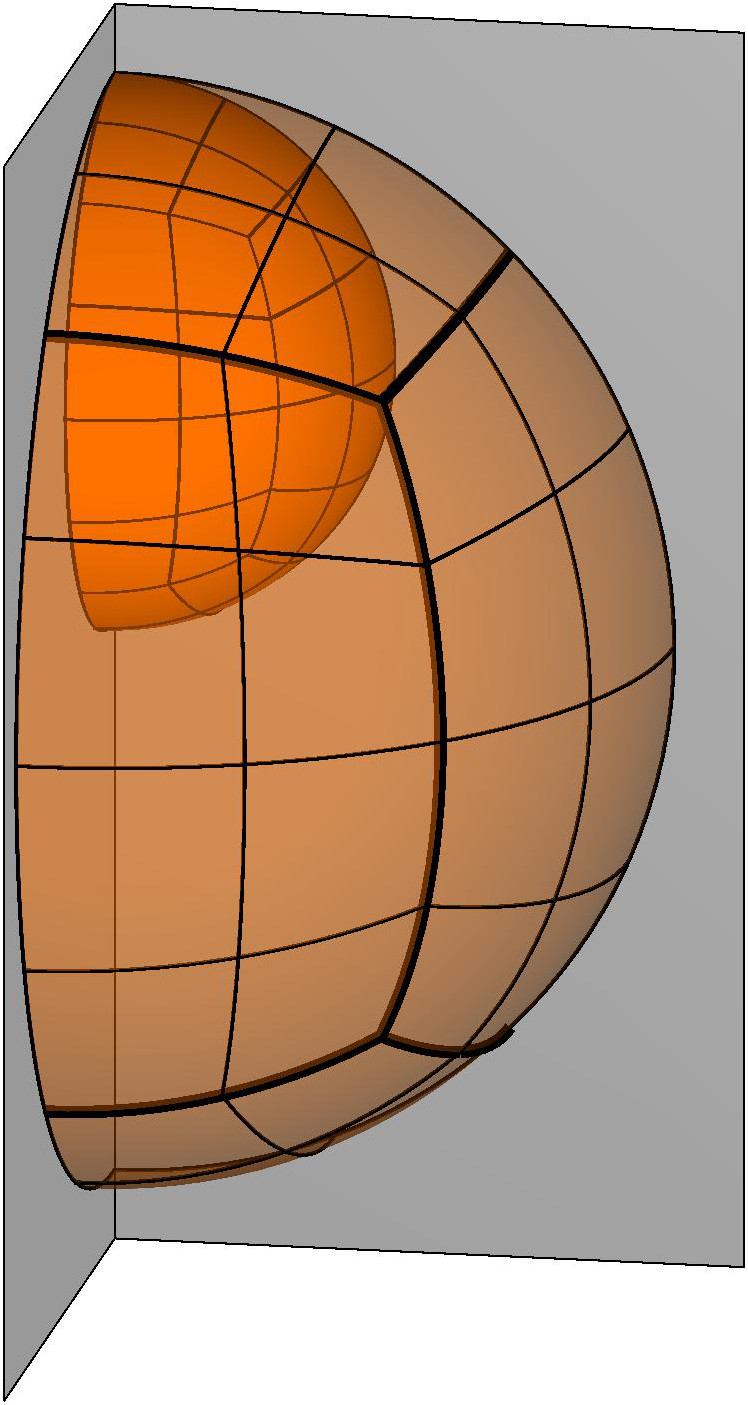}}
	\qquad\quad
		\subfloat[\figlabel{numExViscInflMemSetupLoading}]{
			\begin{tikzpicture}
				\begin{axis}[xlabel={Time $t$ $[T_0]$},ylabel={Stretch $\lambda$ $[-]$},width=0.61\textwidth,height=0.35\textwidth,xmin=0,xmax=1,,grid=both,legend cell align={left},legend pos=north west,ytick={1,1.25,1.5,1.75,2,2.25},ymin=1,ymax=2,legend cell align={left},legend style={nodes={scale=0.75, transform shape}},tick label style={font=\footnotesize}]	
					\addplot[blue,line width=1,domain=0:1,samples=501]{exp(x*ln(2)/1)};
				\end{axis}
			\end{tikzpicture}	
		}
	\caption{Inflated membrane balloon: (a) Initial and deformed configuration (with the bold black lines marking patch interfaces) and (b) imposed stretch over time for $t_\mathrm{end}=1\,T_0$ and $\lambda_\mathrm{end}=2$, see \eqsref{numExViscInflMemStretchFunction}.} \figlabel{numExViscInflMemSetup}
\end{figure}
The initial and current radii are denoted $R$ and $r$, respectively. Likewise, the initial thickness is denoted $T$, while the current thickness is denoted $\tilde{t}$. As shown subsequently, the pressure initially increases but later decreases. Thus, the finite element computation is performed by imposing the enclosed volume $V$ instead of the pressure $p$, see \cite{sauer2014b} for more details. The prescribed stretch $\lambda$ as a function of time $t$ is chosen as
\eqb{l}
	\lambda(t) = \exp\biggl(\dfrac{t}{\tau_\lambda}\biggr)\,, \eqlabel{numExViscInflMemStretchFunction}
\eqe
with the characteristic time
\eqb{l}
	\tau_\lambda:=\dfrac{t_\mathrm{end}}{\ln(\lambda_\mathrm{end})}\,, \eqlabel{numExViscInflMemLoadingRate}
\eqe
where $t_\mathrm{end}$ denotes the end time and $\lambda_\mathrm{end}$ denotes the final stretch, see \figref{numExViscInflMemSetupLoading}. The choice for $\lambda(t)$ in \eqsref{numExViscInflMemStretchFunction} allows for an analytical solution. The resulting volume is then given by
\eqb{l}
	V(t)= \exp\biggl(\dfrac{3\,t}{\tau_\lambda}\biggr)\,V_0\,, \eqlabel{numExViscBalloonVolumeFunction}
\eqe
where the relation $V=\lambda^3\,V_0$ with initial volume $V_0$ has been used. For this example, $\lambda_\mathrm{end}$ is set to $2$, such that the final volume is $V(t_\mathrm{end})=8\,V_0$. The elastic behavior of the rubber membrane is described by the incompressible Neo-Hookean material model from \eqsref{cntViscConstElasticitySigmaABincompressibleNeoHookean}, and the elastic energy in the Maxwell branch is given by \eqsref{cntViscConstElasticityPsiMemClassicalNeoHookean} with $\Lambda=0$.

In analogy to \eqsref{cntViscTotalStress}, the total pressure $p$ can be decomposed into the contributions
\eqb{l}
	p(t) = p_\mrel(t) + p_\mrvisc(t)\,, \eqlabel{numExViscInflMemPressureTotal}
\eqe
where $p_\mrel(t)$ is the pressure function coming from the elastic branch, and $p_\mrvisc(t)$ is the one from the Maxwell branch. These two contributions $p_\bullet(t)$ are derived in detail in \appref{viscAnalyticalSolutionsInflMem}, and they are given by
\eqb{l}
	 p_\mrel(t)=\dfrac{2\,\mu}{R}\,\biggl(\dfrac{1}{\lambda}-\dfrac{1}{\lambda^7}\biggr)=\dfrac{2\,\mu}{R}\,\biggl(\Bigl(\dfrac{V_0}{V}\Bigr)^{\frac{1}{3}}-\Bigl(\dfrac{V_0}{V}\Bigr)^{\frac{7}{3}}\biggr)\,. \eqlabel{numExViscInflMemPressureElastic}
\eqe
and
\eqb{l}
	p_\mrvisc(t)=\dfrac{2\,\mu_1}{R}\,\biggl(\dfrac{1}{\lambda}-\dfrac{1}{\lambda^3\,\hat{a}_\mathrm{ev}}\biggr)\,, \eqlabel{numExViscInflMemPressureViscous}
\eqe
and
\eqb{l}
	\hat{a}_\mathrm{ev}(t):=\dfrac{\mu_1\,\tau_\lambda\,\exp\bigl(-2\,t/\tau_\lambda\bigr)-2\,\eta_\mrs\,\exp\bigl(-\mu_1\,t/\eta_\mrs\bigr)}{\mu_1\,\tau_\lambda-2\,\eta_\mrs}\,. \eqlabel{numExViscInflMemHataEv}
\eqe

First, the convergence of the model with respect to mesh and time step size refinement is investigated. For this, the pressure error
\eqb{l}
	\epsilon_p:=\dfrac{\big|p_\mathrm{num}(t_\mathrm{end})-p_\mathrm{ana}(t_\mathrm{end})\big|}{p_\mathrm{ana}(t_\mathrm{end})}\,, \eqlabel{numExViscInflMemPressureError}
\eqe
is defined. The parameters used for the convergence study are $\mu_1=\mu$, $\eta_\mrs=0.1\,\mu\,T_0$, and $t_\mathrm{end}=1\,T_0$. The time step size $\Delta t$ is chosen to be constant in $[0,t_\mathrm{end}]$. \figref{numExViscInflMemConvergence} clearly shows that the dominating error stems from the time step size, and not from the finite element mesh. Beyond $m\geq2$, an increase of the number of elements for fixed $\Delta t$ does not lead to a further decrease of the error. As the deformation in this example is homogeneous, even a small number of elements is sufficient to accurately capture it. In contrast, the error decreases linearly with a decrease of the time step size $\Delta t$. This convergence rate is also the expected rate for the employed implicit Euler scheme. As the mesh $m=2$ is sufficient to accurately capture the deformation, this mesh is used for all subsequent balloon examples.
\begin{figure}[!ht]
	\setlength{\figwidth}{0.45\textwidth}
	\setlength{\figheight}{0.35\textwidth}
	\centering
	\subfloat[Mesh refinement\figlabel{numExViscInflMemConvergenceMesh}]{
		\begin{tikzpicture}
			\def\cdot{\times}
			\begin{axis}[xmode=log,ymode=log,grid=both,xlabel={Number of elements $n_\mathrm{el}$ $[-]$},ylabel={$\epsilon_p$ $[-]$},width=\figwidth,height=\figheight,
			log basis x = 10,
			xmin=1,xmax=1e5,ymin=3e-7,ymax=0.25e-1,
			tick label style={font=\footnotesize},
			legend style={at={(0,1.2)},anchor=west},legend style={nodes={scale=0.75, transform shape}},legend cell align={left},legend columns=2,
			]
				\addplot[black,line width=1,mark=triangle*,mark size=1.5,mark options={solid}]table [x index = {0}, y index = {1},col sep=comma]{fig/inflMem/inflMemConvergenceMesh.csv};
					\addlegendentry{$\Delta t=10^{-1}\,T_0$};
				\addplot[blue,densely dashed,line width=1,mark=*,mark size=1.5,mark options={solid}]table [x index = {0}, y index = {2},col sep=comma]{fig/inflMem/inflMemConvergenceMesh.csv};
					\addlegendentry{$\Delta t=10^{-2}\,T_0$};
				\addplot[red,densely dotted,line width=1,mark=square*,mark size=1.5,mark options={solid}]table [x index = {0}, y index = {3},col sep=comma]{fig/inflMem/inflMemConvergenceMesh.csv};
					\addlegendentry{$\Delta t=10^{-3}\,T_0$};
				\addplot[cyan,densely dash dot,line width=1,mark=diamond*,mark size=1.5,mark options={solid},restrict x to domain=0:6114]table [x index = {0}, y index = {4},col sep=comma]{fig/inflMem/inflMemConvergenceMesh.csv};
					\addlegendentry{$\Delta t=10^{-4}\,T_0$};
				\addplot[orange,densely dash dot dot,line width=1,mark=pentagon*,mark size=1.5,mark options={solid},restrict x to domain=0:384]table [x index = {0}, y index = {5},col sep=comma]{fig/inflMem/inflMemConvergenceMesh.csv};
					\addlegendentry{$\Delta t=10^{-5}\,T_0$};

			\end{axis}
		\end{tikzpicture}
	}
	\quad
		\subfloat[Time step size refinement\figlabel{numExViscInflMemConvergenceTimeStepSize}]{
		\begin{tikzpicture}
			\def\cdot{\times}
			\begin{axis}[xmode=log,ymode=log,grid=both,xlabel={Time step size $\Delta t$ $[T_0]$},ylabel={$\epsilon_p$ $[-]$},width=\figwidth,height=\figheight,
			tick label style={font=\footnotesize},
			ymin=3e-7,ymax=0.25e-1,
			legend style={at={(0,1.145)},anchor=west},legend style={nodes={scale=0.75, transform shape}},legend cell align={left},legend columns=2,
			]
				\addplot[black,densely dash dot,line width=1,mark=triangle*,mark size=1.5,mark options={solid}]table [x index = {0}, y index = {1},col sep=comma]{fig/inflMem/inflMemConvergenceTimeStepSize.csv};
					\addlegendentry{$m=1$};
				\addplot[blue,solid,line width=1,mark=*,mark size=1.5,mark options={solid}]table [x index = {0}, y index = {2},col sep=comma]{fig/inflMem/inflMemConvergenceTimeStepSize.csv};
					\addlegendentry{$m=2$};
				\addplot[red,densely dashed,line width=1,mark=square*,mark size=1.5,mark options={solid}]table [x index = {0}, y index = {3},col sep=comma]{fig/inflMem/inflMemConvergenceTimeStepSize.csv};
					\addlegendentry{$m=4$};
				\addplot[cyan,densely dotted,line width=1,mark=diamond*,mark size=1.5,mark options={solid}]table [x index = {0}, y index = {4},col sep=comma]{fig/inflMem/inflMemConvergenceTimeStepSize.csv};
					\addlegendentry{$m=8$};
			\addplot[black,solid,line width=1,mark=none,domain=5e-5:3e-4,samples=2]{0.040107095706104*x^1.000269987194659};
			\draw[black,solid,line width=1] (5e-5,2e-6) -- (3e-4,2e-6); \node[anchor=north] at (1.5e-4,2e-6) {\scriptsize $\boldsymbol{1}$};
			\draw[black,solid,line width=1] (3e-4,2e-6) -- ( 3e-4,1.200580642967428e-05); \node[anchor=west] at (3e-4,5e-6) {\scriptsize $\boldsymbol{1}$};
			\end{axis}
		\end{tikzpicture}
	}
	\caption{Inflated membrane balloon: Convergence of the pressure error $\epsilon_p$, see \eqsref{numExViscInflMemPressureError}, over mesh and time step size refinement.} \figlabel{numExViscInflMemConvergence}
\end{figure}
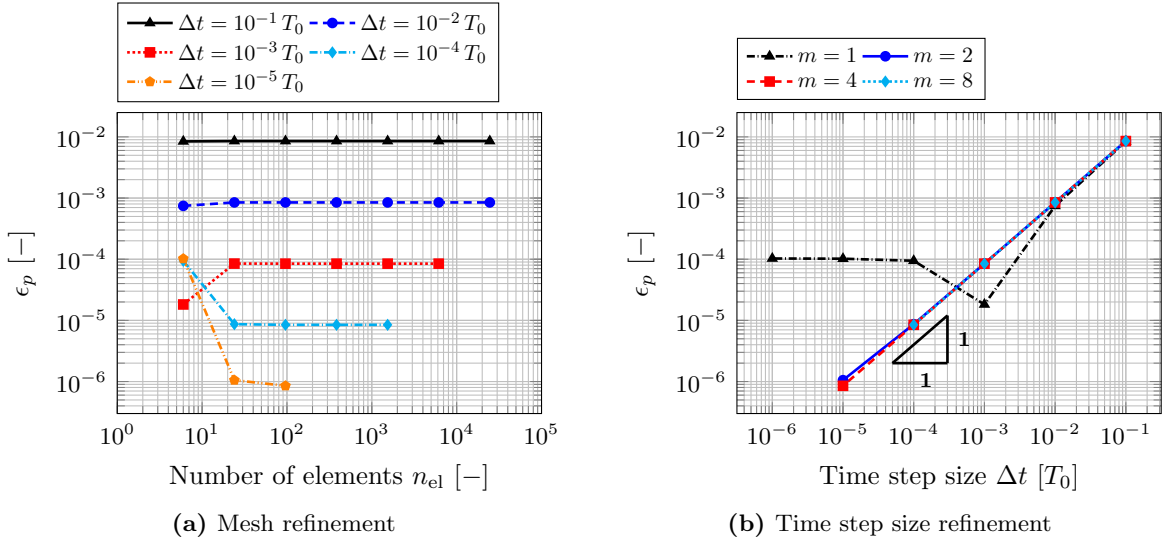

Second, the $p(t)$ relation is shown over a logarithmic time axis in \figref{numExViscInflMemStudyEtaS} for different values of the in-plane shear viscosity $\eta_\mrs$ and fixed $\mu_1=\mu$, and in \figref{numExViscInflMemStudyChi1} for different values of the stiffness ratio $\rchi_1:=\mu_1/\mu$ and fixed $\eta_\mrs=0.5\,\mu\,T_0$. The end time is fixed to $t_\mathrm{end}=1\,T_0$ and $1,000$ time steps are used. The strong nonlinear behavior is captured accurately by the numerical results as \figref{numExViscInflMemStudyEtaSChi1} shows. For increasing values of $\eta_\mrs$ or $\rchi_1$, the magnitude of the pressure increases. The location of the maximum pressure also shifts when those parameters are varied.
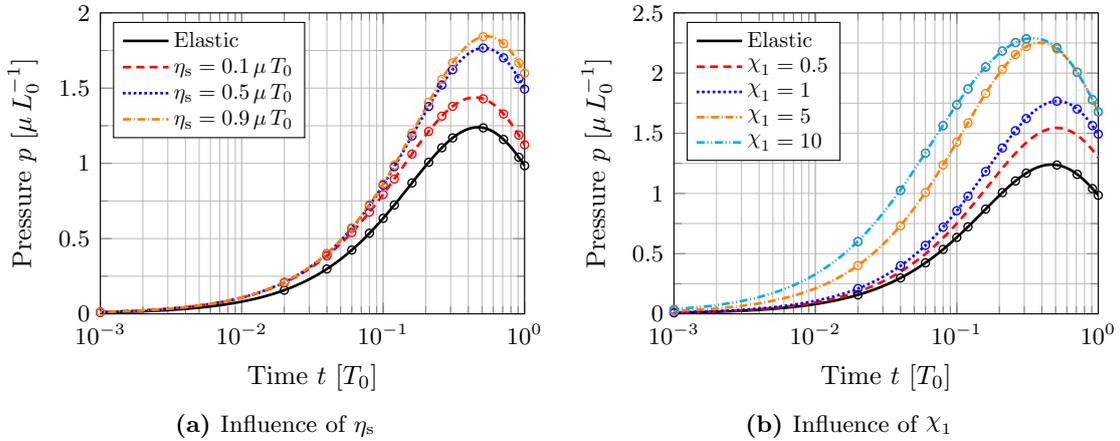
\begin{figure}[!ht]
	\setlength{\figwidth}{0.45\textwidth}
	\setlength{\figheight}{0.35\textwidth}
	\centering
	\subfloat[Influence of $\eta_\mrs$\figlabel{numExViscInflMemStudyEtaS}]{
		\begin{tikzpicture}
			\def\cdot{\times}
			\begin{axis}[xmode=log,grid=both,xlabel={Time $t$ $[T_0]$},ylabel={Pressure $p$ $[\mu\,L_0^{-1}]$},width=\figwidth,height=\figheight,
			xmin=1e-3,xmax=1,ymin=0,ymax=2,minor ytick={0.25,0.75,1.25,1.75},
			tick label style={font=\footnotesize},legend pos = north west,legend style={nodes={scale=0.75, transform shape}},legend cell align={left}
			]
				\addplot[black,solid,line width=1]table [x index = {0}, y index = {1},col sep=comma]{fig/inflMem/inflMemEtaSanalytical.csv};
					\addlegendentry{Elastic};
				\addplot[black,only marks,mark=o,mark size=1.5,line width=.5,forget plot]table [x index = {0}, y index = {1},col sep=comma]{fig/inflMem/inflMemEtaSnumerical.csv};
				\addplot[red,densely dashed,line width=1]table [x index = {0}, y index = {2},col sep=comma]{fig/inflMem/inflMemEtaSanalytical.csv};
					\addlegendentry{$\eta_\mrs=0.1\,\mu\,T_0$};
				\addplot[red,only marks,mark=o,mark size=1.5,line width=.5,forget plot]table [x index = {0}, y index = {2},col sep=comma]{fig/inflMem/inflMemEtaSnumerical.csv};
				\addplot[blue,densely dotted,line width=1]table [x index = {0}, y index = {3},col sep=comma]{fig/inflMem/inflMemEtaSanalytical.csv};
					\addlegendentry{$\eta_\mrs=0.5\,\mu\,T_0$};
				\addplot[blue,only marks,mark=o,mark size=1.5,line width=.5,forget plot]table [x index = {0}, y index = {3},col sep=comma]{fig/inflMem/inflMemEtaSnumerical.csv};
				\addplot[orange,densely dash dot,line width=1]table [x index = {0}, y index = {4},col sep=comma]{fig/inflMem/inflMemEtaSanalytical.csv};
					\addlegendentry{$\eta_\mrs=0.9\,\mu\,T_0$};
				\addplot[orange,only marks,mark=o,mark size=1.5,line width=.5,forget plot]table [x index = {0}, y index = {4},col sep=comma]{fig/inflMem/inflMemEtaSnumerical.csv};				
			\end{axis}
		\end{tikzpicture}
	}
	\subfloat[Influence of $\rchi_1$\figlabel{numExViscInflMemStudyChi1}]{
		\begin{tikzpicture}
			\def\cdot{\times}
			\begin{axis}[xmode=log,grid=both,xlabel={Time $t$ $[T_0]$},ylabel={Pressure $p$ $[\mu\,L_0^{-1}]$},width=\figwidth,height=\figheight,
			xmin=1e-3,xmax=1,ymin=0,ymax=2.5,ytick={0,0.5,1,1.5,2,2.5},minor ytick={0.25,0.75,1.25,1.75,2.25},
			tick label style={font=\footnotesize},legend pos = north west,legend style={nodes={scale=0.75, transform shape}},legend cell align={left}
			]
				\addplot[black,solid,line width=1]table [x index = {0}, y index = {1},col sep=comma]{fig/inflMem/inflMemChi1analytical.csv};
					\addlegendentry{Elastic};
				\addplot[black,only marks,mark=o,mark size=1.5,line width=.5,forget plot]table [x index = {0}, y index = {1},col sep=comma]{fig/inflMem/inflMemChi1numerical.csv};
				\addplot[red,densely dashed,line width=1]table [x index = {0}, y index = {2},col sep=comma]{fig/inflMem/inflMemChi1analytical.csv};
					\addlegendentry{$\rchi_1=0.5$};
				\addplot[red,only marks,mark=o,mark size=1.5,line width=.5,forget plot]table [x index = {0}, y index = {5},col sep=comma]{fig/inflMem/inflMemChi1numerical.csv};
				\addplot[blue,densely dotted,line width=1]table [x index = {0}, y index = {3},col sep=comma]{fig/inflMem/inflMemChi1analytical.csv};
					\addlegendentry{$\rchi_1=1$};
				\addplot[blue,only marks,mark=o,mark size=1.5,line width=.5,forget plot]table [x index = {0}, y index = {3},col sep=comma]{fig/inflMem/inflMemChi1numerical.csv};
				\addplot[orange,densely dash dot,line width=1]table [x index = {0}, y index = {4},col sep=comma]{fig/inflMem/inflMemChi1analytical.csv};
					\addlegendentry{$\rchi_1=5$};
				\addplot[orange,only marks,mark=o,mark size=1.5,line width=.5,forget plot]table [x index = {0}, y index = {4},col sep=comma]{fig/inflMem/inflMemChi1numerical.csv};
				\addplot[cyan,densely dash dot dot,line width=1]table [x index = {0}, y index = {5},col sep=comma]{fig/inflMem/inflMemChi1analytical.csv};
					\addlegendentry{$\rchi_1=10$};
				\addplot[cyan,only marks,mark=o,mark size=1.5,line width=.5,forget plot]table [x index = {0}, y index = {5},col sep=comma]{fig/inflMem/inflMemChi1numerical.csv};
			\end{axis}
		\end{tikzpicture}
	}
	\caption{Inflated membrane balloon: Influence of the in-plane shear viscosity $\eta_\mrs$ and stiffness ratio $\rchi_1:=\mu_1/\mu$ on the pressure $p$. The circles mark the numerical results at various snapshots in time, while the lines show the corresponding analytical results.} \figlabel{numExViscInflMemStudyEtaSChi1}
\end{figure}

Third, the effect of the loading rate is investigated. For this, the end time $t_\mathrm{end}$ is varied in \eqsref{numExViscInflMemLoadingRate}, while the parameters $\eta_\mrs=0.5\,\mu\,T_0$ and $\rchi_1=1$ are fixed. For the temporal integration, $1,000$ time steps are used. As shown in \figref{numExViscInflMemStudyTend}, the total pressure decreases for increasing $t_\mathrm{end}$, and the maximum is shifted to later times. The dissipation $\sD$, see \eqsref{cntViscDissipatedEnergy}, is visualized in \figref{numExViscInflMemDissipation} for three different values of the in-plane shear viscosity. For each value of $\eta_\mrs$, there exists one characteristic time $\tau_\lambda$, see \eqsref{numExViscInflMemLoadingRate}, for which the dissipation is maximal.
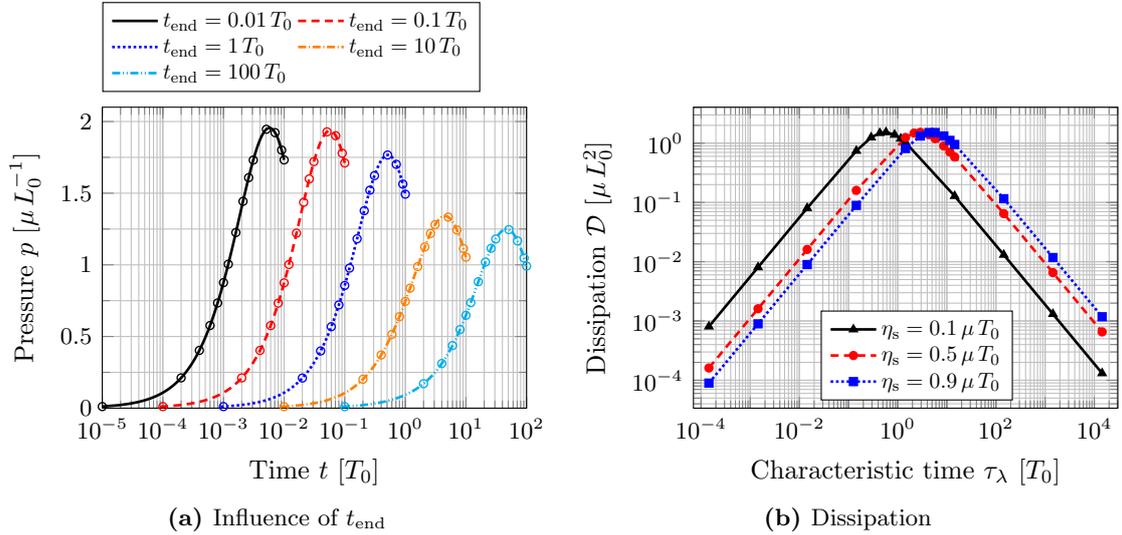
\begin{figure}[!ht]
	\setlength{\figwidth}{0.45\textwidth}
	\setlength{\figheight}{0.35\textwidth}
	\centering
	\subfloat[Influence of $t_\mathrm{end}$\figlabel{numExViscInflMemStudyTend}]{
		\begin{tikzpicture}
			\def\cdot{\times}
			\begin{axis}[xmode=log,grid=both,xlabel={Time $t$ $[T_0]$},ylabel={Pressure $p$ $[\mu\,L_0^{-1}]$},width=\figwidth,height=\figheight,
			xmin=1e-5,xmax=1e2,ymin=0,ymax=2.1,minor ytick={0.25,0.75,1.25,1.75},
			tick label style={font=\footnotesize},
			legend style={at={(0,1.2)},anchor=west},legend style={nodes={scale=0.75, transform shape}},legend cell align={left},legend columns=2,
			]
				\addplot[black,solid,line width=1]table [x index = {0}, y index = {1},col sep=comma]{fig/inflMem/inflMemTendAnalytical.csv};
					\addlegendentry{$t_\mathrm{end}=0.01\,T_0$};
				\addplot[black,only marks,mark=o,mark size=1.5,line width=.5,forget plot]table [x index = {0}, y index = {1},col sep=comma]{fig/inflMem/inflMemTendNumerical.csv};
				\addplot[red,densely dashed,line width=1]table [x index = {2}, y index = {3},col sep=comma]{fig/inflMem/inflMemTendAnalytical.csv};
					\addlegendentry{$t_\mathrm{end}=0.1\,T_0$};
				\addplot[red,only marks,mark=o,mark size=1.5,line width=.5,forget plot]table [x index = {2}, y index = {3},col sep=comma]{fig/inflMem/inflMemTendNumerical.csv};
				\addplot[blue,densely dotted,line width=1]table [x index = {4}, y index = {5},col sep=comma]{fig/inflMem/inflMemTendAnalytical.csv};
					\addlegendentry{$t_\mathrm{end}=1\,T_0$};
				\addplot[blue,only marks,mark=o,mark size=1.5,line width=.5,forget plot]table [x index = {4}, y index = {5},col sep=comma]{fig/inflMem/inflMemTendNumerical.csv};
				\addplot[orange,densely dash dot,line width=1]table [x index = {6}, y index = {7},col sep=comma]{fig/inflMem/inflMemTendAnalytical.csv};
					\addlegendentry{$t_\mathrm{end}=10\,T_0$};
				\addplot[orange,only marks,mark=o,mark size=1.5,line width=.5,forget plot]table [x index = {6}, y index = {7},col sep=comma]{fig/inflMem/inflMemTendNumerical.csv};
				\addplot[cyan,densely dash dot dot,line width=1]table [x index = {8}, y index = {9},col sep=comma]{fig/inflMem/inflMemTendAnalytical.csv};
					\addlegendentry{$t_\mathrm{end}=100\,T_0$};
				\addplot[cyan,only marks,mark=o,mark size=1.5,line width=.5,forget plot]table [x index = {8}, y index = {9},col sep=comma]{fig/inflMem/inflMemTendNumerical.csv};		
			\end{axis}
		\end{tikzpicture}
	}
	\subfloat[Dissipation\figlabel{numExViscInflMemDissipation}]{
		\begin{tikzpicture}
			\def\cdot{\times}
			\begin{axis}[xmode=log,ymode=log,grid=both,xlabel={Characteristic time $\tau_\lambda$ $[T_0]$},ylabel={Dissipation $\sD$ $[\mu\,L_0^{2}]$},width=\figwidth,height=\figheight,
			xmin=7e-5,xmax=3e4,
			xtick={1e-4,1e-3,1e-2,1e-1,1e0,1e1,1e2,1e3,1e4,1e5},
			xticklabels={$10^{-4}$,,$10^{-2}$,,$10^{0}$,,$10^{2}$,,$10^{4}$,},
			tick label style={font=\footnotesize},
			legend style={at={(0.3,0.175)},anchor=west},legend style={nodes={scale=0.75, transform shape}},legend cell align={left}
			]
				\addplot[black,solid,line width=1,mark=triangle*,mark size=1.25,mark options={solid}]table [x index = {0}, y index = {1},col sep=comma]{fig/inflMem/viscBalloonDissipations1.csv};
					\addlegendentry{$\eta_\mrs=0.1\,\mu\,T_0$};
				\addplot[red,densely dashed,line width=1,mark=*,mark size=1.25,mark options={solid}]table [x index = {0}, y index = {1},col sep=comma]{fig/inflMem/viscBalloonDissipations2.csv};
					\addlegendentry{$\eta_\mrs=0.5\,\mu\,T_0$};
				\addplot[blue,densely dotted,line width=1,mark=square*,mark size=1.25,mark options={solid}]table [x index = {0}, y index = {1},col sep=comma]{fig/inflMem/viscBalloonDissipations3.csv};
					\addlegendentry{$\eta_\mrs=0.9\,\mu\,T_0$};
			\end{axis}
		\end{tikzpicture}
	}
	\caption{Inflated membrane balloon: (a) Influence of the end time $t_\mathrm{end}$ on the pressure $p$. Here, the circles mark the numerical results at various snapshots in time, while the lines mark the corresponding analytical results. (b) Dissipation $\sD$, see \eqsref{cntViscDissipatedEnergy}, over the characteristic time $\tau_\lambda$ defined in \eqsref{numExViscInflMemLoadingRate}.} \figlabel{numExViscInflMemStudyTendDissipation}
\end{figure}

\subsection{Pure bending of a flat strip} \seclabel{numExViscPureBend}
This section presents the pure bending of an initially flat shell to verify the formulation for viscoelastic bending. The geometry, boundary and loading conditions are shown in \figref{numExViscPureBendSetup}. At the left and right edge, a distributed bending moment $M$ is applied. Additionally, a displacement is applied on the right edge and a pressure acts on the whole structure. This loading combination ensures that the initially flat sheet is bent into a curved sheet with curvature $\kappa_2$, but not stretched, i.e.~the surface stretches are exactly $\lambda_1=\lambda_2=1$.\footnote{According to thin shell theory, there is a high-order coupling between curvatures and stretches; see, for example, \cite{sauer2017a}.} The curvature is related to the radius $r$ of the deformed shell via $\kappa_2=1/r$. The membrane response in the elastic branch is based on the Neo-Hookean model from \eqsref{cntViscConstElasticityPsiMemClassicalNeoHookean}, and the bending response in the elastic and Maxwell branch is based on the Koiter bending model from \eqsref{cntViscConstElasticityPsiBendKoiter}.
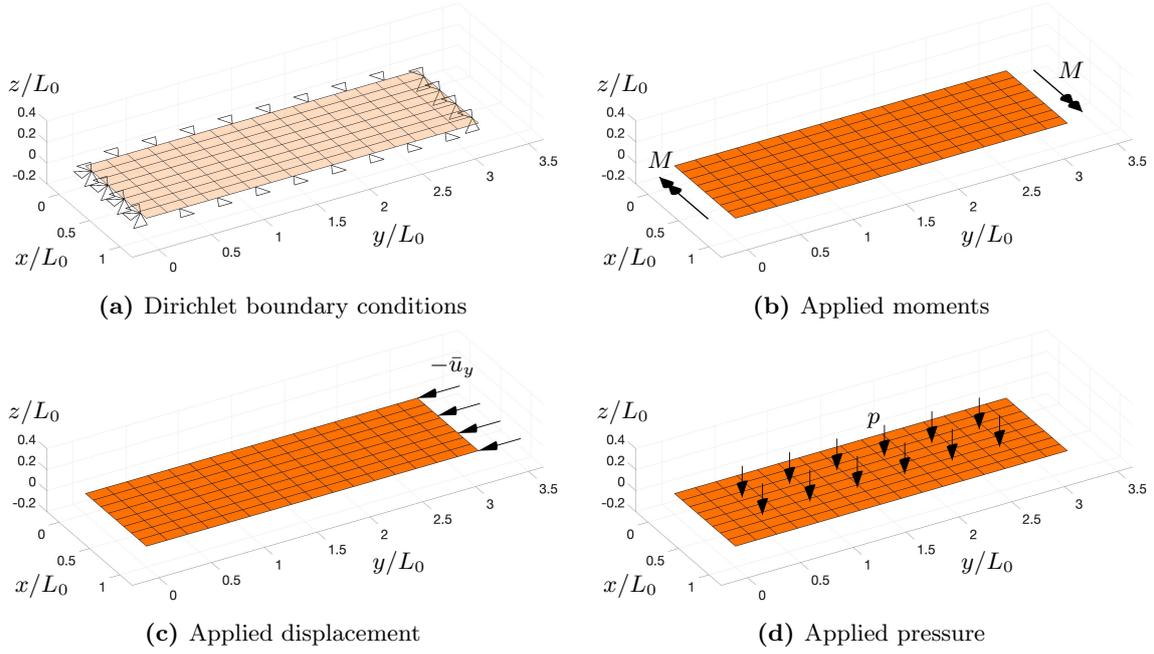
\begin{figure}[!ht]
	\setlength{\figwidth}{0.45\textwidth}
	\centering
	\subfloat[Dirichlet boundary conditions]{
		\begin{tikzpicture}
			\node[align=center] at (0,0) {\includegraphics[width=\figwidth]{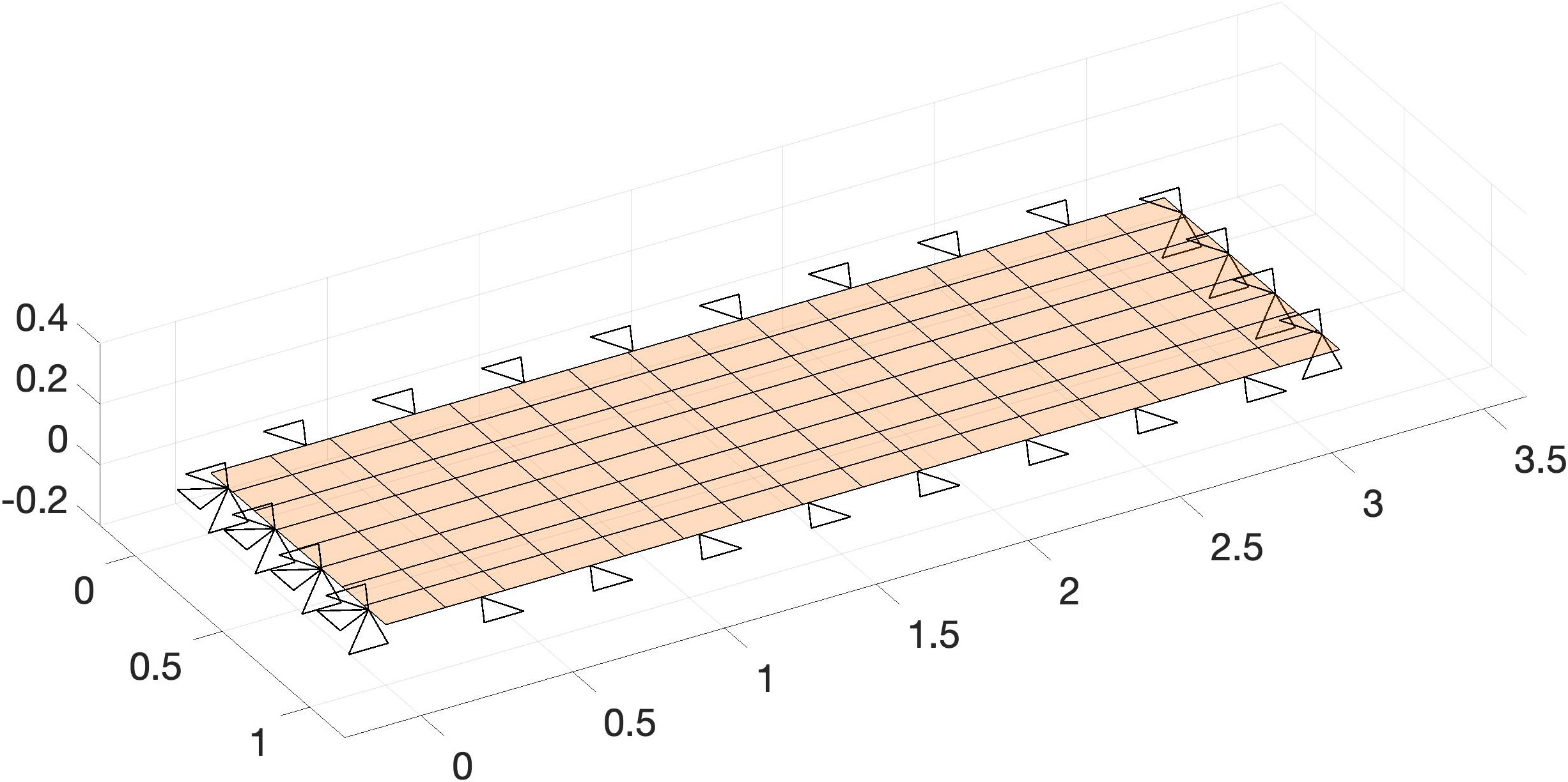}};
			\node[align=center,font=\footnotesize] at (1.5,-1.3) {$y/L_0$};
			\node[align=center,font=\footnotesize] at (-3.2,-1.6) {$x/L_0$};
			\node[align=center,font=\footnotesize] at (-3.3,0.7) {$z/L_0$};
		\end{tikzpicture}
	}
	\subfloat[Applied moments]{
		\begin{tikzpicture}
			\node[align=center] at (0,0) {\includegraphics[width=\figwidth]{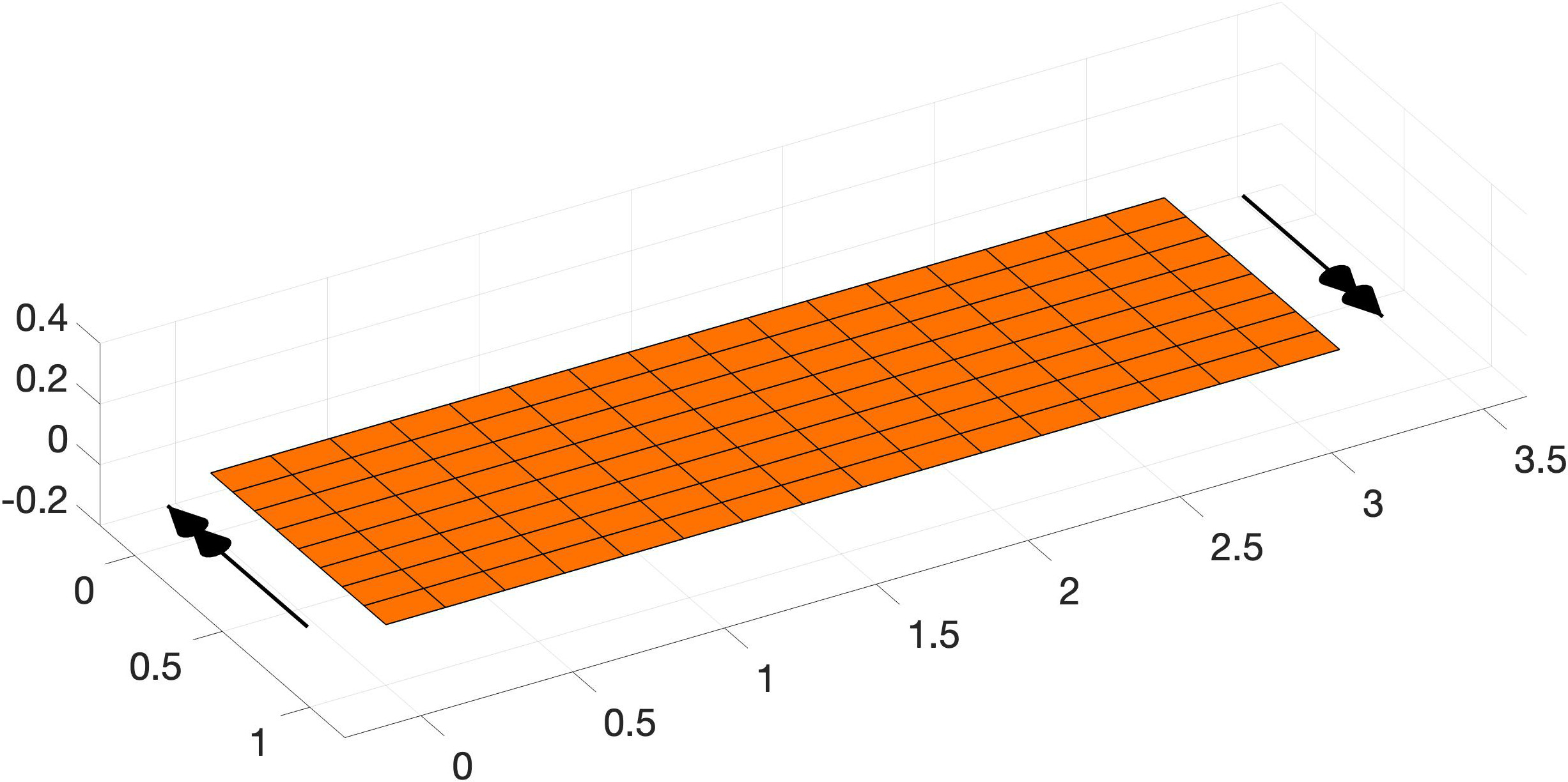}};
			\node[align=center,font=\footnotesize] at (1.5,-1.3) {$y/L_0$};
			\node[align=center,font=\footnotesize] at (-3.2,-1.6) {$x/L_0$};
			\node[align=center,font=\footnotesize] at (-3.3,0.7) {$z/L_0$};
			\node[align=center,font=\footnotesize] at (2.6,0.9) {$M$};
			\node[align=center,font=\footnotesize] at (-2.8,-0.3) {$M$};
		\end{tikzpicture}
	}
	\\\vspace{-4mm}
	\subfloat[Applied displacement]{
		\begin{tikzpicture}
			\node[align=center] at (0,0) {\includegraphics[width=\figwidth]{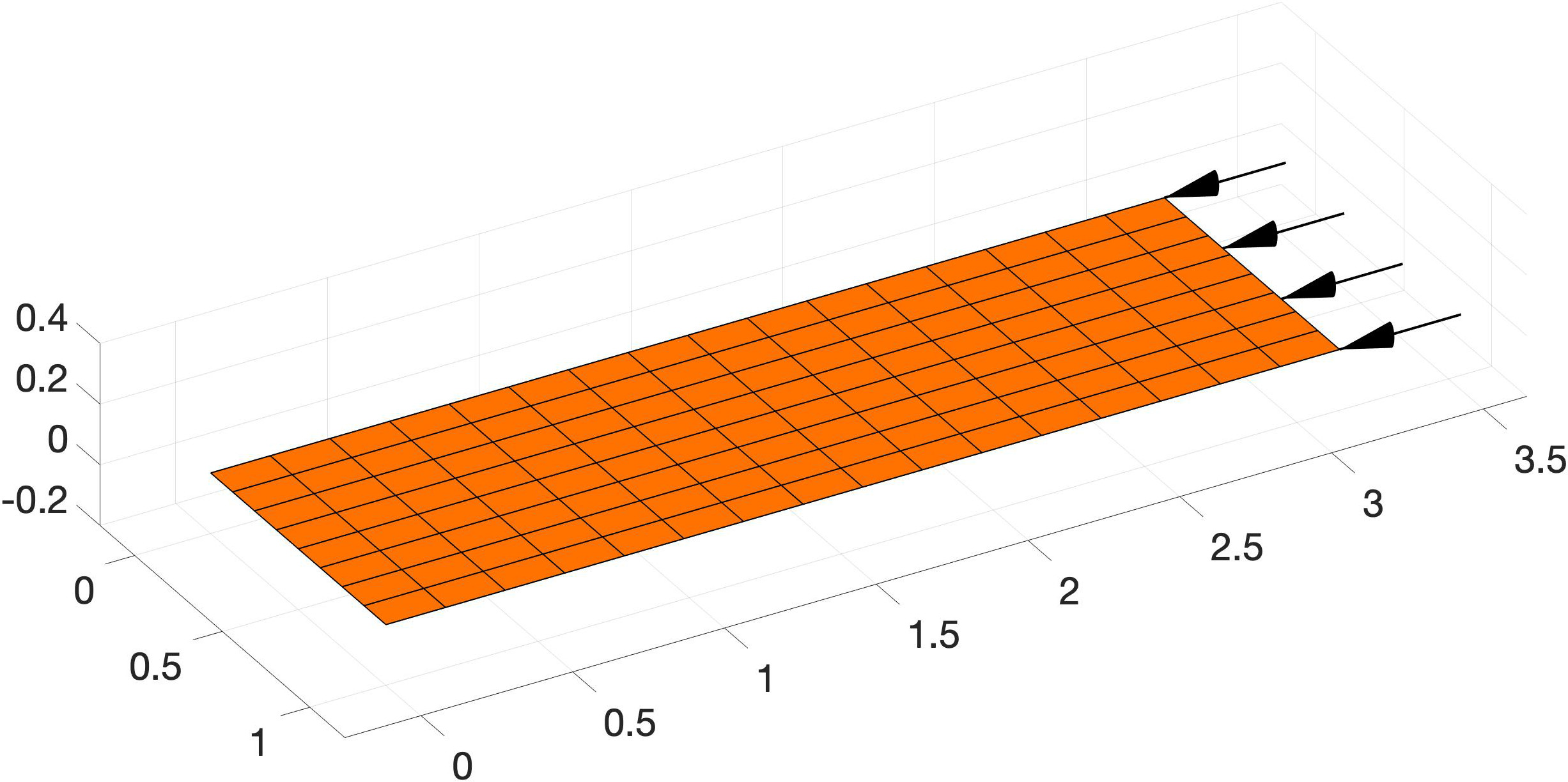}};
			\node[align=center,font=\footnotesize] at (1.5,-1.3) {$y/L_0$};
			\node[align=center,font=\footnotesize] at (-3.2,-1.6) {$x/L_0$};
			\node[align=center,font=\footnotesize] at (-3.3,0.7) {$z/L_0$};
			\node[align=center,font=\footnotesize] at (2.2,1.3) {$-\bar{u}_y$};
		\end{tikzpicture}
	}
	\subfloat[Applied pressure]{
		\begin{tikzpicture}
			\node[align=center] at (0,0) {\includegraphics[width=\figwidth]{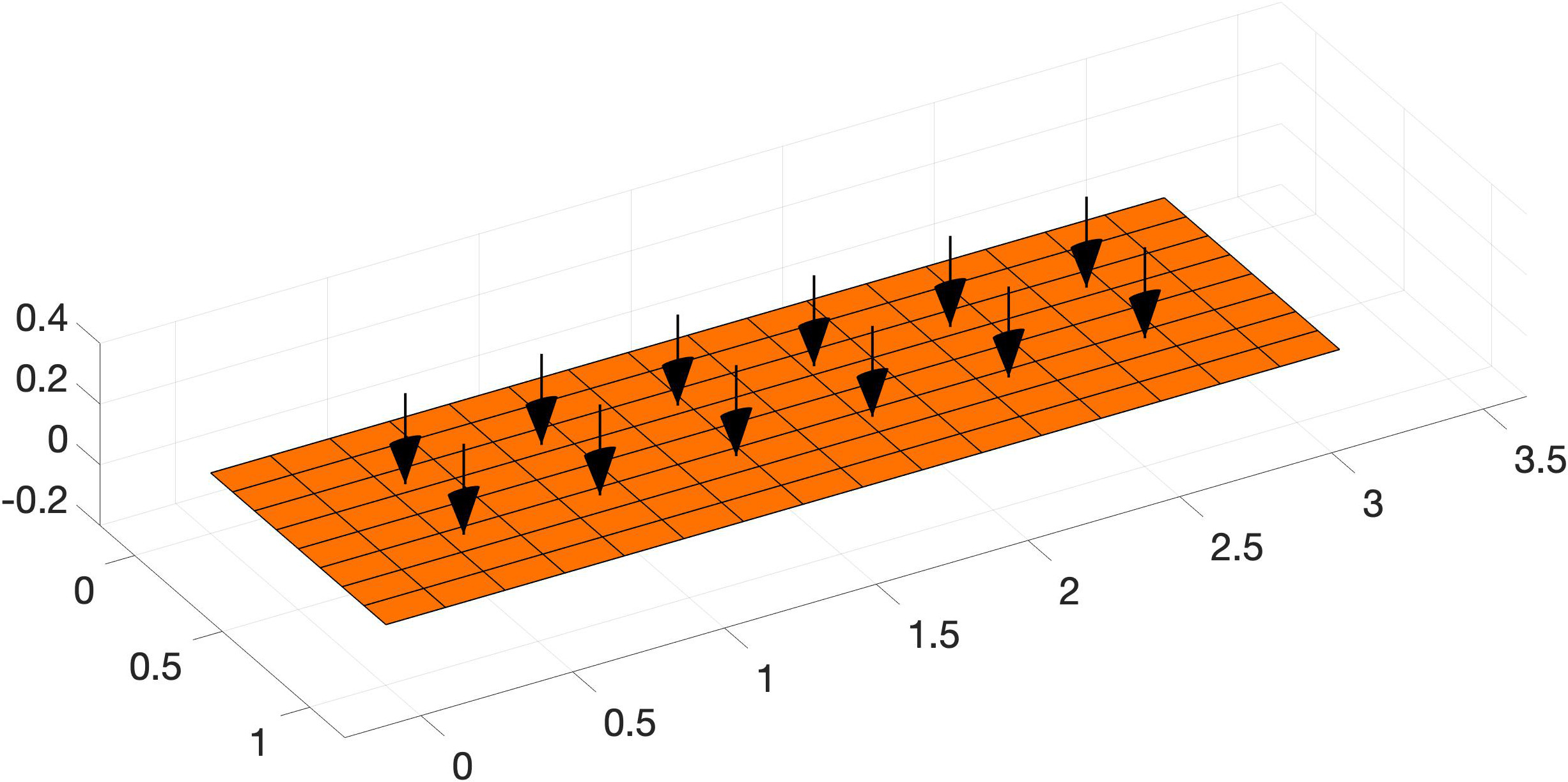}};
			\node[align=center,font=\footnotesize] at (1.5,-1.3) {$y/L_0$};
			\node[align=center,font=\footnotesize] at (-3.2,-1.6) {$x/L_0$};
			\node[align=center,font=\footnotesize] at (-3.3,0.7) {$z/L_0$};
			\node[align=center,font=\footnotesize] at (0,0.6) {$p$};
		\end{tikzpicture}
	}
	\caption{Pure bending of a flat strip: Setup including Dirichlet boundary conditions and three different external loads that are simultaneously applied.} \figlabel{numExViscPureBendSetup}
\end{figure}

Choosing the applied moment
\eqb{l}
	M(t)=M_v\,t\,,\eqlabel{numExViscPureBendM}
\eqe
with constant loading rate $M_v$, the sheet becomes stretch-free for the imposed displacement and pressure
\eqb{l}
	\bar{u}_y(t)=-\biggl(S-\dfrac{2}{\kappa_2(t)}\,\sin\Bigl(\dfrac{S\,\kappa_2(t)}{2}\Bigr)\biggr)\,,\quad\mathrm{and}\quad p(t)=-\Bigl((c+c_1)\,\kappa_2(t)^3-c_1\,\kappa_2(t)^2\,\kappa_2^\mrin(t)\Bigr)\,, \eqlabel{numExViscPureBendDisplacementPressure}
\eqe
where $S=\pi$ denotes the length of the sheet in $y$-direction. \eqsref{numExViscPureBendM} is a chosen rate that allows for an analytical solution as shown in \appref{viscAnalyticalSolutionsPureBend}. The displacement $\bar{u}_y(t)$ in Eq.~(\ref{e:numExViscPureBendDisplacementPressure}.1) is chosen such that a perfect circular arc is obtained for any given curvature $\kappa_2(t)$. The pressure $p(t)$ in Eq.~(\ref{e:numExViscPureBendDisplacementPressure}.2) is required to equilibriate the structure, and it follows from the well-known formula for thin-walled cylindrical pressure vessels, which requires $p(t)=N_2^2(t)/r(t)=N_2^2(t)\,\kappa_2(t)$, where $N_2^2$ denotes the in-plane stress component in $y$-direction. The derivation of the latter is provided in \appref{viscAnalyticalSolutionsPureBend}. This problem can be solved analytically, which leads to the curvatures
\eqb{l}
	\kappa_2^\mrin(t)=\dfrac{M_v\,\tau_b}{c}\,\Bigl(e^{-t/\tau_b}+\dfrac{t}{\tau_b}-1\Bigr)\,, \eqlabel{numExViscPureBendKappa2in}
\eqe
with the characteristic time
\eqb{l}
	\tau_b:=\dfrac{\eta_\mrb\,(c+c_1)}{c\,c_1}\,, \eqlabel{numExViscPureBendcTilde}
\eqe
and
\eqb{l}
	\kappa_2(t)=\dfrac{M(t)+c_1\,\kappa_2^\mrin(t)}{c+c_1}\,, \eqlabel{numExViscPureBendKappa2}
\eqe
see \appref{viscAnalyticalSolutionsPureBend}. The loading rate $M_v$ in \eqsref{numExViscPureBendM} is chosen in a way, such that the final curvature is equal to $\kappa_2^\mathrm{end}:=\kappa_2(t_\mathrm{end})$. Using Eqs.~\eqref{e:numExViscPureBendM} and \eqref{e:numExViscPureBendKappa2in}--\eqref{e:numExViscPureBendKappa2}, this leads to the constant loading rate
\eqb{l}
	M_v:=\dfrac{c+c_1}{t_\mathrm{end}+\dfrac{c_1\,\tau_b}{c}\,\tilde{\kappa}_{2,\mathrm{end}}^\mrin}\,, \eqlabel{numExViscPureBendMv}
\eqe
where $\kappa_2^\mrin(t_\mathrm{end}):=M_v\,\tilde{\kappa}^\mrin_{2,\mathrm{end}}$ with $\tilde{\kappa}^\mrin_{2,\mathrm{end}}:=\tau_b\,\bigl(\exp(-t_\mathrm{end}/\tau_b)+t_\mathrm{end}/\tau_b-1\bigr)/c$.

First, the convergence with respect to the two relative curvature errors
\eqb{l}
	\epsilon_\kappa:=\dfrac{\bigl|\kappa_{2,\mathrm{num}}(t_\mathrm{end})-\kappa_{2,\mathrm{ana}}(t_\mathrm{end})\bigr|}{\kappa_{2,\mathrm{ana}}(t_\mathrm{end})}\,,\quad\mathrm{and}\quad\epsilon_\kappa^\mrin:=\dfrac{\bigl|\kappa_{2,\mathrm{num}}^\mrin(t_\mathrm{end})-\kappa_{2,\mathrm{ana}}^\mrin(t_\mathrm{end})\bigr|}{\kappa_{2,\mathrm{ana}}^\mrin(t_\mathrm{end})}\,, \eqlabel{numExViscPureBendCurvatureErrors}
\eqe
is investigated. The chosen parameters are $\mu=10\,L_0^2/c$, $\Lambda=5\,L_0^2/c$, $c=c_1$, $\eta_\mrb=0.5\,c\,T_0$, $t_\mathrm{end}=1\,T_0$, and $\kappa_2^\mathrm{end}=0.5$, such that $\tau_b=1\,T_0$. \figref{numExViscPureBendConvergence} shows that the errors converge linearly with an increase of the number of elements and a decrease of the time step size.
\begin{figure}[!ht]
	\setlength{\figwidth}{0.45\textwidth}
	\setlength{\figheight}{0.35\textwidth}
	\centering
	\subfloat[Mesh refinement\figlabel{numExViscPureBendConvergenceMesh}]{
		\begin{tikzpicture}
			\def\cdot{\times}
			\begin{axis}[xmode=log,ymode=log,grid=both,xlabel={Number of elements $n_\mathrm{el}$ $[-]$},ylabel={$\epsilon_\kappa$ $[-]$},width=\figwidth,height=\figheight,
			log basis x = 10,
			xmin=5e0,xmax=1e3,ymin=1e-5,ymax=1e-2,
			tick label style={font=\footnotesize},
			legend pos = south west,legend style={nodes={scale=0.75, transform shape}},legend cell align={left},legend columns=1,
			]
				\addplot[black,densely dashed,line width=1,mark=triangle*,mark size=1.5,mark options={solid}]table [x index = {0}, y index = {1},col sep=comma]{fig/pureBend/pureBendConvergenceKappa2.csv};
					\addlegendentry{$\Delta t=10^{-2}\,T_0$};
				\addplot[blue,line width=1,mark=*,mark size=1.5,mark options={solid}]table [x index = {0}, y index = {2},col sep=comma]{fig/pureBend/pureBendConvergenceKappa2.csv};
					\addlegendentry{$\Delta t=10^{-3}\,T_0$};
				\addplot[red,densely dotted,line width=1,mark=square*,mark size=1.5,mark options={solid}]table [x index = {0}, y index = {3},col sep=comma]{fig/pureBend/pureBendConvergenceKappa2.csv};
					\addlegendentry{$\Delta t=10^{-4}\,T_0$};
				\addplot[black,solid,line width=1,mark=none,domain=40:100,samples=2]{0.08*x^-1};
				\draw[black,solid,line width=1] (40,2e-3) -- (100,2e-3); \node[anchor=south] at (70,2e-3) {\scriptsize $\boldsymbol{1}$};
				\draw[black,solid,line width=1] (100,2e-3) -- ( 100,8e-4); \node[anchor=west] at (100,1.4e-3) {\scriptsize $\boldsymbol{1}$};

			\end{axis}
		\end{tikzpicture}
	}
	\quad
		\subfloat[Time step size refinement\figlabel{numExViscPureBendConvergenceTimeStepSize}]{
		\begin{tikzpicture}
			\def\cdot{\times}
			\begin{axis}[xmode=log,ymode=log,grid=both,xlabel={Time step size $\Delta t$ $[T_0]$},ylabel={$\epsilon_\kappa^\mrin$ $[-]$},width=\figwidth,height=\figheight,
			tick label style={font=\footnotesize},
			legend pos=south east,legend style={nodes={scale=0.75, transform shape}},legend cell align={left},legend columns=2,
			]
				\addplot[black,densely dash dot,line width=1,mark=triangle*,mark size=1.5,mark options={solid}]table [x index = {0}, y index = {1},col sep=comma]{fig/pureBend/pureBendConvergenceKappa2in.csv};
					\addlegendentry{$n_\mathrm{el}=8$};
				\addplot[blue,solid,line width=1,mark=*,mark size=1.5,mark options={solid}]table [x index = {0}, y index = {2},col sep=comma]{fig/pureBend/pureBendConvergenceKappa2in.csv};
					\addlegendentry{$n_\mathrm{el}=32$};
				\addplot[red,densely dashed,line width=1,mark=square*,mark size=1.5,mark options={solid}]table [x index = {0}, y index = {3},col sep=comma]{fig/pureBend/pureBendConvergenceKappa2in.csv};
					\addlegendentry{$n_\mathrm{el}=128$};
				\addplot[cyan,densely dotted,line width=1,mark=diamond*,mark size=1.5,mark options={solid}]table [x index = {0}, y index = {4},col sep=comma]{fig/pureBend/pureBendConvergenceKappa2in.csv};
					\addlegendentry{$n_\mathrm{el}=512$};
			\addplot[black,solid,line width=1,mark=none,domain=3e-3:1e-2,samples=2]{0.066666666666667*x^1};
			\draw[black,solid,line width=1] (3e-3,2e-4) -- (1e-2,2e-4); \node[anchor=north] at (6.5e-3,2e-4) {\scriptsize $\boldsymbol{1}$};
			\draw[black,solid,line width=1] (1e-2,2e-4) -- ( 1e-2,6.666666666666666e-04); \node[anchor=west] at (1e-2,3.5e-4) {\scriptsize $\boldsymbol{1}$};
			\end{axis}
		\end{tikzpicture}
	}
	\caption{Pure bending of a flat strip: Convergence of the relative curvature errors, see \eqsref{numExViscPureBendCurvatureErrors}, over mesh and time step size refinement.} \figlabel{numExViscPureBendConvergence}
\end{figure}
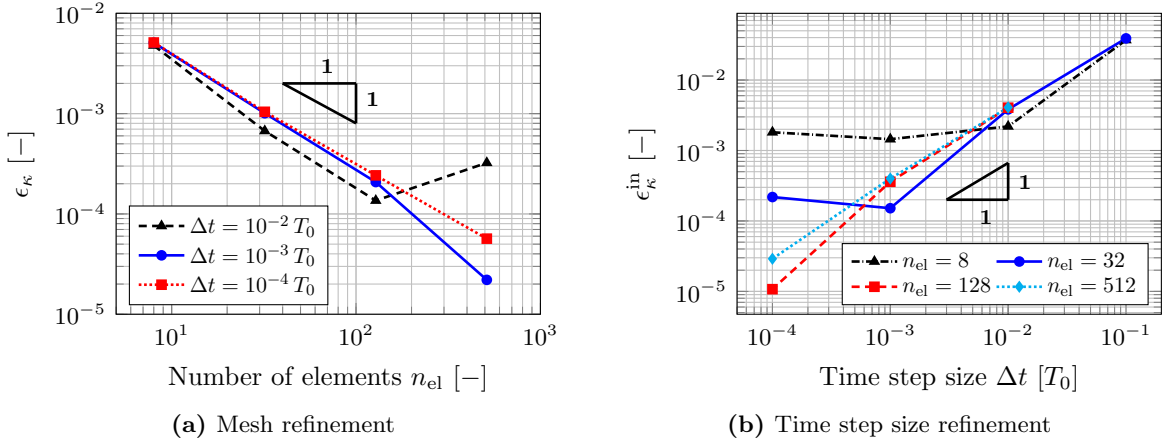

\figref{numExViscPureBendDeformation} shows the deformed surface at various snapshots in time for the parameters $\mu=10\,L_0^2/c$, $\Lambda=5\,L_0^2/c$, $c=c_1$, $\eta_\mrb=0.5\,c\,T_0$, $t_\mathrm{end}=1\,T_0$, $\kappa_2^\mathrm{end}=1$, and $\tau_b=1\,T_0$. The surfaces are colored with the relative curvature error $\epsilon_\kappa$.
\begin{figure}[!ht]
	\setlength{\figwidth}{0.48\textwidth}
	\centering
	\begin{tikzpicture}
		\tikzmath{
			\y2 = -7.975;     
			\y1 = -7.55;     
			\x1 = -1.15;	 
			\x2 = 10.44;	 
		};
			\node[align=center] at (0,1) {\includegraphics[width=\figwidth]{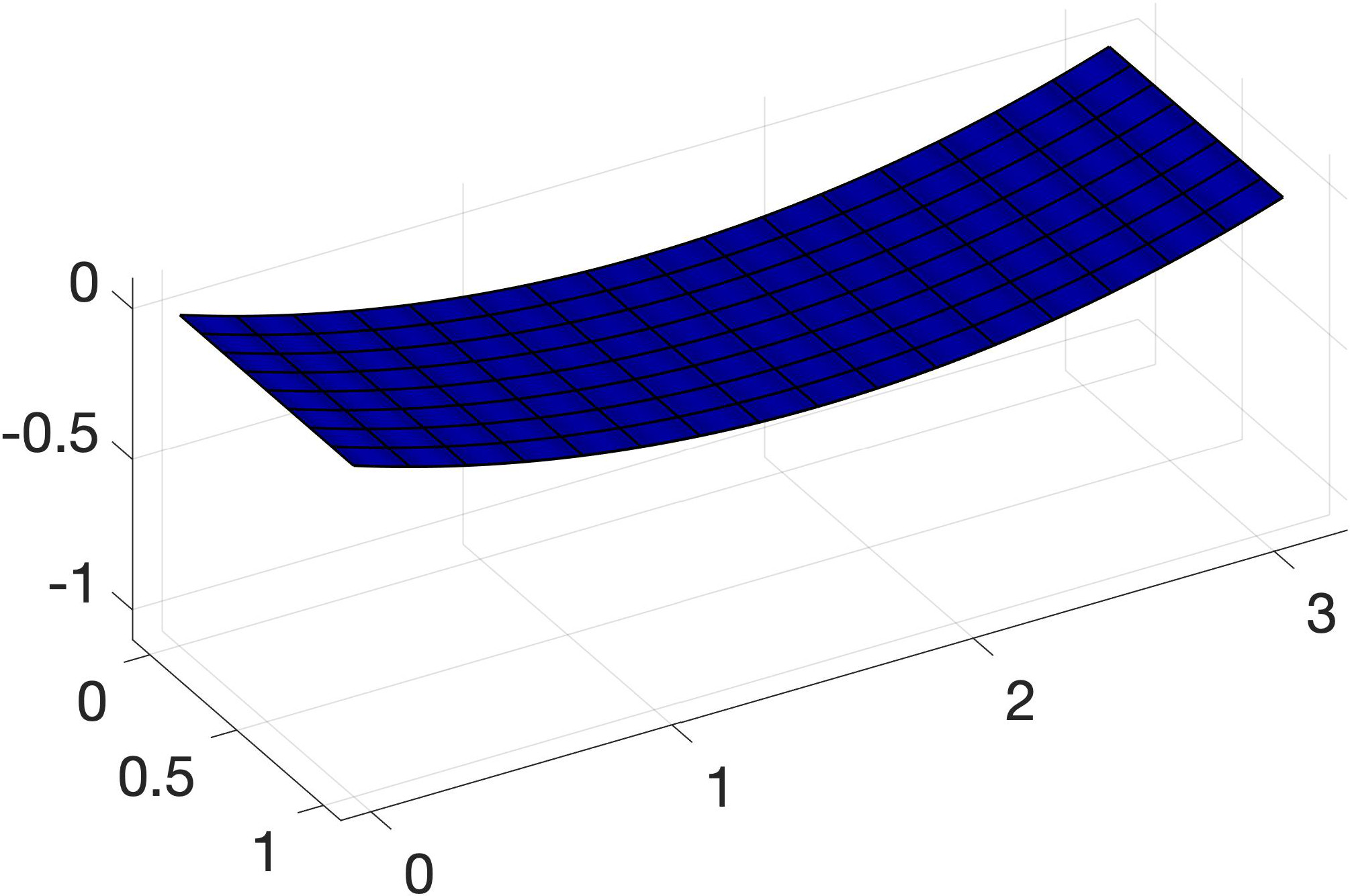}};					
			\node[align=center,font=\footnotesize] at (1.2,-0.85) {$y/L_0$};
			\node[align=center,font=\footnotesize] at (-3.1,-1.3) {$x/L_0$};
			\node[align=center,font=\footnotesize] at (-3.3,2.3) {$z/L_0$};
			\node[align=left,font=\footnotesize] at (-0.5,3) {$t=0.25\,T_0$};
			\node[align=center] at (8,1) {\includegraphics[width=\figwidth]{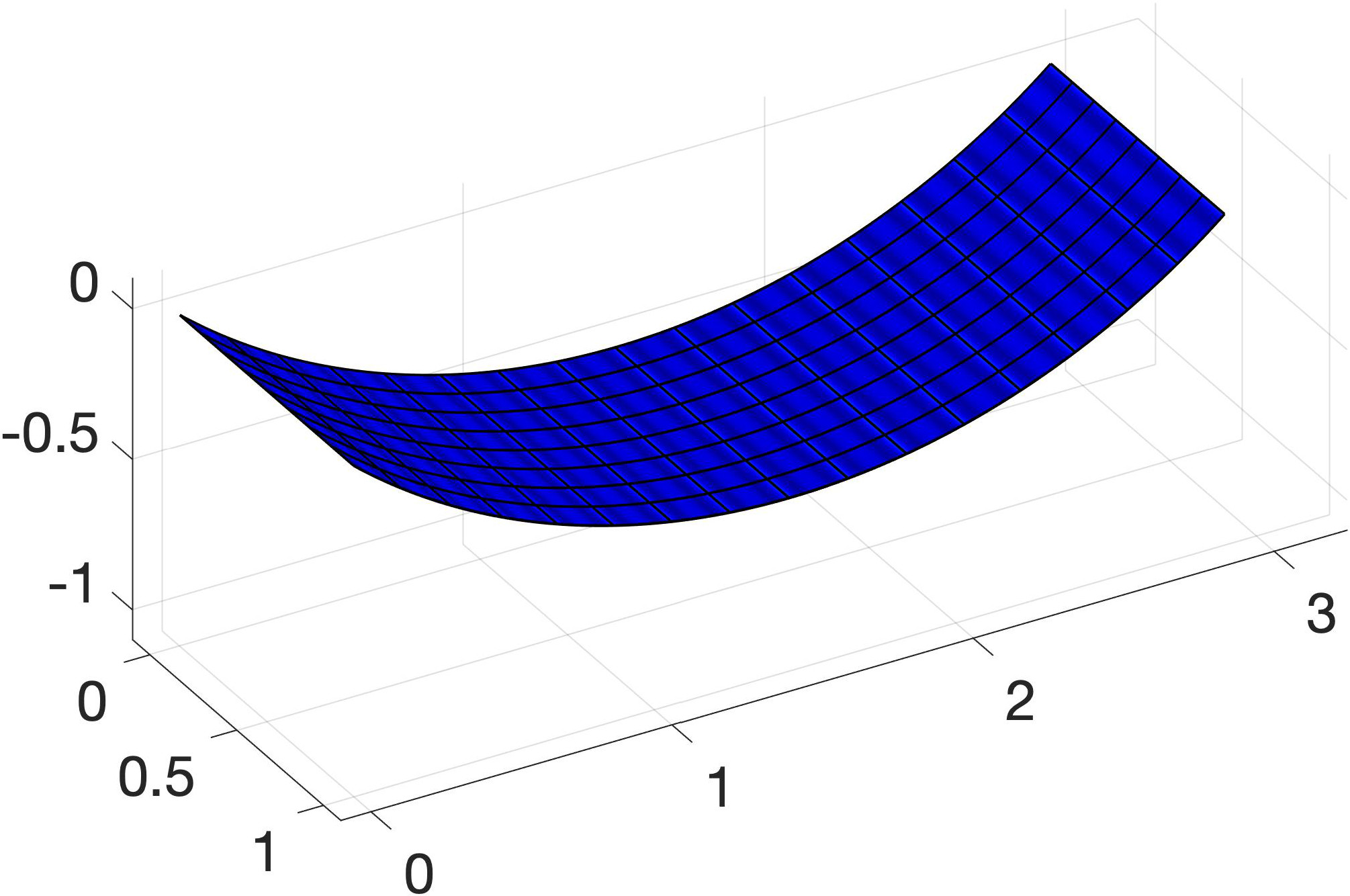}};					
			\node[align=center,font=\footnotesize] at (9.2,-0.85) {$y/L_0$};
			\node[align=center,font=\footnotesize] at (4.9,-1.3) {$x/L_0$};
			\node[align=center,font=\footnotesize] at (4.7,2.3) {$z/L_0$};	
			\node[align=left,font=\footnotesize] at (7.5,3) {$t=0.5\,T_0$};
			\node[align=center] at (0,-4.5) {\includegraphics[width=\figwidth]{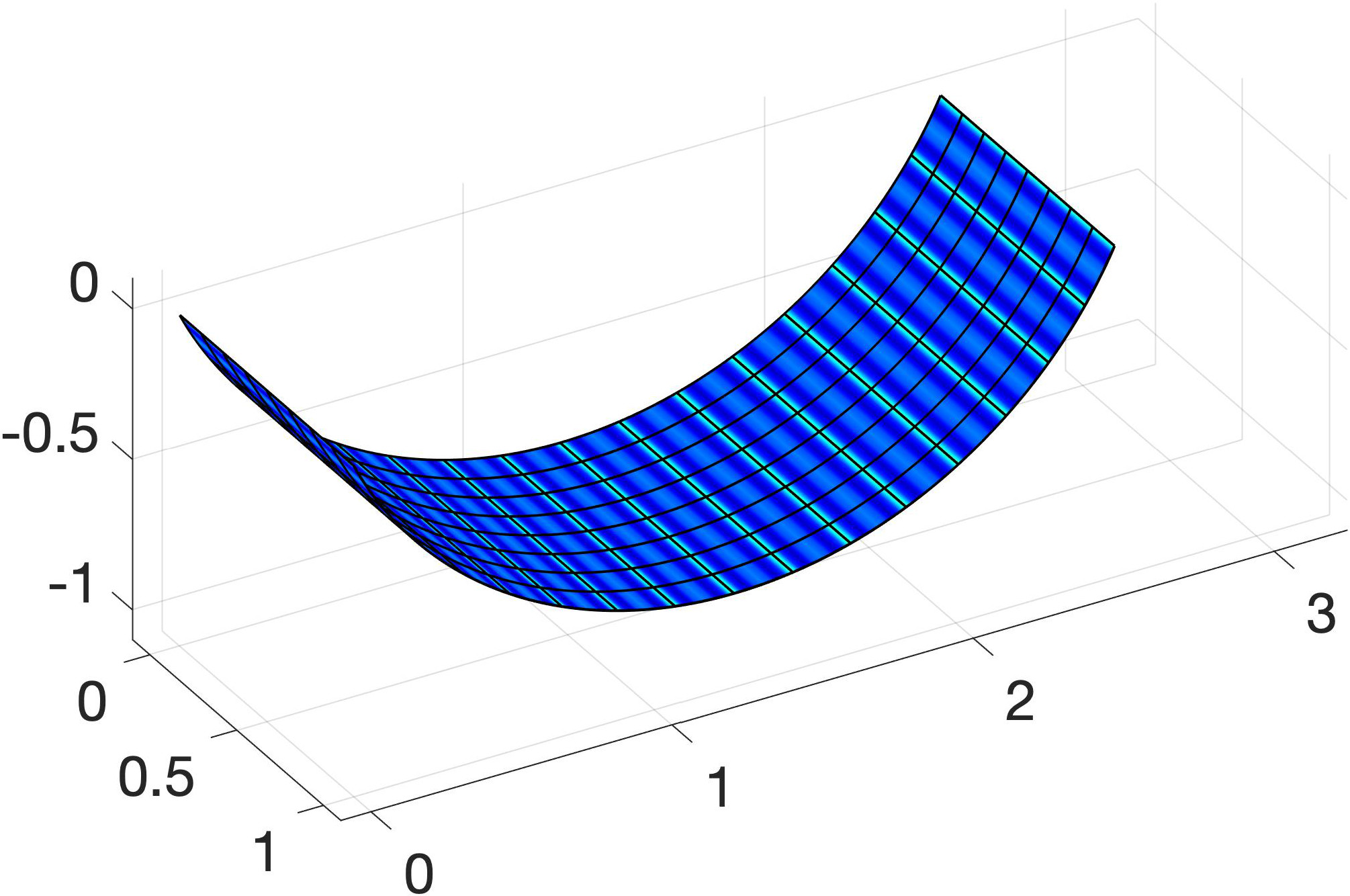}};					
			\node[align=center,font=\footnotesize] at (1.2,-6.35) {$y/L_0$};
			\node[align=center,font=\footnotesize] at (-3.1,-6.8) {$x/L_0$};
			\node[align=center,font=\footnotesize] at (-3.3,-3.2) {$z/L_0$};	
			\node[align=left,font=\footnotesize] at (-0.5,-2.5) {$t=0.75\,T_0$};
			\node[align=center] at (8,-4.5) {\includegraphics[width=\figwidth]{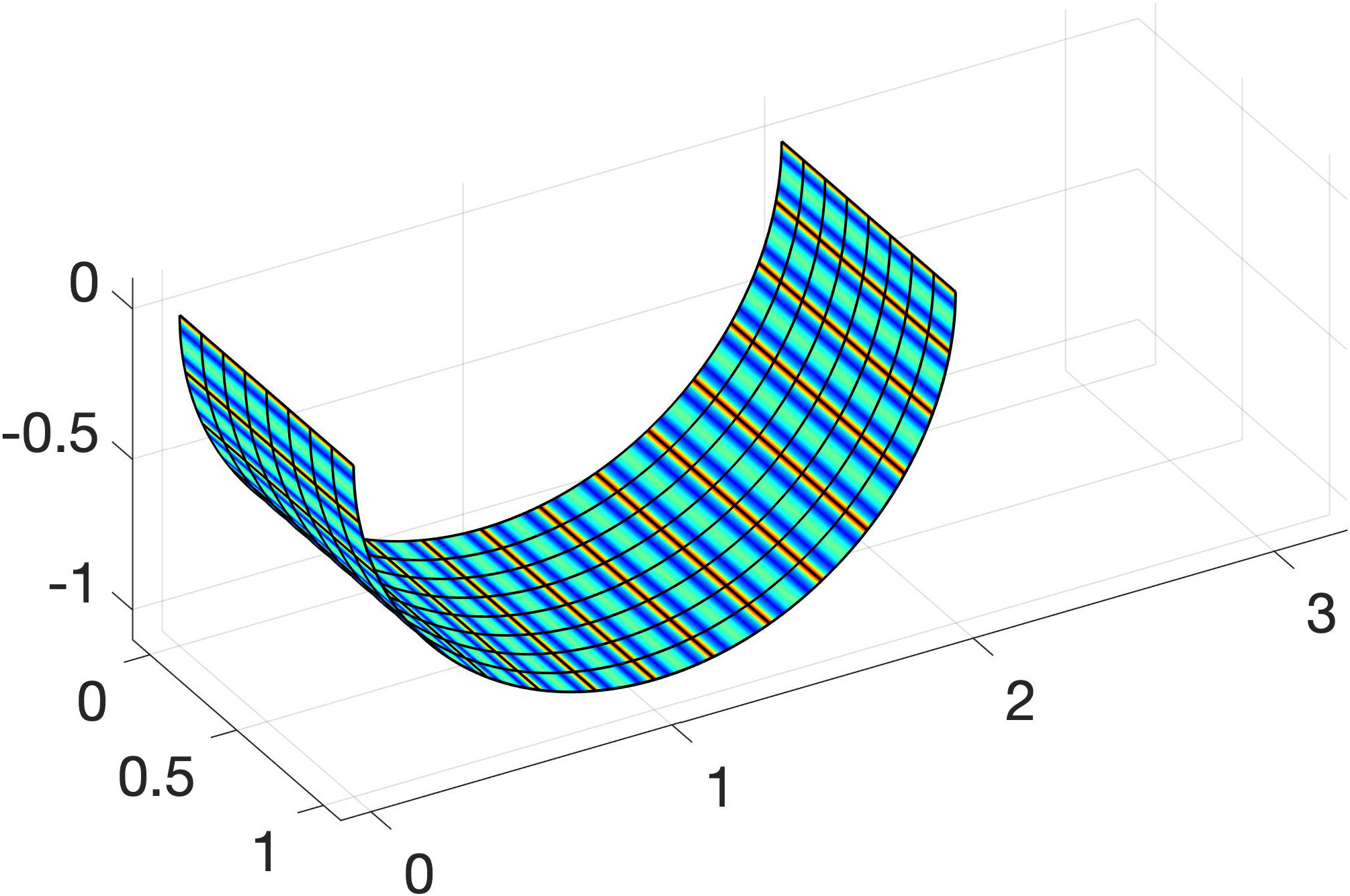}};						
			\node[align=center,font=\footnotesize] at (9.2,-6.35) {$y/L_0$};
			\node[align=center,font=\footnotesize] at (4.9,-6.8) {$x/L_0$};
			\node[align=center,font=\footnotesize] at (4.7,-3.2) {$z/L_0$};
			\node[align=left,font=\footnotesize] at (7.5,-2.5) {$t=1\,T_0$};
			\node[align=center] at (4,\y1) {\includegraphics[width=0.81\textwidth]{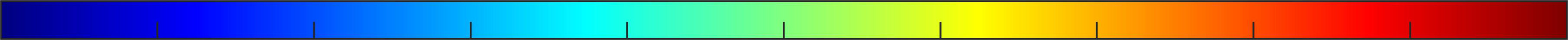}};
			\node[align=center,font=\footnotesize] at ({\x1+(\x2-\x1)/9*0},\y2) {$1$};		
			\node[align=center,font=\footnotesize] at ({\x1+(\x2-\x1)/9*1},\y2) {$2$};
			\node[align=center,font=\footnotesize] at ({\x1+(\x2-\x1)/9*2},\y2) {$3$};		
			\node[align=center,font=\footnotesize] at ({\x1+(\x2-\x1)/9*3},\y2) {$4$};		
			\node[align=center,font=\footnotesize] at ({\x1+(\x2-\x1)/9*4},\y2) {$5$};		
			\node[align=center,font=\footnotesize] at ({\x1+(\x2-\x1)/9*5},\y2) {$6$};		
			\node[align=center,font=\footnotesize] at ({\x1+(\x2-\x1)/9*6},\y2) {$7$};		
			\node[align=center,font=\footnotesize] at ({\x1+(\x2-\x1)/9*7},\y2) {$8$};		
			\node[align=center,font=\footnotesize] at ({\x1+(\x2-\x1)/9*8},\y2) {$9$};		
			\node[align=center,font=\footnotesize] at ({\x1+(\x2-\x1)/9*9},\y2) {$10$};	
			\node[align=center,font=\footnotesize] at (10,-8.4) {$\times10^{-3}$};	
			\node[align=left,font=\footnotesize] at ({\x2+0.8},\y1) {$\epsilon_\kappa$ $[-]$};	
	\end{tikzpicture}
	\caption{Pure bending of a flat stip: Deformed surfaces at various snapshots in time colored by the relative curvature error $\epsilon_\kappa$ see Eq.~(\ref{e:numExViscPureBendCurvatureErrors}.1).} \figlabel{numExViscPureBendDeformation}
\end{figure}
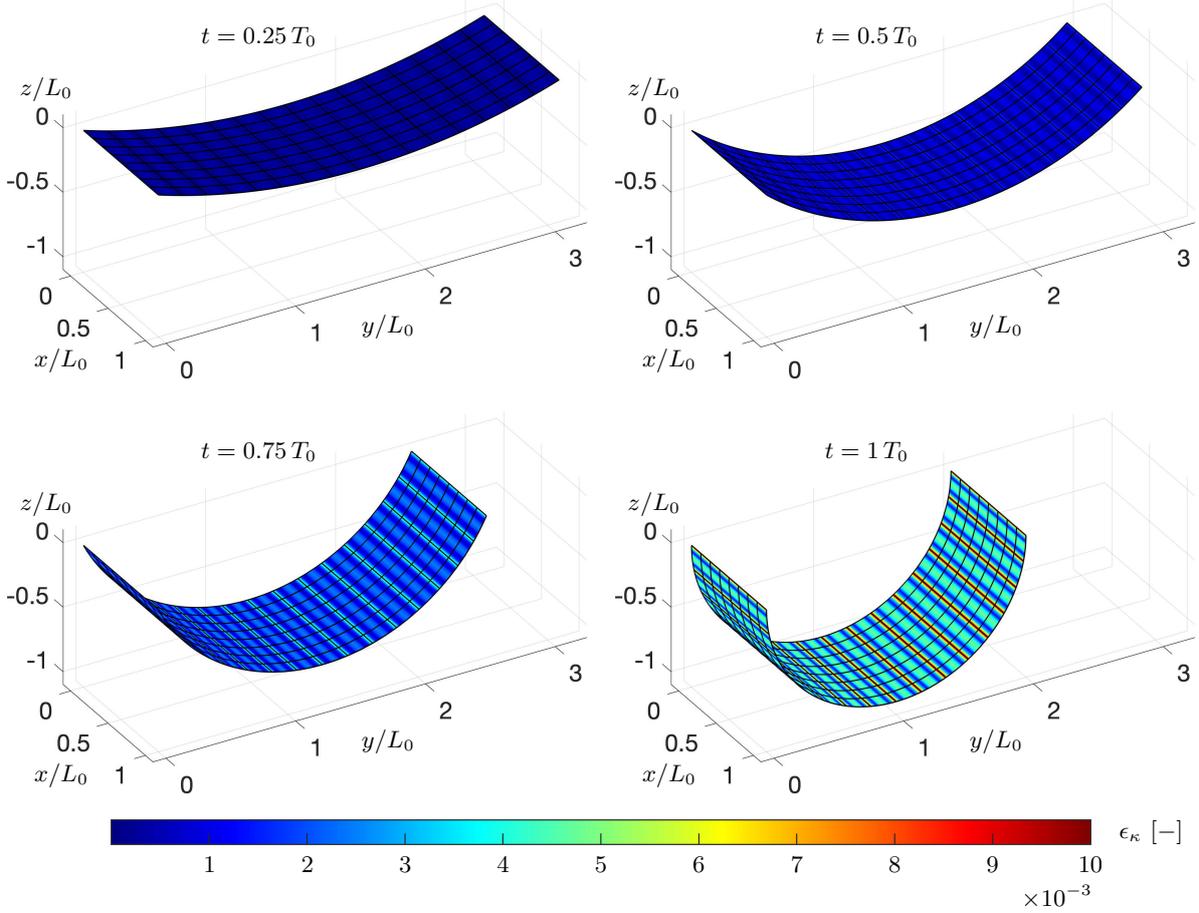
The maximum error of the surface stretches $|\lambda_1-1|$ and $|\lambda_2-1|$ over all time steps are of order $\mathcal{O}(10^{-3})$, which indicates that the chosen boundary and loading conditions in \figref{numExViscPureBendSetup} work well to obtain stretch-free bending deformations. The relative curvature error is of the same magnitude. At $t=t_\mathrm{end}$, the sheet is deformed into a half circle exhibiting large bending deformations. The applied loads over time are visualized in \figref{numExViscPureBendLoading} for this example.
\begin{figure}[!ht]
	\setlength{\figwidth}{0.29\textwidth}
	\setlength{\figheight}{0.29\textwidth}
	\centering
	\subfloat{
		\begin{tikzpicture}
			\def\cdot{\times}
			\begin{axis}[,grid=both,xlabel={Time $t$ $[T_0]$},ylabel={Moment $M$ $[c\,L_0^{-1}]$},width=\figwidth,height=\figheight,
			xmin=0,xmax=1,ymin=0,ymax=1.5,
			xtick={0,0.25,0.5,0.75,1},
			minor ytick={0.25,0.75,1.25},
			tick label style={font=\footnotesize},
			]
				\addplot[blue,line width=1]table [x index = {0}, y index = {1},col sep=comma]{fig/pureBend/pureBendLoadings.csv};

			\end{axis}
		\end{tikzpicture}
	}
	\quad
		\subfloat{
		\begin{tikzpicture}
			\def\cdot{\times}
			\begin{axis}[grid=both,xlabel={Time $t$ $[T_0]$},ylabel={Displacement $\bar{u}_y$ $[L_0]$},width=\figwidth,height=\figheight,
			tick label style={font=\footnotesize},
			xmin=0,xmax=1,ymin=-1.2,ymax=0,
			xtick={0,0.25,0.5,0.75,1},
			minor ytick={-0.25,-0.75},
			]
				\addplot[blue,line width=1]table [x index = {0}, y index = {2},col sep=comma]{fig/pureBend/pureBendLoadings.csv};
			\end{axis}
		\end{tikzpicture}
	}
	\quad
		\subfloat{
		\begin{tikzpicture}
			\def\cdot{\times}
			\begin{axis}[grid=both,xlabel={Time $t$ $[T_0]$},ylabel={Pressure $p$ $[c\,L_0^{-3}]$},width=\figwidth,height=\figheight,
			tick label style={font=\footnotesize},
			xmin=0,xmax=1,
			ymin=-1.5,ymax=0,
			xtick={0,0.25,0.5,0.75,1},
			minor ytick={-0.25,-0.75,-1.25},
			]
				\addplot[blue,line width=1]table [x index = {0}, y index = {3},col sep=comma]{fig/pureBend/pureBendLoadings.csv};

			\end{axis}
		\end{tikzpicture}
	}
	\caption{Pure bending of a flat strip: Imposed loads over time, see also \figref{numExViscPureBendSetup}.} \figlabel{numExViscPureBendLoading}
\end{figure}
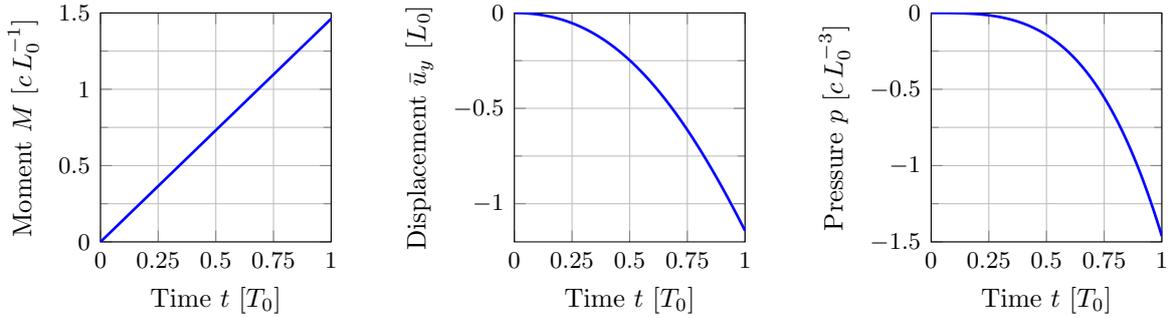

\figref{numExViscPureStudyEtaB} shows the total and inelastic curvatures, $\kappa_2$ and $\kappa_2^\mrin$, respectively, over time for different values of $\eta_\mrb$. While the total curvature is not influenced significantly by $\eta_\mrb$, the curvature of the intermediate configuration shows a strong dependency on $\eta_\mrb$. For smaller values of $\eta_\mrb$, the inelastic curvature increases, i.e.~the intermediate surface is bent more.
\begin{figure}[!ht]
	\setlength{\figwidth}{0.45\textwidth}
	\setlength{\figheight}{0.35\textwidth}
	\centering
	\subfloat{
		\begin{tikzpicture}
			\def\cdot{\times}
			\begin{axis}[grid=both,xlabel={Time $t$ $[T_0]$},ylabel={Curvature $\kappa_2$ $[L_0^{-1}]$},width=\figwidth,height=\figheight,
			xmin=0,xmax=1,ymin=0,ymax=1,minor xtick={0.1,0.3,0.5,0.7,0.9},minor ytick={0.1,0.3,0.5,0.7,0.9},
			tick label style={font=\footnotesize},legend pos = north west,legend style={nodes={scale=0.75, transform shape}},legend cell align={left}
			]
				\addplot[black,solid,line width=1]table [x index = {0}, y index = {1},col sep=comma]{fig/pureBend/pureBendStudyEtabAnalytical.csv};
					\addlegendentry{$\eta_\mrb=0.1\,c\,T_0$};
				%
				\addplot[red,densely dashed,line width=1]table [x index = {0}, y index = {2},col sep=comma]{fig/pureBend/pureBendStudyEtabAnalytical.csv};
					\addlegendentry{$\eta_\mrb=0.5\,c\,T_0$};
				%
				\addplot[blue,densely dotted,line width=1]table [x index = {0}, y index = {3},col sep=comma]{fig/pureBend/pureBendStudyEtabAnalytical.csv};
					\addlegendentry{$\eta_\mrb=0.9\,c\,T_0$};
			\end{axis}
		\end{tikzpicture}
	}
	\subfloat{
		\begin{tikzpicture}
			\def\cdot{\times}
			\begin{axis}[grid=both,xlabel={Time $t$ $[T_0]$},ylabel={Inelastic curvature $\kappa_2^\mrin$ $[L_0^{-1}]$},width=\figwidth,height=\figheight,
			xmin=0,xmax=1,ymin=0,ymax=0.9,minor xtick={0.1,0.3,0.5,0.7,0.9},minor ytick={0.1,0.3,0.5,0.7,0.9},
			tick label style={font=\footnotesize},legend pos = north west,legend style={nodes={scale=0.75, transform shape}},legend cell align={left}
			]
				\addplot[black,solid,line width=1]table [x index = {0}, y index = {4},col sep=comma]{fig/pureBend/pureBendStudyEtabAnalytical.csv};
					\addlegendentry{$\eta_\mrb=0.1\,c\,T_0$};
				\addplot[black,only marks,mark=o,mark size=1.5,line width=.5,forget plot]table [x index = {0}, y index = {4},col sep=comma]{fig/pureBend/pureBendStudyEtabNumerical.csv};
				\addplot[red,densely dashed,line width=1]table [x index = {0}, y index = {5},col sep=comma]{fig/pureBend/pureBendStudyEtabAnalytical.csv};
					\addlegendentry{$\eta_\mrb=0.5\,c\,T_0$};
				\addplot[red,only marks,mark=o,mark size=1.5,line width=.5,forget plot]table [x index = {0}, y index = {5},col sep=comma]{fig/pureBend/pureBendStudyEtabNumerical.csv};
				\addplot[blue,densely dotted,line width=1]table [x index = {0}, y index = {6},col sep=comma]{fig/pureBend/pureBendStudyEtabAnalytical.csv};
					\addlegendentry{$\eta_\mrb=0.9\,c\,T_0$};
				\addplot[blue,only marks,mark=o,mark size=1.5,line width=.5,forget plot]table [x index = {0}, y index = {6},col sep=comma]{fig/pureBend/pureBendStudyEtabNumerical.csv};
			\end{axis}
		\end{tikzpicture}
	}
	\caption{Pure bending of a flat strip: Influence of the parameter $\eta_\mrb$ on the total and inelastic curvatures, $\kappa_2$ and $\kappa_2^\mrin$, respectively. The circles mark the numerical results at various snapshots in time, while the lines show the corresponding analytical results. Given $\kappa_2$ and $\kappa_2^\mrin$, the elastic curvature follows from $\kappa_2=\kappa_2^\mrel+\kappa_2^\mrin$, which is a consequence of \eqsref{cntViscKinCurvatureSplit}.} \figlabel{numExViscPureStudyEtaB}
\end{figure}
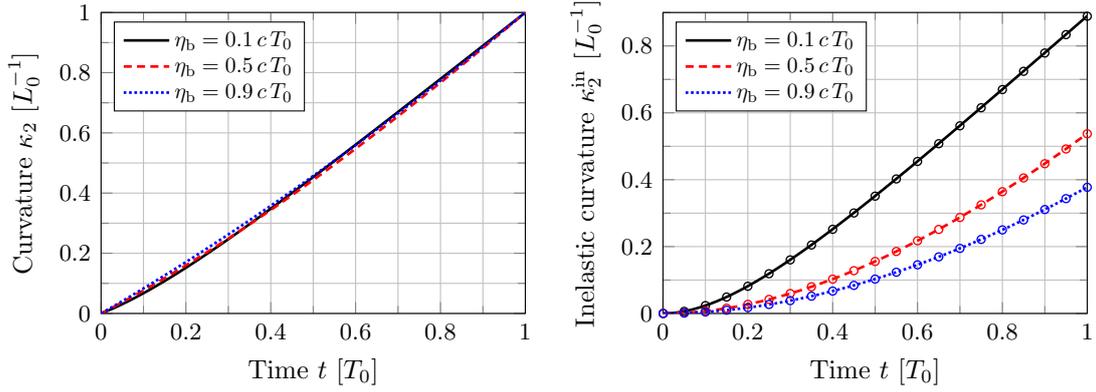

\subsection{Inflated spherical shell} \seclabel{numExViscInflSphr}
This section considers a similar example as in \secref{numExViscInflMem}, but here, also bending resistance is considered. The setup is the same as in \figref{numExViscInflMemSetup} and the parameters $\lambda_\mathrm{end}=4^{1/3}$ and $t_\mathrm{end}=1\,T_0$ are used, see also \eqsref{numExViscInflMemLoadingRate}. The final volume is thus four times as large compared to the initial volume, i.e.~$V(t_\mathrm{end})=4\,V_0$. The employed thin shell formulation requires $C^1$-continuity of the numerical discretization, such that patch constraints need to be enforced along the marked patch interfaces $\Gamma$ in \figref{numExViscInflMemSetupGeometry}. A detailed derivation for the enforcement of patch constraints is provided in \cite{paul2020a}. Here, the Lagrange multiplier method with element-wise constant interpolation is employed to enforce the constraint
\eqb{l}
	\bg_\mrn^\mathrm{planar}=\bn-\tilde\bn=\boldsymbol{0}\,,\quad\forall\,\bx\in\Gamma\,, \eqlabel{numExViscInflSphrPatchConstraint}
\eqe
along the patch interfaces $\Gamma$. Here, $\bn$ and $\tilde\bn$ denote the two surface normals of the elements adjacent to $\Gamma$. Further, the symmetry across the gray marked symmetry planes is enforced by the constraint in \eqsref{numExViscInflSphrPatchConstraint}. In that case, $\tilde{\bn}$ denotes the normal of the symmetry plane.

In the elastic branch, the membrane energy density is given by the incompressible Neo-Hookean material model from \eqsref{cntViscConstElasticitySigmaABincompressibleNeoHookean}, and the bending energy density is given by the Helfrich model from \eqsref{cntViscConstElasticityPsiBendHelfrich}. For the Maxwell branch, the Neo-Hookean model from \eqsref{cntViscConstElasticityPsiMemClassicalNeoHookean} with $\Lambda=0$ and Koiter bending model from \eqsref{cntViscConstElasticityPsiBendKoiter} are employed.

Similar to \secref{numExViscInflMem}, the pressure is composed of two contributions, i.e.
\eqb{l}
	p(t) = p_\mrel(t) + p_\mrvisc(t)\,, \eqlabel{numExViscInflSphrPressureTotal}
\eqe
where $p_\mrel(t)$ is the pressure function coming from the elastic branch, and $p_\mrvisc(t)$ is the one from the Maxwell branch. These two contributions $p_\bullet(t)$ are derived in detail in \appref{viscAnalyticalSolutionsInflSphr}, and they are given by
\eqb{l}
	p_\mrel(t)=\dfrac{2}{R^3}\,\Biggl[\mu\,R\,\biggl(\dfrac{1}{\lambda}-\dfrac{1}{\lambda^7}\biggr)+k\,\biggl(\dfrac{H_0\,R}{\lambda^2}+\dfrac{H_0^2\,R^2}{\lambda}\biggr)\Biggr]\,. \eqlabel{numExViscInflSphrPressureElastic}
\eqe
and
\eqb{l}
	p_\mrvisc(t)=\dfrac{2}{R^3}\,\Biggl[\mu_1\,R^2\,\biggl(\dfrac{1}{\lambda}-\dfrac{1}{\lambda^3\,\hat{a}_\mathrm{ev}}\biggr)+c_1\,\hat{a}_\mathrm{ev}\,\biggl(\dfrac{1}{\lambda}-\dfrac{\hat{b}_\mathrm{ev}}{\lambda^2}\biggr)\Biggr]\,, \eqlabel{numExViscInflSphrPressureViscous}
\eqe
with $\hat{a}_\mathrm{ev}(t)$ given in \eqsref{numExViscInflMemHataEv} and $\hat{b}_\mathrm{ev}(t)$ given by
\eqb{l}
	\hat{b}_\mathrm{ev}(t):=\dfrac{c_1\,\tau_\lambda\,\exp\bigl(t/\tau_\lambda\bigr)+\eta_\mrb\,\exp\bigl(-c_1\,t/\eta_\mrs\bigr)}{\eta_\mrb+c_1\,\tau_\lambda}\,. \eqlabel{numExViscInflSphrHatbEv}
\eqe

First, the convergence of the model with respect to time step size refinement is investigated. The parameters are $\mu=5\,k/R^2$, $\mu_1=\mu$, $c_1=k$, $k^\star=0$, $\eta_\mrs=0.5\,\mu_0\,T_0$, $\eta_\mrb=0.5\,k\,T_0$, $H_0=1/R$, and $t_\mathrm{end}=1\,T_0$. \figref{numExViscInflSphrConvergence} shows the convergence of the pressure error $\epsilon_p$, see \eqsref{numExViscInflMemPressureError}, over the time step size $\Delta t$. As expected for the implicit Euler scheme, the error decreases linearly with decreasing time step size. There is no significant difference between the considered meshes for $m\geq2$, such that the mesh $m=2$ is used for the subsequent examples.
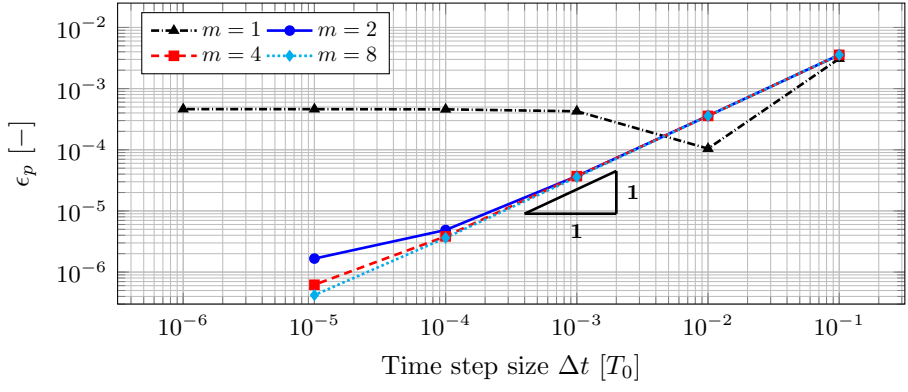
\begin{figure}[!ht]
	\setlength{\figwidth}{0.75\textwidth}
	\setlength{\figheight}{0.35\textwidth}
	\centering
		\begin{tikzpicture}
			\def\cdot{\times}
			\begin{axis}[xmode=log,ymode=log,grid=both,xlabel={Time step size $\Delta t$ $[T_0]$},ylabel={$\epsilon_p$ $[-]$},width=\figwidth,height=\figheight,
			tick label style={font=\footnotesize},
			ymin=3e-7,ymax=0.25e-1,
			legend pos=north west,legend style={nodes={scale=0.75, transform shape}},legend cell align={left},legend columns=2,
			]
				\addplot[black,densely dash dot,line width=1,mark=triangle*,mark size=1.5,mark options={solid}]table [x index = {0}, y index = {1},col sep=comma]{fig/inflSphr/inflSphrConvergenceTimeStepSize.csv};
					\addlegendentry{$m=1$};
				\addplot[blue,solid,line width=1,mark=*,mark size=1.5,mark options={solid}]table [x index = {0}, y index = {2},col sep=comma,]{fig/inflSphr/inflSphrConvergenceTimeStepSize.csv};
					\addlegendentry{$m=2$};
				\addplot[red,densely dashed,line width=1,mark=square*,mark size=1.5,mark options={solid}]table [x index = {0}, y index = {3},col sep=comma,]{fig/inflSphr/inflSphrConvergenceTimeStepSize.csv};
					\addlegendentry{$m=4$};
				\addplot[cyan,densely dotted,line width=1,mark=diamond*,mark size=1.5,mark options={solid}]table [x index = {0}, y index = {4},col sep=comma,]{fig/inflSphr/inflSphrConvergenceTimeStepSize.csv};
					\addlegendentry{$m=8$};
			\addplot[black,solid,line width=1,mark=none,domain=4e-4:2e-3,samples=2]{0.0225*x^1};
			\draw[black,solid,line width=1] (4e-4,9e-6) -- (2e-3,9e-6); \node[anchor=north] at (1e-3,9e-6) {\scriptsize $\boldsymbol{1}$};
			\draw[black,solid,line width=1] (2e-3,9e-6) -- ( 2e-3,4.5e-05); \node[anchor=west] at (2e-3,2e-5) {\scriptsize $\boldsymbol{1}$};
			\end{axis}
		\end{tikzpicture}
	\caption{Inflated spherical shell: Convergence of the pressure error $\epsilon_p$, see \eqsref{numExViscInflMemPressureError}, over time step size refinement.} \figlabel{numExViscInflSphrConvergence}
\end{figure}

For the subsequent results, the parameters as given above are used if not stated otherwise. The time step size is set to $\Delta t=10^{-3}\,T_0$. \figref{numExViscInflSphrStudyEtaSetaB} shows the resulting pressure over time relation for different values of $\eta_\mrs$ and $\eta_\mrb$, and two different values of $H_0\in\{0,1/R\}$. With increasing $\eta_\mrs$ and decreasing $\eta_\mrb$, the total pressure increases. Further, the pressure is larger for $H_0=1/R$ than for $H_0=0$.
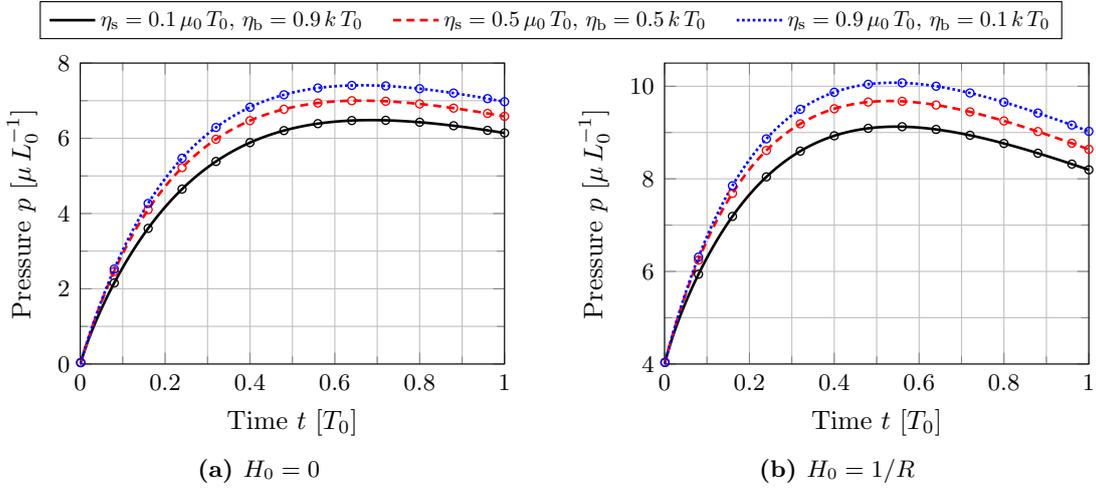
\begin{figure}[!ht]
	\setlength{\figwidth}{0.45\textwidth}
	\setlength{\figheight}{0.35\textwidth}
	\centering
	\begin{tikzpicture}
		\pgfplotsset{width=1\textwidth,height=0.2\textwidth,compat=newest,}
		\begin{axis}[hide axis,xmin=0,xmax=0.00001,ymin=0,ymax=0.00001,legend cell align={left},
					 legend columns=3,legend style={/tikz/every even column/.append style={column sep=2ex}},
					 tick label style={font=\footnotesize},legend style={nodes={scale=0.75, transform shape}},legend cell align={left}]
  			  	\addlegendimage{black,solid,line width=1,}
  			  	\addlegendentry{$\eta_\mrs=0.1\,\mu_0\,T_0$, $\eta_\mrb=0.9\,k\,T_0$};
  			  	\addlegendimage{red,densely dashed,line width=1,}
  			  	\addlegendentry{$\eta_\mrs=0.5\,\mu_0\,T_0$, $\eta_\mrb=0.5\,k\,T_0$};
  			  	\addlegendimage{blue,densely dotted,line width=1,}
  			  	\addlegendentry{$\eta_\mrs=0.9\,\mu_0\,T_0$, $\eta_\mrb=0.1\,k\,T_0$};
  			 \end{axis}
	\end{tikzpicture}
	\\ \vspace{-3mm}
	\subfloat[$H_0=0$]{
		\begin{tikzpicture}
			\def\cdot{\times}
			\begin{axis}[grid=both,xlabel={Time $t$ $[T_0]$},ylabel={Pressure $p$ $[\mu\,L_0^{-1}]$},width=\figwidth,height=\figheight,
			xmin=0,xmax=1,ymin=0,ymax=8,minor xtick={0.1,0.3,0.5,0.7,0.9},minor ytick={1,3,5,7},
			tick label style={font=\footnotesize},
			]
				\addplot[black,solid,line width=1]table [x index = {0}, y index = {3},col sep=comma]{fig/inflSphr/inflSphrEtaSetaBanalytical.csv};
				\addplot[black,only marks,mark=o,mark size=1.5,line width=.5,forget plot]table [x index = {0}, y index = {3},col sep=comma]{fig/inflSphr/inflSphrEtaSetaBnumerical.csv};
				\addplot[red,densely dashed,line width=1]table [x index = {0}, y index = {1},col sep=comma]{fig/inflSphr/inflSphrEtaSetaBanalytical.csv};
				\addplot[red,only marks,mark=o,mark size=1.5,line width=.5,forget plot]table [x index = {0}, y index = {1},col sep=comma]{fig/inflSphr/inflSphrEtaSetaBnumerical.csv};
				\addplot[blue,densely dotted,line width=1]table [x index = {0}, y index = {2},col sep=comma]{fig/inflSphr/inflSphrEtaSetaBanalytical.csv};
				\addplot[blue,only marks,mark=o,mark size=1.5,line width=.5,forget plot]table [x index = {0}, y index = {2},col sep=comma]{fig/inflSphr/inflSphrEtaSetaBnumerical.csv};
			\end{axis}
		\end{tikzpicture}
	}
	\quad
		\subfloat[$H_0=1/R$]{
		\begin{tikzpicture}
			\def\cdot{\times}
			\begin{axis}[grid=both,xlabel={Time $t$ $[T_0]$},ylabel={Pressure $p$ $[\mu\,L_0^{-1}]$},width=\figwidth,height=\figheight,
			xmin=0,xmax=1,ymin=4,ymax=10.5,minor xtick={0.1,0.3,0.5,0.7,0.9},minor ytick={5,7,9,11},
			tick label style={font=\footnotesize},
			]
				\addplot[black,solid,line width=1]table [x index = {0}, y index = {6},col sep=comma]{fig/inflSphr/inflSphrEtaSetaBanalytical.csv};
				\addplot[black,only marks,mark=o,mark size=1.5,line width=.5,forget plot]table [x index = {0}, y index = {6},col sep=comma]{fig/inflSphr/inflSphrEtaSetaBnumerical.csv};
				\addplot[red,densely dashed,line width=1]table [x index = {0}, y index = {4},col sep=comma]{fig/inflSphr/inflSphrEtaSetaBanalytical.csv};
				\addplot[red,only marks,mark=o,mark size=1.5,line width=.5,forget plot]table [x index = {0}, y index = {4},col sep=comma]{fig/inflSphr/inflSphrEtaSetaBnumerical.csv};
				\addplot[blue,densely dotted,line width=1]table [x index = {0}, y index = {5},col sep=comma]{fig/inflSphr/inflSphrEtaSetaBanalytical.csv};
				\addplot[blue,only marks,mark=o,mark size=1.5,line width=.5,forget plot]table [x index = {0}, y index = {5},col sep=comma]{fig/inflSphr/inflSphrEtaSetaBnumerical.csv};
			\end{axis}
		\end{tikzpicture}
	}
	\caption{Inflated spherical shell: Influence of the in-plane shear viscosity $\eta_\mrs$ and out-of-plane viscosity $\eta_\mrb$ on the pressure $p$ for two different values of $H_0$. The circles mark the numerical results at various snapshots in time, while the lines show the corresponding analytical results.} \figlabel{numExViscInflSphrStudyEtaSetaB}
\end{figure}

\subsection{Sagging Scordelis-Lo roof} \seclabel{numExViscScordelisLos}
This section presents a viscoelastic shell that exhibits inhomogeneous deformations. The geometry, loading and boundary conditions are visualized in \figref{numExViscScordelisLoSetup}. The setup corresponds to the Scordelis-Lo roof \citep{macneal1985}.
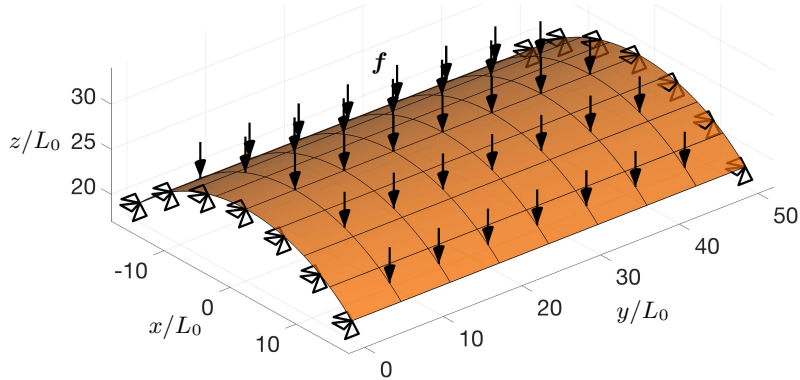
\begin{figure}[!ht]
	\setlength{\figwidth}{0.6\textwidth}
	\centering
		\begin{tikzpicture}
			\node[align=center] at (0,0) {\includegraphics[width=\figwidth]{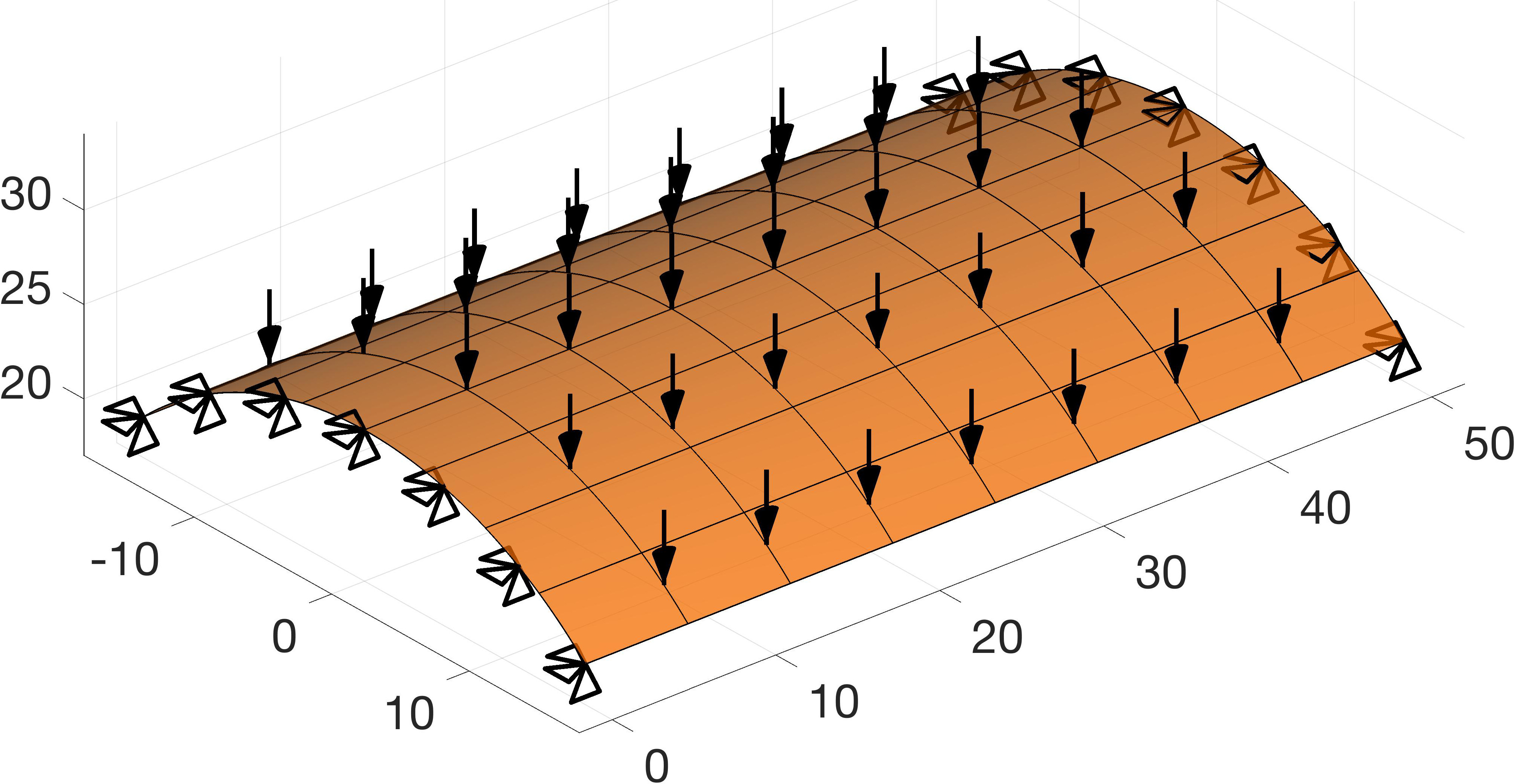}};
			\node[align=center,font=\footnotesize] at (-3.4,-1.8) {$x/L_0$};
			\node[align=center,font=\footnotesize] at (2.75,-1.6) {$y/L_0$};
			\node[align=center,font=\footnotesize] at (-5.25,0.65) {$z/L_0$};
			\node[align=center,font=\footnotesize] at (-0.72,1.7) {$\bff$};
		\end{tikzpicture}
	\caption{Sagging Scordelis-Lo roof: Geometry, loading and boundary conditions.} \figlabel{numExViscScordelisLoSetup}
\end{figure}

The following time-dependent load is applied
\eqb{l}
	\bff(t) = f_v(t)\begin{bmatrix}0\\0\\-1\end{bmatrix}\,,\quad\mathrm{where}\quad f_v(t):=\dfrac{1}{25}\begin{cases}\dfrac{f_0}{t_0}\,t\,, &t\leq t_0\\f_0\,,&t>t_0\end{cases}\,,
\eqe
with $t_0=10\,T_0$ and $f_0=1\,\mu_0/L_0$. The Neo-Hookean material model from \eqsref{cntViscConstElasticityPsiMemNeoHookean} is used for the membrane response in the elastic and Maxwell branches, and the Koiter model from \eqsref{cntViscConstElasticityPsiBendKoiter} is employed for the bending response in both branches. 
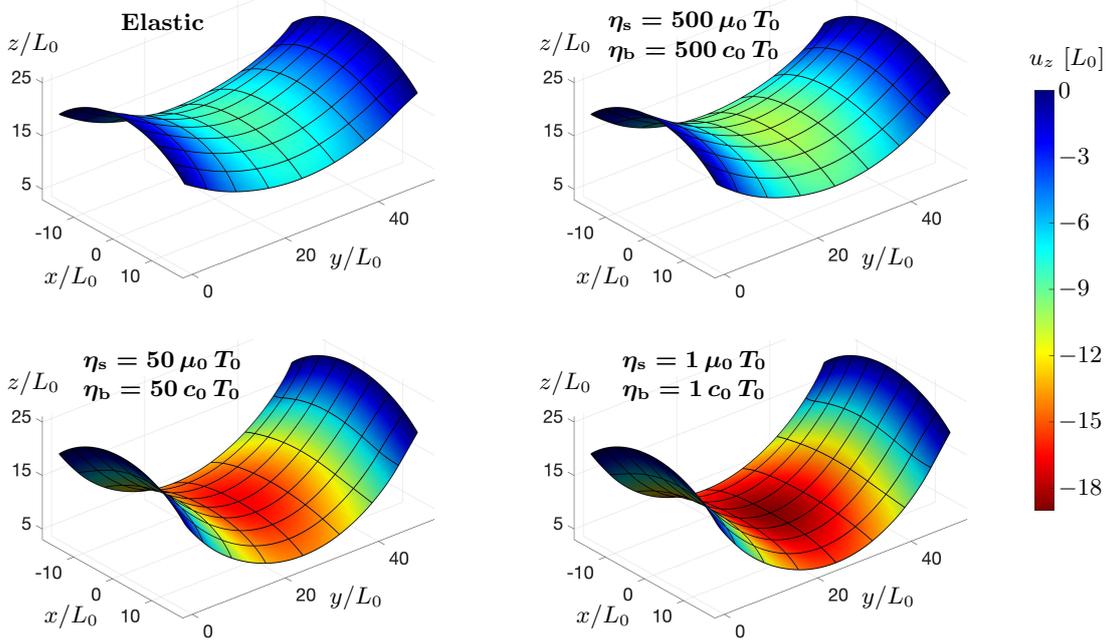
\begin{figure}[!ht]
	\setlength{\figwidth}{0.35\textwidth}
	\centering
	\begin{tikzpicture}
		\tikzmath{
			\y2 = 10.85;     
			\y1 = -7.55;     
			\x1 = -4.49;	 
			\x2 = 0.78;	 
		};
			
			\node[align=center] at (0,0) {\includegraphics[width=\figwidth]{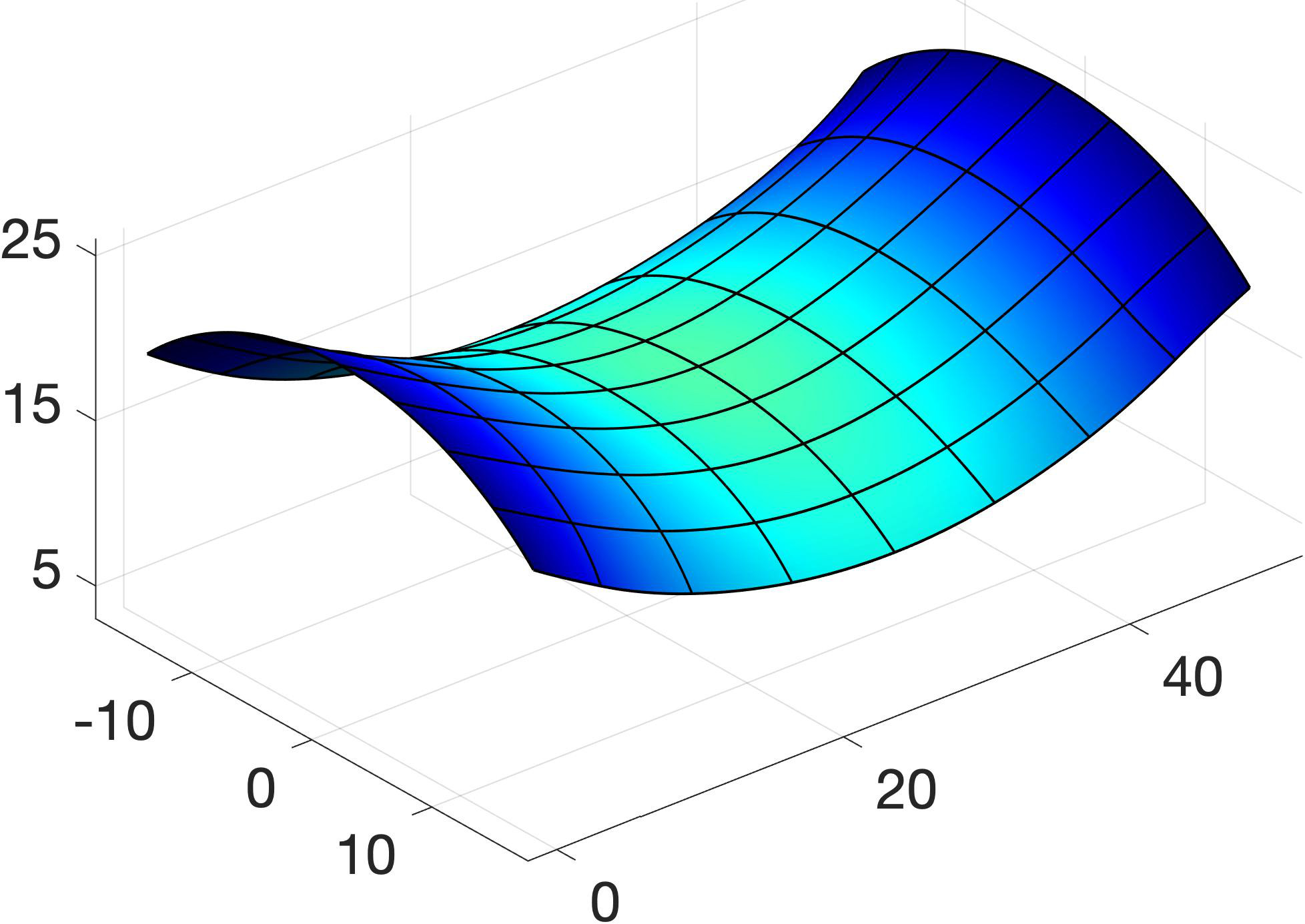}};
			\node[align=center,font=\footnotesize] at (-2,-1.7) {$x/L_0$};
			\node[align=center,font=\footnotesize] at (1.75,-1.45) {$y/L_0$};
			\node[align=center,font=\footnotesize] at (-2.5,1.4) {$z/L_0$};
			\node[align=left,font=\footnotesize] at (-0.8,1.5) {\textbf{Elastic}\\};
			\node[align=center] at (7,0) {\includegraphics[width=\figwidth]{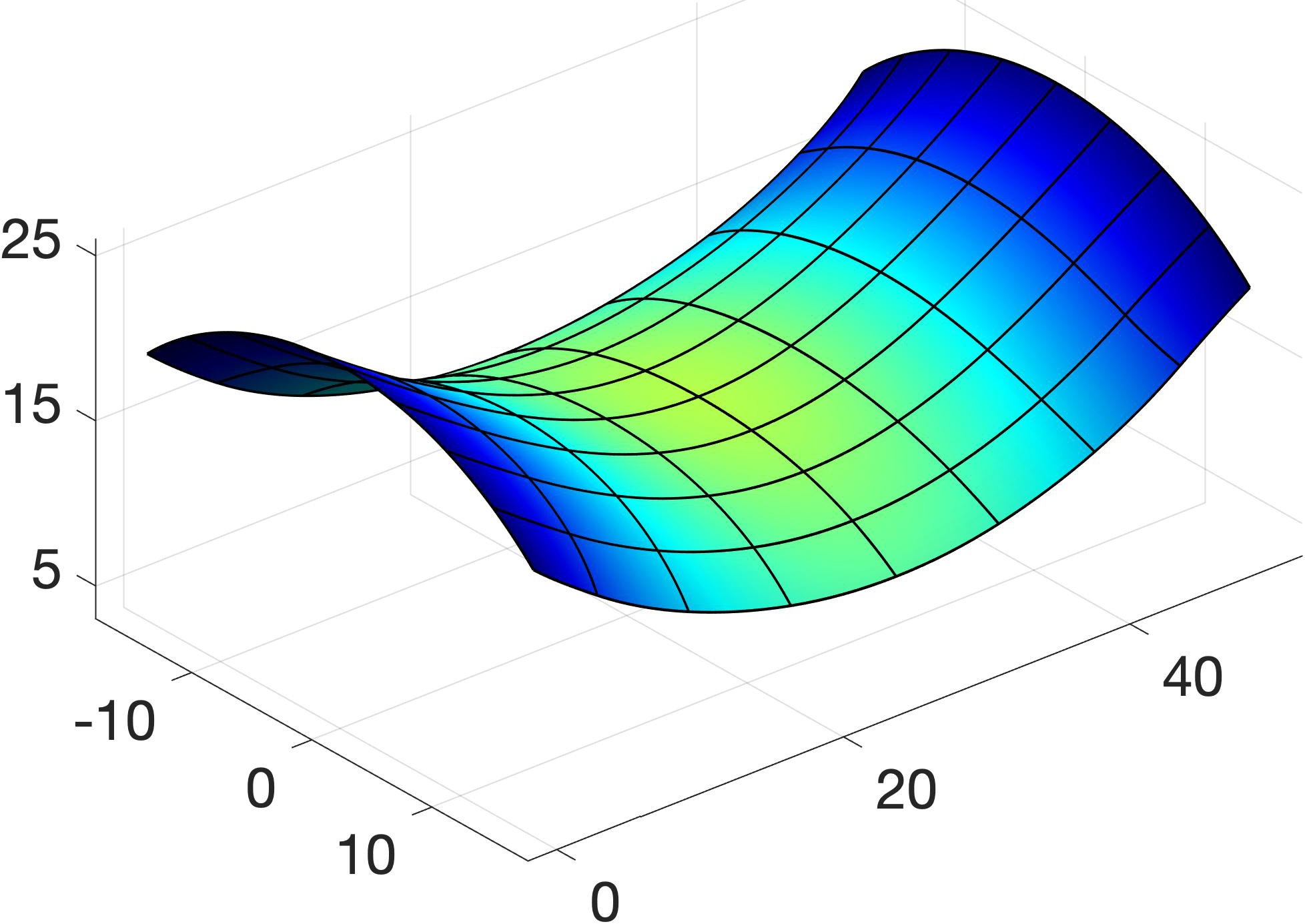}};
			\node[align=center,font=\footnotesize] at (5,-1.7) {$x/L_0$};
			\node[align=center,font=\footnotesize] at (8.75,-1.45) {$y/L_0$};
			\node[align=center,font=\footnotesize] at (4.5,1.4) {$z/L_0$};
			\node[align=left,font=\footnotesize] at (6.2,1.5) {$\boldsymbol{\eta_\mrs=500\,\mu_0\,T_0}$\\$\boldsymbol{\eta_\mrb=500\,c_0\,T_0}$};
			\node[align=center] at (0,-4.5) {\includegraphics[width=\figwidth]{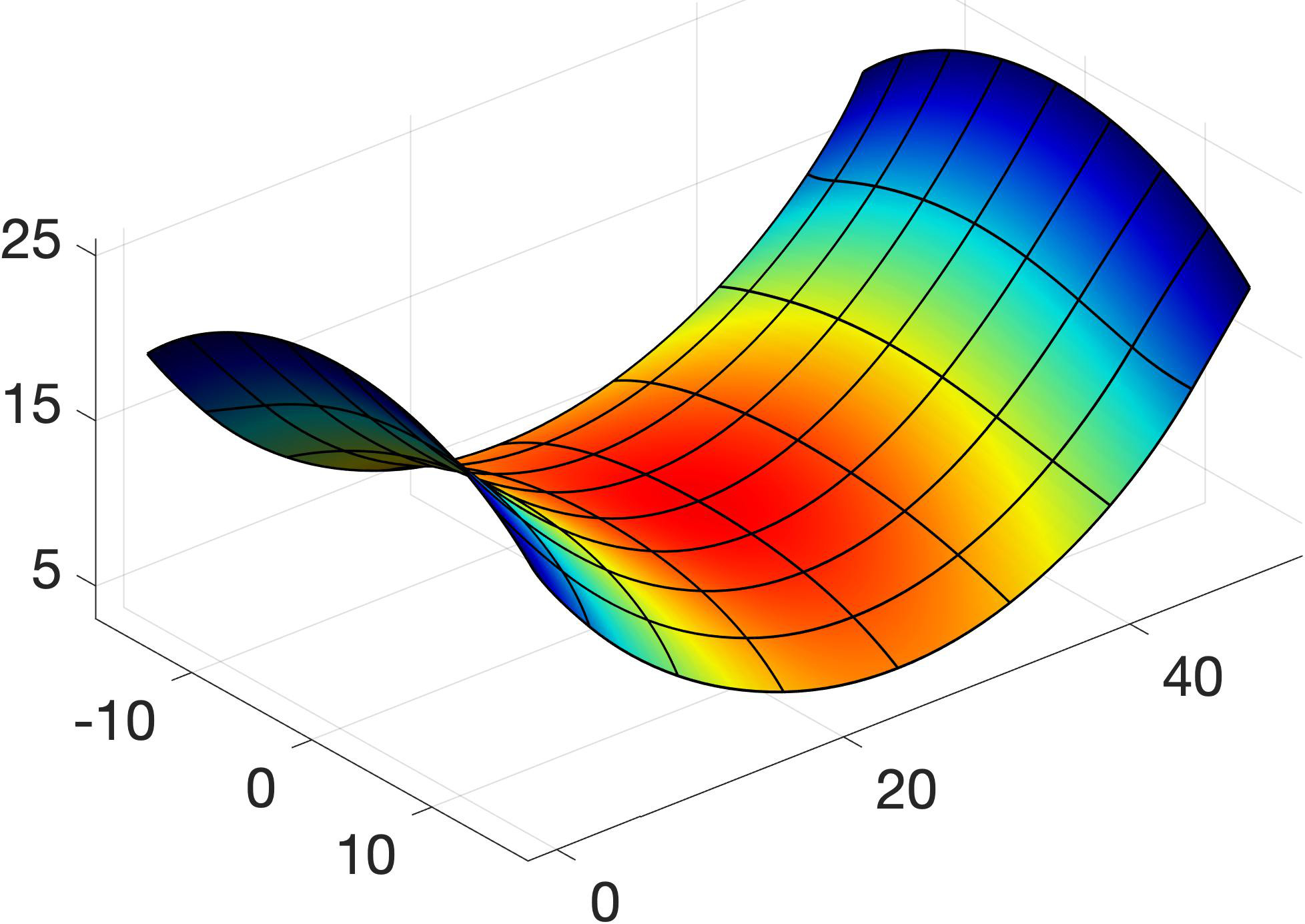}};
			\node[align=center,font=\footnotesize] at (-2,-6.2) {$x/L_0$};
			\node[align=center,font=\footnotesize] at (1.75,-5.95) {$y/L_0$};
			\node[align=center,font=\footnotesize] at (-2.5,-3.1) {$z/L_0$};
			\node[align=left,font=\footnotesize] at (-0.8,-3) {$\boldsymbol{\eta_\mrs=50\,\mu_0\,T_0}$\\$\boldsymbol{\eta_\mrb=50\,c_0\,T_0}$};
			\node[align=center] at (7,-4.5) {\includegraphics[width=\figwidth]{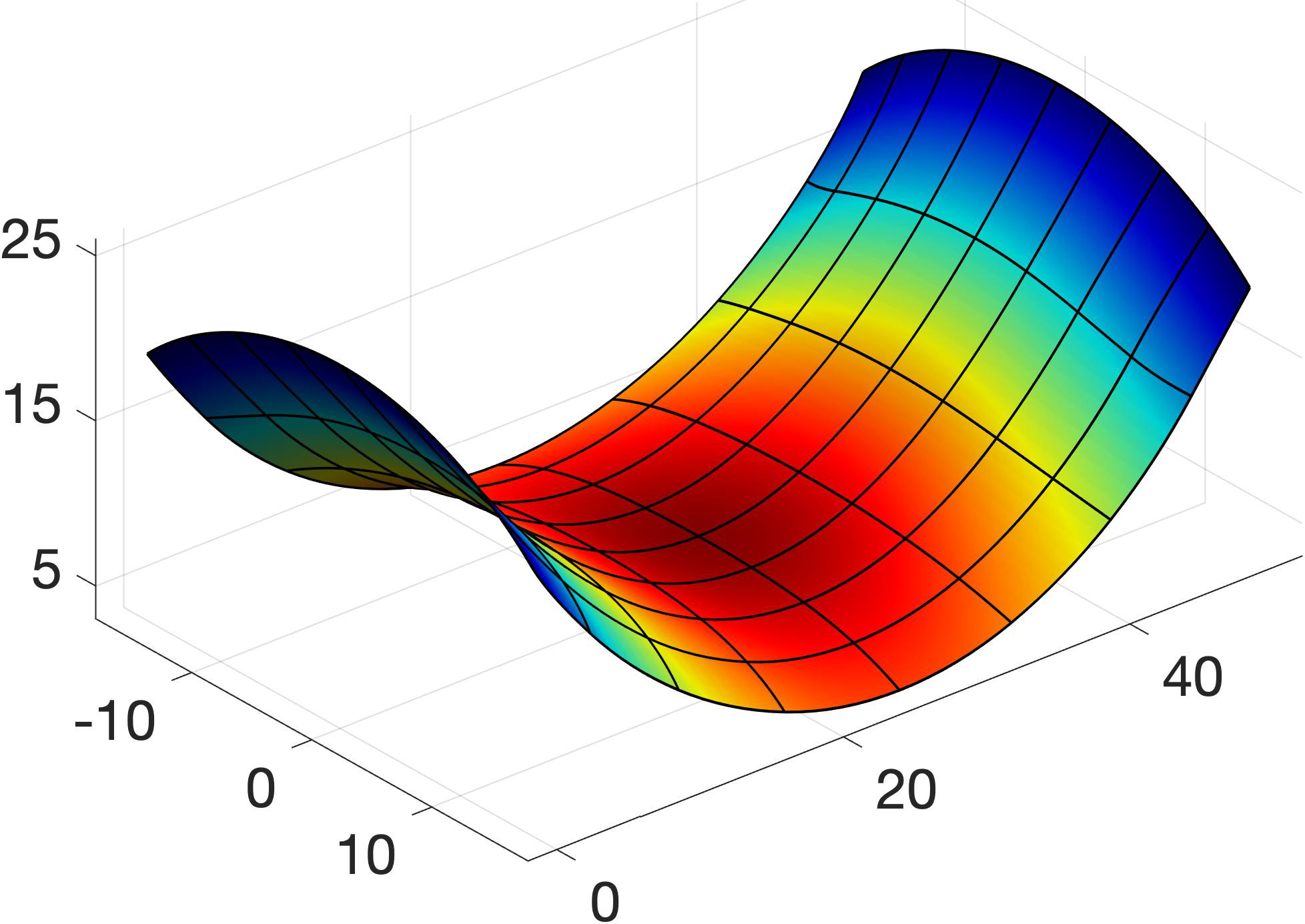}};
			\node[align=center,font=\footnotesize] at (5,-6.2) {$x/L_0$};
			\node[align=center,font=\footnotesize] at (8.75,-5.95) {$y/L_0$};
			\node[align=center,font=\footnotesize] at (4.5,-3.1) {$z/L_0$};
			\node[align=left,font=\footnotesize] at (6.2,-3) {$\boldsymbol{\eta_\mrs=1\,\mu_0\,T_0}$\\$\boldsymbol{\eta_\mrb=1\,c_0\,T_0}$};
			\node[align=center] at (10.8,-2) {\includegraphics[height=0.35\textwidth]{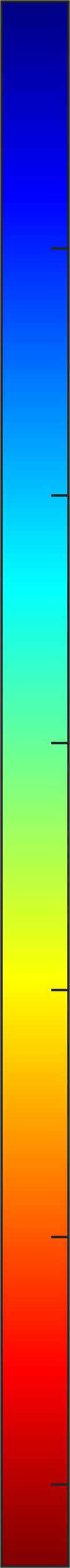}};
			\node[align=left,anchor=west,font=\footnotesize] at (\y2,{\x1+(\x2-\x1)/6*0}) {$-18$};		
			\node[align=left,anchor=west,font=\footnotesize] at (\y2,{\x1+(\x2-\x1)/6*1}) {$-15$};		
			\node[align=left,anchor=west,font=\footnotesize] at (\y2,{\x1+(\x2-\x1)/6*2}) {$-12$};
			\node[align=left,anchor=west,font=\footnotesize] at (\y2,{\x1+(\x2-\x1)/6*3}) {$-9$};		
			\node[align=left,anchor=west,font=\footnotesize] at (\y2,{\x1+(\x2-\x1)/6*4}) {$-6$};		
			\node[align=left,anchor=west,font=\footnotesize] at (\y2,{\x1+(\x2-\x1)/6*5}) {$-3$};		
			\node[align=left,anchor=west,font=\footnotesize] at (\y2,{\x1+(\x2-\x1)/6*6}) {$0$};		
			\node[align=left,font=\footnotesize] at (11.1,1.2) {$u_z$ $[L_0]$};	
	\end{tikzpicture}
	\caption{Sagging Scordelis-Lo roof: Final deformation for the elastic and viscoelastic case for three different values of $\eta_\mrs$ and $\eta_\mrb$. The surfaces are colored by the vertical displacement $u_z$.} \figlabel{numExViscScordelisLoDeformation}
\end{figure}
For the elastic case, the material parameters $\mu=10\,\mu_0$, $K=10\,\mu_0$, $c=10\,\mu_0\,L_0^2$, and $\mu_1=K_1=c_1=0$ are used. For the viscoelastic case, the material parameters $\mu=2\,\mu_0$, $K=2\,\mu_0$, $c=2\,\mu_0\,L_0^2$, $\mu_1=8\,\mu_0$, $K_1=8\,\mu_0$, and $c_1=8\,\mu_0\,L_0^2$ are used. Further, $c_0:=\mu_0\,L_0^2$. The finite element mesh is constructed from $8$ elements in each direction, the end time is $t_\mathrm{end}=50\,T_0$, and $1,000$ time steps are used.

The deformed structure is visualized in \figref{numExViscScordelisLoDeformation} for the elastic and viscoelastic case using different values for $\eta_\mrs$ and $\eta_\mrb$. As shown, the creep is considerably larger for smaller $\eta_\mrs$ and $\eta_\mrb$.

\figref{numExViscScordelisLoDisplacementSurfaceStretch} shows the vertical displacement $u_z$ and the surface stretch $J$ over time. There are large creep deformations, which happen faster for smaller values of $\eta_\mrs$ and $\eta_\mrb$. The decomposition of the surface stretch into its inelastic and elastic components, see Eqs.~\eqref{e:cntViscKinSurfaceStretchSplit}--\eqref{e:cntViscKinSurfaceStretchSplitComponents}, is visualized in \figref{numExViscScordelisLoSurfaceStretchComponents}.
\begin{figure}[!ht]
	\setlength{\figwidth}{0.45\textwidth}
	\setlength{\figheight}{0.35\textwidth}
	\centering
		\begin{tikzpicture}
		\pgfplotsset{width=1\textwidth,height=0.2\textwidth,compat=newest,}
		\begin{axis}[hide axis,xmin=0,xmax=0.00001,ymin=0,ymax=0.00001,legend cell align={left},
					 legend columns=4,legend style={/tikz/every even column/.append style={column sep=2ex}},
					 tick label style={font=\footnotesize},legend style={nodes={scale=0.75, transform shape}},legend cell align={left}]
  			  	\addlegendimage{brown,densely dash dot,line width=1,}
  			  		\addlegendentry{Elastic};
  			  	\addlegendimage{black,solid,line width=1,}
  			  	\addlegendentry{$\eta_\mrs=1\,\mu_0\,T_0$, $\eta_\mrb=1\,c_0\,T_0$};
  			  	\addlegendimage{red,densely dashed,line width=1,}
  			  	\addlegendentry{$\eta_\mrs=50\,\mu_0\,T_0$, $\eta_\mrb=50\,c_0\,T_0$};
  			  	\addlegendimage{blue,densely dotted,line width=1,}
  			  	\addlegendentry{$\eta_\mrs=500\,\mu_0\,T_0$, $\eta_\mrb=500\,c_0\,T_0$};
  			 \end{axis}
	\end{tikzpicture}
	\\ \vspace{-3mm}
	\subfloat[Vertical displacement\figlabel{numExViscScordelisLoDisplacement}]{
		\begin{tikzpicture}
			\def\cdot{\times}
			\begin{axis}[grid=both,xlabel={Time $t$ $[T_0]$},ylabel={Displacement $u_z$ $[L_0]$},width=\figwidth,height=\figheight,
			xmin=0,xmax=50,ymin=-20,ymax=0,minor xtick={5,15,25,35,45},minor ytick={-2.5,-7.5,-12.5,-17.5},
			tick label style={font=\footnotesize},legend pos = north west,legend style={nodes={scale=0.75, transform shape}},legend cell align={left}
			]
				\addplot[brown,densely dash dot,line width=1]table [x index = {0}, y index = {1},col sep=comma]{fig/ScordelisLo/ScordelisLoDisplacements.csv};
				\addplot[black,solid,line width=1]table [x index = {0}, y index = {2},col sep=comma]{fig/ScordelisLo/ScordelisLoDisplacements.csv};
				\addplot[red,densely dashed,line width=1]table [x index = {0}, y index = {3},col sep=comma]{fig/ScordelisLo/ScordelisLoDisplacements.csv};
				\addplot[blue,densely dotted,line width=1]table [x index = {0}, y index = {4},col sep=comma]{fig/ScordelisLo/ScordelisLoDisplacements.csv};
			\end{axis}
		\end{tikzpicture}
	}
	\subfloat[Total surface stretch $J$\figlabel{numExViscScordelisLoSurfaceStretch}]{
		\begin{tikzpicture}
			\def\cdot{\times}
			\begin{axis}[grid=both,xlabel={Time $t$ $[T_0]$},ylabel={Surface stretch $J$ $[-]$},width=\figwidth,height=\figheight,
			xmin=0,xmax=50,ymin=0.97,ymax=1.135,minor xtick={5,15,25,35,45},minor ytick={0.975,1.025,1.075,1.125},
			tick label style={font=\footnotesize},legend pos = north west,legend style={nodes={scale=0.75, transform shape}},legend cell align={left}
			]
				\addplot[brown,densely dash dot,line width=1]table [x index = {0}, y index = {1},col sep=comma]{fig/ScordelisLo/ScordelisLoSurfaceStretches.csv};
				\addplot[black,solid,line width=1]table [x index = {0}, y index = {2},col sep=comma]{fig/ScordelisLo/ScordelisLoSurfaceStretches.csv};
				\addplot[red,densely dashed,line width=1]table [x index = {0}, y index = {3},col sep=comma]{fig/ScordelisLo/ScordelisLoSurfaceStretches.csv};
				\addplot[blue,densely dotted,line width=1]table [x index = {0}, y index = {4},col sep=comma]{fig/ScordelisLo/ScordelisLoSurfaceStretches.csv};
			\end{axis}
		\end{tikzpicture}
	}
	\caption{Sagging Scordelis-Lo roof: (a) Vertical displacements and (b) surface stretches for different values of $\eta_\mrs$ and $\eta_\mrb$ over time, measured at the center of the structure.} \figlabel{numExViscScordelisLoDisplacementSurfaceStretch}
\end{figure}
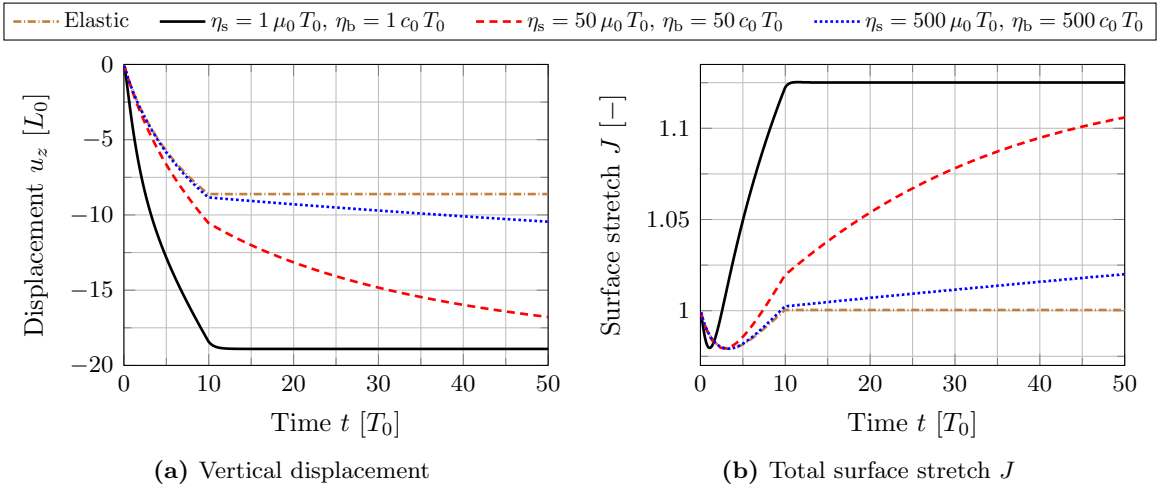
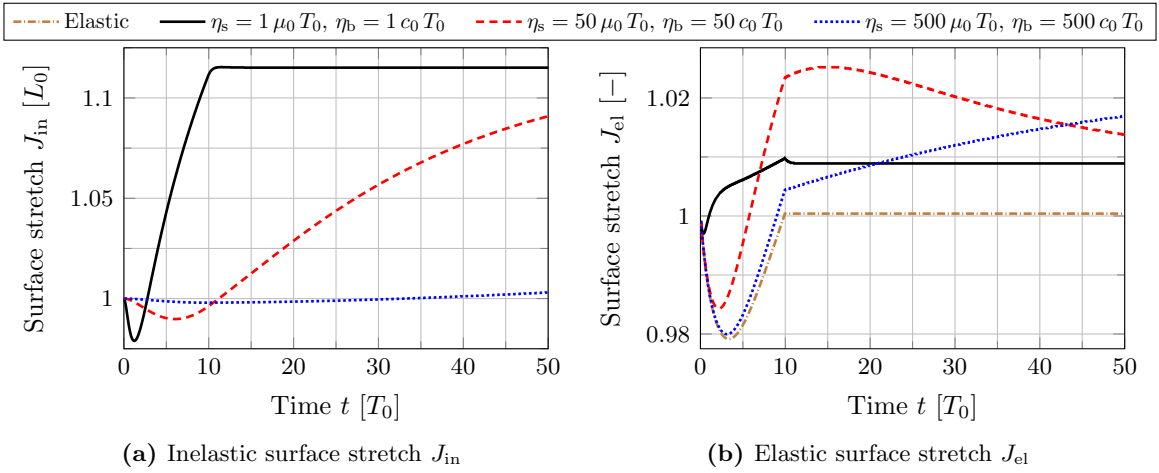
\begin{figure}[!ht]
	\setlength{\figwidth}{0.45\textwidth}
	\setlength{\figheight}{0.35\textwidth}
	\centering
		\begin{tikzpicture}
		\pgfplotsset{width=1\textwidth,height=0.2\textwidth,compat=newest,}
		\begin{axis}[hide axis,xmin=0,xmax=0.00001,ymin=0,ymax=0.00001,legend cell align={left},
					 legend columns=4,legend style={/tikz/every even column/.append style={column sep=2ex}},
					 tick label style={font=\footnotesize},legend style={nodes={scale=0.75, transform shape}},legend cell align={left}]
  			  	\addlegendimage{brown,densely dash dot,line width=1,}
  			  		\addlegendentry{Elastic};
  			  	\addlegendimage{black,solid,line width=1,}
  			  	\addlegendentry{$\eta_\mrs=1\,\mu_0\,T_0$, $\eta_\mrb=1\,c_0\,T_0$};
  			  	\addlegendimage{red,densely dashed,line width=1,}
  			  	\addlegendentry{$\eta_\mrs=50\,\mu_0\,T_0$, $\eta_\mrb=50\,c_0\,T_0$};
  			  	\addlegendimage{blue,densely dotted,line width=1,}
  			  	\addlegendentry{$\eta_\mrs=500\,\mu_0\,T_0$, $\eta_\mrb=500\,c_0\,T_0$};
  			 \end{axis}
	\end{tikzpicture}
	\\ \vspace{-3mm}
	\subfloat[Inelastic surface stretch $J_\mrin$\figlabel{numExViscScordelisLoSurfaceStretchComponentsInelastic}]{
		\begin{tikzpicture}
			\def\cdot{\times}
			\begin{axis}[grid=both,xlabel={Time $t$ $[T_0]$},ylabel={Surface stretch $J_\mrin$ $[L_0]$},width=\figwidth,height=\figheight,
			xmin=0,xmax=50,ymin=0.975,ymax=1.125,minor xtick={5,15,25,35,45},minor ytick={0.975,1.025,1.075,1.125},
			tick label style={font=\footnotesize},legend pos = north west,legend style={nodes={scale=0.75, transform shape}},legend cell align={left}
			]
				\addplot[black,solid,line width=1]table [x index = {0}, y index = {2},col sep=comma]{fig/ScordelisLo/ScordelisLoSurfaceStretchComponents.csv};
				\addplot[red,densely dashed,line width=1]table [x index = {0}, y index = {3},col sep=comma]{fig/ScordelisLo/ScordelisLoSurfaceStretchComponents.csv};
				\addplot[blue,densely dotted,line width=1]table [x index = {0}, y index = {4},col sep=comma]{fig/ScordelisLo/ScordelisLoSurfaceStretchComponents.csv};
			\end{axis}
		\end{tikzpicture}
	}
	\subfloat[Elastic surface stretch $J_\mrel$\figlabel{numExViscScordelisLoSurfaceStretchComponentsElastic}]{
		\begin{tikzpicture}
			\def\cdot{\times}
			\begin{axis}[grid=both,xlabel={Time $t$ $[T_0]$},ylabel={Surface stretch $J_\mrel$ $[-]$},width=\figwidth,height=\figheight,
			xmin=0,xmax=50,ymin=0.9775,ymax=1.0285,minor xtick={5,15,25,35,45},minor ytick={0.99,1.01,1.03},
			tick label style={font=\footnotesize},legend pos = north west,legend style={nodes={scale=0.75, transform shape}},legend cell align={left}
			]
				\addplot[brown,densely dash dot,line width=1]table [x index = {0}, y index = {1},col sep=comma]{fig/ScordelisLo/ScordelisLoSurfaceStretchComponents.csv};
				\addplot[black,solid,line width=1]table [x index = {0}, y index = {5},col sep=comma]{fig/ScordelisLo/ScordelisLoSurfaceStretchComponents.csv};
				\addplot[red,densely dashed,line width=1]table [x index = {0}, y index = {6},col sep=comma]{fig/ScordelisLo/ScordelisLoSurfaceStretchComponents.csv};
				\addplot[blue,densely dotted,line width=1]table [x index = {0}, y index = {7},col sep=comma]{fig/ScordelisLo/ScordelisLoSurfaceStretchComponents.csv};
			\end{axis}
		\end{tikzpicture}
	}
	\caption{Sagging Scordelis-Lo roof: (a) Inelastic and (b) elastic surface stretches for different values of $\eta_\mrs$ and $\eta_\mrb$ over time, measured at the center of the structure. The total surface stretch, shown in \figref{numExViscScordelisLoSurfaceStretch}, satisfies $J=J_\mrin\,J_\mrel$.} \figlabel{numExViscScordelisLoSurfaceStretchComponents}
\end{figure}

In \figref{numExViscScordelisLoMeanCurvatures}, the intermediate mean curvature $\hat{H}$ and mean curvature $H$ are shown over time. The curvature is larger for smaller values of $\eta_\mrs$ and $\eta_\mrb$, which can also be seen in \figref{numExViscScordelisLoDeformation}, e.g.~the creep deformations are larger for small $\eta_\mrs$ and $\eta_\mrb$ in the given time span $t\in[0,t_\mathrm{end}]$.
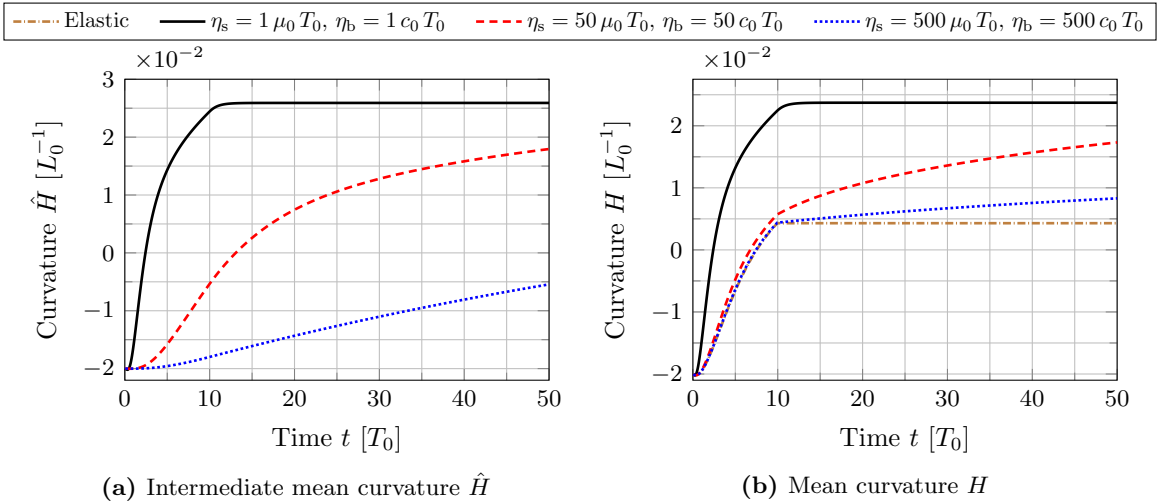
\begin{figure}[!ht]
	\setlength{\figwidth}{0.45\textwidth}
	\setlength{\figheight}{0.35\textwidth}
	\centering
		\begin{tikzpicture}
		\pgfplotsset{width=1\textwidth,height=0.2\textwidth,compat=newest,}
		\begin{axis}[hide axis,xmin=0,xmax=0.00001,ymin=0,ymax=0.00001,legend cell align={left},
					 legend columns=4,legend style={/tikz/every even column/.append style={column sep=2ex}},
					 tick label style={font=\footnotesize},legend style={nodes={scale=0.75, transform shape}},legend cell align={left}]
  			  	\addlegendimage{brown,densely dash dot,line width=1,}
  			  		\addlegendentry{Elastic};
  			  	\addlegendimage{black,solid,line width=1,}
  			  	\addlegendentry{$\eta_\mrs=1\,\mu_0\,T_0$, $\eta_\mrb=1\,c_0\,T_0$};
  			  	\addlegendimage{red,densely dashed,line width=1,}
  			  	\addlegendentry{$\eta_\mrs=50\,\mu_0\,T_0$, $\eta_\mrb=50\,c_0\,T_0$};
  			  	\addlegendimage{blue,densely dotted,line width=1,}
  			  	\addlegendentry{$\eta_\mrs=500\,\mu_0\,T_0$, $\eta_\mrb=500\,c_0\,T_0$};
  			 \end{axis}
	\end{tikzpicture}
	\\ \vspace{-3mm}
	\subfloat[Intermediate mean curvature $\hat{H}$\figlabel{numExViscScordelisLoMeanCurvatureIntermediate}]{
		\begin{tikzpicture}
			\def\cdot{\times}
			\begin{axis}[grid=both,xlabel={Time $t$ $[T_0]$},ylabel={Curvature $\hat{H}$ $[L_0^{-1}]$},width=\figwidth,height=\figheight,
			xmin=0,xmax=50,ymin=-2.2e-2,ymax=0.03,minor xtick={5,15,25,35,45},ytick={-2e-2,-1e-2,0,1e-2,2e-2,3e-2},minor ytick={-1.5e-2,-0.5e-2,0.5e-2,1.5e-2,2.5e-2},
			tick label style={font=\footnotesize},legend pos = north west,legend style={nodes={scale=0.75, transform shape}},legend cell align={left}
			]
				\addplot[black,solid,line width=1]table [x index = {0}, y index = {2},col sep=comma]{fig/ScordelisLo/ScordelisLoMeanCurvatures.csv};
				\addplot[red,densely dashed,line width=1]table [x index = {0}, y index = {3},col sep=comma]{fig/ScordelisLo/ScordelisLoMeanCurvatures.csv};
				\addplot[blue,densely dotted,line width=1]table [x index = {0}, y index = {4},col sep=comma]{fig/ScordelisLo/ScordelisLoMeanCurvatures.csv};
			\end{axis}
		\end{tikzpicture}
	}
	\subfloat[Mean curvature $H$\figlabel{numExViscScordelisLoMeanCurvatureOnS}]{
		\begin{tikzpicture}
			\def\cdot{\times}
			\begin{axis}[grid=both,xlabel={Time $t$ $[T_0]$},ylabel={Curvature $H$ $[L_0^{-1}]$},width=\figwidth,height=\figheight,
			xmin=0,xmax=50,ymin=-2.1e-2,ymax=0.0275,minor xtick={5,15,25,35,45},ytick={-2e-2,-1e-2,0,1e-2,2e-2},minor ytick={-1.5e-2,-0.5e-2,0.5e-2,1.5e-2,2.5e-2},
			tick label style={font=\footnotesize},legend pos = north west,legend style={nodes={scale=0.75, transform shape}},legend cell align={left}
			]
				\addplot[brown,densely dash dot,line width=1]table [x index = {0}, y index = {1},col sep=comma]{fig/ScordelisLo/ScordelisLoMeanCurvatures.csv};
				\addplot[black,solid,line width=1]table [x index = {0}, y index = {5},col sep=comma]{fig/ScordelisLo/ScordelisLoMeanCurvatures.csv};
				\addplot[red,densely dashed,line width=1]table [x index = {0}, y index = {6},col sep=comma]{fig/ScordelisLo/ScordelisLoMeanCurvatures.csv};
				\addplot[blue,densely dotted,line width=1]table [x index = {0}, y index = {7},col sep=comma]{fig/ScordelisLo/ScordelisLoMeanCurvatures.csv};
			\end{axis}
		\end{tikzpicture}
	}
	\caption{Sagging Scordelis-Lo roof: (a) Intermediate mean curvature and (b) mean curvature for different values of $\eta_\mrs$ and $\eta_\mrb$ over time, measured at the center of the structure.} \figlabel{numExViscScordelisLoMeanCurvatures}
\end{figure}

\subsection{Cube encased by a viscoelastic surface} \seclabel{numExViscCube}
This section highlights that the presented formulation for viscoelastic shells can be applied to model boundary viscoelasticity of 3D bodies without any further modifications. The finite element connectivity automatically enforces the coupling between bulk and surface elements and their constitutive behavior. For this, a cube with side length $2\,L_0$ is encased with a viscoelastic surface. Only one eighth of the geometry is modeled by exploiting the symmetry of the problem and the mesh is \textit{a priori} refined towards the sharp edges, see \figref{numExViscSurfaceViscoelasticityDeformedStates}. A similar example is considered by \cite{dortdivanlioglu2021}. In total, $512$ hexahedral elements and $192$ quadrilateral elements are used to discretize the eighth cube and its encasing surface, respectively. Both are discretized by quadratic NURBS shape functions. The material behavior of the bulk material is modeled by a Neo-Hookean material model with elastic energy density and stresses
\eqb{l}
	\tilde\Psi=\dfrac{\tilde{\Lambda}}{2}\,\bigl(\ln\tilde{J}\bigr)^2+\dfrac{\tilde{\mu}}{2}\,\bigl(\tilde{I}_1-3-2\,\ln\tilde{J}\bigr)\,,\quad\mathrm{and}\quad\tilde\bsig=\dfrac{\tilde{\Lambda}}{\tilde{J}}\,\ln\tilde{J}+\dfrac{\tilde{\mu}}{\tilde{J}}\,\bigl(\tilde\bB+\tilde\bI\bigr)\,. \eqlabel{numExViscSurfaceViscosityPsiSigmaBulk}
\eqe
Here, $\tilde\bI$ is the full identity in 3D, $\tilde\bB$ is the left Cauchy-Green tensor in 3D, and $\tilde{I}_1$ and $\tilde{J}$ are the invariants of $\tilde\bB$. A tilde is added to avoid confusion with the corresponding surface quantities presented in \secref{cntViscKin}. The material parameters for the bulk are $\tilde{\Lambda}=5\,\tilde{\Lambda}_0$ and $\tilde{\mu}=5\,\tilde{\Lambda}_0$. The elastic energy density in the elastic branch is given by \eqsref{cntViscConstElasticityPsiMemSurfaceTensionDriven} with prescribed surface tension $\gamma$. Since this model provides no deviatoric surface stiffness, the second part of \eqsref{cntViscConstElasticityPsiMemNeoHookean} with $\mu=1\,\tilde{\Lambda}_0\,L_0$ is added for numerical stabilization. For the Maxwell branch, the spring element from \figref{cntViscConstSrfSurfaceRheology} is omitted, such that the constitutive model resembles a Kelvin model with $\hat{a}_\ab=a_\ab$, $J_\mathrm{in}=J$, and $J_\mrel=1$. The dashpot is chosen to follow model \eqref{e:cntViscSigma1in2}, which in this special case is a pure dilatational model that requires no evolution laws and can be directly computed using $\dot{J}\approx(J-J_n)/\Delta t$, where $J_n$ denotes the surface stretch from the previous time step. The surface tension is imposed over time as
\eqb{l}
	\gamma(t)=\gamma_0\,\begin{cases}t/(2.5\,T_0)\,,&t\leq2.5\,T_0\\1\,,&t>2.5\,T_0\end{cases}\,. \eqlabel{numExViscSurfaceViscosityGammaHatGamma}
\eqe
The end time is $t_\mathrm{end}=10\,T_0$ and $2,500$ time steps are used for the temporal integration.

\figref{numExViscSurfaceViscoelasticityDeformedStates} shows the deformed geometry at $t_\mathrm{end}$ for different values of $\gamma_0$ and fixed $\eta_\mrs=1\,\tilde{\Lambda}_0\,L_0\,T_0$. Energetically, a sphere is the optimal geometry for constant surface tension. Thus, for larger prescribed surface tension, the bulk deforms more into a spherical shape. Due to the stiffness of the bulk material, not a perfect sphere is obtained, but the sharp edges and corners are smoothed out.
\begin{figure}[!ht]
	\setlength{\figwidth}{0.29\textwidth}
	\centering
		\begin{tikzpicture}
			\tikzmath{
				\dx = 5.5;
				\dy = 5.2;
				\dtt = 0.825;
				\dt = 0.47;
				\y1 = -8.7; 
				\y2 = -9.125;     
				\x1 = -0.935;
				\x2 = 11.5225;
				\nl = 7;
			};
			\node[align=center] at (0,0) {\includegraphics[width=\figwidth]{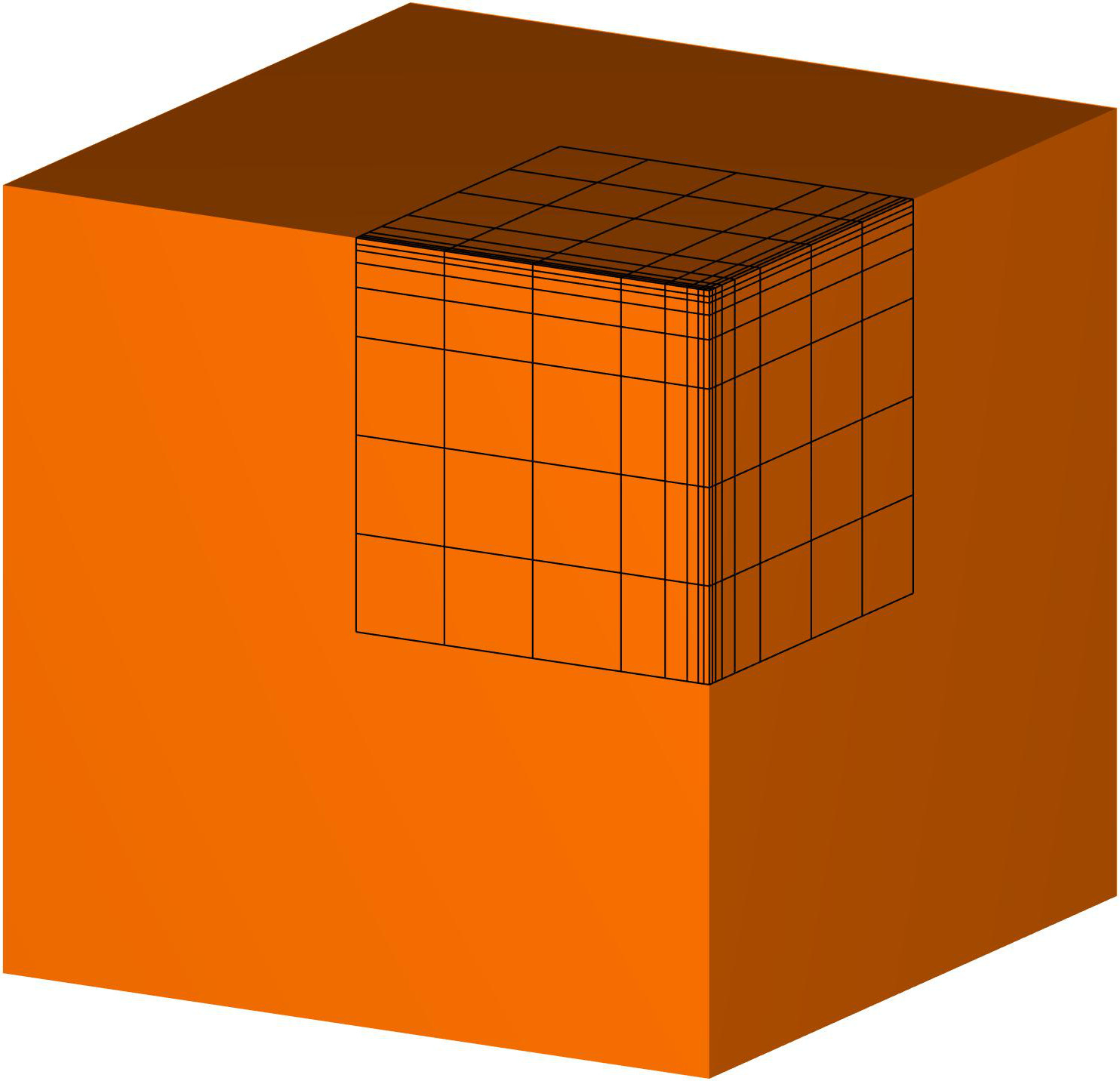}};	
				\node[align=left,font=\footnotesize] at (-\dx+\dtt*\dx,-\dt*\dy) {Initial geometry};
			\node[align=center] at (1*\dx,0) {\includegraphics[width=\figwidth]{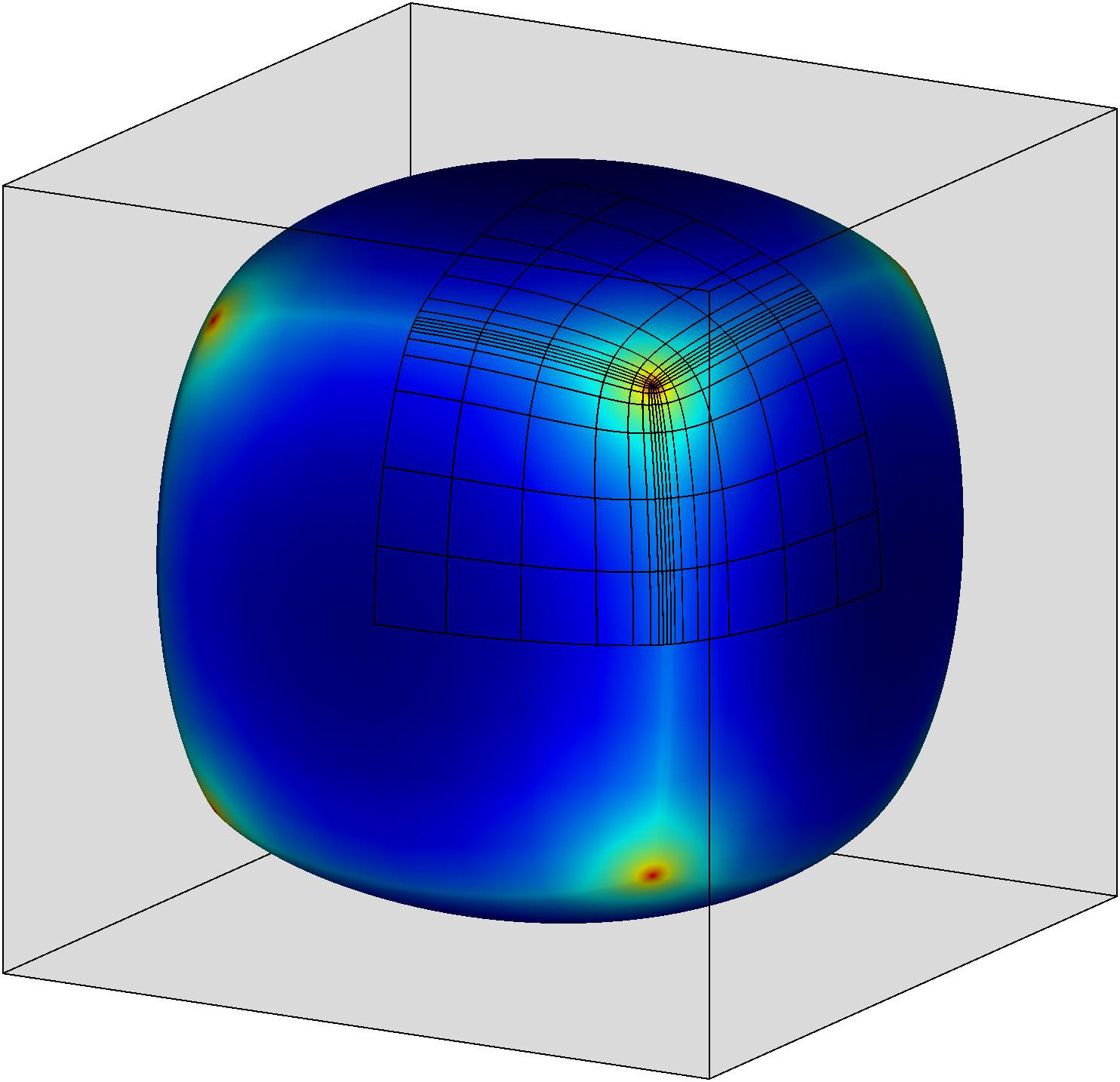}};	
				\node[align=left,font=\footnotesize] at (\dtt*\dx,-\dt*\dy) {$\gamma_0=5\,\tilde{\Lambda}_0\,L_0$};
			\node[align=center] at (2*\dx,0) {\includegraphics[width=\figwidth]{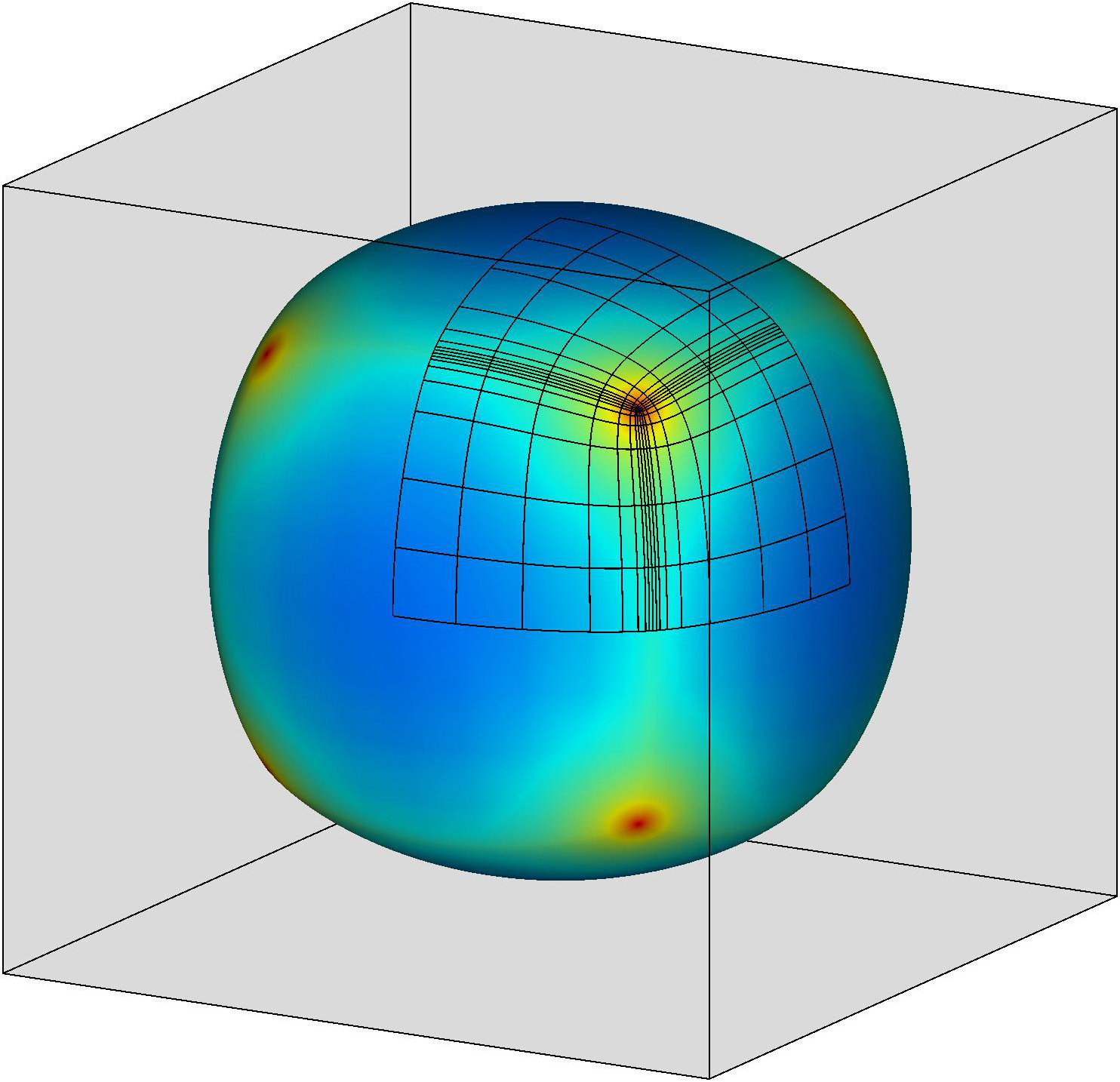}};	
				\node[align=left,font=\footnotesize] at (\dtt*\dx+\dx,-\dt*\dy) {$\gamma_0=10\,\tilde{\Lambda}_0\,L_0$};
			\node[align=center] at (0,-\dy) {\includegraphics[width=\figwidth]{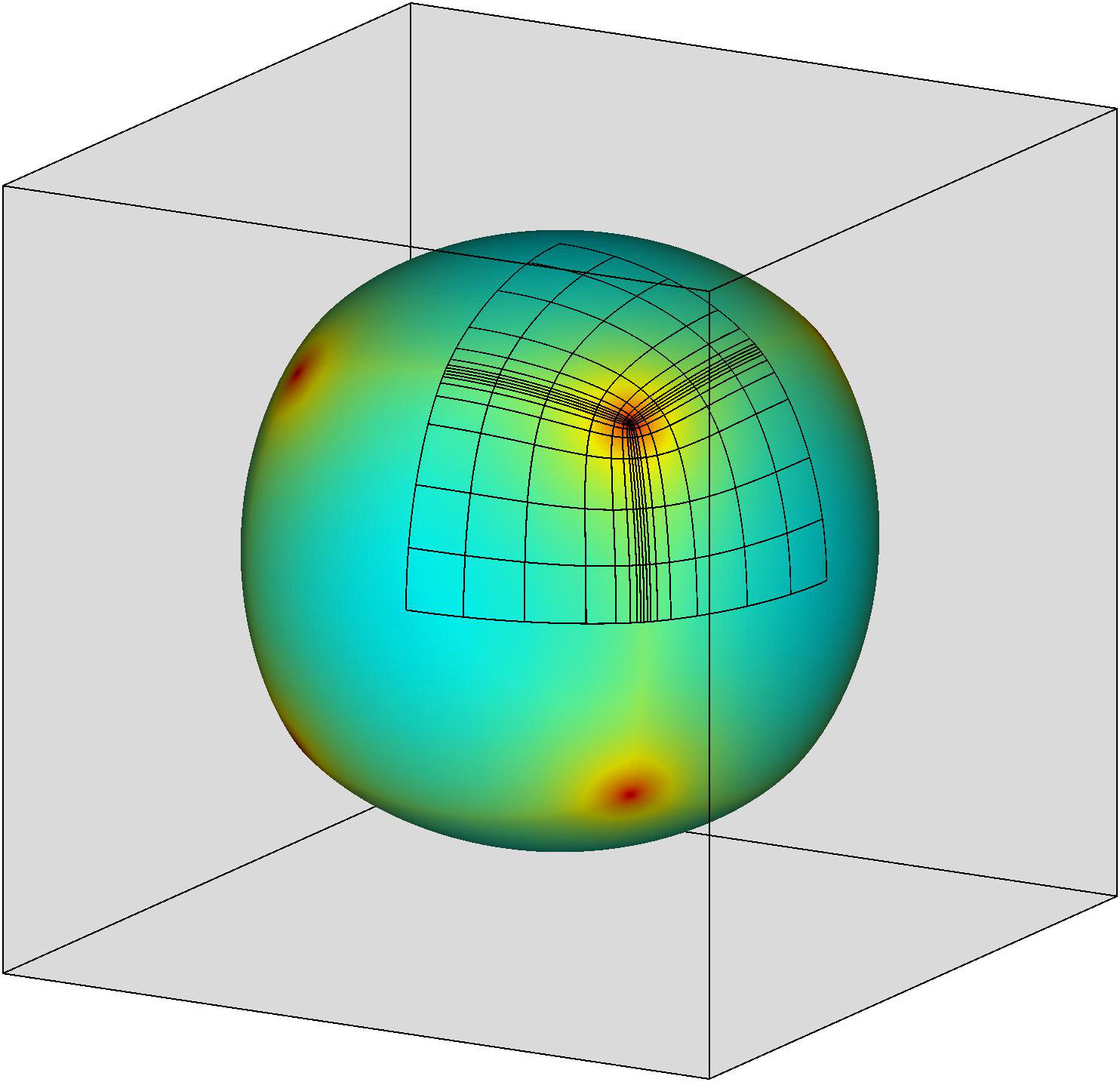}};	
				\node[align=left,font=\footnotesize] at (-\dx+\dtt*\dx,-\dt*\dy-\dy) {$\gamma_0=15\,\tilde{\Lambda}_0\,L_0$};
			\node[align=center] at (1*\dx,-\dy) {\includegraphics[width=\figwidth]{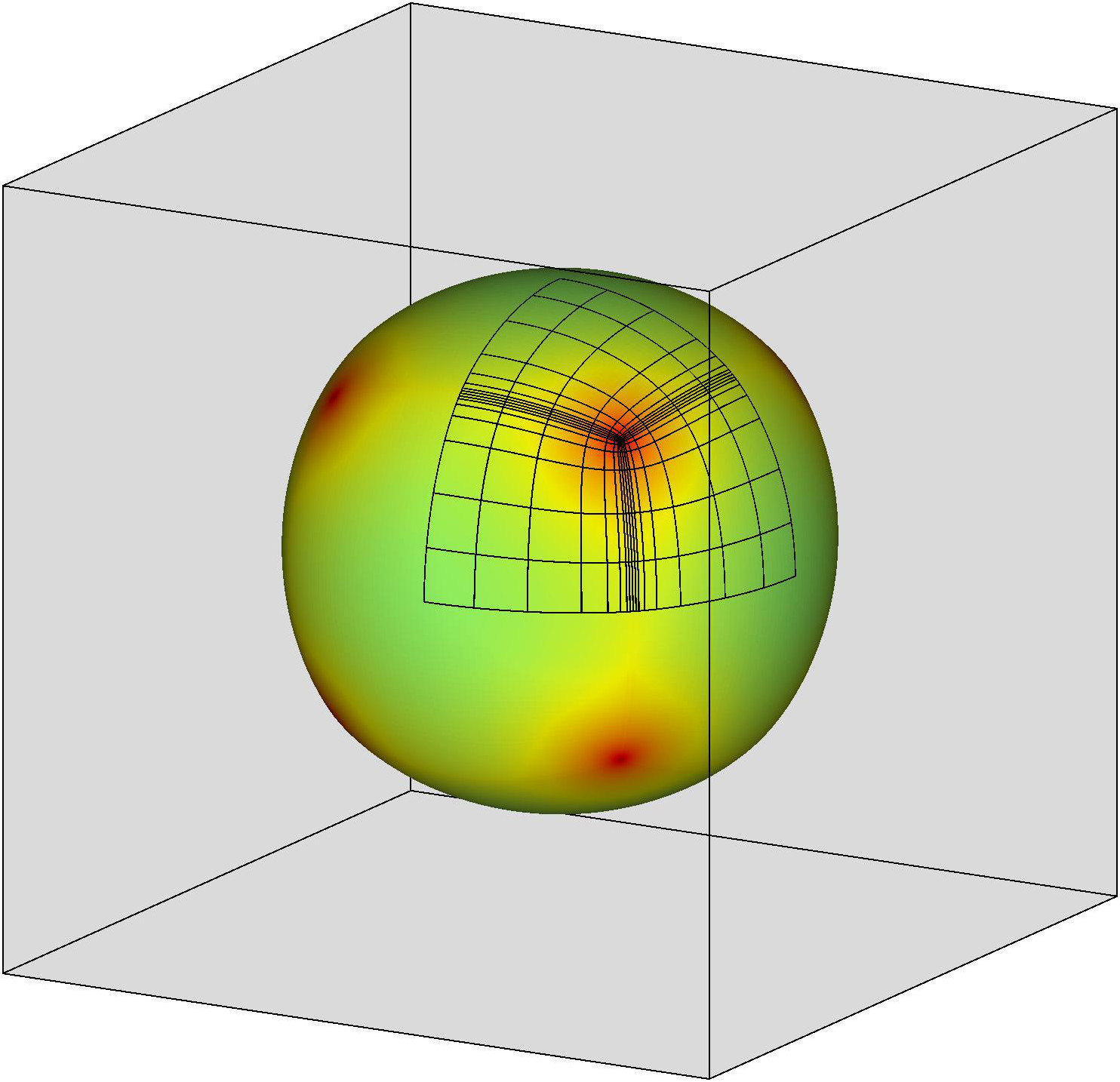}};	
				\node[align=left,font=\footnotesize] at (\dtt*\dx,-\dt*\dy-\dy) {$\gamma_0=25\,\tilde{\Lambda}_0\,L_0$};
			\node[align=center] at (2*\dx,-\dy) {\includegraphics[width=\figwidth]{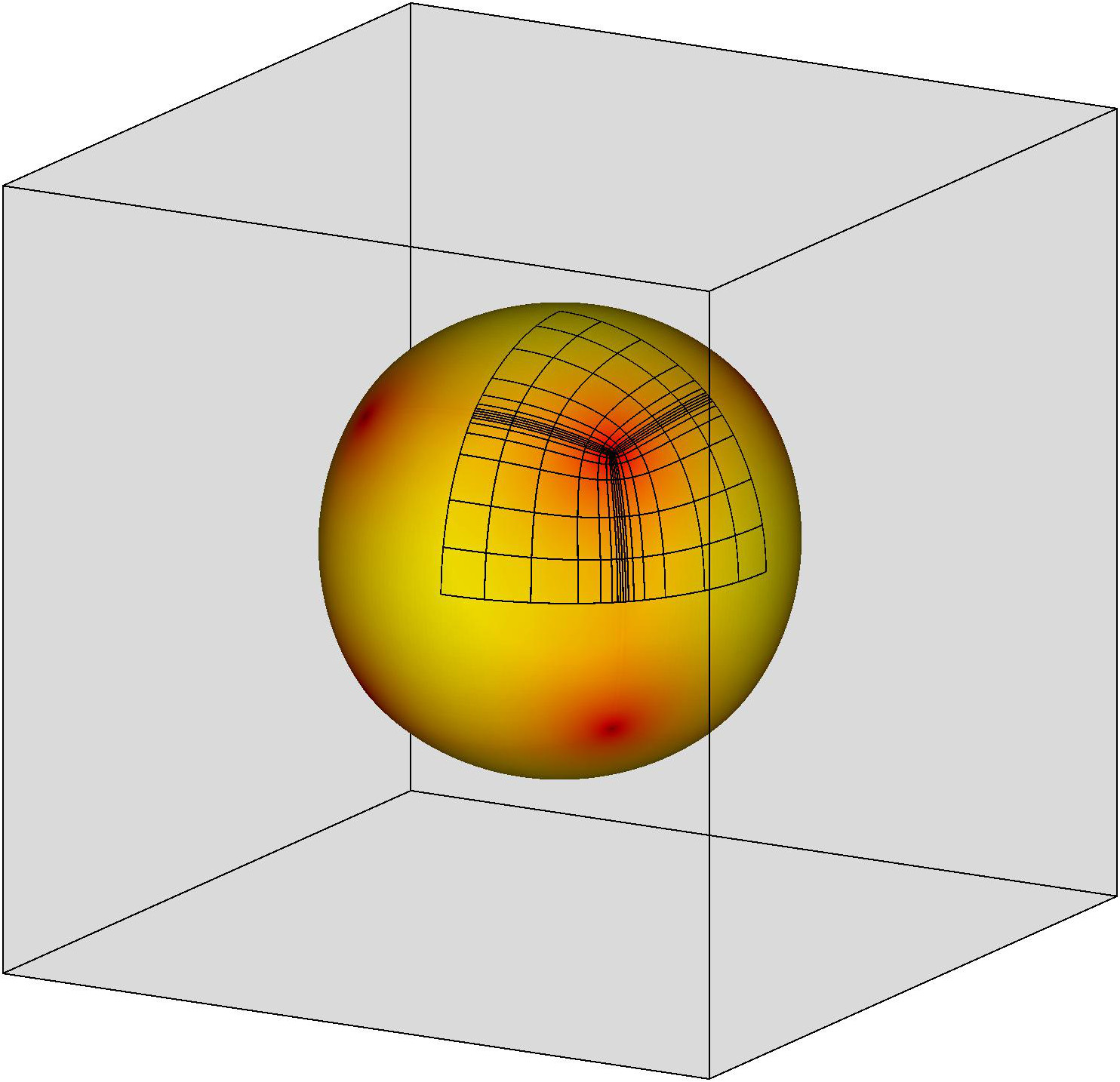}};	
				\node[align=left,font=\footnotesize] at (\dtt*\dx+\dx,-\dt*\dy-\dy) {$\gamma_0=40\,\tilde{\Lambda}_0\,L_0$};
			\node[align=center] at (1*\dx,\y1) {\includegraphics[width=0.81\textwidth]{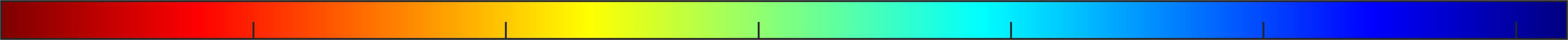}};
			\node[align=center,font=\footnotesize] at ({\x1+(\x2-\x1)/(\nl-1)*0},\y2) {$0$};		
			\node[align=center,font=\footnotesize] at ({\x1+(\x2-\x1)/(\nl-1)*1},\y2) {$0.1$};
			\node[align=center,font=\footnotesize] at ({\x1+(\x2-\x1)/(\nl-1)*2},\y2) {$0.2$};		
			\node[align=center,font=\footnotesize] at ({\x1+(\x2-\x1)/(\nl-1)*3},\y2) {$0.3$};		
			\node[align=center,font=\footnotesize] at ({\x1+(\x2-\x1)/(\nl-1)*4},\y2) {$0.4$};		
			\node[align=center,font=\footnotesize] at ({\x1+(\x2-\x1)/(\nl-1)*5},\y2) {$0.5$};
			\node[align=center,font=\footnotesize] at ({\x1+(\x2-\x1)/(\nl-1)*6},\y2) {$0.6$};		
			\node[align=left,font=\footnotesize] at ({\x2+1},\y1) {$J$ $[-]$};	
		\end{tikzpicture}
	\caption{Cube encased by a viscoelastic surface: Deformed cubes at $t_\mathrm{end}$ for different values of the imposed surface tension $\gamma_0$, colored by the surface stretch $J$, see Eq.~(\ref{e:cntViscKinInvariantsOnS}.2).} \figlabel{numExViscSurfaceViscoelasticityDeformedStates}
\end{figure}
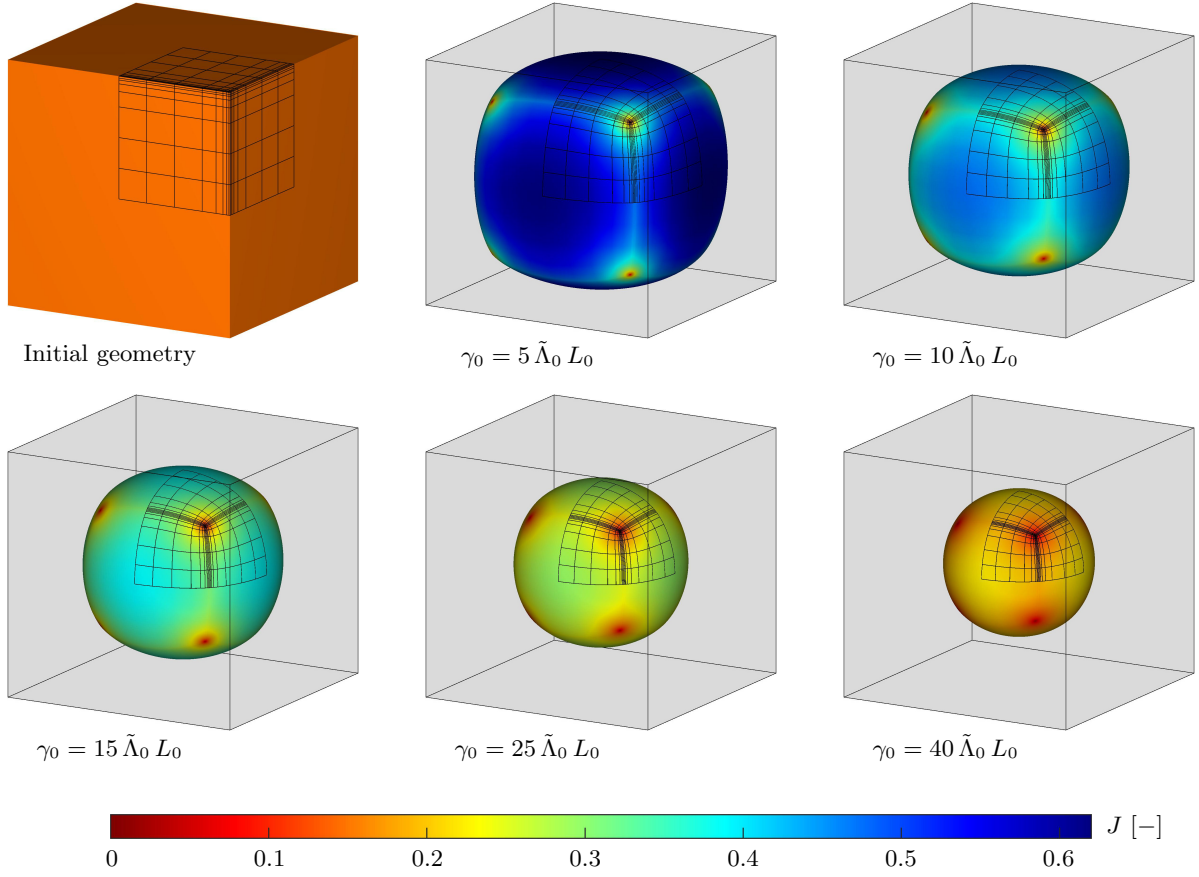

The displacement norm $\norm{\bu}_2=\norm{\bx-\bX}_2$ over time is plotted in \figref{numExViscSurfaceViscoelasticityDisplacementEtas} for different values of the in-plane shear viscosity $\eta_\mrs$ and fixed $\gamma_0=5\,\tilde{\Lambda}_0\,L_0$, and in \figref{numExViscSurfaceViscoelasticityDisplacementGamma} for different values of the surface tension $\gamma_0$ and fixed $\eta_\mrs=1\,\tilde{\Lambda}_0\,L_0\,T_0$. \figref{numExViscSurfaceViscoelasticityDisplacementEtas} shows creep behavior, which becomes more pronounced for increasing $\eta_\mrs$. \figref{numExViscSurfaceViscoelasticityDisplacementGamma} shows that larger values of the surface tension lead to larger deformations, see also \figref{numExViscSurfaceViscoelasticityDeformedStates}.
\begin{figure}[!ht]
	\setlength{\figwidth}{0.45\textwidth}
	\setlength{\figheight}{0.35\textwidth}
	\centering
	\subfloat[Influence of $\eta_\mrs$ for  $\gamma_0=5\,\tilde{\Lambda}_0\,L_0$\figlabel{numExViscSurfaceViscoelasticityDisplacementEtas}]{
		\begin{tikzpicture}
			\def\cdot{\times}
			\begin{axis}[grid=both,xlabel={Time $t$ $[T_0]$},ylabel={Displacement $\norm{\bu}_2$ $[L_0]$},width=\figwidth,height=\figheight,
			xmin=0,xmax=10,ymin=0,ymax=0.1,minor xtick={1,3,5,7,9},ytick={0,0.025,0.05,0.075,0.1},minor ytick={0.0125,0.0375,0.0625,0.0875},			
			tick label style={font=\footnotesize},legend style={at={(0,1.2)},anchor=west},legend style={nodes={scale=0.75, transform shape}},legend cell align={left}, legend columns=2,
			yticklabel style={/pgf/number format/fixed,/pgf/number format/precision=5},scaled y ticks=false]
				\addplot[black,solid,line width=1]table [x index = {0}, y index = {12},col sep=comma]{fig/cube/cubeDisplacementNorm.csv};
					\addlegendentry{Elastic};
				\addplot[red,densely dashed,line width=1]table [x index = {0}, y index = {1},col sep=comma]{fig/cube/cubeDisplacementNorm.csv};
					\addlegendentry{$\eta_\mrs=1\,\tilde{\Lambda}_0\,L_0\,T_0$};
				\addplot[blue,densely dotted,line width=1]table [x index = {0}, y index = {2},col sep=comma]{fig/cube/cubeDisplacementNorm.csv};
					\addlegendentry{$\eta_\mrs=10\,\tilde{\Lambda}_0\,L_0\,T_0$};
				\addplot[cyan,densely dash dot,line width=1]table [x index = {0}, y index = {3},col sep=comma]{fig/cube/cubeDisplacementNorm.csv};
					\addlegendentry{$\eta_\mrs=20\,\tilde{\Lambda}_0\,L_0\,T_0$};
				\addplot[orange,densely dash dot dot,line width=1]table [x index = {0}, y index = {4},col sep=comma]{fig/cube/cubeDisplacementNorm.csv};
					\addlegendentry{$\eta_\mrs=30\,\tilde{\Lambda}_0\,L_0\,T_0$};
			\end{axis}
		\end{tikzpicture}
	}
	\subfloat[Influence of $\gamma_0$ for $\eta_\mrs=1\,\tilde{\Lambda}_0\,L_0\,T_0$\figlabel{numExViscSurfaceViscoelasticityDisplacementGamma}]{
		\begin{tikzpicture}
			\def\cdot{\times}
			\begin{axis}[grid=both,xlabel={Time $t$ $[T_0]$},ylabel={Displacement $\norm{\bu}_2$ $[L_0]$},width=\figwidth,height=\figheight,
			xmin=0,xmax=10,ymin=0,ymax=0.45,minor xtick={1,3,5,7,9},ytick={0,0.1,0.2,0.3,0.4},minor ytick={0.05,0.15,0.25,0.35},
			tick label style={font=\footnotesize},legend style={at={(0,1.2)},anchor=west},legend style={nodes={scale=0.75, transform shape}},legend cell align={left}, legend columns=2,
			]
				\addplot[black,solid,line width=1]table [x index = {0}, y index = {1},col sep=comma]{fig/cube/cubeDisplacementNorm.csv};
					\addlegendentry{$\gamma_0=5\,\tilde{\Lambda}_0\,L_0$};
				\addplot[red,densely dashed,line width=1]table [x index = {0}, y index = {5},col sep=comma]{fig/cube/cubeDisplacementNorm.csv};
					\addlegendentry{$\gamma_0=10\,\tilde{\Lambda}_0\,L_0$};
				\addplot[blue,densely dotted,line width=1]table [x index = {0}, y index = {6},col sep=comma]{fig/cube/cubeDisplacementNorm.csv};
					\addlegendentry{$\gamma_0=15\,\tilde{\Lambda}_0\,L_0$};
				\addplot[cyan,densely dash dot,line width=1]table [x index = {0}, y index = {8},col sep=comma]{fig/cube/cubeDisplacementNorm.csv};
					\addlegendentry{$\gamma_0=25\,\tilde{\Lambda}_0\,L_0$};
				\addplot[orange,densely dash dot dot,line width=1]table [x index = {0}, y index = {11},col sep=comma]{fig/cube/cubeDisplacementNorm.csv};
					\addlegendentry{$\gamma_0=40\,\tilde{\Lambda}_0\,L_0$};
			\end{axis}
		\end{tikzpicture}
	}
	\caption{Cube encased by a viscoelastic surface: Displacement norm $\norm{\bu}_2$ for different values of the in-plane shear viscosity $\eta_\mrs$ and surface tension $\gamma_0$, measured at the central point of the top surface.} \figlabel{numExViscSurfaceViscoelasticityDisplacement}
\end{figure}
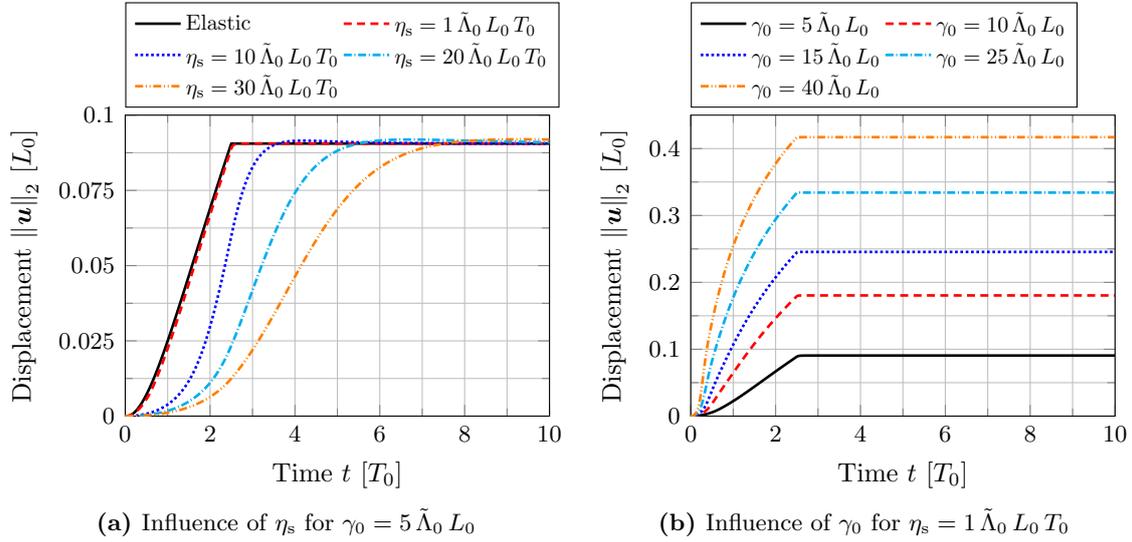

\section{Conclusion} \seclabel{concl}
This work presents a computational formulation to model isotropic finite strain viscoelasticity for membranes, thin shells, and boundaries of 3D bodies. The material behavior is modeled based on the generalized viscoelastic solid, for which a multiplicative split of the surface deformation gradient is employed. The implementation of membrane and bending viscosity is verified by several numerical examples and ideal convergence rates are obtained in all cases. The chosen examples capture large deformations and standard viscoelasticity behavior of thin shells, as well as boundary viscoelasticity of 3D bodies.

This work demonstrates that the previously developed multiplicative split of the surface deformation gradient works robustly and accurately in finite strain computations. The employed direct surface formulation and its decomposition of the elastic energy density into membrane and bending parts makes the constitutive modeling more flexible. For example, one can consider viscous material behavior only for membrane deformations but not for bending, or vice versa. At the same time, the formulation allows to use known 3D material models. The presented formulation also allows to describe boundary viscoelasticity of 3D bodies. Thus, the proposed formulation unifies boundary, membrane and shell viscoelasticity. The use of rotation-free finite elements and isogeometric shape functions increases efficiency and accuracy in comparison to classical FE discretizations. This increase is greatest when used in conjunction with direct surface-based constitutive models such as are provided here.

In order to avoid problems stemming from ill-conditioned parametrizations noted in Remark~\ref{r:reformulatingODEs}, the ODEs for $\hat{a}^\ab$ can be reformulated, e.g.~as ODEs for $I_1^\mrel$, $J_\mrin$, and $J_\mrel$. Also, constant area for the inelastic deformation can be enforced by coupling the evolution laws to the constraint $J_\mrin=1$, which can be directly plugged into the ODEs to eliminate one of them, or enforced by a Lagrange multiplier approach, for instance.

As the viscoelastic framework introduces a time scale, the quasi-static shell framework can also be extended to incorporate inertia, e.g.~to investigate the influence of the viscous effects and inertia on each other. Further, temperature can be introduced in the formulation in order to model the change of temperature as a cause of dissipation. The elastic and viscoelastic material properties can then also be dependent on the temperature.

\section*{Acknowledgments}
The authors acknowledge funding by the Deutsche Forschungsgemeinschaft (DFG, German Research Foundation) – 333849990/GRK2379 (IRTG Modern Inverse Problems). Simulations were partly performed with computing resources granted by RWTH Aachen University under project rwth0917.

\bigskip
\appendix
\section*{Appendix}

\section{Linearization} \applabel{viscLinearization}

\subsection{Auxiliary derivatives} \applabel{viscLinearizationAuxDervs}
For the linearization in \appref{viscLinearizationImplicitEuler}, the derivatives of Eqs.~(\ref{e:cntViscKinSurfaceStretchSplitComponents}.1), \eqref{e:cntViscKinFirstInvariantElastic} and (\ref{e:cntViscKinMeanCurvatures}.1) w.r.t.~$\hat{a}^\ab$ and $\hat{b}_\ab$ are required. They are given by
\eqb{l}
	\dpa{I_1^\mrel}{\hat{a}^\ab}=a_\ab\,,\quad\dpa{J_\mrel}{\hat{a}^\ab}=\dfrac{1}{2}\,J_\mrel\,\hat{a}_\ab\,,\quad\dfrac{\partial\hat{H}}{\partial\hat{a}^\ab}=\dfrac{1}{2}\,\hat{b}_\ab\,,\quad\mathrm{and}\quad\dfrac{\partial\hat{H}}{\partial\hat{b}_\ab}=\dfrac{1}{2}\,\hat{a}^\ab\,, \eqlabel{cntViscDerivativesInvariants}
\eqe
where $\partial\bigl(\det[\hat{a}^\ab]\bigr)/\partial t=\det[\hat{a}^\ab]\,\hat{a}_\gd\,\dot{\hat{a}}^\gd$ has been used.

For the linearization of the stresses in the Maxwell branch, see \appref{viscLinearizationFEM}, the derivatives of $J_\mrel$, $J_\mrin$, and $I_1^\mrel$ w.r.t.~$a_\gd$ are required, see Eqs.~\eqref{e:cntViscKinSurfaceStretchSplitComponents}--\eqref{e:cntViscKinFirstInvariantElastic}. They are given by
\eqb{l}
	\dpa{I_1^\mrel}{a_\gd}=\hat{a}^\gd+\dpa{\hat{a}^\ez}{a_\gd}\,a_\ez\,, \eqlabel{varLinLinearizationViscoelasticShellsAuxiliaryDerivativeI1elForLinearization}
\eqe
and
\eqb{lll}
	\dpa{J_\mrin}{a_\gd}=\dpa{J_\mrin}{\hat{a}^\ez}\,\dpa{\hat{a}^\ez}{a_\gd}\is\dfrac{1}{\sqrt{\det[A_\ab]}}\,\dpa{}{\hat{a}^\ez}\biggl(\dfrac{1}{\sqrt{\det[\hat{a}^\ab]}}\biggr)\,\dpa{\hat{a}^\ez}{a_\gd}\\[5mm]
	\is-\dfrac{1}{2}\,\dfrac{1}{\sqrt{\det[A_\ab]}}\,\dfrac{1}{\bigl(\sqrt{\det[\hat{a}^\ab]}\bigr)^{3/2}}\,\dpa{}{\hat{a}^\ez}\Bigl(\det[\hat{a}^\ab]\Bigr)\,\dpa{\hat{a}^\ez}{a_\gd}\\[5mm]
	\is-\dfrac{1}{2}\,J_\mrin\,\hat{a}_\ez\,\dpa{\hat{a}^\ez}{a_\gd}\,, \eqlabel{varLinLinearizationViscoelasticShellsAuxiliaryDerivativesJinForLinearization}
\eqe
as $\partial\bigl(\det[\hat{a}^\ab]\bigr)/\partial\hat{a}^\ez=\det[\hat{a}^\ab]\,\hat{a}_\ez$. Likewise,
\eqb{lll}
	\dpa{J_\mrel}{a_\gd}\is\sqrt{\det[\hat{a}^\ab]}\,\dpa{\bigl(\sqrt{\det[a_\ab]}\bigr)}{a_\gd}+\sqrt{\det[a_\ab]}\,\dpa{\bigl(\sqrt{\det[\hat{a}^\ab]}\bigr)}{a_\gd}\\[5mm]
	\is\dfrac{1}{2}\,\dfrac{\sqrt{\det[\hat{a}^\ab]}}{\sqrt{\det[a_\ab]}}\,\dpa{\bigl(\det[a_\ab]\bigr)}{a_\gd}+\dfrac{1}{2}\,\dfrac{\sqrt{\det[a_\ab]}}{\sqrt{\det[\hat{a}^\ab]}}\,\dpa{\bigl(\det[\hat{a}^\ab]\bigr)}{\hat{a}^\ez}\,\dpa{\hat{a}^\ez}{a_\gd}\\[5mm]
	\is\dfrac{1}{2}\,J_\mrel\,a^\gd+\dfrac{1}{2}\,J_\mrel\,\hat{a}_\ez\,\dpa{\hat{a}^\ez}{a_\gd}\,, \eqlabel{varLinLinearizationViscoelasticShellsAuxiliaryDerivativesJelForLinearization}
\eqe
as $\partial\bigl(\det[a_\ab]\bigr)/\partial a_\gd=\det[a_\ab]\,a^\gd$. The derivatives of the mean curvature $\hat{H}$, see Eq.~(\ref{e:cntViscKinMeanCurvatures}.1), w.r.t. $a_\gd$ and $b_\gd$ are given by
\eqb{l}
	\dfrac{\partial\hat{H}}{\partial a_\gd} = \dfrac{1}{2}\,\biggl(\dfrac{\partial\hat{a}^\ez}{\partial a_\gd}\,\hat{b}_\ez+\hat{a}^\ez\,\dfrac{\partial\hat{b}_\ez}{\partial a_\gd}\biggr)\,,\quad\mathrm{and}\quad\dfrac{\partial\hat{H}}{\partial b_\gd} = \dfrac{1}{2}\,\biggl(\dfrac{\partial\hat{a}^\ez}{\partial b_\gd}\,\hat{b}_\ez+\hat{a}^\ez\,\dfrac{\partial\hat{b}_\ez}{\partial b_\gd}\biggr)\,. \eqlabel{varLinLinearizationViscoelasticShellsAuxiliaryDerivativeshatHForLinearization}
\eqe
For the derivatives in Eqs.~\eqref{e:varLinLinearizationViscoelasticShellsAuxiliaryDerivativeI1elForLinearization}--\eqref{e:varLinLinearizationViscoelasticShellsAuxiliaryDerivativeshatHForLinearization}, the derivatives of $\hat{a}^\ez$ and $\hat{b}_\ez$ w.r.t.~$a_\gd$ and $b_\gd$ need to be computed, due to the elimination noted in Remark~\ref{r:eliminationHataabHatbab}. The derivation of these derivatives depends on the employed material model and is presented in \appref{viscLinearizationFEM}. 
Further, the following derivatives are used in the subsequent sections \citep{sauer2018}
\eqb{rll}
	a^{\ab\gd}\dis\dfrac{\partial a^\ab}{\partial a_\gd}=-\dfrac{1}{2}\,\bigl(a^{\alpha\gamma}\,a^{\beta\delta}+a^{\alpha\delta}\,a^{\beta\gamma}\bigr)\,,\\[4mm]
	\hat{a}^{\ab\gd}\dis\dfrac{\partial\hat{a}^\ab}{\partial\hat{a}_\gd}=-\dfrac{1}{2}\,\bigl(\hat{a}^{\alpha\gamma}\,\hat{a}^{\beta\delta}+\hat{a}^{\alpha\delta}\,\hat{a}^{\beta\gamma}\bigr)\,,\quad\mathrm{and}\\[4mm]
	b^{\ab\gd}\dis\dfrac{\partial b^\ab}{\partial a_\gd}=-\dfrac{1}{2}\,\bigl(a^{\alpha\gamma}\,b^{\beta\delta}+b^{\alpha\gamma}\,a^{\beta\delta}+a^{\alpha\delta}\,b^{\beta\gamma}+b^{\alpha\delta}\,a^{\beta\gamma}\bigr)\,,\eqlabel{varLinLinearizationViscoelasticShellsAuxiliaryDerivativesAabgdAndHatAabgd}
\eqe
from which $\partial b^\ab/\partial b_\gd=-a^{\ab\gd}$ follows.

\subsection{Linearization for the implicit Euler scheme} \applabel{viscLinearizationImplicitEuler}
For the time integration of the evolution laws based on the implicit Euler scheme, the derivatives in \eqsref{cmpViscAssembledEqnsLin} are required for the different material models. Subsequently, those derivatives are reported.

\subsubsection{Koiter membrane model}
The derivative of \eqsref{cmpViscDiscretizedODEKoiter} w.r.t.~$\hat{a}^\ab$ is given by
\eqb{rll}
\dpa{\hat{\mrg}_\mrs^\ab}{\hat{a}^\gd}\is\dfrac{1}{\Delta t}\,\delta^\alpha_\gamma\,\delta^\beta_\delta + \dfrac{\Lambda_1}{2\,\eta_\mrs}\,\Bigl(\hat{a}^\ab\,a_\gd+\bigl(I_1^{\mrel}-2\bigr)\,\delta^\alpha_\gamma\,\delta^\beta_\delta\Bigr) + \dfrac{\mu_1}{\eta_\mrs}\,\Bigl(a_{\delta\epsilon}\,\bigl(\delta^\alpha_\gamma\,\hat{a}^{\beta\epsilon}+\delta^\beta_\gamma\,\hat{a}^{\alpha\epsilon}\bigr)-\delta^\alpha_\gamma\,\delta^\beta_\delta\Bigr)\,,
\eqe
with Kronecker delta $\delta^\alpha_\beta$.

\subsubsection{Neo-Hookean membrane model with dilatational/deviatoric split}
The derivative of \eqsref{cmpViscDiscretizedODENeoHookean} w.r.t.~$\hat{a}^\ab$ is given by
\eqb{rll}
\dpa{\hat{\mrg}_\mrs^\ab}{\hat{a}^\gd}\is\dfrac{1}{\Delta t}\,\delta^\alpha_\gamma\,\delta^\beta_\delta + \dfrac{K_1}{2\,\eta_\mrs}\,J_{\mrel}^ 2\,a^\ab\,\hat{a}_\gd+\dfrac{\mu_1}{2\,\eta_\mrs\,J_{\mrel}}\,\Bigl(2\,\delta^\alpha_\gamma\,\delta^\beta_\delta-a^\ab\,a_\gd-\hat{a}^\ab\,\hat{a}_\gd+\dfrac{1}{2}\,I_1^{\mrel}\,a^\ab\,\hat{a}_\gd\Bigr)\,,
\eqe
with $\bigl[\hat{a}_\ab\bigr]=\bigl[\hat{a}^\ab\bigr]^{-1}$ and Kronecker delta $\delta^\alpha_\beta$.

\subsubsection{Incompressible Neo-Hookean membrane model}
The derivative of \eqsref{cmpViscDiscretizedODEIncompressibleNeoHookean} w.r.t.~$\hat{a}^\gd$ is given by
\eqb{l}
	\dpa{\hat{\mrg}_\mrs^\ab}{\hat{a}^\gd}=\dfrac{1}{\Delta t}\,\delta^\alpha_\gamma\,\delta^\beta_\delta + \dfrac{\mu_1}{\eta_\mrs}\,\biggl(\delta^\alpha_\gamma\,\delta^\beta_\delta+\dfrac{a^\ab}{J_{\mrel}^2}\,\hat{a}_\gd\biggr)\,,
\eqe
with $\bigl[\hat{a}_\ab\bigr]=\bigl[\hat{a}^\ab\bigr]^{-1}$ and Kronecker delta $\delta^\alpha_\beta$.

\subsubsection{Membranes with constant surface tension}
The derivative of \eqsref{cmpviscDiscretizedODEsurfaceTensionDriven} w.r.t.~$\hat{a}^\gd$ is given by
\eqb{l}
	\dpa{\hat{\mrg}_\mrs^\ab}{\hat{a}^\gd}=\dfrac{1}{\Delta t}\,\delta^\alpha_\gamma\,\delta^\beta_\delta + \dfrac{\hat\gamma}{2\,\eta_\mrs}\,J_{\mrel}\,a^\ab\,\hat{a}_\gd\,,
\eqe
with $\bigl[\hat{a}_\ab\bigr]=\bigl[\hat{a}^\ab\bigr]^{-1}$ and Kronecker delta $\delta^\alpha_\beta$.

\subsubsection{Helfrich bending model}
The derivatives of Eqs.~\eqref{e:cmpViscDiscretizedODEHelfrichBendingGs}--\eqref{e:cmpViscDiscretizedODEHelfrichBendingGb} w.r.t.~$\hat{a}^\gd$ and $\hat{b}_\gd$ are given by
\eqb{rll}
	\dpa{\hat{\mrg}_\mrs^\ab}{\hat{a}^\gd}\is\dfrac{1}{\Delta t}\,\delta^\alpha_\gamma\,\delta^\beta_\delta + \dfrac{k_1\,J_{\mrel}}{\eta_\mrs}\,\Biggl[\biggl(\dfrac{1}{2}\Delta H^2\,a^\ab-\Delta H\,b^\ab\biggr)\,\hat{a}_\gd-\biggl(\Delta H\,a^\ab-b^\ab\biggr)\,\hat{b}_\gd\Biggr]\,,\\[5mm]
	\dpa{\hat{\mrg}_\mrs^\ab}{\hat{b}_\gd}\is-\dfrac{k_1\,J_{\mrel}}{\eta_\mrs}\,\Bigl(\Delta H\,a^\ab-b^\ab\Bigr)\,\hat{a}^\gd\,, \eqlabel{viscLinearizationImplicitEulerHelfrichGs}
\eqe
and
\eqb{rll}
	\dpa{\hat{\mrg}^\mrb_\ab}{\hat{a}^\gd}\is-\dfrac{k_1\,J_{\mrel}}{\eta_\mrb}\,a_\ab\,\Bigl(\Delta H\,\hat{a}_\gd-\hat{b}_\gd\Bigr)\,,\\[5mm]
	\dpa{\hat{\mrg}^\mrb_\ab}{\hat{b}_\gd}\is\dfrac{1}{\Delta t}\,\delta^\alpha_\gamma\,\delta^\beta_\delta + \dfrac{k_1\,J_{\mrel}}{2\,\eta_\mrb}\,a_\ab\,\hat{a}^\gd\,, \eqlabel{viscLinearizationImplicitEulerHelfrichGb}
\eqe
with $\bigl[\hat{a}_\ab\bigr]=\bigl[\hat{a}^\ab\bigr]^{-1}$, Kronecker delta $\delta^\alpha_\beta$, and $\Delta H:=H-\hat{H}$.

\subsection{Linearization for the finite element method} \applabel{viscLinearizationFEM}
Subsequently, the additional contributions in the tangent matrices, see Eqs.~\eqref{e:cmpViscTangentsFE}--\eqref{e:cmpViscMaterialTangents}, coming from the Maxwell branch are derived for the employed material models. 

\subsubsection{Koiter membrane model}
Given the stresses in \eqsref{cntViscConstElasticitySigmaABkoiter},\footnote{with the modifications mentioned in Remark~\ref{r:hatPsiMem}\label{footnote:linearization}} the contribution in the linearized weak form is
\eqb{rll}
	c_1^{\ab\gd}&=2\,\dpa{\tau_{1(\mrel)}^\ab}{a_\gd}\\[5mm]
	&=\Lambda_1\,\dpa{J_\mrin}{a_\gd}\,\bigl(I_1^\mrel-2\bigr)\,\hat{a}^\ab+\Lambda_1\,J_\mrin\,\Bigl(\dpa{I_1^\mrel}{a_\gd}\,\hat{a}^\ab+\bigl(I_1^\mrel-2\bigr)\,\dpa{\hat{a}^\ab}{a_\gd}\Bigr)\\[5mm]
	&\quad+2\,\mu_1\,\dpa{J_\mrin}{a_\gd}\,\bigl(\hat{a}^{\alpha\epsilon}\,a_\ez\,\hat{a}^{\beta\zeta}-\hat{a}^\ab\bigr)\\[5mm]
	&\quad+2\,\mu_1\,J_\mrin\,\Bigl(\dpa{\hat{a}^{\alpha\epsilon}}{a_\gd}\,a_\ez\,\hat{a}^{\beta\zeta}+\dfrac{1}{2}\,\bigl(\hat{a}^{\alpha\gamma}\,\hat{a}^{\beta\delta}+\hat{a}^{\alpha\delta}\,\hat{a}^{\beta\gamma}\bigr)+\hat{a}^{\alpha\epsilon}\,a_\ez\,\dpa{\hat{a}^{\beta\zeta}}{a_\gd}-\dpa{\hat{a}^\ab}{a_\gd}\Bigr)\,. \eqlabel{varLinLinearizationViscoelasticShellsAuxiliaryCabgdKoiter}
\eqe
Still, the derivative $\partial\hat{a}^\ab/\partial a_\gd$ needs to be found to evaluate \eqsref{varLinLinearizationViscoelasticShellsAuxiliaryCabgdKoiter}. For this, the nonlinear equations in \eqsref{cmpViscDiscretizedODEKoiter}, which need to be solved for $\hat{a}^\ab$ within the element routine at each quadrature point and each time step, need to be differentiated w.r.t.~$a_\gd$. The auxiliary arrays
\eqb{l}
	\hat\ma_\mathrm{con}:=\begin{bmatrix}\hat{a}^{11}\\\hat{a}^{12}\\\hat{a}^{21}\\\hat{a}^{22}\end{bmatrix}\,,\quad\hat\ma_\mathrm{co}:=\begin{bmatrix}\hat{a}_{11}\\\hat{a}_{12}\\\hat{a}_{21}\\\hat{a}_{22}\end{bmatrix}\,,\quad\ma_\mathrm{con}:=\begin{bmatrix}a^{11}\\a^{12}\\a^{21}\\a^{22}\end{bmatrix}\,,\quad\mathrm{and}\quad\ma_\mathrm{co}:=\begin{bmatrix}a_{11}\\a_{12}\\a_{21}\\a_{22}\end{bmatrix}\,, \eqlabel{varLinLinearizationViscoelasticShellsAuxiliaryArrays}
\eqe
are defined to simplify the notation. The required derivatives $\partial\hat{a}^\ab/\partial a_\gd$ can then be extracted from the $(4\times4)$-matrix
\eqb{l}
	\dfrac{\partial\hat\ma_\mathrm{con}}{\partial\ma_\mathrm{co}}=\begin{bmatrix}
		\nicefrac{\partial\hat{a}^{11}}{\partial a_{11}}&\nicefrac{\partial\hat{a}^{11}}{\partial a_{12}}&\nicefrac{\partial\hat{a}^{11}}{\partial a_{21}}&\nicefrac{\partial\hat{a}^{11}}{\partial a_{22}}\\
		\nicefrac{\partial\hat{a}^{12}}{\partial a_{11}}&\nicefrac{\partial\hat{a}^{12}}{\partial a_{12}}&\nicefrac{\partial\hat{a}^{12}}{\partial a_{21}}&\nicefrac{\partial\hat{a}^{12}}{\partial a_{22}}\\
		\nicefrac{\partial\hat{a}^{21}}{\partial a_{11}}&\nicefrac{\partial\hat{a}^{21}}{\partial a_{12}}&\nicefrac{\partial\hat{a}^{21}}{\partial a_{21}}&\nicefrac{\partial\hat{a}^{21}}{\partial a_{22}}\\
		\nicefrac{\partial\hat{a}^{22}}{\partial a_{11}}&\nicefrac{\partial\hat{a}^{22}}{\partial a_{12}}&\nicefrac{\partial\hat{a}^{22}}{\partial a_{21}}&\nicefrac{\partial\hat{a}^{22}}{\partial a_{22}}
	\end{bmatrix}\,, \eqlabel{varLinLinearizationViscoelasticShellsAuxiliarydhatacondaco}
\eqe
which follows from the solution of
\eqb{l}
	\Biggl[\biggl(\dfrac{\eta_\mrs}{\Delta t}-\mu_1+\dfrac{\Lambda_1}{2}\,\bigl(I_1^\mrel-2\bigr)\biggr)\,\mI_4+\dfrac{\Lambda_1}{2}\,\hat\ma_\mathrm{con}\,\ma_\mathrm{co}^\mrT+\mu_1\,\tilde\mC\Biggr]\,\dfrac{\partial\hat\ma_\mathrm{con}}{\partial\ma_\mathrm{co}}=-\dfrac{\Lambda_1}{2}\,\hat\ma_\mathrm{con}\,\hat\ma_\mathrm{con}^\mrT+\mu_1\,\dfrac{\partial\hat\ma_\mathrm{con}}{\partial\hat\ma_\mathrm{co}}\,, \eqlabel{varLinLinearizationViscoelasticShellsAuxiliaryEquationSystemKoiter}
\eqe
where $\mI_4$ denotes the $(4\times4)$-identity matrix. The derivative $\partial\hat\ma_\mathrm{con}/\partial\hat\ma_\mathrm{co}$ follows from Eq.~(\ref{e:varLinLinearizationViscoelasticShellsAuxiliaryDerivativesAabgdAndHatAabgd}.2) and it is arranged in analogy to \eqsref{varLinLinearizationViscoelasticShellsAuxiliarydhatacondaco}. The matrix $\tilde\mC$ is given by
\eqb{l}
	\renewcommand*{\arraystretch}{1.25}
	\tilde\mC:=\begin{bmatrix}2\,c_1^1&2\,c_2^1&0&0\\c_1^2&c_2^2&c_1^1&c_2^1\\c_1^2&c_2^2&c_1^1&c_2^1\\0&0&2\,c_1^2&2\,c_2^2\end{bmatrix}\,,
\eqe
with $c^\alpha_\gamma:=\hat{a}^\ab\,a_{\beta\gamma}$.

\begin{remark} \label{r:efficientLinearization}
	Note that the computation of the derivatives $\partial\hat{a}^\ab/\partial a_\gd$ can become more efficient by exploiting the symmetries of $\hat{a}^\ab$ and $a_\ab$, e.g.~$\hat{a}^{12}=\hat{a}^{21}$ and $a_{12}=a_{21}$. The $(4\times4)$-matrix in \eqsref{varLinLinearizationViscoelasticShellsAuxiliarydhatacondaco} can then be reduced to a $(3\times3)$-matrix, such that also the dimension of the linear system of equations in \eqsref{varLinLinearizationViscoelasticShellsAuxiliaryEquationSystemKoiter} decreases.
\end{remark}

\subsubsection{Neo-Hookean membrane model}
Considering the stresses in \eqsref{cntViscConstElasticitySigmaABclassicalNeoHookean} with $\Lambda=0$,\footnoteref{footnote:linearization} the corresponding linearization is
\eqb{l}
	c_1^{\ab\gd}=2\,\dpa{\tau_{1(\mrel)}^\ab}{a_\gd}=2\,\mu_1\,\biggl[\dpa{J_\mrin}{a_\gd}\,\bigl(\hat{a}^\ab-a^\ab\bigr)+J_\mrin\,\Bigl(\dpa{\hat{a}^\ab}{a_\gd}-a^{\ab\gd}\Bigr)\biggr]\,, \eqlabel{varLinLinearizationViscoelasticShellsAuxiliaryCabgdNeoHookeanSimple}
\eqe
where
\eqb{l}
	\dpa{\hat{a}^\ab}{a_\gd} = \dfrac{\mu_1\,\Delta t}{\eta_\mrs+\mu_1\,\Delta t}\,a^{\ab\gd}\,, \eqlabel{varLinLinearizationViscoelasticShellsAuxiliaryDhataabDagd}
\eqe
see also \eqsref{cmpviscDiscretizedODENeoHookeanSimple2}. Note that if $\eta_\mrs/\Delta t\rightarrow0$, the derivative in \eqsref{varLinLinearizationViscoelasticShellsAuxiliaryDhataabDagd} approaches $a^{\ab\gd}$, and for $\eta_\mrs/\Delta t\rightarrow\infty$ it approaches zero.

\subsubsection{Neo-Hookean membrane model with dilatational/deviatoric split}
Given the stress in \eqsref{cntViscConstElasticitySigmaABneoHookean},\footnoteref{footnote:linearization} the contribution in the linearized weak form is
\eqb{rll}
	c_1^{\ab\gd}=2\,\dpa{\tau^\ab_{1(\mrel)}}{a_\gd}&=K_1\,\biggl[\dpa{J_\mrin}{a_\gd}\,\bigl(J_\mrel^2-1\bigr)\,a^\ab+2\,J\,\dpa{J_\mrel}{a_\gd}\,a^\ab+J_\mrin\,\bigl(J_\mrel^2-1\bigr)\,a^{\ab\gd}\biggr]\\[5mm]
	&\quad+\mu_1\,\biggl[-\dfrac{J_\mrin}{J_\mrel^2}\,\dpa{J_\mrel}{a_\gd}\,\bigl(2\,\hat{a}^\ab-I_1^\mrel\,a^\ab\bigr)+\dfrac{1}{J_\mrel}\,\dpa{J_\mrin}{a_\gd}\,\bigl(2\,\hat{a}^\ab-I_1^\mrel\,a^\ab\bigr)\\[5mm]
	&\qquad\qquad+\dfrac{J_\mrin}{J_\mrel}\,\Bigl(2\,\dpa{\hat{a}^\ab}{a_\gd}-\dpa{I_1^\mrel}{a_\gd}\,a^\ab-I_1^\mrel\,a^{\ab\gd}\Bigr)\biggr]\,. \eqlabel{varLinLinearizationViscoelasticShellsAuxiliaryCabgdNeoHookean}
\eqe
Still, the derivative $\partial\hat{a}^\ab/\partial a_\gd$ needs to be found to evaluate \eqsref{varLinLinearizationViscoelasticShellsAuxiliaryCabgdNeoHookean}. For this, the nonlinear equations from \eqsref{cmpViscDiscretizedODENeoHookean}, which need to be solved for $\hat{a}^\ab$ in the element routine at each quadrature point and each time step, need to be differentiated w.r.t.~$a_\gd$. Similar to \eqsref{varLinLinearizationViscoelasticShellsAuxiliaryEquationSystemKoiter}, the required derivatives $\partial\hat{a}^\ab/\partial a_\gd$ can be extracted from the $(4\times4)$-matrix $\partial\hat\ma_\mathrm{con}/\partial\ma_\mathrm{co}$, see \eqsref{varLinLinearizationViscoelasticShellsAuxiliarydhatacondaco}, which follows from the solution of 
\eqb{l}
\Biggl[\biggl(\dfrac{2\,\eta_\mrs}{\Delta t}+\dfrac{2\,\mu_1}{J_\mrel}\biggr)\,\mI_4+\biggl(K_1\,J_\mrel^2+\dfrac{\mu_1\,I_1^\mrel}{2\,J_\mrel}\biggr)\,\ma_\mathrm{con}\,\hat\ma_\mathrm{co}^\mrT-\dfrac{\mu_1}{J_\mrel}\Bigl(\hat\ma_\mathrm{con}\,\hat\ma_\mathrm{co}^\mrT+\ma_\mathrm{con}\,\ma_\mathrm{co}^\mrT\Bigr)\Biggr]\,\dfrac{\partial\hat\ma_\mathrm{con}}{\partial\ma_\mathrm{co}}\\[4mm]
\qquad\qquad=-K_1\,\biggl(J_\mrel^2\,\ma_\mathrm{con}\,\ma_\mathrm{con}^\mrT+\bigl(J_\mrel^2-1\bigr)\,\dfrac{\partial\ma_\mathrm{con}}{\partial\ma_\mathrm{co}}\biggr)\\[4mm]
\qquad\qquad\qquad\qquad+\dfrac{\mu_1}{J_\mrel}\,\biggl(\ma_\mathrm{con}\,\hat\ma_\mathrm{con}^\mrT+I_1^\mrel\,\dfrac{\partial\ma_\mathrm{con}}{\partial\ma_\mathrm{co}}+\hat\ma_\mathrm{con}\,\ma_\mathrm{con}^\mrT-\dfrac{I_1^\mrel}{2}\,\ma_\mathrm{con}\,\ma_\mathrm{con}^\mrT\biggr)\,, \eqlabel{varLinLinearizationViscoelasticShellsAuxiliaryEquationSystemNeoHookean}
\eqe
where $\mI_4$ denotes the $(4\times4)$-identity matrix. The derivative $\partial\ma_\mathrm{con}/\partial\ma_\mathrm{co}$ follows from Eq.~(\ref{e:varLinLinearizationViscoelasticShellsAuxiliaryDerivativesAabgdAndHatAabgd}.1) and it is arranged in analogy to \eqsref{varLinLinearizationViscoelasticShellsAuxiliarydhatacondaco}.

\subsubsection{Incompressible Neo-Hookean membrane model}
Given the stresses in \eqsref{cntViscConstElasticitySigmaABincompressibleNeoHookean},\footnoteref{footnote:linearization}  the contribution in the linearized weak form is
\eqb{l}
	c_1^{\ab\gd}=2\,\dpa{\tau^\ab_{1(\mrel)}}{a_\gd}=2\,\mu_1\,\biggl[\dfrac{\partial J_\mrin}{\partial a_\gd}\,\Bigl(\hat{a}^\ab-\dfrac{a^\ab}{J_\mrel^2}\Bigr)+J_\mrin\,\Bigl(\dfrac{\partial\hat{a}^\ab}{\partial a_\gd}-\dfrac{a^{\ab\gd}}{J_\mrel^2}+2\,\dfrac{a^\ab}{J_\mrel^3}\,\dfrac{\partial J_\mrel}{\partial a_\gd}\Bigr)\biggr]\,. \eqlabel{varLinLinearizationViscoelasticShellsAuxiliaryCabgdIncompressibleNeoHookean}
\eqe
The required derivative $\partial\hat{a}^\ab/\partial a_\gd$ is extracted from the $(4\times4)$-matrix $\partial\hat\ma_\mathrm{con}/\partial\ma_\mathrm{co}$, see \eqsref{varLinLinearizationViscoelasticShellsAuxiliarydhatacondaco}, which follows from the solution of
\eqb{l}
	\Biggl[\biggl(\dfrac{\eta_\mrs}{\Delta t}+\mu_1\biggr)\,\mI_4+\dfrac{\mu_1}{J_\mrel^2}\,\ma_\mathrm{con}\,\hat\ma_\mathrm{co}^\mrT\Biggr]\,\dfrac{\partial\hat\ma_\mathrm{con}}{\partial\ma_\mathrm{co}}=\dfrac{\mu_1}{J_\mrel^2}\,\biggl(\dfrac{\partial\ma_\mathrm{con}}{\partial\ma_\mathrm{co}}-\ma_\mathrm{con}\,\ma_\mathrm{con}^\mrT\biggr)\,,
\eqe
where $\mI_4$ denotes the $(4\times4)$-identity matrix. The derivative $\partial\ma_\mathrm{con}/\partial\ma_\mathrm{co}$ follows from Eq.~(\ref{e:varLinLinearizationViscoelasticShellsAuxiliaryDerivativesAabgdAndHatAabgd}.1) and it is arranged in analogy to \eqsref{varLinLinearizationViscoelasticShellsAuxiliarydhatacondaco}.

\subsubsection{Membranes with constant surface tension}
Given the stresses in \eqsref{cntViscConstElasticitySigmaABsurfaceTensionDriven},\footnoteref{footnote:linearization}  the contribution in the linearized weak form is
\eqb{l}
	c_1^{\ab\gd}=2\,\dpa{\tau^\ab_{1(\mrel)}}{a_\gd}=\hat\gamma\,J\,\bigl(a^\ab\,a^\gd+2\,a^{\ab\gd}\bigr)\,. \eqlabel{varLinLinearizationViscoelasticShellsAuxiliaryCabgdSurfaceTensionDriven}
\eqe

\subsubsection{Koiter bending model}
For the moment components given in \eqsref{cntViscConstElasticityMabKoiter},\footnoteref{footnote:linearization}  the material tangent is given by
\eqb{l}
	f^{\ab\gd}_1=\dpa{M^\ab_{0(1)(\mrel)}}{b_\gd}=J_\mrin\,\hat{f}^{\ab\gd}\,\dfrac{\eta_\mrb}{\eta_\mrb+c_1\,\Delta t}\,,
\eqe
as the derivative $\partial\hat{b}_\ab/\partial b_\gd$ follows in analogy to \eqsref{varLinLinearizationViscoelasticShellsAuxiliaryDhataabDagd}.

\subsubsection{Helfrich bending model}
Given the stress and moment components in Eqs.~\eqref{e:cntViscConstElasticitySigmaABhelfrich}--\eqref{e:cntViscConstElasticityMabHelfrich} with $k^\star=0$,\footnoteref{footnote:linearization}  the material tangents in \eqsref{cmpViscTangentsFE} follow as
\eqb{rll}
	c^{\ab\gd}_1 \is 2\,k_1\,J\,\biggl(\Bigl(\dfrac{1}{2}\Delta H^2\,a^\ab-\Delta H\,b^\ab\Bigr)\,a^\gd\\[2mm]
	&&\qquad\qquad\qquad\qquad+\Bigl(\Delta H\,a^\ab-b^\ab\Bigr)\,\Delta H_a^\gd+\Delta H^2\,a^{\ab\gd}-2\,\Delta H\,b^{\ab\gd}\biggr)\,, \\[4mm]
	d^{\ab\gd}_1 \is k_1\,J\,\biggl(\Bigl(\Delta H\,a^\ab-b^\ab\Bigr)\,\Delta H_b^\gd + 2\,\Delta H\,a^{\ab\gd}\biggr)\,, \\[4mm]
	e^{\ab\gd}_1 \is k_1\,J\,\Bigl(\Delta H\,a^\ab\,a^\gd+\Delta H_a^\gd\,a^\ab+2\,\Delta H\,a^{\ab\gd}\Bigr)\,, \\[4mm]
	f^{\ab\gd}_1 \is \dfrac{k_1\,J}{2}\,\Delta H_b^\gd\,a^\ab\,,
\eqe
where $\Delta H:=H-\hat{H}$ and
\eqb{l}
	\Delta H_a^\gd:=-b^\gd-\dfrac{\partial\hat{a}^\ez}{\partial a_\gd}\,\hat{b}_\ez-\hat{a}^\ez\,\dfrac{\partial\hat{b}_\ez}{\partial a_\gd}\,,\quad\mathrm{and}\quad\Delta H_b^\gd:=a^\gd-\dfrac{\partial\hat{a}^\ez}{\partial b_\gd}\,\hat{b}_\ez-\hat{a}^\ez\,\dfrac{\partial\hat{b}_\ez}{\partial b_\gd}\,. \eqlabel{varLinLinearizationViscoelasticShellsHelfrichDerivativeMeanCurvature}
\eqe
Still, the derivatives of $\hat{a}^\ez$ and $\hat{b}_\ez$ w.r.t.~$a_\gd$ and $b_\gd$ are required. For this, the nonlinear algebraic equations from Eqs.~\eqref{e:cmpViscDiscretizedODEHelfrichBendingGs}--\eqref{e:cmpViscDiscretizedODEHelfrichBendingGb} are differentiated w.r.t.~$a_\gd$ and $b_\gd$. In analogy to \eqsref{varLinLinearizationViscoelasticShellsAuxiliaryArrays}, the auxiliary arrays
\eqb{l}
	\hat{\mathbf{b}}_\mathrm{con}:=\begin{bmatrix}\hat{b}^{11}\\\hat{b}^{12}\\\hat{b}^{21}\\\hat{b}^{22}\end{bmatrix}\,,\quad\hat{\mathbf{b}}_\mathrm{co}:=\begin{bmatrix}\hat{b}_{11}\\\hat{b}_{12}\\\hat{b}_{21}\\\hat{b}_{22}\end{bmatrix}\,,\quad\mathbf{b}_\mathrm{con}:=\begin{bmatrix}b^{11}\\b^{12}\\b^{21}\\b^{22}\end{bmatrix}\,,\quad\mathrm{and}\quad\mathbf{b}_\mathrm{co}:=\begin{bmatrix}b_{11}\\b_{12}\\b_{21}\\b_{22}\end{bmatrix}\,, \eqlabel{varLinLinearizationViscoelasticShellsAuxiliaryArrays2}
\eqe
are defined to simplify the notation. The required derivatives are then obtained by solving
\eqb{l}
	\renewcommand*{\arraystretch}{1.25}
	\begin{bmatrix}\partial\hat{\mg}_\mrs/\partial\ma_\mathrm{co}\\\partial\hat{\mg}_\mrs/\partial\mathbf{b}_\mathrm{co}\\\partial\hat{\mg}^\mrb/\partial\ma_\mathrm{co}\\\partial\hat{\mg}^\mrb/\partial\mathbf{b}_\mathrm{co}\end{bmatrix}=\boldsymbol{0}\,,
\eqe
for the unknown derivatives. The resulting linear equation system is given by
\eqb{l}
	\renewcommand*{\arraystretch}{1.25}
	\begin{bmatrix}\mA_{11}&\boldsymbol{0}&\mA_{13}&\boldsymbol{0}\\\boldsymbol{0}&\mA_{22}&\boldsymbol{0}&\mA_{24}\\\mA_{31}&\boldsymbol{0}&\mA_{33}&\boldsymbol{0}\\\boldsymbol{0}&\mA_{42}&\boldsymbol{0}&\mA_{44}\end{bmatrix}\,\begin{bmatrix}\partial\hat\ma_\mathrm{con}/\partial\ma_\mathrm{co}\\\partial\hat\ma_\mathrm{con}/\partial\mathbf{b}_\mathrm{co}\\\partial\hat{\mathbf{b}}_\mathrm{co}/\partial\ma_\mathrm{co}\\\partial\hat{\mathbf{b}}_\mathrm{co}/\partial\mathbf{b}_\mathrm{co}\end{bmatrix}=\begin{bmatrix}\mr_1\\\mr_2\\\mr_3\\\mr_4\end{bmatrix}\,,
\eqe
with the $(4\times4)$-tangent matrix blocks
\eqb{rllrrl}
	\mA_{11} \dis \dfrac{\eta_\mrs}{k_1\,\Delta t\,J_\mrel}\,\mI_4+\mcc_\mathrm{con}\,\hat{\ma}_\mathrm{co}^\mrT-\md_\mathrm{con}\,\hat{\mathbf{b}}_\mathrm{co}^\mrT\,,&\quad
	\mA_{13} \dis -\md_\mathrm{con}\,\hat{\ma}_\mathrm{con}^\mrT\,,\\[4mm]
	\mA_{22} \dis \dfrac{\eta_\mrs}{k_1\,\Delta t\,J_\mrel}\,\mI_4-\md_\mathrm{con}\,\hat{\mathbf{b}}_\mathrm{co}^\mrT\,,&
	\mA_{24} \dis -\md_\mathrm{con}\,\hat{\ma}_\mathrm{con}^\mrT\,,\\[4mm]
	\mA_{31} \dis -\dfrac{1}{2}\,\Bigl(\Delta H\,\ma_\mathrm{co}\,\hat{\ma}_\mathrm{co}^\mrT-\ma_\mathrm{co}\,\hat{\mathbf{b}}_\mathrm{co}^\mrT\Bigr)\,,&
	\mA_{33} \dis \dfrac{\eta_\mrb}{k_1\,\Delta t\,J_\mrel}\,\mI_4 + \dfrac{1}{2}\,\ma_\mathrm{co}\,\hat{\ma}_\mathrm{con}^\mrT\,,\\[4mm]
	\mA_{42} \dis \dfrac{1}{2}\,\ma_\mathrm{co}\,\hat{\mathbf{b}}_\mathrm{co}^\mrT\,,&
	\mA_{44} \dis \dfrac{\eta_\mrb}{k_1\,\Delta t\,J_\mrel}\,\mI_4+\dfrac{1}{2}\,\ma_\mathrm{co}\,\hat{\ma}_\mathrm{con}^\mrT\,,
\eqe
and the $(4\times4)$-right-hand-side blocks
\eqb{rll}
	\mr_1 \dis -\mcc_\mathrm{con}\,\ma_\mathrm{con}^\mrT+\md_\mathrm{con}\,\mathbf{b}_\mathrm{con}^\mrT-\Delta H^2\,\dfrac{\partial\ma_\mathrm{con}}{\partial\ma_\mathrm{co}}+2\,\Delta H\,\dfrac{\partial\mathbf{b}_\mathrm{con}}{\partial\mathbf{b}_\mathrm{co}}\,,\\[4mm]
	\mr_2 \dis -\md_\mathrm{con}\,\ma_\mathrm{con}^\mrT+2\,\Delta H\,\dfrac{\partial\mathbf{b}_\mathrm{con}}{\partial\mathbf{b}_\mathrm{co}}\,,\\[4mm]
	\mr_3 \dis \dfrac{1}{2}\,\Delta H\,\ma_\mathrm{co}\,\ma_\mathrm{con}^\mrT-\dfrac{1}{2}\,\ma_\mathrm{co}\,\mathbf{b}_\mathrm{con}^\mrT+\Delta H\,\mI_4\,,\\[4mm]
	\mr_4 \dis \dfrac{1}{2}\,\ma_\mathrm{co}\,\ma_\mathrm{con}^\mrT\,,
\eqe
where
\eqb{l}
	\mcc_\mathrm{con}:=\dfrac{1}{2}\,\Bigl(\Delta H^2\,\ma_\mathrm{con}-2\,\Delta H\,\mathbf{b}_\mathrm{con}\Bigr)\,,\quad\mathrm{and}\quad\md_\mathrm{con}:=\Delta H\,\ma_\mathrm{con}-\mathbf{b}_\mathrm{con}\,.
\eqe
Here, the $(4\times4)$-identity matrix is denoted $\mI_4$ and the derivatives $\partial\ma_\mathrm{con}/\partial\ma_\mathrm{co}$, $\partial\mathbf{b}_\mathrm{con}/\partial\ma_\mathrm{co}$ and $\partial\mathbf{b}_\mathrm{con}/\partial\mathbf{b}_\mathrm{co}$ follow from \eqsref{varLinLinearizationViscoelasticShellsAuxiliaryDerivativesAabgdAndHatAabgd} and they are arranged in analogy to \eqsref{varLinLinearizationViscoelasticShellsAuxiliarydhatacondaco}.

\section{Analytical solutions} \applabel{viscAnalyticalSolutions}

\subsection{Inflated membrane balloon} \applabel{viscAnalyticalSolutionsInflMem}
This appendix derives the analytical solution for the inflated spherical membrane example from \secref{numExViscInflMem}, see also Eqs.~\eqref{e:numExViscInflMemPressureTotal}--\eqref{e:numExViscInflMemPressureViscous}. The two contributions $p_\mrel(t)$ and $p_\mrvisc(t)$ are derived subsequently.

\subsubsection{Pressure from the elastic branch} \applabel{viscAnalyticalSolutionsInflMemPressureElastic}
The elastic behavior of rubber can be described by the incompressible Neo-Hookean material model from \eqsref{cntViscConstElasticitySigmaABincompressibleNeoHookean}. For the inflated balloon, the current radius $r$ is related to the initial radius $R$ via the stretch $\lambda$, i.e.~$r=\lambda\,R$. As $\dif\bx=\bF\,\dif\bX$, the surface deformation gradient becomes $\bF=\lambda\,\bi$. The left surface Cauchy-Green tensor then becomes $\bB=\bF\bF^\mrT=\lambda^2\,\bi$. The surface stretch, which is related to the area change, i.e.~$\dif a=J\,\dif A$, is given by $J=\lambda^2$. Finally, the incompressibility of the material results in the relation $\tilde{t}=T/J$ between the current and initial thickness.

Using these kinematic relations, \eqsref{cntViscConstElasticitySigmaABincompressibleNeoHookean}, and the plane stress condition $\sigma_{33}=0$, the stress tensor becomes
\eqb{l}
	\tilde\bsig=\dfrac{\bsig}{\tilde{t}}+\sigma_{33}\,\bigl(\bn\otimes\bn\bigr)=\dfrac{1}{\tilde{t}}\,\dfrac{\mu}{J}\,\Bigl(\bB-\dfrac{\bi}{J^2}\Bigr)=\dfrac{1}{\tilde{t}}\,\dfrac{\mu}{\lambda^2}\,\Bigl(\lambda^2\,\bi-\dfrac{\bi}{\lambda^4}\Bigr)=\dfrac{\mu}{\tilde{t}}\,\Bigl(1-\dfrac{1}{\lambda^6}\Bigr)\,\bi\,. \eqlabel{numExViscInflMemIncompressibleNeoHookSigmaTildeAnalytical}
\eqe
As derived in \cite{needleman1977}, the in-plane normal stress in \eqsref{numExViscInflMemIncompressibleNeoHookSigmaTildeAnalytical} is equal to $\tilde\sigma=p_\mrel\,r/2/\tilde{t}$, which leads to the pressure-stretch or pressure-volume relation from \eqsref{numExViscInflMemPressureElastic}. The surface tension, see Eq.~(\ref{e:cntViscConstGammaNabDev}.1), is then given by $\gamma=\mu\,(1-\lambda^{-6})$.

\subsubsection{Solution of the evolution laws} \applabel{viscAnalyticalSolutionsInflMemODE}
To determine the Maxwell stresses, the intermediate configuration needs to be fully specified. For this, the evolution laws 
\eqb{l}
	\dot{\hat{a}}^\ab+\dfrac{\mu_1}{\eta_\mrs}\,\hat{a}^\ab=\dfrac{\mu_1}{\eta_\mrs}\,a^\ab(t)\,,\quad \hat{a}^\ab(t=0)=A^\ab\,, \eqlabel{numExViscInflMemODE}
\eqe
need to be solved, see \eqsref{cntViscMemODEneoHookeanSimple}. Based on the kinematic relations
\eqb{l}
	\hat{a}_\ab=J_\mrin\,A_\ab\,,\qquad\hat{a}^\ab=\dfrac{A^\ab}{J_\mrin}\,,\qquad a_\ab=J\,A_\ab\,,\qquad a^\ab=\dfrac{A^\ab}{J}\,, \eqlabel{numExViscInflMemKinematicRelations}
\eqe
the contravariant surface metric can be written as $a^\ab=A^\ab/\lambda^2$. Thus, \eqsref{numExViscInflMemODE} represent linear, inhomogeneous, first-order ODEs, which are solved with variation of constants. The solution is composed of an homogeneous (`$\mathrm{hom}$') and particular (`$\mathrm{p}$') part, i.e.
\eqb{l}
	\hat{a}^\ab(t)=\hat{a}^\ab_\mathrm{hom}(t)+\hat{a}^\ab_\mathrm{p}(t)\,. \eqlabel{numExViscInflMemHataabComposition}
\eqe
The homogeneous part is given by
\eqb{l}
	\hat{a}^\ab_\mathrm{hom}(t)=c^\ab_0\exp\Bigl(-\dfrac{\mu_1}{\eta_\mrs}\,t\Bigr)\,,\quad c^\ab_0=\mathrm{const.}\,. \eqlabel{numExViscInflMemHataabHomogeneous}
\eqe
The particular part is now assumed to take the form $\hat{a}^\ab_\mathrm{p}(t)=c^\ab(t)\,\exp(-\mu_1/\eta_\mrs\,t)$, where $c^\ab(t)$ denotes an unknown function. Plugging this $\hat{a}^\ab_\mathrm{p}$ into the ODEs in  \eqsref{numExViscInflMemODE} and applying some algebraic manipulations yields
\eqb{l}
	 \dot{c}^\ab(t)=\dfrac{\mu_1}{\eta_\mrs}\,a^\ab(t)\,\exp\Bigl(\dfrac{\mu_1}{\eta_\mrs}\,t\Bigr)\,, \eqlabel{numExViscInflMemODEc}
\eqe
which resembles ODEs for the unknown $c^\ab(t)$. The solution is given by
\eqb{l}
	c^\ab(t)=\ds\int\dfrac{\mu_1}{\eta_\mrs}\,a^\ab(\tilde{t})\,\exp\Bigl(\dfrac{\mu_1}{\eta_\mrs}\,\tilde{t}\Bigr)\,\dif\tilde{t}+c_1^\ab\,,\quad c_1^\ab=\mathrm{const.}\,. \eqlabel{numExViscInflMemC}
\eqe
Plugging the homogeneous and particular solution into \eqsref{numExViscInflMemHataabComposition} then results in the solution
\eqb{l}
	\hat{a}^\ab(t)=\ds\exp\Bigl(-\dfrac{\mu_1}{\eta_\mrs}\,t\Bigr)\,\biggl[c_2^\ab+\dfrac{\mu_1}{\eta_\mrs}\,\int\limits_0^t a^\ab(\tilde{t})\,\exp\Bigl(\dfrac{\mu_1}{\eta_\mrs}\,\tilde{t}\Bigr)\biggr]\,, \eqlabel{numExViscInflMemGeneralSolution}
\eqe
where the constant $c^\ab_2=A^\ab$ follows from the initial condition in \eqsref{numExViscInflMemODE}. The integral in \eqsref{numExViscInflMemGeneralSolution} can then be explicitly computed using  $a^\ab=A^\ab/\lambda^2$ with $\lambda(t)$ given in \eqsref{numExViscInflMemStretchFunction}. Defining $\hat{a}^\ab(t):=A^\ab\,\hat{a}_\mathrm{ev}(t)$, with $\hat{a}_\mathrm{ev}(t)$ given in \eqsref{numExViscInflMemHataEv}, the intermediate surface metric is given as the initial surface metric multiplied with an evolution function.

\subsubsection{Pressure from the Maxwell branch}\applabel{viscAnalyticalSolutionsInflMemPressureViscous}
For the Maxwell branch, the elastic energy density shown in \eqsref{cntViscConstElasticityPsiMemClassicalNeoHookean} with $\Lambda=0$ is considered. The resulting stresses are given in \eqsref{cntViscConstElasticitySigmaABclassicalNeoHookean}. Plugging the kinematic relations from \eqsref{numExViscInflMemKinematicRelations} into the analytical solution for the contravariant surface metric in the intermediate configuration, $\hat{a}^\ab$, from \appref{viscAnalyticalSolutionsInflMemODE} yields
\eqb{l}
	\dfrac{A^\ab}{J_\mrin}=A^\ab\,\hat{a}_\mathrm{ev}(t)\,. \eqlabel{numExViscInflMemEquationForJel}
\eqe
Multiplying \eqsref{numExViscInflMemEquationForJel} with the surface stretch $J$, eliminating $A^\ab$, and using $J=J_\mrel\,J_\mrin$ results in the required relation for the elastic surface stretch, i.e.
\eqb{l}
	J_\mrel = J\,\hat{a}_\mathrm{ev}(t)\,, \eqlabel{numExViscInflMemJel}
\eqe
with $\hat{a}_\mathrm{ev}(t)$ given in \eqsref{numExViscInflMemHataEv}.

Based on the kinematic relations and the analytical solution for $\hat{a}^\ab$, the stresses in \eqsref{cntViscConstElasticitySigmaABclassicalNeoHookean} become
\eqb{l}
	\sigma_{1(\mrel)}^\ab=\dfrac{\mu_1}{J\,\hat{a}_\mathrm{ev}(t)}\,\bigl(A^\ab\,\hat{a}_\mathrm{ev}(t)-a^\ab\bigr)\,.
\eqe
In tensor notation, this becomes
\eqb{l}
	\bsig_{1(\mrel)}=\sigma_{1(\mrel)}^\ab\,\bigl(\ba_\alpha\otimes\ba_\beta\bigr)=\dfrac{\mu_1\,\bigl(\bB\,\hat{a}_\mathrm{ev}(t)-\bi\bigr)}{J\,\hat{a}_\mathrm{ev}(t)}=\dfrac{\mu_1\,\bigl(\lambda^2\,\hat{a}_\mathrm{ev}(t)-1\bigr)}{\lambda^2\,\hat{a}_\mathrm{ev}(t)}\,\bi=\mu_1\,\biggl(1-\dfrac{1}{\lambda^2\,\hat{a}_\mathrm{ev}(t)}\biggr)\,\bi\,,\eqlabel{numExViscInflMemSigma1AB}
\eqe
where the kinematic relations for $\bB$ and $J$ have been inserted. Similar to \eqsref{numExViscInflMemIncompressibleNeoHookSigmaTildeAnalytical}, the in-plane normal stress is compared to $\tilde\sigma=p\,r/2/\tilde{t}$ to obtain the pressure-stretch relation, see \eqsref{numExViscInflMemPressureViscous}. The total pressure in the system is then given by the sum of Eqs.~\eqref{e:numExViscInflMemPressureElastic} and \eqref{e:numExViscInflMemPressureViscous}, see also \eqsref{numExViscInflMemPressureTotal}. The surface tension w.r.t.~the intermediate configuration, see also Eq.~(\ref{e:cntViscConstGammaNabDev}.1), is then given by $\hat\gamma=\mu_1\,(1-\lambda^{-2}\,\hat{a}_\mathrm{ev}^{-1})$.

\subsection{Pure bending of a flat strip} \applabel{viscAnalyticalSolutionsPureBend}
In this section, the analytical solution from Eqs.~\eqref{e:numExViscPureBendKappa2in}--\eqref{e:numExViscPureBendKappa2} is derived. The initially flat sheet of dimension $L\times S$ is deformed to a curved sheet with radius $r=1/\kappa_2$. The boundary and loading conditions are chosen such that the surface stretches are equal to one, i.e.~$\lambda_1=\lambda_2=1$. The derivation follows the elastic case considered in \cite{sauer2017a}.  The surface is parametrized by the coordinates $\xi\in[0,L]$ and $\eta\in[0,S]$, i.e.~$\xi$ points along the $x$-direction in \figref{numExViscPureBendSetup} and $\eta$ points along the $y$-direction. The reference and current surface can be described via the mappings
\eqb{l}
	\bX(\xi,\eta) = \xi\,\be_1+\eta\,\be_2\,, \eqlabel{numExViscPureBendMappingX}
\eqe
and
\eqb{l}
	\bx(\xi,\eta) = \xi\,\be_1+r\,\sin\theta\,\be_2+r\,\bigl(1-\cos\theta\bigr)\,\be_3\,, \eqlabel{numExViscPureBendMappingx}
\eqe
with $\theta:=\kappa_2\,\eta$. Based on these mappings, the tangent vectors follow as $\bA_1=\be_1$, $\bA_2=\be_2$, $\ba_1=\be_1$, and $\ba_2=\cos\theta\,\be_2+\sin\theta\,\be_2$, and the surface normal follows as $\bn=-\sin\theta\,\be_2+\cos\theta\,\be_3$. The surface metrics are then given by
\eqb{l}
	\bigl[A_\ab\bigr]=\bigl[A^\ab\bigr]=\bigl[a_\ab\bigr]=\bigl[a^\ab\bigr]=\begin{bmatrix}1&0\\0&1\end{bmatrix}\,, \eqlabel{numExViscPureBendSurfaceMetrics}
\eqe
such that the surface stretch is $J=1$ by construction. The curvature tensor components are given by
\eqb{l}
	\bigl[b_\ab\bigr]=\bigl[b^\alpha_\beta\bigr]=\bigl[b^\ab\bigr]=\begin{bmatrix}0&0\\0&\kappa_2\end{bmatrix}\,, \eqlabel{numExViscPureBendCurvatureTensors}
\eqe
such that the mean and Gaussian curvature follow as $H=\kappa_2/2$ and $\kappa=0$, respectively. Similar to Eqs.~\eqref{e:numExViscPureBendSurfaceMetrics}--\eqref{e:numExViscPureBendCurvatureTensors}, the intermediate surface metric and curvature tensor components are
\eqb{l}
	\bigl[\hat{a}_\ab\bigr]=\bigl[\hat{a}^\ab\bigr]=\begin{bmatrix}1&0\\0&1\end{bmatrix}\,,\quad\mathrm{and}\quad\bigl[\hat{b}_\ab\bigr]=\bigl[\hat{b}^\ab\bigr]=\begin{bmatrix}0&0\\0&\kappa_2^\mrin\end{bmatrix}\,, \eqlabel{numExViscPureBendIntermediateQuantities}
\eqe
with inelastic curvature $\kappa_2^\mrin$. The intermediate mean curvature follows from $\hat{H}=\kappa_2^\mrin/2$. Further, $J_\mrin=J_\mrel=1$.

Based on the Neo-Hookean and Koiter bending material models for the elastic and Maxwell branch, see Eqs.~\eqref{e:cntViscConstElasticityPsiMemClassicalNeoHookean} and \eqref{e:cntViscConstElasticityPsiBendKoiter}, the stresses are given by $\sigma^{11}=\sigma^{12}=\sigma^{21}=\sigma^{22}=0$, as $\lambda_1=\lambda_2=\lambda_1^\mrel=\lambda_2^\mrel=1$. The moment components are $M^{11}=M^{12}=M^{21}=0$ and
\eqb{l}
	M^{22} = c\,\kappa_2+c_1\,\bigl(\kappa_2-\kappa_2^\mrin\bigr)\,. \eqlabel{numExViscPureBendMomentComponents}
\eqe
The resulting in-plane stress components $N^\ab=\sigma^\ab+b^\alpha_\gamma\,M^{\gamma\beta}$ follow as
\eqb{l}
	N^{11}=N^{12}=N^{21}=0\,,\quad\mathrm{and}\quad 	N^{22}=c\,\kappa_2^2+c_1\,\bigl(\kappa_2^2-\kappa_2\,\kappa_2^\mrin\bigr)\,. \eqlabel{numExViscPureBendNab}
\eqe
As $[a_\ab]$ is equal to the identity matrix, see \eqsref{numExViscPureBendSurfaceMetrics}, it follows that $[N^\alpha_\beta]=[N^\ab]$. Considering a cut at $\theta$, which is perpendicular to the normal $\bnu=\ba_2$, the distributed bending moment on this cut can be computed as 
\eqb{l}
	M=M^\ab\,\nu_\alpha\,\nu_\beta=c\,\kappa_2+c_1\,\bigl(\kappa_2-\kappa_2^\mrin\bigr)\,, \eqlabel{numExViscPureBendMoment}
\eqe
such that the curvature $\kappa_2(t)$ follows as given in \eqsref{numExViscPureBendKappa2}.

Still, the inelastic curvature $\kappa_2^\mrin(t)$ is unknown. Using Eqs.~\eqref{e:cntViscBendODEKoiter}, \eqref{e:numExViscPureBendKappa2} and \eqref{e:numExViscPureBendIntermediateQuantities}, the ODE for $\kappa_2^\mrin$ is given by
\eqb{l}
	\dot{\kappa}_2^\mrin=\dfrac{c_1}{\eta_\mrb\,(c+c_1)}\,\bigl(M-c\,\kappa_2^\mrin\bigr)\,, \eqlabel{numExViscPureBendODE}
\eqe
with initial condition $\kappa_2^\mrin(t=0)=0$. Using the definition from \eqsref{numExViscPureBendcTilde}, the ODE is rewritten in the form
\eqb{l}
	\dot{\kappa}_2^\mrin+\dfrac{1}{\tau_b}\,\kappa_2^\mrin=\dfrac{1}{c\,\tau_b}\,M(t)\,. \eqlabel{numExViscPureBendODE2}
\eqe
In analogy to \appref{viscAnalyticalSolutionsInflMemODE}, this ODE can be solved based on the variation of parameters. The solution reads
\eqb{l}
	\kappa_2^\mrin(t)=\ds\dfrac{1}{c\,\tau_b}\,e^{-t/\tau_b}\int_0^tM(\tilde{t})\,e^{\tilde{t}/\tau_b}\,\dif \tilde{t}\,. \eqlabel{numExViscPureBendKappa2inGeneral}
\eqe
Using the moment profile $M(t)$ from \eqsref{numExViscPureBendM}, the integral in \eqsref{numExViscPureBendKappa2inGeneral} can be analytically solved such that $\kappa_2^\mrin(t)$ is obtained as given in \eqsref{numExViscPureBendKappa2in}.

\subsection{Inflated spherical shell} \applabel{viscAnalyticalSolutionsInflSphr}
This section derives the analytical solution for the inflated spherical shell example from \secref{numExViscInflSphr}, see also Eqs.~\eqref{e:numExViscInflSphrPressureTotal}--\eqref{e:numExViscInflSphrPressureViscous}. The derivation of the elastic pressure, $p_\mrel(t)$, can be found in \cite{sauer2017b}, but it is repeated here for completeness.

\subsubsection{Pressure from the elastic branch} \applabel{viscAnalyticalSolutionsInflSphrPressureElastic}
Given the material models from Eqs.~\eqref{e:cntViscConstElasticitySigmaABincompressibleNeoHookean} and \eqref{e:cntViscConstElasticityPsiBendHelfrich}, the in-plane traction components are given by
\eqb{l}
	N^\ab_\mrel=N_a^\mrel\,a^\ab+N_b^\mrel\,b^\ab\,, \eqlabel{numExViscInflSphrNabCompositionElastic}
\eqe
with
\eqb{l}
	N_a^\mrel:=\mu\,\biggl(1-\dfrac{1}{J^3}\biggr)+k\,\bigl(H-H_0\bigr)^2\,, \quad\mathrm{and\quad} N_b^\mrel:=-k\,\bigl(H-H_0\bigr)\,. \eqlabel{numExViscInflSphrNabElastic}
\eqe
Based on the surface parametrization
\eqb{l}
	\bx(\phi,\theta)=\begin{bmatrix}r\,\cos\phi\,\sin\theta\\r\,\sin\phi\,\sin\theta\\-r\,\cos\theta\end{bmatrix}\,, \eqlabel{numExViscInflSphrSurfaceParametrization}
\eqe
the following relations can be found
\eqb{l}
	\bigl[a^\ab\bigr]=\dfrac{1}{r^2}\,\begin{bmatrix}1/\sin^2\theta&0\\0&1\end{bmatrix}\,,\quad b^\ab=-a^\ab/r\,,\quad\mathrm{and}\quad H=-\dfrac{1}{r}\,. \eqlabel{numExViscInflSphrKinematicRelations}
\eqe
The traction vector $\bT=\bT^\alpha\,\nu_\alpha$ on a cut orthogonal to $\bnu$ then becomes
\eqb{l}
	\bT=\bigl(N^\ab_\mrel\,\ba_\beta+S^\alpha\,\bn\bigr)\,\nu_\alpha = \biggl[N_a^\mrel-\dfrac{N_b^\mrel}{r}\biggr]\,\bnu\,, \eqlabel{numExViscInflSphrT}
\eqe
where the kinematic relations from \eqsref{numExViscInflSphrKinematicRelations} have been used. Similar to \appref{viscAnalyticalSolutionsInflMem}, the in-plane component $T_\nu:=N_a^\mrel-N_b^\mrel/r$ needs to equilibriate the current pressure according to \citep{needleman1977}
\eqb{l}
	p_\mrel=\dfrac{2\,T_\nu}{r}=\dfrac{2}{r}\,\biggl[N_a^\mrel-\dfrac{N_b^\mrel}{r}\biggr]\,. \eqlabel{numExViscInflSphrPressureTractionRelation}
\eqe
Plugging \eqsref{numExViscInflSphrNabElastic} into \eqsref{numExViscInflSphrPressureTractionRelation} and using the kinematic relations $J=\lambda^2$, $r=\lambda\,R$ and $H=-1/r$ yields the elastic pressure given in \eqsref{numExViscInflSphrPressureElastic}.

\subsubsection{Solution of the evolution laws} \applabel{viscAnalyticalSolutionsInflSphrODE}
To determine the Maxwell moments, the intermediate configuration needs to be fully specified. The intermediate surface metric is the same as derived in \appref{viscAnalyticalSolutionsInflMemODE}, i.e.~$\hat{a}^\ab(t)=A^\ab\,\hat{a}_\mathrm{ev}(t)$ and $J_\mrel(t)=J(t)\,\hat{a}_\mathrm{ev}(t)$ with $\hat{a}_\mathrm{ev}(t)$ given in \eqsref{numExViscInflMemHataEv}. The bending of the intermediate surface follows from the solution of the evolution laws
\eqb{l}
	\dot{\hat{b}}_\ab+\dfrac{c_1}{\eta_\mrb}\,\hat{b}_\ab=\dfrac{c_1}{\eta_\mrb}\,b_\ab(t)\,,\quad\hat{b}_\ab(t=0)=B_\ab\,, \eqlabel{numExViscInflSphrODE}
\eqe
see \eqsref{cntViscBendODEKoiter}. Based on the surface parametrization in \eqsref{numExViscInflSphrSurfaceParametrization}, the relation $b_\ab(t)=\lambda(t)\,B_\ab$ can be derived. As \eqsref{numExViscInflSphrODE} resembles first-order, linear, inhomogeneous ODEs, the same approach as in \appref{viscAnalyticalSolutionsInflMemODE} can be used in order to solve for $\hat{b}_\ab$. The solution is then given by
\eqb{l}
	\hat{b}_\ab(t)=B_\ab\,\hat{b}_\mathrm{ev}(t)\,, \eqlabel{numExViscInflSphrhatbab}
\eqe
with $\hat{b}_\mathrm{ev}(t)$ given in \eqsref{numExViscInflSphrHatbEv}. Note that the function $\lambda(t)$ from \eqsref{numExViscInflMemStretchFunction} has been inserted to obtain a closed-form solution for $\hat{b}_\mathrm{ev}(t)$.

\subsubsection{Pressure from the Maxwell branch} \applabel{viscAnalyticalSolutionsInflSphrPressureViscous}
Given the material models in Eqs.~\eqref{e:cntViscConstElasticityPsiMemClassicalNeoHookean} and \eqref{e:cntViscConstElasticityPsiBendKoiter} with $\Lambda=0$, the Maxwell stresses and moments follow as
\eqb{l}
	\sigma_{1(\mrel)}^\ab=\dfrac{\mu_1}{J_\mrel}\,\bigl(\hat{a}^\ab-a^\ab\bigr)\,,\quad\mathrm{and}\quad M_{1(\mrel)}^\ab=\dfrac{1}{J_\mrel}\,\hat{f}^{\ab\gd}\,\kappa_\gd^\mrel\,. \eqlabel{numExViscInflSphrSigmaM1}
\eqe
Using the kinematic relations from \eqsref{numExViscInflMemKinematicRelations}, $b_\ab=\lambda\,B_\ab$ and $B^\ab=\lambda^3\,b^\ab$ yields
\eqb{l}
	\sigma_{1(\mrel)}^\ab=\mu_1\,\biggl(1-\dfrac{1}{J_\mrel}\biggr)\,a^\ab\,,\quad\mathrm{and}\quad M_{1(\mrel)}^\ab=c_1\,\hat{a}_\mathrm{ev}\,\bigl(\lambda^2-\lambda\,\hat{b}_\mathrm{ev}\bigr)\,b^\ab\,, \eqlabel{numExViscInflSphrSigmaM2}
\eqe
with $\hat{a}_\mathrm{ev}(t)$ given in \eqsref{numExViscInflMemHataEv} and $\hat{b}_\mathrm{ev}(t)$ given in \eqsref{numExViscInflSphrHatbEv}. Similar to \eqsref{numExViscInflSphrNabCompositionElastic}, the in-plane traction components for the Maxwell branch are given by
\eqb{l}
	N^\ab_\mrvisc=N_a^\mrvisc\,a^\ab+N_b^\mrvisc\,b^\ab\,, \eqlabel{numExViscInflSphrNabCompositionViscous}
\eqe
where
\eqb{l}
	N_a^\mrvisc:=\mu_1\,\biggl(1-\dfrac{1}{\lambda^2\,\hat{a}_\mathrm{ev}}\biggr)\,, \quad\mathrm{and\quad} N_b^\mrvisc:=\dfrac{c_1}{R}\,\hat{a}_\mathrm{ev}\,\bigl(\hat{b}_\mathrm{ev}-\lambda\bigr)\,. \eqlabel{numExViscInflSphrNabViscous}
\eqe
Using \eqsref{numExViscInflSphrPressureTractionRelation} and replacing the elastic traction components with the viscous traction components from \eqsref{numExViscInflSphrNabViscous} yields the pressure $p_\mrvisc(t)$ as given in \eqsref{numExViscInflSphrPressureViscous}.

\bigskip
\bibliographystyle{apalike}
\bibliography{visc}

\end{document}